\documentclass[twoside,phd]{iitkgp}

\usepackage[utf8]{inputenc} 
\usepackage[T1]{fontenc}   

\usepackage{graphicx}
\usepackage{paralist, tabularx}
\usepackage[centertags]{amsmath}
\usepackage{csquotes}
\usepackage{enumerate}
\usepackage{rotating}
\usepackage{times}
\usepackage{cite}
\usepackage[normalem]{ulem}
\usepackage[table,xcdraw]{xcolor}
\usepackage{bbold}
\usepackage{bm}
\usepackage{graphicx}
\usepackage{caption}
\usepackage{subcaption}
\usepackage{color}
\usepackage{tcolorbox}
\usepackage{multirow}
\usepackage{hhline}
\usepackage{booktabs}
\usepackage{float}
\usepackage{amsfonts,amssymb,amsthm}
\usepackage{chngcntr}
\usepackage{adjustbox}
\usepackage{hhline}
\usepackage{pifont}
\usepackage{algorithm}
\usepackage{algpseudocode}
\usepackage{amsmath}
\usepackage{parskip}

 \usepackage{algpseudocode}
\usepackage{bbm} 
\usepackage{bigstrut}

\DeclareMathOperator*{\argmax}{arg\,max}
\DeclareMathOperator*{\argmin}{arg\,min}

\newcommand{\EMPTOPTAU}{\mbox{$\hat{S}^{\tau_{\delta}}_m$}}
\newcommand{\EMPWORSTTAU}[1]{(\EMPTOPTAU)^c}

\newcommand{\HA}[1]{\text{H}^{\varepsilon}(\mu)}

\newcommand{\mathscr}[1]{\mathcal{#1}}
\newcommand{\maxm}[1]{%
  \ifthenelse{\isempty{#1}}%
    {\overset{m}{\max}}
    {\underset{#1}{\overset{m}{\max}}\, }
}
\newcommand{\minm}[1]{%
  \ifthenelse{\isempty{#1}}%
    {\overset{m}{\min}}
    {\underset{#1}{\overset{m}{\min}}\, }
}
\newcommand{\argmaxm}[1]{%
  \ifthenelse{\isempty{#1}}%
    {\overset{m}{\argmax}}
    {\underset{#1}{\overset{m}{\argmax}}\, }
}
\newcommand{\argminm}[1]{%
  \ifthenelse{\isempty{#1}}%
    {\overset{m}{\argmin}}
    {\underset{#1}{\overset{m}{\argmin}}\, }
}
\newcommand{\maxmset}[1]{%
  \ifthenelse{\isempty{#1}}%
    {\overset{[m]}{\max}}
    {\underset{#1}{\overset{[m]}{\max}}\, }
}
\newcommand*{\colorboxed}{}
\def\colorboxed#1#{%
  \colorboxedAux{#1}%
}
\newcommand*{\colorboxedAux}[3]{%
  \begingroup
    \colorlet{cb@saved}{.}%
    \color#1{#2}%
    \boxed{%
      \color{cb@saved}%
      #3%
    }%
  \endgroup
}

\newcommand{\minmset}[1]{%
  \ifthenelse{\isempty{#1}}%
    {\overset{[m]}{\min}}
    {\underset{#1}{\overset{[m]}{\min}}\, }
}
\newcommand{\argmaxmset}[1]{%
  \ifthenelse{\isempty{#1}}%
    {\overset{[m]}{\argmax}}
    {\underset{#1}{\overset{[m]}{\argmax}}\, }
}
\newcommand{\argminmset}[1]{%
  \ifthenelse{\isempty{#1}}%
    {\overset{[m]}{\argmin}}
    {\underset{#1}{\overset{[m]}{\argmin}}\, }
}

\usepackage{color}

\usepackage{soul}
\setul{}{1pt}
\usepackage{epigraph}

\epigraphsize{\small}
\usepackage{etoolbox}
\makeatletter
\patchcmd{\epigraph}{\@epitext{#1}}{\itshape\@epitext{#1}}{}{}
\makeatother

\usepackage{lscape}
\usepackage[bottom]{footmisc}
\usepackage[titletoc]{appendix}
\newcolumntype{P}[1]{>{\centering\arraybackslash}p{#1}}
\newcolumntype{M}[1]{>{\centering\arraybackslash}m{#1}}

\usepackage{fancyhdr}
  \renewcommand{\chaptermark}[1]{\markboth{\chaptername \ \thechapter \ \ #1}{}}

\DeclareCaptionFormat{suggested}{\singlespace \textbf{#1}\textbf{#2}#3 \doublespace}
\let\markeverypar\everypar
\newtoks\everypar
\everypar\markeverypar
\markeverypar{\the\everypar\looseness=-1\relax}

\usepackage{cite}

\usepackage{makeidx}
\makeindex

\usepackage[hyphens]{url}
\usepackage[bookmarksnumbered,pdfpagelabels=true,plainpages=false,colorlinks=true,
            linkcolor=black,citecolor=black,urlcolor=blue]{hyperref}

\usepackage{comment}
\usepackage{tikz}
\usepackage{tikz-cd}

\usepackage{moresize}

\usepackage{color}
\definecolor{blue}{RGB}{17,220,247}
\definecolor{purple}{RGB}{163,115,250}
\definecolor{caribbeangreen}{rgb}{0.0, 0.8, 0.6}

\definecolor{GREEN}{RGB}{84,130,53}

\usepackage{soul}

\newcommand{\todokd}[1]{\textcolor{red}{#1}}
\newcommand{\xhdr}[1]{\noindent{{\bf #1.}}}

\newcommand{\rev}[1]{\textcolor{black}{ #1}}

\newcommand{\iclrmodel}{\textbf{\textit{TGDMat}}}
\newcommand{\uaimodel}{\textbf{\textit{CrysMMNet}}}
\newcommand{\aaaimodel}{\textbf{\textit{CrysGNN}}}
\newcommand{\npjmodel}{\textbf{\textit{CrysXPP}}}

\newcommand{\Ecal}{\mathcal{E}}

\newcommand{\Gcal}{\mathcal{G}}

\newcommand{\Mcal}{\mathcal{M}}

\newcommand{\Pcal}{\mathcal{P}}

\newcommand{\Scal}{{\mathcal{S}}}

\newcommand{\Vcal}{\mathcal{V}}

\newcommand{\EE}{\mathbb{E}} 
\usepackage{pgfplots}
\pgfplotsset{compat=1.15}
\usepgfplotslibrary{
  statistics,
  colorbrewer,
  groupplots,
}
\usetikzlibrary{
  patterns,
  shapes.geometric,
  decorations.text,
  matrix,
  fit,
  backgrounds,
  positioning,
}
\tikzset{
  fignode/.style={
    outer sep=0.25em,
  }
}
\tikzset{
  framedfignode/.style={
    outer sep=0.25em,
    inner sep=0.5em,
    rounded corners,
    draw,
  }
}
\colorlet{plotColorNeutral}{gray}
\definecolor{plotColor1}{HTML}{f61a1c}
\definecolor{plotColor2}{HTML}{377eb8}
\definecolor{plotColor3}{HTML}{4daf4a}
\definecolor{plotColor4}{HTML}{984ea3}
\definecolor{plotColor5}{HTML}{FFFFCB}
\definecolor{plotColor6}{HTML}{1e90ff}
\colorlet{plotColorNeutral*}{plotColorNeutral!40}
\colorlet{plotColor1*}{plotColor1!60}
\colorlet{plotColor2*}{plotColor2!60}
\colorlet{plotColor3*}{plotColor3!60}
\colorlet{plotColor4*}{plotColor4!60}
\colorlet{plotColor5*}{plotColor5!60}
\colorlet{plotColor6*}{plotColor6!60}
\pgfplotsset{
    colormap={greenred}{HTML=(4daf4a) HTML=(e41a1c)},
    colormap={redgreen}{HTML=(e41a1c) HTML=(4daf4a)}
}

  \Year{2026}
  \Month{January}
  \Author{\textbf{Kishalay Das}}
  \degree{Doctor of Philosophy}

\TitleTop{\textbf{Robust and Efficient AI Frameworks for Scalable }}
\TitleBottom{\textbf{Material Design and Property Prediction}}

\Advisor{Prof. Niloy Ganguly and Prof. Pawan Goyal}

\Approval{
\singlespace
\hspace{9cm}
Date:\hspace{.8cm}$/ \ \ \ \ \  \ /$ 26  \\  \\
Certified that the thesis entitled {\bf ``Robust and Efficient AI Frameworks for Scalable Material Design and Property Prediction''} 
submitted by Kishalay Das to the Indian Institute
of Technology, Kharagpur, for the award of the degree of Doctor of Philosophy
has been accepted by the external examiners and that the student has successfully
defended the thesis in the viva-voce examination held today.

\vspace{0.5in}

\vspace{1in}
\noindent
(Member of DSC)~~~~~\hfill(Member of DSC)~~~~~\hfill(Member of DSC)

\vspace{0.3in}
\vspace{0.7in}
\noindent
(Supervisor)\hspace{2.5cm}  ~~~~~~~(Joint Supervisor)

\vspace{0.3in}

\vspace{0.7in}
\noindent
(External Examiner)\hspace{2cm}(Chairman)\hfill ~~~~~~~
}

\Certificate{

\noindent%
{\em This is to certify that the thesis entitled} {\bf {\em``Robust and Efficient AI Frameworks for Scalable Material Design and Property Prediction''}}, {\em submitted by Kishalay Das to the Indian Institute of Technology, 
Kharagpur, for the partial fulfillment of the award of the degree of Doctor of Philosophy in Computer Science and Engineering, is a record of bona fide 
research work carried out by him under my supervision and guidance.}
{\em The thesis in my opinion, is worthy of consideration for the award of the degree of Doctor of Philosophy in 
accordance with the regulations of the Institute. To the best of my knowledge, the results embodied in this thesis 
have not been submitted to any other University or Institute for the award of any other Degree or Diploma.} 

\signaturebox{Pawan Goyal\\ Professor\\ Department of Computer \\Science and Engineering,\\ IIT Kharagpur}
\signaturebox{Niloy Ganguly\\ Professor\\ Department of Computer \\Science and Engineering,\\ IIT Kharagpur}

\datebox~~~~~~~ 
}

\Declaration{
\noindent
I certify that
\begin{enumerate}

\item[a.]   the work contained in this thesis is original and has been done by me under the guidance of my supervisor.
\item[b.]   the work has not been submitted to any other Institute for any degree or diploma.
\item[c.]  I have followed the guidelines provided by the Institute in preparing the \mbox{thesis}.
\item[d.]   I have conformed to the norms and guidelines given in the Ethical Code of Conduct of the Institute.
\item[e.]   whenever I have used materials (data, theoretical analysis, figures, and text) from other sources, I have given due credit to them by citing them in the text of the thesis and giving their details in the references. Further, I have taken permission from the copyright owners of the sources, whenever necessary.

\end{enumerate}

\vspace{0.6in}

\hfill Kishalay Das ~ ~ ~
}

\Acknowledgments{
	\color{black}
I would like to express my heartfelt gratitude to my supervisors, Dr. Pawan Goyal and Dr. Niloy Ganguly, for their unwavering support and invaluable guidance throughout my doctoral journey. Their mentorship went far beyond academic oversight; they gave me the freedom to pursue research problems of my own choosing and supported me unconditionally at every step. When we started working on AI for Material Design, we all were new to the domain. Their mentorship has significantly enhanced my understanding, critical thinking, problem formulation, and providing me with the tools necessary to comprehend and address the various challenges inherent in the research process. Their constant encouragement and belief in my abilities were instrumental in helping me reach this academic milestone.

I am deeply grateful to my external collaborators, Dr. Satadeep Bhattacharjee and Dr. Bidisha Samanta, for their insightful suggestions and stimulating discussions. Our collaborative brainstorming sessions greatly contributed to the development of my research ideas and played a pivotal role in shaping this thesis.

I extend my sincere thanks to the members of my Doctoral Scrutiny Committee (DSC), Dr. Soumya Kanti Ghosh, Dr. Sourangshu Bhattacharya, Dr. Ayan Chaudhury, and Dr. Plaban Kumar Bhowmik, for their constructive feedback and thoughtful suggestions.

I also extend my gratitude to the administrative and technical staff of the CSE department for their timely assistance and support. Special thanks to Hoimonti Di, always ready to help with any administrative matters, and to Suman Da, for his prompt and diligent support with server and infrastructure issues, ensuring uninterrupted progress in my research.

I am also grateful to all members of the Complex Networks Research Group for their continuous support and guidance throughout my academic journey. I would like to specially acknowledge Abhilash, Ankan, Anurag, Arpan, Arijit, Bishal, Gunjan, Kiran, Punyajoy, Rajdeep Da, Soumyadeep, Sayantan, Shounak, Siddharth, Soham, Subhendu, Sugand, Sujeet, and Paramita Di—thank you for the shared experiences that have enriched both my personal and academic growth.

Finally, I wish to express my deepest gratitude to my family, whose unwavering love and support have been the foundation of my journey. My wife, Dr. Uttirna, has been my constant companion, standing steadfastly by my side through every challenge and triumph. My uncle, Mr. Swapan Das, my aunt, Mrs. Gita Das, and my brother, Snehasis Das, have been my pillars of encouragement, always ready with a listening ear and words of comfort. Above all, my parents, Mrs. Anjana Das and Mr. Mukunda Das, have been my guiding light, supporting me with endless love, believing in me even when I doubted myself, and providing everything I needed to pursue my dreams. From the decision to leave my job at TCS to prepare for GATE, to choosing a Ph.D. after M.Tech, none of this would have been possible without their sacrifices, faith, and constant encouragement. Every achievement I have reached is a testament to their love, guidance, and unwavering belief in me, words can never fully express the depth of my gratitude. 

 Kishalay Das\\
IIT Kharagpur, India\\

}

\Biography {
\noindent Kishalay Das received his B.Tech degree in Computer Science and Engineering from the Meghnad Saha Institute of Technology, Kolkata, in 2012. He then worked as a Software Engineer at Accenture, Tata Consultancy Services, and Indian Space Research Organization (ISRO). Subsequently, he joined the Department of Computer Science and Automation (CSA) at the Indian Institute of Science (IISc), Bangalore, where he completed his M.Tech in 2020. In Jan 2022, he started pursuing his Ph.D. at the Department of Computer Science and Engineering, Indian Institute of Technology, Kharagpur. His research interests include Graph Representation Learning and Generative Modeling, with a strong focus on AI4Science applications.

\singlespace
\begin{center}
	\vspace{0.3cm}
	{\bfseries {\large Publications from the Thesis}}
	\vspace{0.3cm}
\end{center}

\thispagestyle{empty}

\begin{enumerate}

    \item {\bf Kishalay Das}, Bidisha Samanta, Pawan Goyal, Seung-Cheol Lee, Satadeep Bhattacharjee, Niloy Ganguly. "CrysXPP: An Explainable Property Predictor for Crystalline Material." NPJ Computational Materials Journal, 2022.

    \item {\bf Kishalay Das}, Bidisha Samanta, Pawan Goyal, Seung-Cheol Lee, Satadeep Bhattacharjee, Niloy Ganguly. "CrysGNN: Distilling pre-trained knowledge to enhance property prediction for crystalline materials." AAAI Conference on Artificial Intelligence (AAAI), 2023.

    \item {\bf Kishalay Das}, Pawan Goyal, Seung-Cheol Lee, Satadeep Bhattacharjee, Niloy Ganguly. "CrysMMNet: Multimodal Representation for Crystal Property Prediction." Conference on Uncertainty in Artificial Intelligence (UAI), 2023.

    \item {\bf Kishalay Das}, Subhojyoti Khastagir, Pawan Goyal, Seung-Cheol Lee, Satadeep Bhattacharjee, Niloy Ganguly. "Periodic Materials Generation using Text-Guided Joint Diffusion Model." International Conference on Learning Representations (ICLR), 2025.
		
\end{enumerate}

}

\Abstract{
\color{black}
Screening 3D periodic structures and their atomic compositions to identify novel crystal materials with specific chemical properties remains a long-standing challenge in the materials design community. These materials have been fundamental to key innovations such as the development of batteries, solar cells, semiconductors, quantum computers, etc.~\cite{kohn1965self,desiraju2002cryptic,butler2018machine}. A major challenge in materials design is discovering novel crystal structures with desired chemical properties that are both stable and physically realizable. Typically, this is done through a two-step process: first, the generation of 3D periodic structures for potential new candidate materials (Generative Task). Next, various chemical properties are evaluated (Discriminative Task) to confirm their validity and filter out the best ones, which are further sent to wet lab for synthesizing. Historically, Density Functional Theory (DFT) was an effective tool for both the tasks, however it is both resource-intensive and time-consuming. However, in recent times AI based techniques have demonstrated great potential to solve both the tasks. Against this backdrop, in this thesis, we focus on developing robust and efficient AI frameworks targeting both generative and discriminative tasks for material design.
\\\\
Specifically, in the first work of this thesis, we focus on the property prediction task and leverage the \textit{pretraining-finetuning paradigm} to reduce the reliance on large volumes of property-tagged data for training deep property prediction models in crystal materials. To this end, we developed a deep learning framework, \npjmodel{}~\cite{das2022crysxpp}, which addresses this challenge through the design of an autoencoder, CrysAE. Trained in an unsupervised manner on property-untagged crystal graphs, CrysAE effectively captures essential structural and chemical information. These learned representations are then transferred to the property prediction model via transfer learning. This approach significantly lowers the requirement for large property-tagged datasets and achieves low prediction errors across various crystal property prediction tasks.
\\\\
In our second work, we further scale this idea by developing a \textit{deep pre-trained GNN framework} for crystal materials. We propose \aaaimodel{}~\cite{das2023crysgnn}, which is trained on a large corpus of unlabeled data in a self-supervised way that captures (a) atomic connectivity, (b) atomic properties, and (c) graph-level similarity. CrysGNN is designed as a versatile framework that can be seamlessly integrated with any state-of-the-art property predictor for crystal materials. Using knowledge distillation, we transfer essential structural and chemical information from CrysGNN to downstream tasks. This integration enables property predictors to leverage prior knowledge, providing a scalable and robust solution for crystal property prediction.
\\\\
 In our third work, we focus on learning a more robust and enriched representation of crystal materials by using \textit{multi-modal data} i.e graph structure and textual description of materials. Towards that goal, we first curate the textual dataset of popular materials databases (Graph-based), containing textual descriptions of each material of those databases. We propose \uaimodel{}~\cite{das2023crysmmnet}, a simple multi-modal framework for crystal materials that fuse both graph structural and textual representations to generate a more enriched multi-modal representation of materials. We observe that CrysMMNet outperforms all the baseline models across diverse sets of properties on popular benchmark datasets.
\\\\
Finally, in the final part of the thesis, we focus on the generative task, i.e, discovering 3D periodic structure of a new material and explore a \textit{text-guided joint diffusion model} for material generation. We propose \iclrmodel{}, that conducts joint diffusion on lattices, atom types, and coordinates, enhancing its ability to accurately capture the crystal geometry. Additionally, we incorporate global structural knowledge through textual descriptions at each denoising step improves \iclrmodel{}’s ability to generate plausible materials with valid and stable structures. Our proposed framework not only enhances the performance on material generation tasks on benchmark tasks, also bridges the gap between natural language understanding and material structure generation.
\\\\
Overall, this research advances the development of a robust, efficient, and scalable AI framework for materials design, targeting both the generation of novel materials and the prediction of their diverse properties. The proposed methods hold broad applicability within the AI for materials design community and pave the way toward accelerating the discovery and design of materials with tailored properties.
\color{black}

\vspace{3 mm}
}

\fussy

\begin{document}


 \frontmatter


\makepreliminarypages

\singlespace

 \tableofcontents
 \clearemptydoublepage

%

 \listoffigures
 \clearemptydoublepage

 \listoftables
 \clearemptydoublepage


 \onehalfspace

\mainmatter
\addtolength{\parskip}{0.7\baselineskip}

\abovedisplayskip=13pt
\belowdisplayskip=13pt

\clearemptydoublepage
\chapter{\textcolor{black}{Introduction}}
\label{chap:intro}
\rev{In recent times, AI for Science has made rapid progress by leveraging advances in machine learning to accelerate scientific discovery, reduce computational cost, and enable exploration of previously intractable problem spaces. Data-driven and foundation models are increasingly complementing first-principles simulations by providing fast and reasonably accurate predictions across physics, chemistry, materials science, and biology. The integration of AI with domain knowledge, high-throughput simulations, and experimental pipelines is gradually shifting scientific workflows toward more automated, scalable, and closed-loop discovery paradigms.
\\
Within this broader context, materials design and synthesis represent one of the core challenges in scientific discovery. Technological progress is fundamentally driven by the discovery and development of new materials, with advances in energy storage and conversion, electronics, catalysis, healthcare, transportation, and infrastructure all relying on materials with carefully engineered properties~\cite{butler2018machine,desiraju2002cryptic}. In particular, crystalline materials form the backbone of modern civilization, enabling applications ranging from semiconductors and carbon capture to advanced energy storage technologies.
\\
A key component of this pipeline is the computational design and characterization of stable or metastable crystal structures, which enables the systematic prioritization of promising candidates for experimental synthesis and validation. The speed and efficiency of this discovery process play a decisive role in determining the pace of technological breakthroughs. Consequently, a central question in contemporary materials design is: \textbf{\textit{“How can we discover valid and stable materials whose properties meet the design requirements of a target application?”}} This challenge, addressed through materials screening, has become a cornerstone of modern materials discovery.}

Despite its importance, the discovery of novel 3D crystal materials with desired chemical properties remains a long-standing open problem in the materials design community, as progress has been significantly constrained by the immense vastness of the chemical and structural space. A central challenge lies in identifying crystal structures that are not only stable and physically realizable but also tailored to meet the specific property requirements of target applications.
\\\\
A typical material discovery pipeline generally involves two key phases:
\begin{itemize}
    \item \textbf{Generation Phase:} In this stage, the objective is to generate 3D periodic structures of novel materials. Given a dataset of known stable materials, the task is to design new candidates with valid and stable periodic structures. Typically, in this step, around 10,000 candidate materials are generated or screened.
    \item \textbf{Prediction Phase:} Once candidate structures are generated, their validity is assessed by estimating their chemical properties using a pre-trained property prediction model. In this step, various properties of the candidate materials are  predicted, and the candidates are filtered based on whether the predicted values meet the desired requirements. Finally, only a small subset, usually about 5 to 10 promising materials, is forwarded to the wet lab for experimental synthesis and validation.
\end{itemize}

\begin{figure}[ht!]
	\centering
	\boxed{\includegraphics[width=\columnwidth]{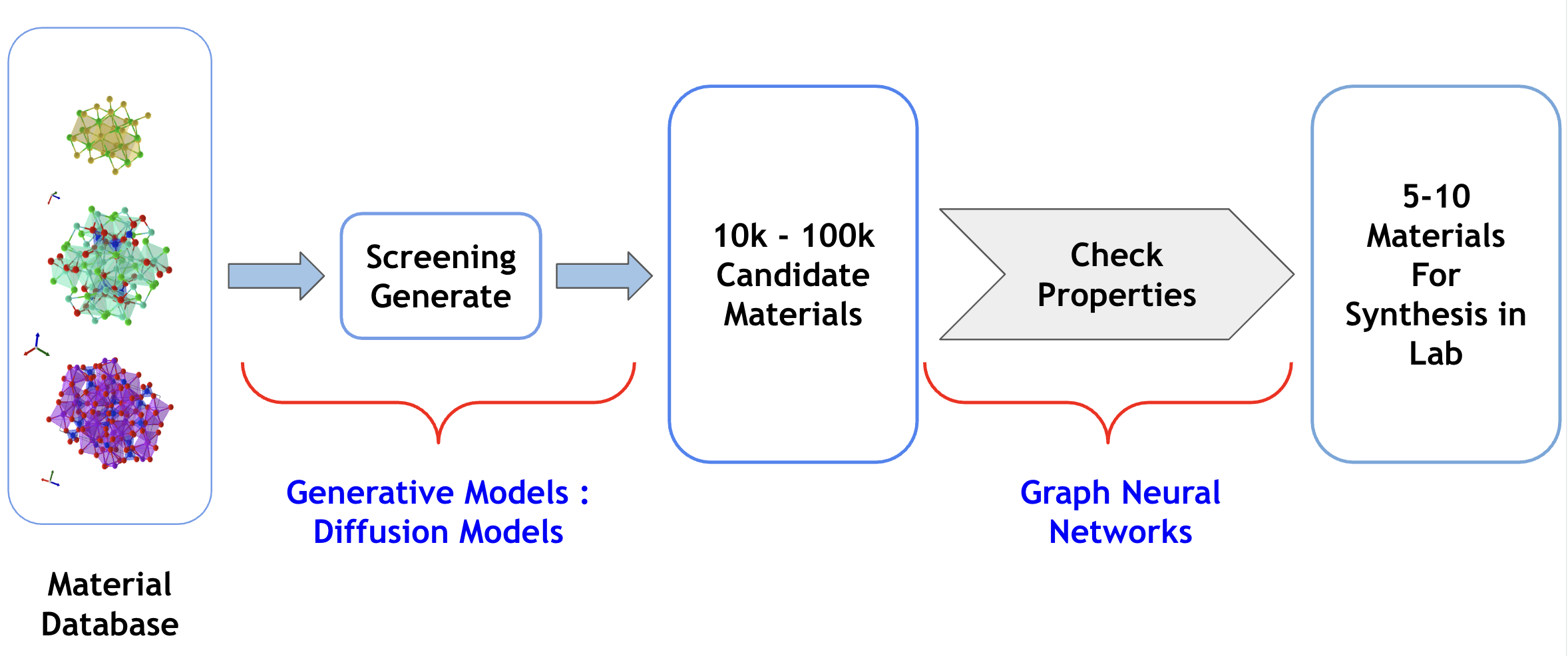}}
	\caption{A typical material design pipeline. In the first step, around 10,000 candidate materials are generated or screened. Their various properties are then predicted, and the candidates are filtered based on whether the predicted values meet the desired requirements. Finally, only a small subset, usually about 5 to 10 promising materials, is forwarded to the wet lab for experimental synthesis and validation.}
	\label{fig:text}
\end{figure}

Traditionally, materials discovery has relied on human insight, following a cycle of experimentation driven by hypothesis. In this approach, scientists propose potential structures based on their physical and chemical understanding, and later verify them through laboratory synthesis and characterization~\cite{schmidt2019recent}. This process demands long iterative cycles and was constrained by the limited number of candidate materials that can be evaluated. Further this approach has been significantly advanced through computational techniques, like Density Functional Theory (DFT)~\cite{kohn1965self} to compute the energy at each iteration. This process is typically guided by optimization algorithms, such as random search, Bayesian optimization, and others, that iteratively explore the energy landscape in search of stable states corresponding to local minima. However, DFT is computationally expensive and resource-intensive, with its cost scaling unfavorably with increasing system size and complexity. This limitation makes it impractical for accelerating the rapid discovery and generation of new materials. With recent progress of deep generative models, the majority of research in novel material discovery~\cite{noh2019inverse,hoffmann2019data,long2021constrained,kim2020generative,ren2020inverse,zhao2021high} has shifted focus to learn the data distribution directly from the training data consisting of stable material structures. Most recent advancements in  equivariant diffusion models~\cite{hoogeboom2022equivariant,bao2022equivariant}, have opened up a promising trajectory for the generation of novel three-dimensional periodic structures of crystal materials. Specifically, majority of the state-of-the-art diffusion models for material discovery~\cite{xie2021crystal,luo2023towards,jiao2023crystal} follow a sequential process: they first inject noise to destroy stable material structure (data $\rightarrow$ noise) and then learn to denoise the corrupted data through the reverse procedure (noise $\rightarrow$ data) using equivariant denoising network, which eventually helps to sample new stable material structure. 
\\\\
Fast and accurate prediction of different material properties is another important task in material science to validate different properties of newly generated materials, thus facilitating rapid screening of large material search spaces~\cite{meredig2014combinatorial,jha2018elemnet,choudhary2018machine}. Density functional theory (DFT)~\cite{orio2009density} is an effective tool to estimate several chemical properties which are directly related to the unique \emph{functionals of the electronic density}  of a given ground state conformation of solids and molecules. As it involves a large set of approximations, for crystals with large number of atoms, the calculation requires substantial computation costs; hence makes the screening process inefficient. With the advent of machine learning, the data-driven  approaches~\cite{tom2011materials,gaultois2016perspective,lu2018accelerated,gomez2016design,xue2016accelerated,xie2018crystal} have gained popularity and have potential to mitigate the computation issue. The existing techniques either use handcrafted feature based descriptors~\cite{seko2015prediction,xue2016accelerated,isayev2017universal,ghiringhelli2015big,isayev2015materials} or deep graph neural network (GNN)~\cite{xie2018crystal,sanyal2018mt,chen2019graph,louis2020global,cheng2021geometric,banjade2021structure,jin2020hierarchical,qiao2020orbnet,ye2020symmetrical}  to generate a representation from the 3D crystal structures. In specific, Graph neural network (GNN) based approaches are getting popular recently for their ability to encode graph information in an enriched representation space. Recent state-of-the-art GNN models~\cite{xie2018crystal,chen2019graph,louis2020graph, Wolverton2020,schmidt2021crystal,choudhary2021atomistic,hsu2021efficient,das2022crysxpp,yan2022periodic} construct multi-edge graphs for a 3D material structure where they create edges between nearby atoms within a pre-specified distance threshold in 3D space and apply GNN model to learn representations of crystal structures that are optimized for downstream property prediction tasks.
\\\\
In this thesis, we investigate \textbf{robust and efficient machine learning approaches for learning enriched representations of crystal materials}, with the goal of accelerating their discovery and design. We explore (1) Deep graph pretraining methods to improve crystal property prediction; (2) Develop multi-modal crystal representations by integrating both graph and text data; and (3) Incorporate textual guidance into joint diffusion models to enable conditional generation of more stable and valid materials.

\color{black}
\section{Challenges and Motivation}\label{sec:motivationThesis}
In this section, we first explore the challenges associated with both crystal material generation and property prediction, which further motivate the development of more efficient and robust machine learning methods aimed at mitigating these challenges and enhancing performance in both tasks.
\\\\
\noindent \textbf{Crystal Material Generation.}
Recent advances in generative modeling, particularly equivariant diffusion and flow matching models ~\cite{bao2022equivariant,hoogeboom2022equivariant}, have opened up a promising trajectory for generating novel 3D periodic structures of crystal material.  Although these models demonstrate the potential to generate stable materials, they possess several inherent limitations. 
\textbf{(1) Limitations in Joint Modeling of Crystal Structures:} \rev{As per our current knowledge, existing SOTA models} learn the joint distribution of atom coordinates, types, and lattice structure of the material through an end-to-end diffusion network. Existing models like CDVAE~\cite{xie2021crystal} and SyMat~\cite{luo2023towards} learn lattice parameters and atom types separately using a VAE model and further use a score network to learn the conditional distribution of atom coordinates given atom types and lattice. DiffCSP~\cite{jiao2023crystal}, on the other hand, focuses primarily on the structure prediction task, where it assumes atom types are given and predicts the stable crystal structure (lattice and coordinates). 
\textbf{(2) Challenges in Capturing Global Structural Knowledge:} Furthermore, these models use SE(3)-equivariant GNNs as backbone denoising networks, which largely rely on message passing around the local neighborhood of the atoms. Hence, they fail to incorporate global structural knowledge into the diffusion process, which can enhance the diffusion performance. 
\textbf{(3) Need for Conditional Crystal Generation:} Finally, these models are unconditional by design. From initial noisy structures without any external constraints, they generate stable crystal structures, which are distributionally similar to structures of the training dataset. This setup may have limited utility in real-world scenarios, as it lacks a mechanism for users to specify criteria for the material to be generated.  In a realistic setup, users would want to specify certain key details about the target material, like the chemical formula, space group, crystal symmetry, bond lengths, chemical properties, etc. as input to the diffusion model, which the generated structure must then match. 
\\
These limitations of existing state-of-the-art models motivate us to pursue the ambitious objective of generating crystal materials under realistic settings, where the goal is to produce stable periodic structures based on provided textual descriptions. To achieve this, we propose incorporating textual guidance into the diffusion process, leveraging material descriptions to steer the reverse denoising trajectory toward generating novel and stable materials that align with the specified requirements.
\\\\
\noindent \textbf{Crystal Property Prediction.} In the recent past, we have witnessed a surge of interest in developing machine learning models ~\cite{seko2015prediction,pilania2015structure,lee2016prediction,de2016statistical,seko2017representation,isayev2017universal,ward2017including,lu2018accelerated,im2019identifying} for fast and accurate property prediction of crystalline materials. In particular, graph neural network (GNN) models have gained significant prominence for their ability to encode complex graph-structured information into enriched representation spaces. State-of-the-art GNN approaches ~\cite{xie2018crystal,chen2019graph,louis2020graph,Wolverton2020,schmidt2021crystal,choudhary2021atomistic} typically construct multi-edge graphs for 3D material structures by connecting nearby atoms within a predefined distance threshold in 3D space. These graphs are then processed using GNNs to learn crystal structure representations that are optimized for downstream property prediction tasks. However, they suffer from the following major limitations: 
\textbf{(A) Scarcity of Labeled Data:} Existing models are supervised in nature, possessing large trainable parameters like typical deep neural networks. Hence, a large amount of property-labeled data is needed to train these models, which makes it challenging for many material properties where we do not have enough property-tagged data. 
\textbf{(B) Lack of Interpretability: } Current methods lack interpretability and algorithmic transparency for their results, limiting their utility in material science applications. Therefore, it is necessary to explore and provide the reasons behind a prediction for any given property. 
\textbf{(C) DFT Error Bias: }Also, as available experimental data for the various properties are small and less diverse~\cite{kubaschewski1993materials,bracht1995properties,turns1995understanding} as they are very expensive to curate, existing models are trained using data gathered from the DFT calculations~\cite{Materials_project,kirklin2015open,Wolverton}. As DFT data often differ from experimental ground truth training with DFT-only data may incorporate the inaccuracies of DFT in the prediction. The current methods can hardly remove or mitigate such  DFT error bias.
\textbf{(D) Dependency on Domain Knowledge: } The architectural innovations of these models come from incorporating specific domain knowledge into a deep encoding module.  However, as different properties expressed by crystal materials are a complex function of different inherent structural and chemical properties of the constituent atoms, it is extremely difficult to explicitly incorporate them into the encoder architecture. 
\textbf{(E) Lack of Pre-trained Graph Model: } While pre-training has been effective in language and vision domains, it remains an open question how to effectively use pre-training on graph datasets like crystals, which will be robust and task agnostic.
\textbf{(F) Lack of Global Knowledge: }Existing models rely on a single modality of crystal data, i.e, crystal graph structure, and fail to incorporate crucial global periodic structural information, which can aid the property prediction accuracy.
\\
These challenges motivate us to focus on developing machine learning algorithms capable of learning more robust and enriched representations of crystal materials, with the aim of improving property prediction accuracy and alleviating the aforementioned limitations.

\rev{\noindent \textbf{Accuracy \& Cost Trade-off Between DFT and AI-Based Models.}
Density Functional Theory (DFT)~\cite{kohn1965self} provides high-fidelity predictions of material properties by explicitly solving quantum-mechanical equations. However, this accuracy comes at a significant computational cost. A typical DFT calculation scales as $\mathcal{O}(N^3)$ with respect to the number of electrons or basis functions, and often requires minutes to hours per structure on high-performance computing (HPC) resources. For large-scale materials screening involving $10^4 \text{–} 10^6$ candidate structures, the total computational cost becomes prohibitive in terms of CPU/GPU hours, energy consumption, and monetary expense.
\\
In contrast, AI-based surrogate models, once trained, provide predictions with constant-time inference, typically requiring milliseconds per structure on a single GPU or CPU. Although the initial training cost can range from tens to hundreds of GPU-hours, this cost is amortized over all subsequent predictions. As a result, AI models enable orders-of-magnitude reduction in computational cost when deployed at scale.
\\
Quantitatively, prior studies ~\cite{xie2018crystal,chen2019graph,Wolverton2020,schmidt2021crystal,choudhary2021atomistic} have shown that AI models trained on DFT-labeled datasets can achieve Mean Absolute Errors (MAE) within $2\% \text{–} 10\%$ of DFT accuracy for key material properties such as formation energy, band gap, and elastic constants. In exchange, computational cost is reduced by approximately $10^4 \text{–} 10^6 \times$ compared to performing fresh DFT calculations for each structure.
\\
For example, predicting formation energy using DFT may require ~1–10 CPU-hours per material, whereas an AI model can generate the same prediction in < 0.01 seconds, resulting in negligible runtime cost. Even accounting for training overhead, the break-even point is reached after predicting only a few thousand structures.
\\
Importantly, the AI-based approach does not aim to replace DFT entirely, but rather to approximate DFT-level trends with high fidelity, making it well-suited for high-throughput screening, candidate prioritization, and inverse design. Final shortlisted candidates can still be validated using DFT, ensuring that the overall pipeline retains accuracy while drastically reducing computational expense.
\\
Thus, the trade-off is favorable: a modest loss in accuracy (typically a few eV/atom or a few percent error) yields several orders-of-magnitude savings in computational cost, making AI-based methods a scalable and practical alternative for large-scale materials discovery.}

\color{black}
\section{Research Objectives}\label{Sec:RQs}

\begin{table*}[!ht]
    \centering
    \resizebox{1.0\textwidth}{!}{
    \begin{tabular}{p{4cm} | p{4.5cm} | p{5cm} | p{4.5cm} | p{4.5cm}}
    \hline
        \textbf{Objective} & \textbf{Chapter 3} & \textbf{Chapter 4} & \textbf{Chapter 5} & \textbf{Chapter 6}\\ \hline
         Chapter Title & CrysXPP: An Explainable Property Predictor for Crystalline Materials & CrysGNN: Distilling Pre-trained Knowledge to Enhance Property Prediction for Crystalline Materials & CrysMMNet: Multimodal Representation for Crystal Property Prediction & TGDMat: Periodic Materials Generation using Text-Guided Joint Diffusion Model\\ \hline
         Framework  & Explainable Property Predictor & Pre-training GNN & Multi-modal GNN & Text Guided Diffusion\\ \hline
         Input Type   & Multi-Graph & Multi-Graph & Multi-Graph + Text & 3D graphs + Text\\ \hline
         Impact / Utility  & Improved property prediction performance & Improved property prediction performance, Dataset contribution  & Improved property prediction performance, Dataset contribution & Improved generation performance, Dataset contribution \\ \hline
         Evaluation Tasks  & Crystal Property Prediction & Crystal Property Prediction & Crystal Property Prediction & Random Material Generation (Gen), Crystal Structure Prediction (CSP)\\ \hline
    \end{tabular}}
    \caption{Overview of the research objectives and how they are addressed by various approaches developed in this thesis.}
    \label{tab:research-objectives}
\end{table*}

The primary objective of this thesis is to develop efficient and robust machine learning frameworks for crystal material discovery and property prediction. Building on the motivations and challenges discussed in the preceding section, the specific goals of this work are as follows:

\begin{itemize}

    \item \textbf{Explainable Property Predictor for Crystalline Materials: }  One fundamental problem in crystal property prediction is \textit{scarcity of labeled data and lack of interpretability in their predictions}. We leverage a transfer learning-based unsupervised framework to develop an explainable property predictor \npjmodel{} that mitigates these issues and enhanced the property prediction performance.

    \item \textbf{GNN Pretraining Model for Crystal Property Prediction:} We extend the idea of GNN pretraining by developing a deep pre-trained GNN framework \aaaimodel{} for crystal materials, trained on a large corpus of unlabeled data. By leveraging a self-supervised pretraining paradigm at both the node and graph levels, the model effectively captures local atomic chemical features as well as global structural information, thereby enhancing the performance of downstream property prediction tasks.

    \item \textbf{Crystal Multi-Modal Representation: } One of the fundamental limitations of existing crystal property predictors is that they \textit{rely on a single modality}, i.e, crystal graph structure, which limits the expressive power of these models. To overcome this, we develop \uaimodel{} where we incorporate textual information alongside graph-structured data to learn multi-modal representations of crystal materials. These enriched and more robust representations lead to improved property prediction performance.

     \item \textbf{Text Guided Joint Diffusion Model for Periodic Material Generation: }Crystal materials can be modeled by a minimal \textit{unit cell}, which contains three components: atom type ($\mathbf{A}$), atomic coordinates  ($\mathbf{X}$), and the lattice structure ($\mathbf{L}$). We develop \iclrmodel{}, a text-guided diffusion model for material generation, jointly diffusing lattices, atom types, and coordinates with textual knowledge at each step. It outperforms baselines in CSP and generation tasks, lowers computational cost, and shows strong real-world generative ability.
\end{itemize}
In the next section, we will describe how the work done in this thesis achieves the research objectives stated above. For reference, Table~\ref{tab:research-objectives} presents the names of our approaches (that have been developed in this thesis) and the research objectives addressed by each approach. 

\color{black}
\section{Contributions of the Thesis}\label{sec:contributionThesis}
In this section, we summarize the key contributions of the thesis that tackle the challenges outlined in Section~\ref {sec:motivationThesis}.
\subsection{Explainable Property Predictor for Crystalline Materials}
Our first contribution is leveraging a transfer learning-based unsupervised framework to develop an explainable property predictor. A fundamental challenge in materials design is the fast and accurate prediction of crystal material properties. While existing GNN models achieve high precision in property prediction, they face several key limitations, including \textit{scarcity of labeled data, lack of interpretability, and DFT error bias}. To mitigate these issues and learn a more robust and enriched crystal representation for property prediction, we first leverage a transfer learning-based unsupervised framework to develop an explainable property predictor, \textbf{Crys}tal e\textbf{X}plainable \textbf{P}roperty \textbf{P}redictor (\npjmodel{}). It is built upon CrysAE, an auto-encoder-based architecture that is trained with all available (untagged) crystal data. This leads to the deep encoding module capturing all the important structural and chemical information of the constituent atoms of the crystal graph. The learned information is leveraged to build the property predictor, \npjmodel{}, where the knowledge-enhanced encoder helps to produce a high-quality representation of a candidate crystal. Consequently, the property predictor provides superior performance even when trained with a small amount of property-tagged data. Further, we introduce a feature selector that helps to provide an explanation by highlighting the subset of the atomic features responsible for the manifestation of a property of the given crystal. The node features are first passed through a feature selector, which is a trainable weight vector that selects a weighted subset of important node-level features for a given crystal property of interest.
\\
Through extensive analysis of a popular inorganic crystal dataset across seven properties, we show that \npjmodel{} can achieve the lowest error compared to other alternative baselines; the improvement is particularly significant when only a small amount of tagged data is available for training. Further, with appropriate case studies, we show that the feature selection module can effectively provide explanations of the importance of different features towards prediction, which are in sync with domain knowledge. Our novel contributions in this work are as follows:
\begin{itemize}
    \item We propose an explainable property predictor, \npjmodel{}, built on a transfer learning–based unsupervised framework. The underlying autoencoder (CrysAE) learns enriched structural and chemical representations from large-scale unlabeled crystal data, enabling robust property prediction with limited labeled samples.
    \item We design a trainable feature selector that identifies the most relevant atomic-level features for a given property. This module not only improves predictive accuracy but also provides interpretable explanations consistent with domain knowledge.
    \item Through extensive experiments on a widely used inorganic crystal dataset spanning seven properties, \npjmodel{} demonstrates significantly lower prediction error than strong baselines, particularly in low-data regimes, while also offering interpretability.
\end{itemize}

\subsection{GNN Pretraining Model for Crystal Property Prediction}
Our second contribution is developing a large deep pretrained GNN model for crystal property prediction.
We extend the idea of pretraining to address the challenges of \textit{domain knowledge dependency and the lack of a pre-trained graph models in materials science}. We propose a graph pretraining method that captures (a) atomic connectivity, (b) atomic features, and (c) graph similarity from a large collection of unlabeled data. To this end, we curate a new large-scale crystal dataset consisting of 800 K crystal graphs and train a deep pretraining framework, \aaaimodel{}, on it. \aaaimodel{} learns crystal graph representations by incorporating both node-level (atom) and graph-level (crystal) losses. At the node level, we pretrain the GNN to reconstruct node features and their connectivity in a self-supervised manner. At the graph level, we employ supervised and contrastive learning to capture structural similarities between crystal graphs using space group and crystal system information. Finally, we retrofit the pretrained \aaaimodel{} model into state-of-the-art property predictors to enhance learning and improve performance. To achieve this, we employ knowledge distillation to transfer essential structural and chemical information from the pretrained model into the property prediction process.
With rigorous experimentation on two popular benchmark materials datasets, we demonstrate that distilling information from \aaaimodel{} into various property predictors yields substantial performance improvements for both GNN-based architectures and the complex ALIGNN model. The improvements range from 4.19\% to 16.20\% over several highly optimized SOTA models. We also conduct ablation studies to analyze the impact of different pretraining losses on SOTA model performance and observe significant gains when combining both node- and graph-level pretraining, compared to using either individually. Furthermore, employing both supervised and contrastive graph-level pretraining enables learning more robust and expressive graph representations, thereby enhancing property predictor performance, particularly in sparse data scenarios. Additionally, since property-tagged datasets are theoretically derived (DFT-based), they inherently introduce bias, which propagates into the property predictors. We find that training with a small amount of experimental data substantially reduces this DFT bias. Key contributions of this work are as follows:
\begin{itemize}
    \item We curate a new dataset of 800K crystal graphs and develop \aaaimodel{}, a large deep pretrained GNN model for crystal property prediction. The framework jointly learns atomic- and crystal-level representations using self-supervised node reconstruction and graph-level supervised + contrastive learning.
    \item We retrofit \aaaimodel{} into state-of-the-art property predictors through knowledge distillation, transferring structural and chemical knowledge to enhance learning. This yields substantial improvements (4.19\%–16.20\%) across diverse GNN architectures, including ALIGNN, particularly in low-data regimes.
    \item Through rigorous experiments and ablation studies, we demonstrate the effectiveness of combining node- and graph-level pretraining losses. Further, we show that incorporating limited experimental data helps mitigate DFT-induced bias, leading to more robust and accurate property prediction.
    
\end{itemize}

\subsection{Multi-Modal Representation for Crystal Property Prediction}
Our third contribution addresses the reliance of crystal property predictors on a single data modality i.e., crystal graph structure and proposes learning a more robust and enriched representation by integrating multi-modal data, i.e., graph structures and textual descriptions of materials. A key advantage of using textual descriptions is that they provide diverse periodic structural information useful for estimating different crystal properties, which is difficult to explicitly encode into graph structures. We first curate textual datasets from two popular graph-based materials databases, Materials Project (MP) and JARVIS-DFT, containing textual descriptions of each material. For this, we employ Robocrystallographer~\cite{ganose2019robocrystallographer} to generate descriptions of global crystal structures, including space group numbers, crystal symmetries, rotational information, component orientations, and heterostructure information. We propose \uaimodel{}, a simple multi-modal framework for crystal materials consisting of two components: a Graph Encoder and a Text Encoder. Given a material, the Graph Encoder processes its graph structure using a GNN-based approach to capture local neighborhood structural information around atoms and learn a crystal-level representation. In parallel, the Text Encoder, a transformer-based model, encodes global structural knowledge from the material’s textual description and generates a textual representation. Finally, both graph-based and text-based representations are fused to form a richer multimodal representations of materials, capturing both global and local structural knowledge, thereby improving property prediction accuracy. Extensive experiments show \uaimodel{} outperforms all the popular state-of-the-art baselines across ten diverse sets of properties on two popular datasets. Further, we conduct ablation studies to demonstrate the expressiveness and robustness of textual representation on different crystal GNN encoders and show performance gain across all the properties using both global and local information combined as textual knowledge, compared to only global or local knowledge separately. Key contributions of this work as follows:
\begin{itemize}
    \item We propose \uaimodel{}, a framework that integrates graph structures and textual descriptions of materials to capture both local (atomic neighborhood) and global (crystal symmetry, orientation, and structural) information, leading to richer crystal representations.
    \item We curate large-scale textual datasets for crystals from Materials Project and JARVIS-DFT using Robocrystallographer, enabling the incorporation of diverse structural information (space groups, symmetries, orientations, heterostructures) not easily encoded in graph structures.
    \item Through extensive experiments on two benchmark datasets across ten crystal properties, \uaimodel{} consistently outperforms state-of-the-art baselines. Ablation studies further validate that combining global and local structural knowledge via multimodal fusion improves robustness and predictive accuracy.
\end{itemize}

\subsection{Text-Guided Joint Diffusion Model for Periodic Materials Generation} 
Our final contribution of this thesis is to develop a generative AI tool for periodic material generation. Crystal materials can be modeled by a minimal \textit{unit cell}, which gets repeated infinite times in 3D space on a regular lattice to form the periodic crystal structure \todokd{[cite]}. Formally, a crystal material $\mathbf{M}$ can be represented using three core components: the atom type matrix ($\mathbf{A}$), the coordinate matrix ($\mathbf{X}$), and the lattice matrix ($\mathbf{L}$). Recently, the equivariant diffusion models~\cite{jiao2023crystal,luo2023towards,xie2021crystal} have demonstrated great potential to generate stable 3D periodic structures of new crystal materials. However, these models have several inherent limitations. (1) None of the existing SOTA models learns the joint distribution of atom coordinates, types, and lattice structure in an end-to-end manner. CDVAE~\cite{xie2021crystal} and SyMat~\cite{luo2023towards} separately model lattice parameters and atom types with a VAE, then use a score network for atom coordinates. DiffCSP~\cite{jiao2023crystal} assumes atom types are given and predicts only lattice and coordinates. (2) These models rely on SE(3)-equivariant GNNs as denoising backbones, which focus on local message passing and fail to capture global structural knowledge that could improve diffusion. (3) Finally, they are unconditional: starting from noisy inputs, they generate crystals similar to the training distribution but without user control. In practice, users would want to specify constraints such as chemical formula, space group, symmetry, bond lengths, or desired properties, which the generated structures must satisfy. In this work, we propose \iclrmodel{}, a novel \textit{\underline{T}ext-\underline{G}uided joint \underline{D}iffusion Model for \underline{Mat}erial Generation} that mitigates the limitations mentioned above and enhances the generation capability. Our proposed model jointly generates $\mathbf{A}$, $\mathbf{X}$, and $\mathbf{L}$ using a periodic SE(3)-equivariant denoising diffusion model. This design ensures that the learned generative process adheres to the periodic E(3) invariance properties inherent to crystal structures. Building on this foundation, we further incorporate textual conditioning into the reverse diffusion process. This allows the model to generate crystal structures that align with given textual descriptions, effectively guiding the denoising trajectory using semantic cues. Such text guidance introduces a powerful modality for controllable generation and opens up new possibilities for materials discovery based on natural language specifications. Comprehensive experiments on standard benchmark datasets across multiple tasks demonstrate that TGDMat consistently outperforms existing baseline methods by a significant margin. In particular, for the structure prediction task (CSP), \iclrmodel{} achieves superior performance with just a single generated sample, highlighting the strength of text-guided generation. Moreover, in the de novo generation task (Gen), TGDMat outperforms all baselines, including their text-fusion variants, underscoring the effectiveness of our joint diffusion paradigm. Importantly, the integration of textual information not only improves generative performance but also leads to reduced training and inference costs, showcasing the practical benefits of our approach in terms of both efficiency and accuracy. Our novel contributions in this work are as follows:
\begin{itemize}
    \item To the best of our knowledge, we are the first to explore text-guided diffusion for material generation. Our proposed \iclrmodel{} bridges the gap between natural language understanding and material structure generation.
    \item Unlike prior models, \iclrmodel{} conducts joint diffusion on lattices, atom types, and coordinates, enhancing its ability to accurately capture the crystal geometry. Additionally, incorporating global structural knowledge through textual descriptions at each denoising step improves TGDMat’s ability to generate plausible materials with valid and stable structures.
    \item Extensive experiments using popular datasets on benchmark tasks show \iclrmodel{} outperforms baseline models by a good margin. Notably, in the CSP task, with just one generated sample, \iclrmodel{} outperforms all baseline models, highlighting the importance of text-guided diffusion. Moreover, in the generation task, \iclrmodel{} outperforms all baselines and their text-fusion variants, showcasing the effectiveness of the joint diffusion paradigm.
    \item Fusing textual knowledge reduces the overall computational cost for both training and inference of the diffusion model. Moreover, when applied to real-world custom text prompts by experts, \iclrmodel{} demonstrates rich generative capability under general textual conditions.
\end{itemize}

\color{black}
\section{Organization}\label{Sec: Organization}
The entire thesis consists of seven chapters. A brief overview of the chapters of the thesis is as follows:
\begin{itemize}

    \item \textbf{Chapter ~\ref{chap:intro}:} presents the background and objectives of the thesis in detail, followed by an explanation of its contributions.

	\item \textbf{Chapter~\ref{chap: related}:} This chapter provides a detailed literature review of existing research aligned with the primary objectives of the thesis. For each objective, relevant studies are examined, and our contributions are positioned within the field. While key ideas are revisited in later chapters, this chapter serves as a foundational overview, establishing the background that underpins the work presented in this thesis.

    \item \textbf{Chapter~\ref{chap: crysxpp}:} This chapter, containing the first work of my thesis, is titled {\it CrysXPP: An Explainable Property Predictor for Crystalline Materials}. This work addresses \textit{scarcity of labeled data, lack of interpretability, and DFT error bias} issues of crystal property prediction. It proposes \npjmodel{}, a transfer learning–based unsupervised framework for explainable crystal property prediction. Built on CrysAE, an autoencoder trained on unlabeled crystal data, the model learns enriched structural and chemical representations that enable superior prediction even with limited labeled samples. To ensure interpretability, a feature selector highlights the most influential atomic features responsible for specific properties. \npjmodel{} outperforms alternative baselines, with significant gains in low-data settings, while its explanations align well with domain knowledge.

    \item \textbf{Chapter~\ref{chap: crysgnn}:} This chapter, containing the second work of my thesis, is titled {\it CrysGNN: Distilling pre-trained knowledge to enhance property prediction for crystalline materials}. This work addresses the challenges of \textit{domain knowledge dependency and the lack of a pre-trained graph model in materials science}. It proposes \aaaimodel{}, a large pretrained GNN for crystal property prediction, trained on a newly curated dataset of 800 K crystal graphs. The framework jointly learns atomic- and crystal-level representations through self-supervised node reconstruction and graph-level supervised/contrastive learning. By distilling knowledge from \aaaimodel{} into state-of-the-art property predictors, we achieve performance gains across diverse models, including ALIGNN. Extensive experiments and ablation studies confirm that combining node- and graph-level pretraining yields robust representations, while incorporating small amounts of experimental data helps mitigate DFT-induced bias.

    \item \textbf{Chapter~\ref{chap: crysmmnet}:} This chapter, containing the third work of my thesis, is titled {\it CrysMMNet: Multimodal Representation for Crystal Property Prediction}. This work addresses the reliance of crystal property predictors on a single data modality i.e. crystal graph structure, and proposes learning a more robust and enriched representation by integrating multi-modal data. It proposes \uaimodel{}, which combines a graph encoder (GNN-based) to learn local structural representations with a text encoder (transformer-based) to capture global structural knowledge. Experiments on the Materials Project (MP) and JARVIS-DFT datasets show that \uaimodel{} outperforms state-of-the-art baselines across ten crystal property prediction tasks. Ablation studies further highlight the value of textual information, demonstrating consistent performance gains when combining global and local knowledge.

    \item \textbf{Chapter~\ref{chap: textgudedgen}:} This chapter, containing the final work of my thesis, is titled {\it TGDMat: Periodic Materials Generation using Text-Guided Joint Diffusion Model}. This work introduces \iclrmodel{}, a text-guided joint diffusion model for periodic material generation. It jointly learns atom types, coordinates, and lattice parameters within a periodic E(3)-equivariant framework, overcoming limitations of prior models. By incorporating textual conditioning, \iclrmodel{} enables controllable crystal generation and outperforms state-of-the-art baselines in both structure prediction and de novo generation, while also reducing computational costs.

    \item \textbf{Chapter~\ref{chap: conclusion}:} The final chapter of this thesis summarizes the key contributions: spanning explainable property prediction, large-scale pretraining, multimodal learning, and text-guided material generation, and outlines promising future research directions toward scalable, controllable, and foundation models for periodic materials.

\end{itemize}
\color{black}

%
\clearemptydoublepage
\chapter{\textcolor{black}{Related Work}}
\chaptermark{\textcolor{black}{Background Study}}
\label{chap: related}

\section{Crystal Property Prediction.}
Historically, Density Functional Theory (DFT)~\cite{orio2009density} has been an effective tool for estimating several properties of crystals by examining the electronic density of their atomic configurations. However, DFT calculations require substantial computational costs; hence, they pose a significant challenge for novel materials discovery in the vast materials space.
Recently, thanks to the advances in deep learning, data-driven~\cite{seko2015prediction,pilania2015structure,lee2016prediction,de2016statistical,seko2017representation,isayev2017universal,ward2017including,lu2018accelerated,im2019identifying} techniques have become quite popular to establish relationships between crystal atomic structures and their properties effectively. Graph neural networks have emerged as highly promising models in various domains of computer science, showcasing significant potential in many real-world applications, including social networks \cite{hamilton2017inductive,chen2018fastgcn,dai2018learning}, recommender systems \cite{berg2017graph,ying2018graph}, hyper-networks \cite{yadati2019hypergcn,bandyopadhyay2020hypergraph}, chemical and biological networks \cite{duvenaud2015convolutional,gilmer2017neural}, etc.
Recently, graph neural network (GNN) based approaches have been very effective in encoding structural information of crystal materials into an enriched embedding space so that they can predict different crystal properties with high accuracy.
CGCNN \cite{xie2018crystal} is the first proposed model for crystal property prediction, which constructs a multi-edge graph from the 3D crystal structure and applies a GNN model to encode the neighborhood structural information around an atom.
Following CGCNN, there are a lot of subsequent studies~\cite{chen2019graph,louis2020graph,Wolverton2020,schmidt2021crystal}, where authors proposed different variants of GNN architectures for effective crystal representation learning. Through multiple layers of graph convolutions, these models can implicitly encode many-body interactions. Further, ALIGNN~\cite{choudhary2021atomistic} explicitly captures many-body interactions by incorporating bond angles and local geometric distortions into the GNN encoding module to enhance property prediction accuracy and became SOTA for a wide range of properties. 
Recently, the transformer-based architecture Matformer~\cite{yan2022periodic} is proposed to learn the periodic graph representation of the material, which is invariant to periodicity and can capture repeating patterns explicitly. Matformer marginally improves the performance compared to ALIGNN; however, it is much faster than it. Recently, PotNet~\cite{lin2023efficient} is proposed, which models inter-atomic potentials directly based on physical principles. 

Moreover, data scarcity is a known challenge in this field, and to address it, several graph pre-training strategies have been introduced. CrysXPP~\cite{das2022crysxpp} is the first pre-trained/fine-tune model, which uses an autoencoder to train on a large volume of untagged crystal graphs (unsupervised pre-training), and the learned knowledge is transferred to initialize the encoder of a property predictor, which is further fine-tuned with property-specific tagged data. Following this work, CrysGNN~\cite{das2023crysgnn} proposed large-scale pre-training with a huge corpus of curated unlabelled data and incorporates the idea of knowledge distillation to distill the important information from the pre-trained model and inject it into the property prediction process.

\section{Graph Pre-training Strategies}
There are attempts to pre-train GNNs to extract graph and node-level representations.~\cite {hu2020pretraining} develops an effective pre-training strategy for GNNs, where they perform both node-level and graph-level pre-training on GNNs to capture domain specific knowledge about nodes and edges, in addition to global graph-level knowledge. GPT-GNN~\cite{hu2020gpt} proposes a self-supervised attributed graph generation task to pre-train a GNN model, which captures structural and semantic properties of the graph.~\cite{qiu2020gcc} presents GCC, which leverages the idea of contrastive learning to design the graph pre-training task as subgraph instance discrimination, to capture universal network topological properties across multiple networks. Another recent work is GraphCL~\cite{you2020graph}, which proposes different graph augmentation methods and maximises the agreement between two augmented views of the same graph via a contrastive loss.

\section{Diffusion Models}
The fundamental idea of the diffusion model, as initially proposed by~\cite{sohl2015deep}, is to gradually corrupt data with diffusion noise and learn a neural model to recover back data from noise. The idea of diffusion was further developed in two broad categories: (1) \textit{Score Matching Network}~\cite{song2019generative,song2020improved} and (2) \textit{Denoising Diffusion Probabilistic Models (DDPM)}~\cite{ho2020denoising}. In recent times, diffusion models have emerged as a powerful new family of deep generative models, achieving remarkable performance records across numerous applications such as image synthesis~\cite{dhariwal2021diffusion,ramesh2022hierarchical,rombach2022high}, molecular conformer generation~\cite{shi2021learning,xu2022geodiff}, molecular graph generation~\cite{liu2021graphebm}, protein folding~\cite{luo2022antigen,wu2021se}, etc. 
\section{Conditional Diffusion Models}
The initial DDPM model~\cite{ho2020denoising} demonstrated unconditional diffusion models for image generation, where the output cannot be directed towards a desired characteristic or property. In guided diffusion models, the sampling process can be steered by a prompt, which can be a textual description of the desired output, reference image, or any other type of media.

In the field of image generation by diffusion models, Ramesh et al~\cite{ramesh2022hierarchical} came up with a text-guided diffusion model called DALL·E 2, which showed how textual prompts can be used to steer the sampling process. While training the model, both the image and its textual description are encoded and mapped together, and the encoding of the prompt is used to generate the image during sampling. Another way of guiding the diffusion process using a separate classifier model was shown by~\cite{dhariwal2021diffusion}. They trained a classifier on the noisy images and used the gradient of the classifier to guide the sampling process. In the classifier-free setting, ~\cite{ho2022classifierfree} trained two diffusion models, one guided and one unguided, and combined the resulting score estimated during sampling to get the desired outcome. OpenAI's CLIP~\cite{radford2021learning} further improved the relevance of the generated image to the given prompt by scoring the correctness of the generated image given the textual prompt.

Similar efforts have been made in the field of molecular generative models. The shortcomings of SMILES-based autoregressive models were addressed by TGM-DLM~\cite{gong2024text} by utilizing diffusion models. This necessitates a two-step process, a text-guided generation phase, where the SMILES representation is generated from Gaussian noise with the help of a textual description, and a correction phase, where necessary corrections are made for the correctness of the SMILES string format. This is one of the drawbacks of the SMILES string format, which was later addressed by 3M-Diffusion~\cite{zhu20243mdiffusion}, where they have generated molecular graphs from a given textual description.

\section{Crystal Material Generation}
In the past, there were limited efforts in creating novel periodic materials, with researchers concentrating on generating the atomic composition of periodic materials while largely neglecting the 3D structure. With the advancement of generative models, the majority of the research focuses on using popular generative models like VAEs or GANs to generate 3D periodic structures of materials, however, they either represent materials as three-dimensional voxel images~\cite{court20203,hoffmann2019data,long2021constrained,noh2019inverse}  and generate images to depict material structures (atom types, coordinates, and lattices), or they directly encode material structures as embedding vectors~\cite{kim2020generative,ren2020inverse,zhao2021high}. However, these models neither incorporate stability in the generated structure nor are invariant to Euclidean and periodic transformations. 

\xhdr{Diffusion Models}
In recent times, equivariant diffusion models~\cite{xie2021crystal,luo2023towards,jiao2023crystal,yang2023scalable,jiao2024space,miller2024flowmm} have become the leading method for generating stable crystal materials, thanks to their capability to utilize the physical symmetries of periodic material structures. In specific, state-of-the-art models like CDVAE~\cite{xie2021crystal} and SyMat~\cite{luo2023towards} integrate a variational autoencoder (VAE) and a powerful score-based decoder network, work directly with the atomic coordinates of the structures, and use an equivariant graph neural network to ensure euclidean and periodic invariance. However, both CDVAE and SyMat first predict the lattice parameters and atomic composition using the VAE model and subsequently update the coordinates using score based diffusion model. Subsequent models, such as DiffCSP~\cite{jiao2023crystal} and MatterGen~\cite{zeni2023mattergen}, adopt a joint diffusion framework to simultaneously learn atomic composition, fractional coordinates, and lattice parameters. This approach effectively captures the crystal data distribution, significantly improving the performance of generative models in both crystal structure prediction (CSP) and generation tasks. 
These models represent 3D material structures as graphs and utilize SE(3)-equivariant graph neural networks (EGNN), adapted for periodicity, as denoising networks for the reverse diffusion process. This ensures both euclidean and periodic invariance in the learned distribution. 

In contrast, UniMat~\cite{yang2023scalable} introduces a unified crystal representation using a four-dimensional tensor, where atomic coordinate positions are stored within their corresponding atom entries in the periodic table. To model this representation, a probabilistic diffusion model is trained on the 4D tensor, employing interleaved attention and convolution layers as the denoising network. This unified approach effectively addresses the challenge of jointly modeling discrete atom types and continuous atomic coordinates.

Lately, there have been a few efforts to develop text-guided diffusion models for text-conditional material generation. Models such as TGDMat~\cite{das2025periodic} and Chemeleon~\cite{park2025exploration} incorporate contextual representations, leveraging pretrained models like MatSciBERT or CLIP, into a GNN-based denoising network. This enables the generation of valid and stable periodic materials that align with the conditions specified in textual descriptions.

\xhdr{Latent Diffusion Models}
A key limitation of existing diffusion-based approaches is their operation in a high-dimensional feature space, where they attempt to model the joint distribution over atom types, fractional atomic coordinates, and lattice structures. This joint distribution is inherently multimodal, with each component exhibiting distinct statistical characteristics. Fractional coordinates, which follow a wrapped normal distribution, are often modeled using score-based generative methods~\cite{song2020score}; atom types, being categorical, are typically handled with discrete diffusion models such as D3PM~\cite{austin2021structured}; and lattice structures are usually modeled via denoising diffusion probabilistic models (DDPMs)~\cite{ho2020denoising}. Capturing these heterogeneous components jointly requires complex denoising architectures and a large number of diffusion steps to generate high-quality crystals. As a result, these models entail substantial computational cost for both training and inference, limiting their applicability in resource-constrained settings. Furthermore, when applied to larger datasets like MPTS-52, which feature more intricate compositional spaces, existing approaches often struggle to scale and fail to achieve strong performance. To address this challenge, it is crucial to represent these complex structures in meaningful lower-dimensional spaces.  A few of the recent works, like CrysLDM~\cite{khastagir2025crysldm}, ADiT~\cite{joshi2025all} utilize a latent diffusion model to address the above limitations, reducing sampling time for crystal material generation. Operating in the latent space, these offer unique advantages in generative modeling complexity over existing feature-domain diffusion models, making it more efficient in terms of time and resource consumption costs.

\xhdr{Symmetry Aware Generation} While most diffusion models learn to freely position each atom in a unit cell, enforcing invariance to space group symmetry within a crystal greatly reduces the number of degrees of freedom that the model must predict.
DiffCSP++ \cite{jiao2024space} extends DiffCSP by constraining atomic coordinates and lattice parameters to obey a given space group. Lattices are parametrized such that the generated crystals belong to the correct lattice system, and atomic fractional coordinates are constrained to a set of Wyckoff positions for that space group. These Wyckoff positions are taken from templates drawn from the training data.
SymmCD~\cite{levysymmcd} jointly learns the fractional coordinates along with the Wyckoff position of each atom, using a representation of the site symmetry corresponding to each Wyckoff position.
By only modeling one atom for each crystallographic orbit, SymmCD and DiffCSP++ are able to greatly reduce the dimensionality of the generative task. 
WyCryst \cite{zhu2024wycryst} and Wyckoff Transformer \cite{kazeev2024wyckofftransformer} introduce similar space group constrained generative approaches for VAEs and autoregressive transformers, respectively.

\xhdr{Flow matching} Flow matching (FM)~\cite{lipman2023flow, albergo2023building, liu2023flow} has recently emerged as a compelling alternative to diffusion-based approaches for crystalline material generation. In contrast to diffusion models, which progressively denoise a sample initialized from a Gaussian prior, flow matching directly learns a time-dependent velocity field that continuously transports an arbitrary base distribution into the target distribution of stable crystals. The flexibility in choosing the base distribution, combined with a \emph{simulation-free} training objective defined on simple geometries \cite{chen2024flow}, enables the introduction of strong inductive biases into FM-based generative models. The first work applying these principles was FlowMM  \cite{miller2024flowmm}. This work proposed a representation that explicitly respects global rotational and translational symmetries, as well as periodic boundary conditions, and defined base distributions with equivariant flows to guarantee symmetry invariance. Within this framework, crystals were represented differently for two settings: Crystal Structure Prediction (CSP) and De Novo Generation (DNG). Both formulations describe crystals on distinct manifolds of valid configurations. The unit cell is parameterized by six rotation-invariant real numbers (three lengths and three angles), while atomic positions are specified as fractional coordinates with respect to the unit cell basis. For CSP, atom types were provided as one-hot encodings; for DNG, they had to be generated. Since no flow matching model at the time could directly generate discrete data, FlowMM employed an “analog bit” binary encoding that was later discretized during sampling \cite{chen2023analog}. Across both CSP and DNG, the method achieved strong performance while substantially reducing the computational cost of sampling.

\xhdr{GFlowNets}
Generative Flow Networks (GFlowNets) \cite{bengio2021gflownet, bengio2023gflownetfoundations} provide a unique generative framework inspired by reinforcement learning \cite{tiapkin2024gfnmaxentrl, deleu2024gfnmaxentrl}. They treat generation as a sequential decision-making process, where policy networks iteratively construct molecules or materials within a predefined environment. In Crystal-GFN \cite{milaai4science2023crystalgfn}, the model first samples a space group, followed by atomic composition, and finally lattice parameters to assemble a complete crystal structure. Throughout the generation process, physico-chemical and symmetry constraints are strictly enforced by masking infeasible actions. For materials discovery, the reward function typically reflects target properties such as formation energy, band gap, or mechanical stress. Crystal-GFN has demonstrated its effectiveness by producing crystals with low formation energy, specific band gaps, and high cell density. Extensions of this approach have also been applied to design reticular frameworks for CO\textsubscript{2} capture \cite{cipcigan2024reticulargflownet}.

\xhdr{Large Language Models}
Large language models (LLMs) are increasingly adapted in the natural sciences as versatile priors for reasoning over sequences, graphs, and spatial data \cite{zhang2024comprehensive}, and this trend has recently extended to materials generation. By encoding crystal structures as textual descriptions of unit cells and atomic positions, token-based language models have demonstrated performance that rivals—and in some cases surpasses—specialized geometric approaches \cite{gruver2024fine}.

\color{black}

\clearemptydoublepage
\chapter{CrysXPP: An Explainable Property Predictor for Crystalline Materials}
\chaptermark{CrysXPP [NPJCompMat-2022]}
\label{chap: crysxpp}

\noindent\fbox{%
\parbox{\textwidth}{%
Work of this chapter are based on the following publication:  
\\
\textit{CrysXPP: An Explainable Property Predictor for Crystalline Material.}
\\
\underline{Kishalay Das}, Bidisha Samanta, Pawan Goyal, Seung-Cheol Lee, Satadeep Bhattacharjee and Niloy Ganguly.
\\
NPJ Computational Materials Journal, 2022.}}
\vspace{1em}

\section{Introduction} 
Several machine learning techniques~\cite{seko2015prediction,xue2016accelerated,isayev2017universal,ghiringhelli2015big,isayev2015materials,xie2018crystal,sanyal2018mt,chen2019graph} have recently been developed to provide fast and accurate predictions of various properties of crystalline materials, thereby enabling the efficient screening of large material search spaces~\cite{meredig2014combinatorial,jha2018elemnet,choudhary2018machine}.
The existing approaches rely either on handcrafted feature-based descriptors~\cite{seko2015prediction,xue2016accelerated,isayev2017universal,ghiringhelli2015big,isayev2015materials} or on deep graph neural networks (GNNs)~\cite{xie2018crystal,sanyal2018mt,chen2019graph,louis2020global,cheng2021geometric,banjade2021structure,jin2020hierarchical,qiao2020orbnet,ye2020symmetrical}, which learn representations directly from the 3D conformations of crystal structures.
Generating handcrafted features requires substantial domain knowledge and manual intervention, which inherently limits these methods. In contrast, deep learning approaches eliminate the need for explicit feature design by automatically learning structure–property relationships, making them a highly attractive alternative.
\\\\
Graph neural network (GNN)-based approaches have recently gained popularity due to their ability to encode graph information into an enriched representation space. Orbital-based GNNs \cite{qiao2020orbnet, ye2020symmetrical} use symmetry-adapted atomic orbital features to predict different molecular properties. Though orbital-based GNNs predict molecular properties well, they are not an excellent choice for capturing complicated periodic structures such as crystals since they describe the nature of the electron distribution particularly close to atoms. On the other hand, motif-centric GNNs \cite{banjade2021structure, jin2020hierarchical} convert motif substructures of a crystal as a node and encode their interconnections for a large set of crystalline compounds using an unsupervised learning algorithm. Although these methods demonstrate improvements in property prediction tasks for metal oxides, their applicability remains limited, as they overlook the atomic configurations within motif substructures, which play a crucial role. In a different line of work, CGCNN~\cite{xie2018crystal} and MTCGCNN~\cite{sanyal2018mt} construct convolutional neural networks directly on 2D crystal graphs derived from 3D crystal structures. Further, GATGNN~\cite{louis2020global} applies the concept of graph attention networks to crystal graphs, enabling the model to learn the relative importance of different atomic bonds, while MEGNet~\cite{chen2019graph} incorporates global state attributes to enhance quantitative structure–state–property relationship prediction in materials. As this class of methods aims to capture the information of any crystal graph just from the connectivity and atomic features, we contribute in this promising direction. 
\\\\
Like other large deep neural network models, GNN-based architectures introduce a substantial number of trainable parameters. As a result, to estimate these parameters correctly for better accuracy, a huge amount of tagged training data is required, which is not always available for all the crystal properties.  Therefore, developing a deep learning model that can be effectively trained with limited labeled data would be highly valuable for inferring diverse properties of crystalline materials. Also since the available experimental data for various properties are often limited and lack diversity~\cite{kubaschewski1993materials,bracht1995properties,turns1995understanding}, these models are typically trained on datasets generated from DFT calculations~\cite{Materials_project,kirklin2015open, Wolverton}. Since DFT data often deviates from experimental ground truth due to limitations in accurately describing the many-body ground state, particularly for properties such as band gaps~\cite{bandgap} or van der Waals interactions~\cite{van}, training models solely on DFT-generated data may propagate these inaccuracies into the predictions. Moreover, in most of the cases, the existing property predictors are trained to predict a specific property. Hence, the generated descriptor or embeddings of any crystal are specific to a given property. It prevents them from sharing common structural information relevant to multiple properties. Though multi-task learning setup achieves information sharing across properties~\cite{sanyal2018mt}, it works well only for properties that are correlated with each other~\cite{zhang2021survey}. Last but not least, the existing  neural network based  methods~\cite{meredig2014combinatorial,ward2016general,jha2018elemnet,wu2018moleculenet,sanyal2018mt,xie2018crystal,louis2020global,chen2019graph,cheng2021geometric,banjade2021structure,jin2020hierarchical,qiao2020orbnet,ye2020symmetrical}  hardly provide any explanation for their results. This lack of interpretability and algorithmic transparency limits their practical utility in materials science. Hence, it is essential to develop approaches that not only make accurate predictions but also provide clear reasoning behind the predicted properties.
\\\\
In this paper, we propose an explainable deep property predictor for crystalline materials ~\npjmodel{} \footnote{The source code and dataset for CrysXPP are available at \url{https://github.com/kdmsit/crysxpp}}. It is built upon  CrysAE, an auto-encoder based architecture, trained with a large amount of easily available crystal data, that is, property-agnostic structural information of the crystal graph. This leads to the deep encoding module capturing all the important structural and basic chemical information of the constituent atoms (nodes)  of the crystal graph. The learned information is leveraged to build the property predictor, ~\npjmodel{}, where the {\em knowledgeable} encoder helps to  produce high quality representation of a candidate crystal. Consequently, the property predictor provides superior performance (better than all the competing baselines) even when trained with a small amount of property-tagged data, thus largely mitigating the need for having a huge amount of dataset tagged with a specific property. The structural information learned in the encoding model of an auto-encoder is robust and can remove the error bias introduced by DFT by fine-tuning the system with a small amount of experimental data, whenever available. Further, we introduce a feature selector that helps to provide an explanation by highlighting the subset of the atomic features responsible for the manifestation of a chemical property of the given crystal.\\
Through extensive analysis of an inorganic crystal dataset across seven properties, we show that our method can achieve the lowest error compared to other alternative baselines; the improvement is particularly significant when only a small amount of tagged data is available for training. We have further shown that \npjmodel{} is effective towards removing error bias due to DFT tagged data by incorporating a small amount of experimental data in the training set for both formation energy and band gap. Finally, with appropriate case studies, we show that the feature selection module can effectively provide explanations of the importance of different features towards prediction, which are in sync with the domain knowledge.

\section{Methodology}
\label{sec:method}
\begin{figure}[ht!]
	\centering
	\boxed{\includegraphics[clip,width=\textwidth]{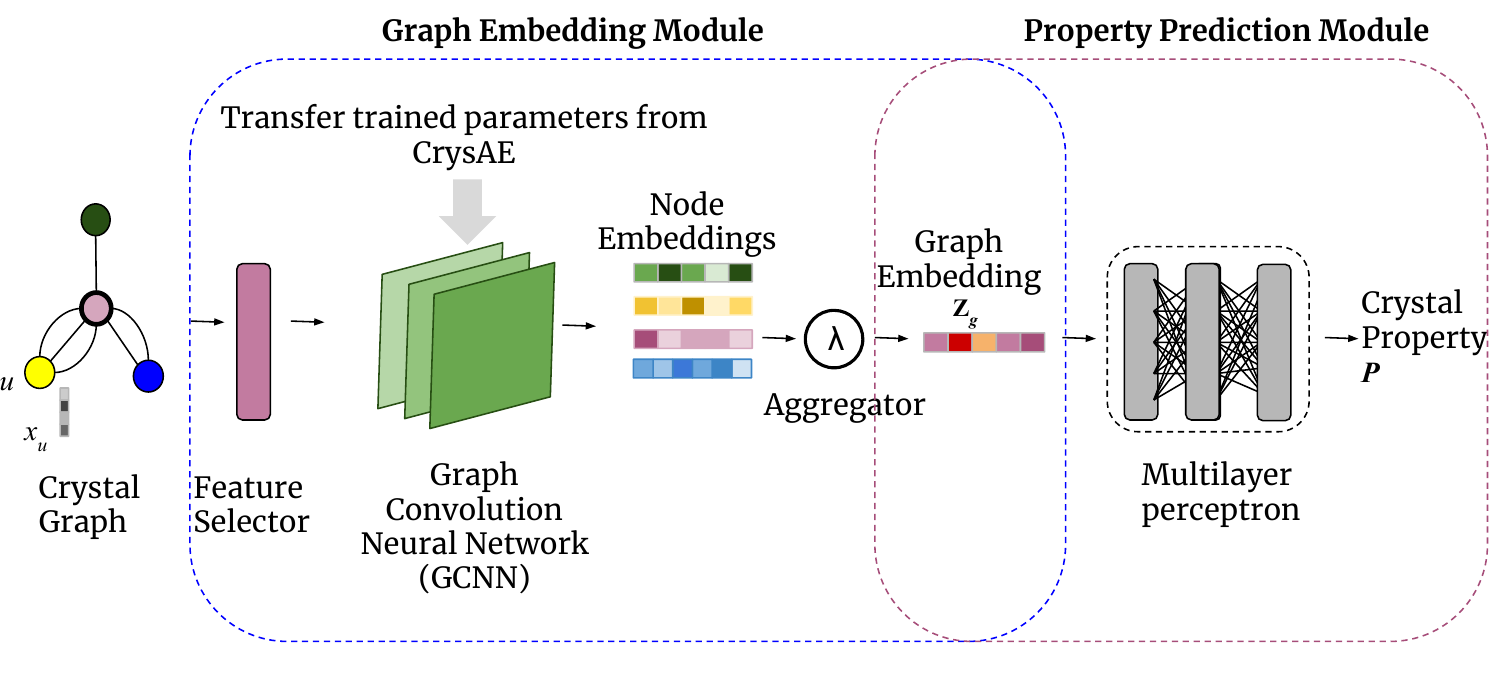}}
	\caption{The architecture of the Crystal eXplainable Property Predictor (\npjmodel{}) consists of two main components: (a) a multilayer Graph Convolutional Neural Network (GCNN) serving as the graph embedding module, and (b) a multilayer perceptron functioning as the property prediction module. Given the graph structure and node features, the GCNN produces an embedding for each graph. This embedding is then passed to the property predictor, a deep regression module, which outputs the predicted property value.}
	\label{fig:crysxpp_architecture}
\end{figure}
\subsection{Model Architecture: Overview}
In this section we discuss in more detail the key technical contributions \rev{towards designing a deep explainable crystal property predictor}, followed by the training process and implementation details.\\
We propose \textbf{Crys}tal e\textbf{X}plainable \textbf{P}roperty \textbf{P}redictor (\npjmodel{}), which realizes a crystalline material as a graph structure (say \textit{$\Gcal$})  and predicts the value of a property (eg. formation energy) given the crystal graph structure. As depicted in  Fig.\ref{fig:crysxpp_architecture}, \npjmodel{} comprises two building blocks (a). a property prediction module
and (b). a graph embedding module. In the graph embedding module we have a crystal graph  encoder based on  graph convolution neural network (GCNN) \cite{xie2018crystal}, which takes a crystal graph structure along with node and edge feature information as input and returns an embedding corresponding to each node as output. We consider nine different atomic properties (check Table \ref{tab:feature_desc}) as node features and the weights of those node features are determined by the  feature selector layer. Moreover,  the graph embedding module needs to capture the structural and chemical properties of the underlying crystal, hence one can use the huge amount of available crystal information (irrespective of the property) to train  the graph convolution network. 
For this at first we separately train the GCNN as a part (the encoder) of CrysAE (Fig.\ref{fig:Enc_architecture}); and the weights thereby obtained are used  as an initialization of the GCNN of \npjmodel{}. The structural information learned in the encoding model of CrysAE and duly transferred to the GCNN of \npjmodel{} makes \npjmodel{} more robust. \\
Our overall model architecture is essentially composed of the  following two modules:
\begin{itemize}
	\item \textbf{Auto encoder (CrysAE): } \\$q_{\bm{\theta}}: (\bm{\Vcal}, \bm{\Ecal}, \bm{\mathcal{X}}, \bm{\mathcal{F}}) \to \bm{\mathcal{Z}}$;
	$p_{\bm{\phi}}:  \bm{\mathcal{Z}} \to (\bm{\Vcal}, \bm{\Ecal}, \bm{\mathcal{X}}, \bm{\mathcal{F}})$
	\item \textbf{Property predictor (\npjmodel{}): }\\ $p_{\bm{\zeta}, \bm{\theta^\prime}, \bm{\psi}}: \bm{\mathcal{X}} \to_{\bm{\zeta}} \bm{\mathcal{X}}^\prime; (\bm{\Vcal}, \bm{\Ecal}, \bm{\mathcal{X}}^\prime, \bm{\mathcal{F}}) \to_{\bm{\theta^\prime}} \bm{\mathcal{Z}}; \bm{\mathcal{Z}} \to_{\bm{\psi}} \Pcal$
\end{itemize}
In the above characterization, $\bm{\theta}, \bm{\phi}, \bm{\zeta}, \bm{\theta}^\prime$ and $\bm{\psi}$ are the trainable parameters of the respective modules. Here $\bm{\theta}$ and $\bm{\phi}$ are the parameters for the encoder and decoder respectively of the CrysAE. $\bm{\zeta}$ is the trainable parameter of feature selector $\Scal$, $\bm{\theta}^\prime$ is the parameter of the encoder  and $\bm{\psi}$ is the parameter of the multi layer perceptron of \npjmodel{} model. We initialize $\bm{\theta}^\prime:=\bm{\theta}$ i.e, we first train the autoencoder and then the parameters of the encoder of CrysAE are transferred to the \npjmodel{}. \\\\
\begin{table}
	\centering
	\small
	\setlength{\tabcolsep}{4pt}
	\scalebox{0.9}{
	\begin{tabular}{|c|c|c|c|}
		\hline
		\textbf{Features} & \textbf{Unit} &\textbf{Range}  & \textbf{Feature Dimension} \\
		\hline
		Group Number & - & 1,2, ..., 18 & 18  \\
		\hline
		Period Number & - & 1,2, ..., 9 & 9\\
		\hline
		Electronegativity & - & 0.5-4.0 & 10 \\
		\hline
		Covalent Radius & pm & 25-250 & 10 \\
		\hline
		Valence Electrons & - &1,2, ..., 12 & 12\\
		\hline
		First Ionization Energy & eV & 1.3–3.3 & 10\\
		\hline
		Electron Affinity & eV & -3–3.7 & 10 \\
		\hline
		Block & - & s, p, d, f & 4 \\
		\hline
		Atomic Volume & $cm^{3} /mol$ & 1.5–4.3 & 10 \\
		\hline
	\end{tabular}
	}
	\caption{\rev{Crystalline materials are realized  as crystal graph structures $\bm{\mathcal{D}} = \{\Gcal_i =(\bm{\mathcal{V}}_i, \bm{\mathcal{E}}_i, \bm{\mathcal{X}}_i, \bm{\mathcal{F}}_i ) \}$. In this table we present the description of different atomic features $\mathcal{X}_i$ and their unit, range and dimensions. All of them together forms 92 dimensional node feature vector $\mathcal{X}_i$.}}
    \centering
	\label{tab:feature_desc}
\end{table}
\begin{figure}[h]
	\centering
	\vspace*{-1mm}
	\boxed{\includegraphics[width=0.8\columnwidth]{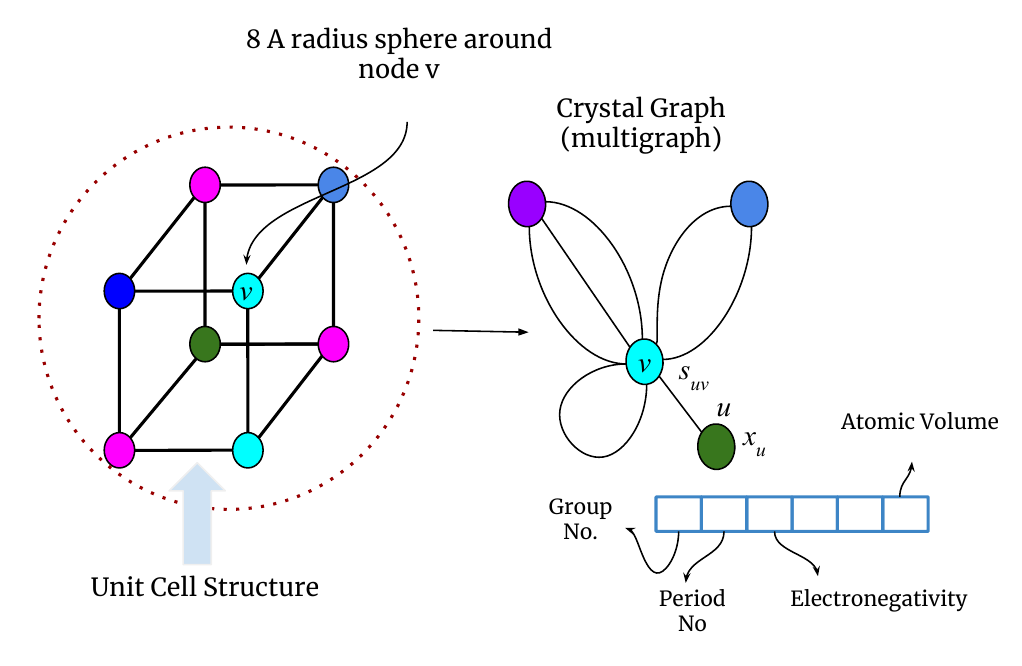}}
	\caption{\rev{Multi-graph Representation from Crystal 3D Structure. A crystal is modeled as an undirected weighted multi-graph, where nodes represent atoms in a unit cell and edges capture interatomic interactions arising from periodic repetition within a cutoff radius (r). Multiple edges between a node pair ((u,v)) account for neighbors in different repeating cells, each weighted by the corresponding bond length. Node features encode atomic properties, and a maximum of (K) edges per node pair is allowed.}}
	\label{fig:3d_crsytal_rep}
\end{figure}
\xhdr{Crystal Representation}
Our model realizes crystalline materials  as crystal graph structures $\bm{\mathcal{D}} = \{\Gcal_i =(\bm{\mathcal{V}}_i, \bm{\mathcal{E}}_i, \bm{\mathcal{X}}_i, \bm{\mathcal{F}}_i ) \}$ as proposed in~\cite{xie2018crystal}. 
Crystals have a repeating structure as depicted in Fig.\ref{fig:Enc_architecture}, where a unit cell gets repeated across all the three dimensions. Hence, unlike simple graphs, the $\Gcal_i$ is an undirected weighted multi-graph where $\bm{\mathcal{V}}_i$ denotes a set of nodes (atoms) present in a unit cell of the crystal structure and $\bm{\mathcal{E}}_i = \{(u,v,k_{uv})\}$ denotes a multi-set of node pairs and the number of edges  between them.  $k_{uv}$ edges between a pair of nodes ($u,v$) indicate that $v$ is present in $k_{uv}$ repeating cells within $r$ radius from $u$ ($r$ is a hyper-parameter). 
$\bm{\mathcal{X}}_i$ represents node features i.e.  features that uniquely identify the chemical properties such as atomic volume, electron affinity, etc. of an atom as described in Table \ref{tab:feature_desc}. Lastly, $\bm{\mathcal{F}}_i$ corresponds to a multi-set of edge weights. We denote $\bm{\mathcal{F}}_i =\{\{s^k\}_{(u,v)} |  (u,v) \in \bm{\mathcal{E}}_i\}$ where $s^k$ denotes the $k^{th}$ bond length between the node pair $(u,v)$.  Between any pair of nodes, a maximum of $K$ edges are possible where $K$ is empirically determined. The bond length helps to specify the relative distance of an atom from its neighboring atoms. We use this graphical abstraction of a crystal as this can effectively embed the periodicity (indicated by the number of bonds) along with relative positioning for each atom in a simpler way, which otherwise was difficult to capture. 

\rev{\xhdr{Choice of radius $r$} In the multi-graph construction, the cutoff radius $r$ of the sphere around each atom determines the number of neighboring atoms considered. We set r = 8 Å following prior works such as CGCNN and MT-CGCNN. Empirically, using smaller values of r (e.g., 2 - 4 Å) results in a sparser graph and faster GCN computation, but significantly degrades property prediction performance. Conversely, increasing r to 10 Å or higher leads to much denser graphs, substantially increasing computational cost while yielding only marginal performance gains. Therefore, r = 8 Å represents a well-balanced trade-off between accuracy and efficiency.}

Next, we formally define the autoencoder (CrysAE) and property predictor (CrysXPP).\\\\
\subsection{Auto Encoder: CrysAE}
We build \textbf{Crys}tal \textbf{A}uto \textbf{E}ncoder (CrysAE) which composes of a simple encoder followed by an appropriate decoder to facilitate the overall training in order to learn necessary information in the encoding mechanism. 
\begin{figure*}[ht!]
	\centering
	\boxed{\includegraphics[clip,width=\columnwidth]{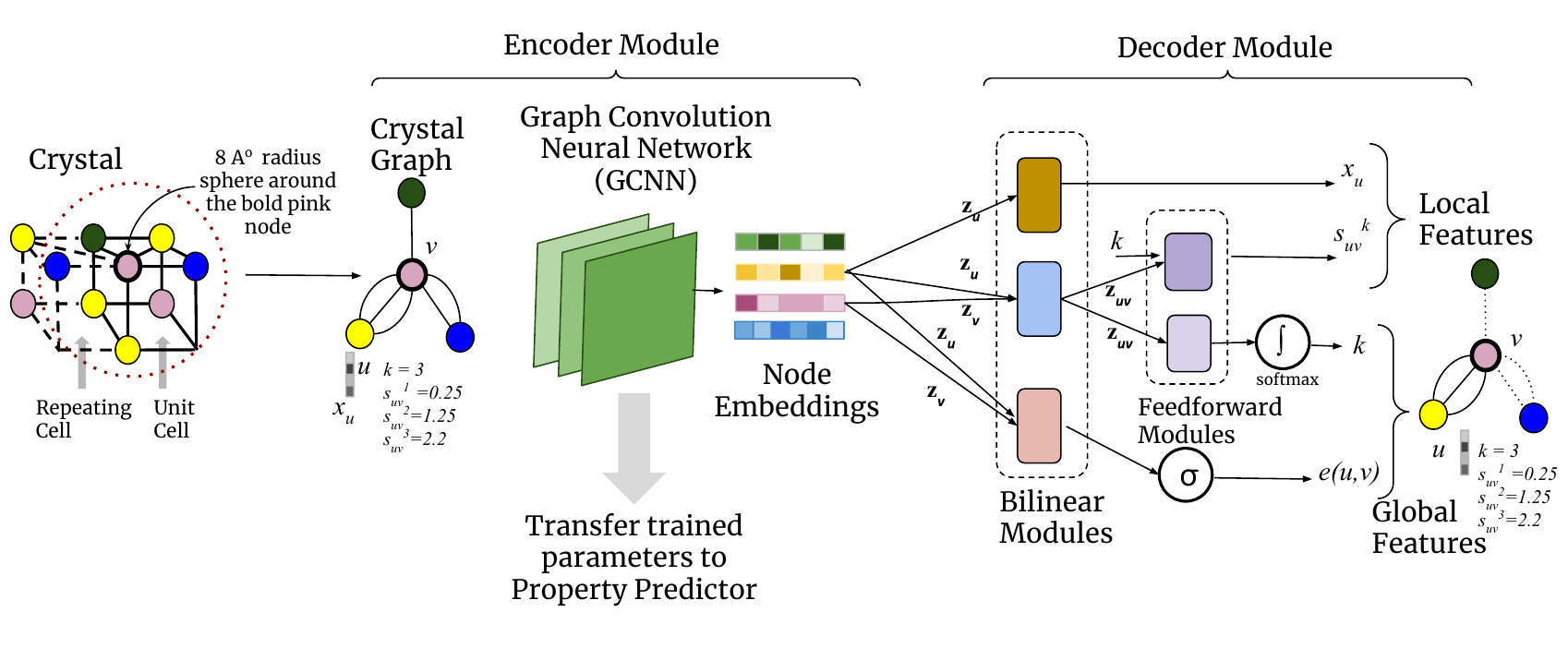}}
	\caption{\rev{The architecture of the Crystal Auto Encoder (CrysAE) module which comprises multilayer graph convolution network as the encoder and a set of decoding modules for reconstructing different local and global features. A periodic crystal is represented as an undirected weighted multi-graph by connecting atoms within a cutoff radius across repeating unit cells. The encoder uses a graph convolutional neural network (GCNN) to transform atomic and bond features into latent node embeddings. The decoder reconstructs local and global graph structure, including edge existence, edge multiplicity, and bond lengths, using bilinear and feedforward modules. The learned encoder representations are transferable to downstream crystal property prediction tasks.}}
	\label{fig:Enc_architecture}
\end{figure*}\\
\xhdr{Encoder}
We extend the crystal graph encoder proposed by Xie et.al.~\cite{xie2018crystal} to encode the chemical and structural information of a crystal graph $\Gcal$. Specifically, we encode $L$-hop neighbouring information of each node as: 
\small{
\begin{align}\label{eq:enc}
&\bm{h}_{(u,v)_k}^{l} = \bm{z}_u^{l} \oplus \bm{z}_v^{l} \oplus {s^k_{(u,v)}}\\
&\bm{z}_u^{l+1}  = \bm{z}_u^{l} + \sum\limits_{v,k} \sigma ( {\bm{h}^l_{(u,v)}}_k \bm{W}_c^{(l)} +  \bm{b}_c^{(l)}) \odot
g( {\bm{h}^l_{(u,v)}}_k \bm{W}_s^{(l)}+\bm{b}_s^{(l)}) \notag
\end{align}
}
where $\bm{z}_u^l$ denotes the embedding of node $u$ after $l$ hop neighbor information aggregation. The embedding of a node $u$ is initialized to a transformed node feature vector, i.e. it is a function of the atom $u$'s chemical features as $\bm{z}_u^0 := \bm{x}_u \bm{W}_x$ where $\bm{W}_x$ is the trainable parameter of the transformation network and $ \bm{x}_u $ is the input node feature vector. ${s^k_{(u,v)}} \in \bm{\mathcal{F}}_u $ represents the length of the $k^{th}$ edge between nodes $u$ and $v$. The $\oplus$ operator denotes concatenation and $\odot$ denotes element-wise multiplication. $\bm{W}_c^{(l)},\bm{W}_s^{(l)},\bm{b}_c^{(l)},\bm{b}_s^{(l)}$ are the convolution weight matrix, self weight matrix, convolution bias, self bias of $l^{th}$ hop convolution, respectively. $\sigma$ is a \rev{non-linear transformation (sigmoid) function} and it is used to generate a squeezed real value in [0,1] indicating the edge importance and $g$ is a feed forward network. 
After neighborhood aggregation we accumulate local information at each node which can be represented as $\bm{z}_u :=\bm{z}_u^L$. Subsequently we generate a graph level global information $\bm{\mathcal{Z}} = \{\bm{z}_1,..., \bm{z}_{|\bm{\mathcal{V}}|}\}$.
We do not aggregate the node embeddings further to prevent information loss in autoencoder.
We denote the set of trainable parameters for this encoder as $\bm{\theta}$ for future reference.
\\\\
\xhdr{Decoder}
We design an effective decoder that helps the encoder to transform the desired information in the representation vector space of $\bm{\mathcal{Z}}$. The decoder plays an inevitable role in learning the local and global structure as well as chemical features which are extremely useful. As mentioned earlier the global chemical features i.e. the crystal properties are a function of the local chemical environment and the overall conformation of the repeating  crystal cell structure; hence, we carefully design the decoder which can reconstruct two important features that induce the local chemical environment. They are (a) the node features i.e chemical properties of individual atoms and (b) local connectivity i.e the relative position of the nodes with respect to their neighbors. 
Precisely we reconstruct these information as below:
\begin{align}
&z_{uv} = \bm{z}_u^T \bm{W}_f \bm{z}_v + b_f \\
&\hat{s_{(u,v)}^k} = \begin{cases}
\gamma_s(z_{uv}\odot k) & \text{if } \gamma_s(.) > 0 \\
0 & \text{otherwise}
\end{cases}
\label{eq:l1}\\
&\bm{\mathcal{\hat X}}_{u} = \bm{W}_x^{T}\bm{z}_u + \bm{b}_x \label{eq:l2}
\end{align}
Eqs.~\ref{eq:l1} and~\ref{eq:l2} correspond to reconstructing the node property or atom's chemical property and a node's position relative to it's neighbors as we intend to achieve in (a) and (b) respectively. $z_{uv}$ is a combined transformed embedding of nodes $u$ and $v$ and $\gamma_s$ is a feed forward network which generates a real number corresponding to the length of the bonds. \\
Further we reconstruct the global structure i.e (c) the connectivity and periodicity of the crystal structures as below
\begin{align}
&(u,v) \sim p(e=(u,v)) = \sigma(\bm{z}_u^{T}\bm{W}_e\bm{z}_v + b_e ) \label{eq:g1}\\
&k_{(u,v)} = \arg \max_k \frac{e^{\gamma_k(z_{uv}, k)}}{\sum_k e^{\gamma_k(\bm{z}_{uv}, k)} } 
\label{eq:g2}
\end{align}
Here, $\bm{W}_e, b_e$ are trainable weight and bias associated with the bilinear edge reconstruction module, respectively. $\sigma$ is a squashing factor which provides a  value between $[0,1]$ denoting the edge probability. Similarly $\bm{W}_f, b_f$ are the trainable weight and bias parameters associated with the intermediate bi-linear transformation module, respectively. $\gamma_k$ represents a feed forward neural network that generates a $K$ length logit vector. We use a softmax to determine the exact number of edges present in the graph. Please note that though Eqs.~\ref{eq:g2},~\ref{eq:l1} correspond to global and local information respectively, they are heavily dependent upon each other, i.e the number of bonds and bond length both depend on the two end nodes (atoms) information. Hence, we design a coupled embedding $z_{uv}$ which is shared by both the modules. We denote the set of parameters in the decoder as $\bm{\phi}$.
\\\\
\xhdr{Training of auto-encoder} We learn the trainable parameters of both encoder and decoder by minimizing the reconstruction loss of different global and local structural and chemical features defined in Eqs.~\ref{eq:l1}-\ref{eq:g2}.
We minimize the cross-entropy loss of the predicted global features and node features along with mean squared loss of the edge weight or bond length in the  following objective: 
\begin{align}
&\EE_{\Gcal \sim \bm{\mathcal{D}}} -\sum_{(u,v) \in \Ecal} \big[\text{log}\ p(e=(u,v)) + \text{log} \ p(k_{(u,v)})\big]  \nonumber \\
& -\sum_{u \in \Vcal}\text{log}\ p(\bm{\mathcal{\hat X}}_{u}) + \sum_{(u,v) \in \Ecal} \sum_{k \in [1,\dots,K]} (s^k_{(u,v)} - \hat{s^k_{(u,v)}})^2 
\label{eq:aetrain}
\end{align}
where $p(.)$ denotes the probability of any event.
Thus by minimizing the reconstruction loss we not only fine tune parameters of decoder but efficiently train the encoder to generate a rich $\bm{\mathcal{Z}}$ which facilitates decoder operations.
\subsection{Property predictor: CrysXPP}
Next, we design a property predictor specific to a property that can take the advantage of the structural information that is learned by the encoder as described above. We generate a graph level representation using the same graph encoder module as described in Eq.\ref{eq:enc}, thus in a way transferring the rich encoded knowledge to the property predictor. Next, we use \rev{a symmetric aggregation function (Mean Pooling)} to generate a single vector as graph representation $\bm{\mathcal{Z}}_g$. Thus the obtained representation of the graph is  invariant of the node orderings. Then the obtained representation is fed to a multilayer perceptron which predicts the value of the properties. More formally the property predictor can be characterized as:
\begin{align}
&\bm{\mathcal{Z}}_g  = \Lambda(\bm{z}_1\dots, \bm{z}_{|\Vcal|}) \label{eq:pool}\\
& \Pcal = \Mcal_{\bm{\psi}}(\bm{\mathcal{Z}}_g) \label{eq:prop}
\end{align}
Here, $\Lambda$ is the aggregation function which is symmetric. $\Mcal$ denotes a multilayer perceptron that has a trainable  parameter set $\bm{\psi}$. 
\\\\
\xhdr{Feature Selection}
\rev{The node features are first passed through a feature 
selector which is a trainable weight vector that selects a weighted subset of important node level features $\bm{\mathcal{X}}^\prime$ for a given property of interest $\Pcal$.} $\bm{\mathcal{X}}^\prime$ forms input to the encoder.
\begin{align}
&\bm{\mathcal{X}}^\prime = \Scal_{\bm{\zeta}}(\bm{\mathcal{X}}) ;
\bm{\mathcal{Z}}_g = \Lambda(\text{Encoder}_{\bm{\theta^\prime}}(\bm{\mathcal{V}}, \bm{\mathcal{E}}, \bm{\mathcal{X}}^\prime, \bm{\mathcal{F}})) \nonumber 
\end{align}
In the above set of characteristic equations, $\Scal$ is the feature selector and $\bm{\zeta}$ is its trainable weight. We will show in section \ref{sec:explain} how the weights chosen by the feature selection layer help us to explain the role of a node feature in the manifestation of a particular crystal property. 
\\\\
\xhdr{Training of \npjmodel{}} We train the property predictor after the autoencoder. We initialize the trainable parameter  $\bm{\theta}^\prime:=\bm{\theta}$ where $\bm{\theta}$ is trained in the autoencoding module. Thus we first transfer the trained information such that the property predictor benefits from the inductive bias already learned by training the autoencoder.\\ 
\rev{We use a LASSO~\cite{li2006lasso} regression to impose sparsity on the feature selector layer. Intuitively, if some atomic features ($\bm{\mathcal{X}}^\prime$) are crucial to predict a chemical property of the crystal,  the corresponding feature selector value will be high and conversely, if some feature is not so important, the corresponding feature selector value will be negligible.}  Hence, along with the property prediction loss we also consider the LASSO regression loss as formally represented below:
\begin{align}\label{eq:proppred}
\min_{\bm{\zeta}, \bm{\theta^\prime}, \bm{\psi}}  (\hat{\Pcal}-\Pcal)^2 + \lambda_{1} * \vert{\bm{\zeta}} \vert_{L_1} 
\end{align}
where $\bm{\zeta}$ denotes the trainable parameters of feature selector $\Scal$ and  $\lambda_{1}$ is a hyper parameter which controls the degree of the regularization imposed. Before reporting the results, we  briefly discuss about the dataset and the baselines used for comparison.

\section{Results}
\label{sec:results}
\subsection{Dataset}
Following prior work~\cite{xie2018crystal}, we have used the Materials Project database~\cite{MP} for our experiments which consists of $\sim$36,835 crystalline materials and is diverse in structure having materials with 87 different types of atoms, seven different lattice systems and 216 space groups. The unit cell of any crystal can have a maximum of 200 atoms. We consider nine properties for each atom which were used to construct the feature vector of each node ~\cite{xie2018crystal}. The details of the properties are given in Table~\ref{tab:feature_desc}. We convert them to categorical values if they are already not in that form. The dataset also provides DFT calculated target property values for the crystal structures. Experiments were done on a smaller training set than the original baseline papers.
\subsection{Comparison with similar baseline algorithms}
We compare the performance of \npjmodel{} with four state-of-the-art algorithms for crystal property prediction. These selected competing methods are  varied in terms of input data processing and working paradigms as described below:
\begin{enumerate}
	\item \textbf{CGCNN}~\cite{xie2018crystal}: This  method generates crystal graphs from inorganic crystal materials and builds a graph convolution based supervised model for predicting various properties of the crystals.
	
	\item \textbf{MT-CGCNN}~\cite{sanyal2018mt} : This model uses the graph convolution based encoding as proposed in the previous model. Moreover, it incorporates multitask learning to jointly predict multiple properties of a single material.
	
	\item \textbf{MEGNET}~\cite{chen2019graph}: Here authors improved the CGCNN model further by introducing global state attributes including  temperature, pressure, entropy etc for quantitative structure-state-property relationship prediction in materials. Doing so they found that the crystal embeddings in MEGNet model encode periodic chemical trends. Further to address the issue of data limitation the 
	embeddings from a MEGNet model trained on formation energies is transferred and used to improve the accuracy of ML models for the band gap and elastic moduli.
	
	\item \textbf{ELEMNET}~\cite{jha2019enhancing}: This work does not specifically consider any structural properties of the crystal graph, rather it considers only the compositional atoms. It uses deep feed-forward networks to implicitly capture the effect of atoms on each other. It uses transfer learning to mitigate the error bias of DFT tagged data.
	
	\item \textbf{GATGNN}~\cite{louis2020graph}: In this work authors have incorporated a graph neural network with multiple graph-attention layers (GAT) and a
    global attention layer, which can learn efficiently the importance of different complex bonds shared among the atoms within each atom’s local neighborhood.
\end{enumerate}
For all the baselines we have used the hyper parameters as mentioned in the original papers.
\subsection{Evaluation criteria}
We predict seven different properties of crystals in our experiments. Out of these, four are crystal state properties, namely, (a)  Formation Energy, (b) Band Gap, (c) Fermi Energy, (d) Magnetic Moment, and three are elastic properties, namely, (e) Bulk Moduli, (f) Shear Moduli, and (g) Poisson Ratio. All of these properties significantly depend on the details of  the crystal structure except Magnetic Moment which is more dependent on the atomic/node specifications as the magnetic moment arises from the unpaired \textit{d or f} electrons in an atom. Also the size of the moment depends on the local environments ~\cite{Lars,S1}. Moreover, we have very little DFT tagged data for Magnetic Moment and Band Gap.\\
We  focus on three different evaluation criteria as described below:
\begin{enumerate}
	\item \textit{How effective is the property predictor?} We evaluate its performance, particularly when trained on a limited amount of DFT-tagged data.
    \item \textit{How robust is the structural encoding?} We examine whether the structural encoding helps mitigate noise introduced by DFT-calculated properties.
    \item \textit{How effective is the explanation?} We validate the generated explanations against established domain knowledge.
\end{enumerate}
\begin{table*}[ht!]
	\centering
	\small
	\setlength{\tabcolsep}{4pt}
	\resizebox{1.0\textwidth}{!}{
		\begin{tabular}{c c c c c c c c c}
		\toprule
		& \textbf{Property}& \textbf{Unit} & \textbf{CGCNN}  & \textbf{MTCGCNN } & \textbf{MEGNet} & \textbf{GATGNN} &  \textbf{Elemnet} & \textbf{\npjmodel{}} \\
		\hline
		\midrule
		\multirow{4}{*}{\rotatebox[origin=c]{90}{\shortstack{State \\ Properties}}}
		& Formation Energy   & eV/atom & 0.127 & 0.112 (0.147) & 0.142 & 0.164 & 0.098\textbf{*}  & \textbf{0.086} \\
		& Band Gap  & eV & 0.503 & 0.497 (0.518) & 0.498 & 0.489\textbf{*} & 0.491 & \textbf{0.467} \\ 
		& Fermi Energy  & eV & 0.528 & 0.503\textbf{*} (0.601) & 0.533 & 0.533 &  0.588 & \textbf{0.471} \\
		& Magnetic Moment  & $\mu_B$ & 1.21 & 1.16 (1.22) & 1.19 & 1.09 & \textbf{0.96} & 1.03\textbf{*}  \\
		\hline\hline
		\multirow{3}{*}{\rotatebox[origin=c]{90}{\shortstack{Elastic \\ Properties}}}
		& Bulk Moduli  & log(GPa) & 0.09 & 0.09 (0.09) & 0.105 &0.088\textbf{*} & 0.1057 & \textbf{0.08} \\
		& Shear Moduli  & log(GPa) & 0.125\textbf{*} & 0.120 (0.078) & 0.187 & 0.123 &  0.148 & \textbf{0.105}  \\
		& Poisson Ratio  &  - & 0.04 & 0.037\textbf{*} (0.039)& 0.041 & 0.039 &  0.039 & \textbf{0.035} \\
		\bottomrule
	\end{tabular}
    }
	\caption{Summary of the prediction performance [Mean Absolute Error (MAE)] of different properties trained on 20\%  data and evaluated on $80\%$ of the data. The best performance is highlighted in bold and second best with \textbf{*}. We report MAE jointly training most correlated property (average on all property pairs) for MTCGCNN.}
	
	\label{tab:mt_mape}

\end{table*}
\subsection{Effectiveness of Property Predictor}
\begin{figure*}[h!]
	\centering
	\includegraphics[width=\columnwidth]{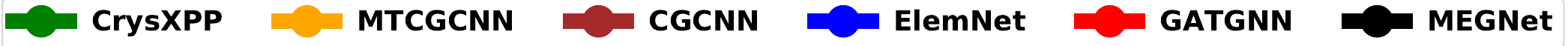}
	\\
	\vspace*{-1mm}
	\subfloat[Formation Energy]{
		\includegraphics[width=0.3\columnwidth, height=40mm]{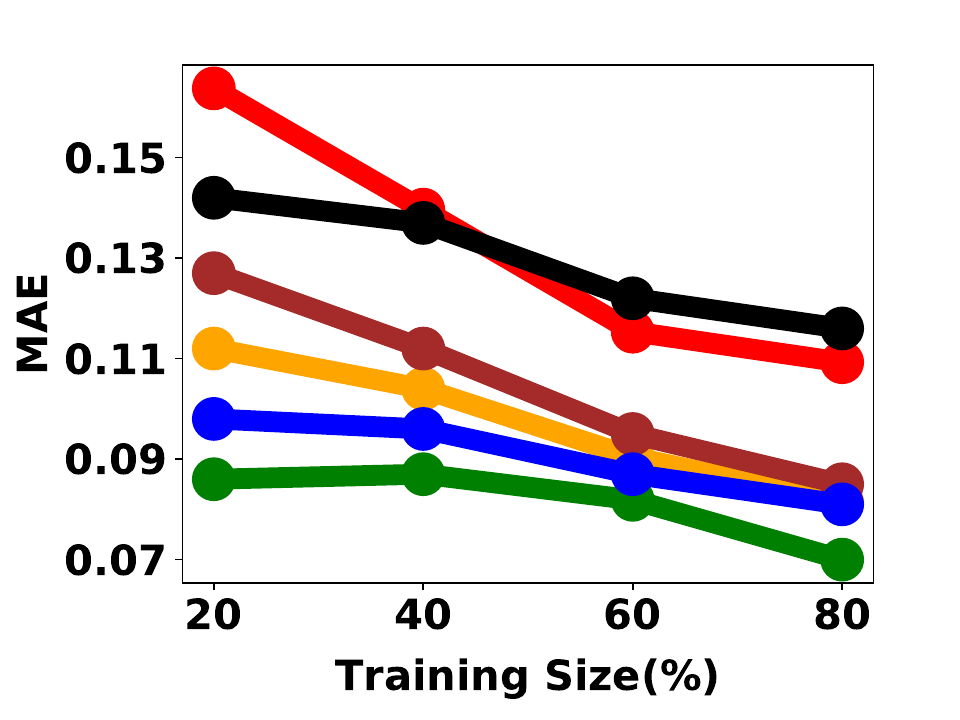}}
	\subfloat[Band Gap]{
		\includegraphics[width=0.3\columnwidth, height=40mm]{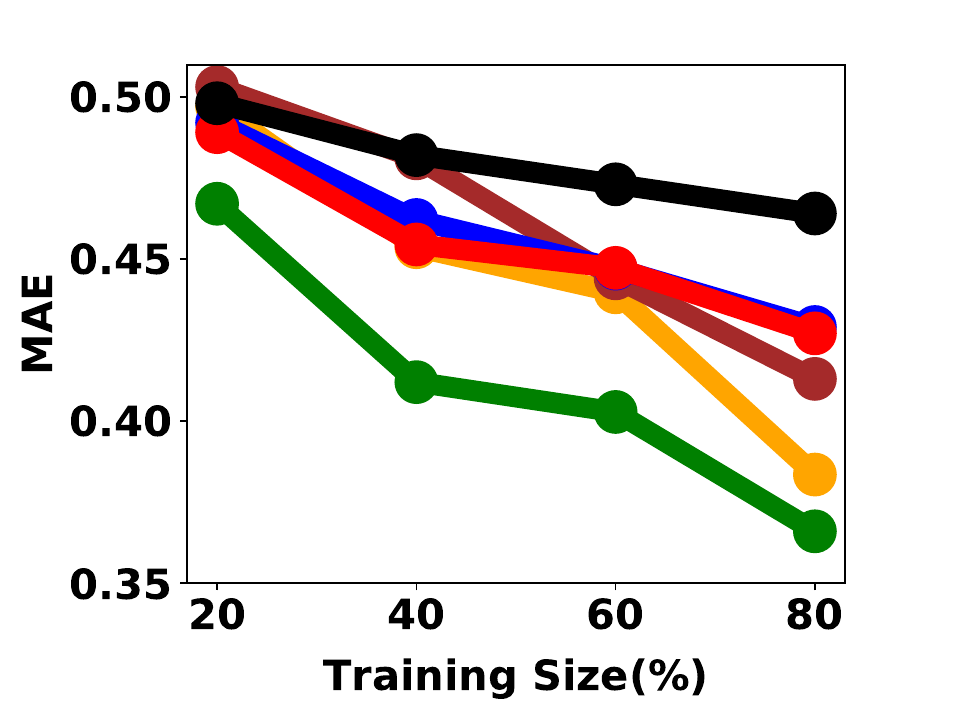}}
	\subfloat[Fermi Energy]{
		\includegraphics[width=0.3\columnwidth, height=40mm]{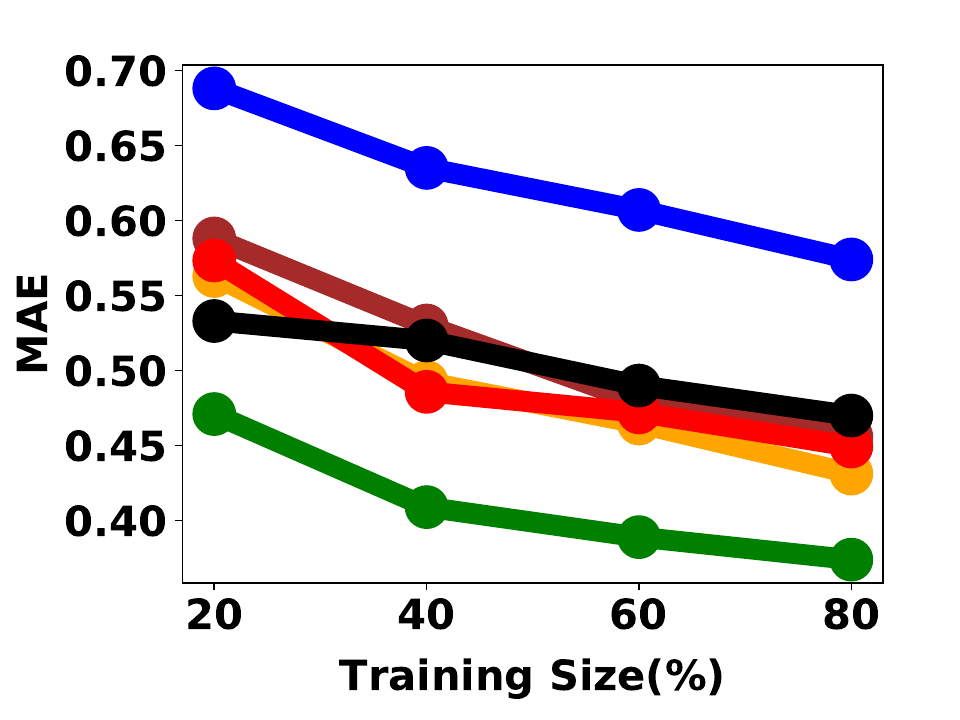}}\\
	\subfloat[Magnetic Moment]{
		\includegraphics[width=0.3\columnwidth, height=40mm]{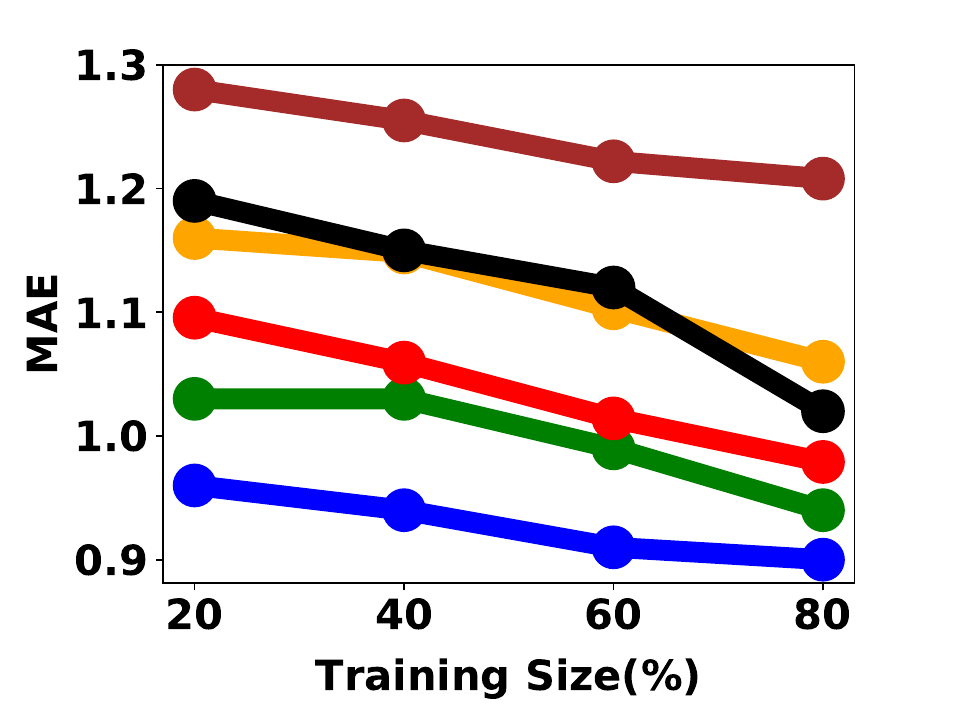}}
	\subfloat[Shear Moduli]{
		\includegraphics[width=0.3\columnwidth, height=40mm]{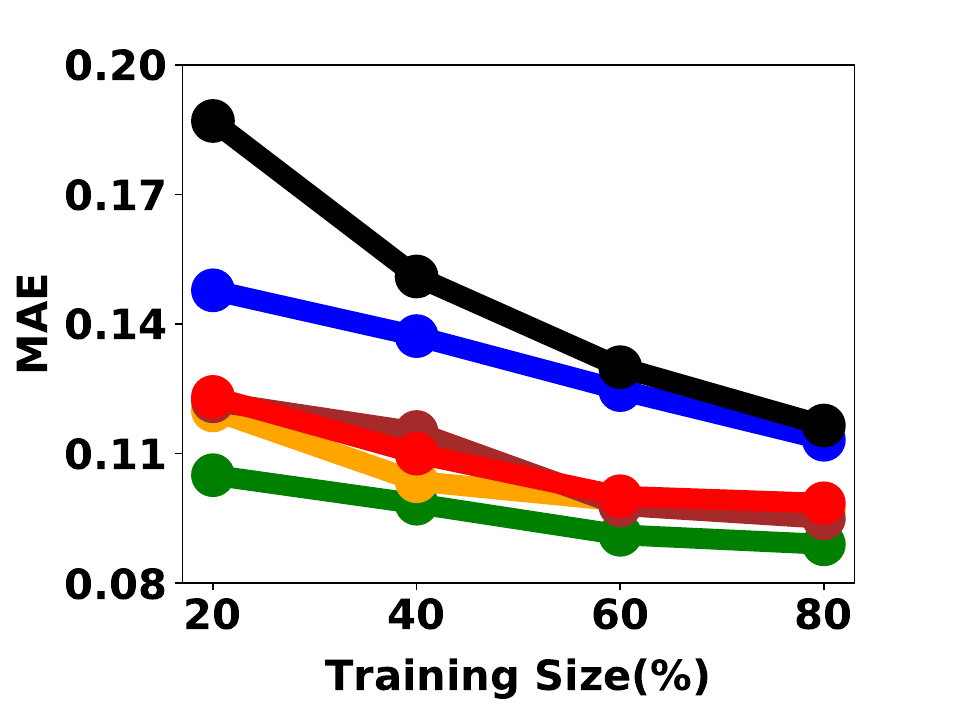}}
	\subfloat[Bulk Moduli]{
		\includegraphics[width=0.3\columnwidth, height=40mm]{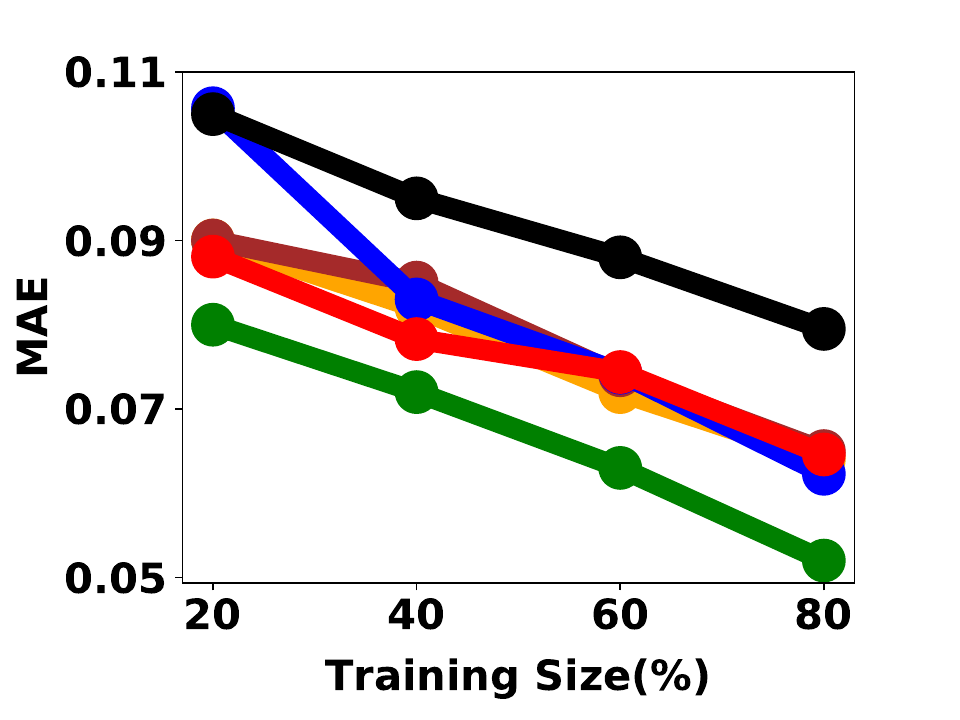}}\\
	\subfloat[Poisson Ratio]{
		\includegraphics[width=0.3\columnwidth, height=40mm]{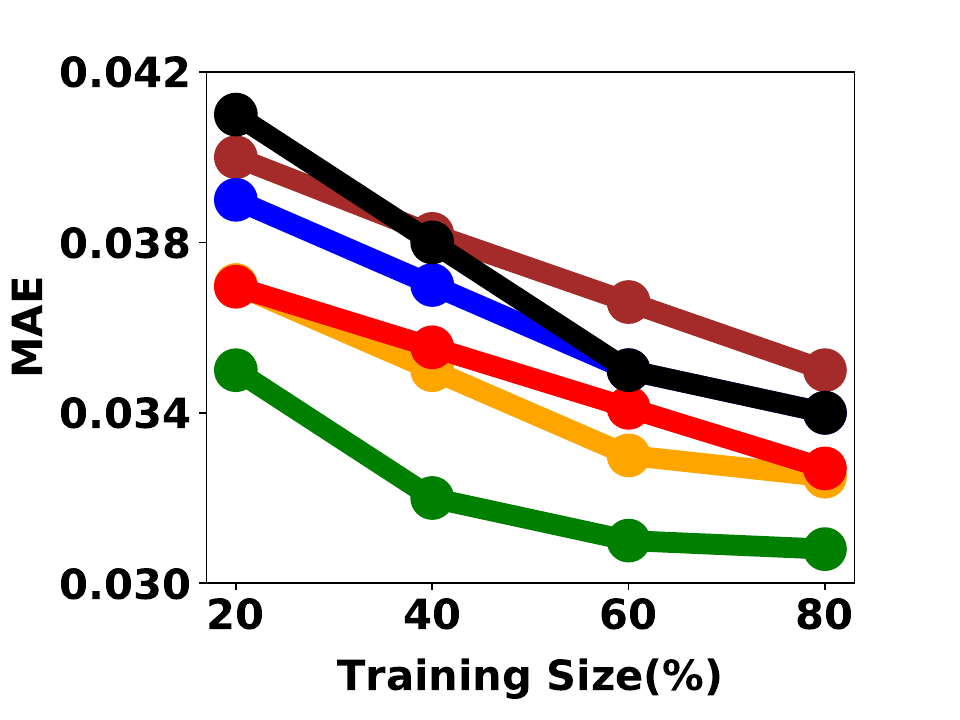}}
	
	\caption{\rev{Variation of Mean Absolute Error (MAE) with the increase in training instances from 20\% to 80\%. We observe a monotonic decrease of MAE between predicted and  DFT calculated vales for most of the models where \npjmodel{} yields consistently smaller MAE and maintains the leadership position for all the properties expect Magnetic Moment. This shows the robustness of our model to be able to perform consistently across a diverse set of properties with varied training instances. The MAE margin between \npjmodel{} and closest competitor (which is variable across properties), however, reduces  as training size increases.}}
	\label{fig:results_on_more_train_data}
\end{figure*}

We first train the autoencoder with all untagged crystal graph present in the dataset, which captures all the structural information of the crystal graphs. Next for a given property of interest, we train the property predictor with 20\% of the available DFT tagged data and test on the rest. We report the 10 fold cross validation results.\\\\
\xhdr{Metric} We report Mean Absolute Error (MAE) to compare the performance of the participating methods. MAE is defined as $ \frac{1}{|\bm{\mathcal{D}}|}\sum_{\Gcal \in \bm{\mathcal{D}}}\big|{\Pcal_\Gcal - \hat \Pcal_\Gcal} \big|$, where $\Pcal_\Gcal$ is the property value calculated by DFT and $\hat \Pcal_\Gcal$ is the predicted value of a graph $\Gcal$.\\\\
\xhdr{Results}
In Table~\ref{tab:mt_mape} we report the MAE for \npjmodel{} as well as other alternatives on seven property values. We observe that \npjmodel{} outperforms every baseline across all the properties except magnetic moment. For MTCGCNN we report two values: the MAE obtained while jointly predicting the most correlated property, and the average MAE across all possible combinations (in bracket). It is interesting to note that its performance significantly degrades if the other property is not correlated with the current property of interest. A careful inspection reveals that for elastic properties, graph neural network based methods perform better than that of Elemnet. Elemnet only considers the composition of the crystal and ignores the global structural information, whereas these properties heavily depend on the crystal structure. In contrast, for Magnetic Moment the local information is important and hence, Elemnet performs the best and \npjmodel{} is the second best method. For the rest of the crystal state based properties, there is no consistent second best method. However, \npjmodel{} is a clear winner with a considerable margin which is due to the fact that the property predictor benefits from the structural knowledge transferred from the autoencoder.
\\\\
\xhdr{Behaviour with increase in tagged data}
Further, we check the robustness of \npjmodel{}, by increasing  the percentage of tagged training data for property prediction. We  report the behavior of \npjmodel{} as well as other baselines in Fig.\ref{fig:results_on_more_train_data} for all the properties. We observe a monotonic decrease of MAE between predicted and  DFT calculated vales for most of the models where \npjmodel{} yields consistently smaller MAE and maintains the leadership position for all the properties expect Magnetic Moment.
This shows the robustness of our model to be able to perform consistently across a diverse set of properties with varied training instances. The MAE margin between \npjmodel{} and closest competitor (which is variable across properties), however, reduces  as training size increases. For Magnetic Moment, the local chemical information is more vital, hence ElemNet, which concentrate more on local chemical information, shows the best performance.
\subsection{ Removal of DFT error bias }
An important aspect of the crystal property prediction is that since we rely only on DFT data for training, we would be limited by the inaccuracies of DFT. In this section, we investigate with a system where we further fine tune the model with a small amount of available experimental data and check whether the system can remove the error propagated due to DFT.
\\\\
\xhdr{Calculation setup}
We consider a property predictor initially trained on crystals with DFT-derived property values. We then fine tune it with a limited amount of experimental data; we perform it for two different properties, namely, Band Gap and Formation Energy. For Formation Energy, we use 1,500 instances available at~\cite{kirklin2015open} and use different percentages of the data to fine-tune the model parameters. For Band Gap, we collect 20 experimental instances out of which we randomly pick 10 instances to fine-tune the parameters and report the prediction value for the rest. 
\begin{table}[]
    \centering
    \small
    \setlength{\tabcolsep}{3pt}
    \resizebox{1.0\textwidth}{!}{
    \begin{tabular}{l c c c c c c}
    \toprule
	\textbf{Experiment Settings} & \textbf{CGCNN} & \textbf{MTCGCNN} & \textbf{GATGNN} & \textbf{MEGNet} & \textbf{ElemNet} & \textbf{\npjmodel{}}\\
	\hline
	\midrule
	\textbf{\vtop{\hbox{\strut Train on 20\% DFT }\hbox{\strut Test on Full Experimental Data}}}  & 0.24 & 0.74 & 0.30 & 0.28 & \textbf{0.215}  & 0.22 \\
	\hline
	\textbf{\vtop{\hbox{\strut Train on 20\% DFT, 20 \% Experimental Data }\hbox{\strut Test on 80 \% Experimental Data}}} & 0.21 & 0.24 & 0.23 & 0.23 & 0.16\textbf{*} & \textbf{0.15} (0.206)  \\
	\hline
	\textbf{\vtop{\hbox{\strut Train on 80\% DFT, 20 \% Experimental Data }\hbox{\strut Test on 80 \% Experimental Data}}} & 0.16 & 0.22 & 0.19 & 0.18 & 0.1344\textbf{*} & \textbf{0.1319} (0.195)  \\
	\hline
	\textbf{\vtop{\hbox{\strut Train on 80\% DFT, 80 \% Experimental Data }\hbox{\strut Test on 20 \% Experimental Data}}} & 0.12 & 0.15 & 0.13 & 0.125 & 0.0905\textbf{*} & \textbf{0.0892} (0.174) \\
	\hline
    \bottomrule
    \end{tabular}
	}
    \caption{Mean Absolute Error (MAE) of predicting experimental values after fine tuning different methods with different percentages of experimental data {for formation energy}. MAE of the experiment where we replace the experimental data with the same amount of DFT data to train \npjmodel{}, is provided in the bracket. The closest prediction is marked in bold and second best with \textbf{*}.
    }
    \label{tab:fe_dft}
\end{table} 
\begin{table}[ht]
	\centering
	\small
	\setlength{\tabcolsep}{4pt}
	\scalebox{0.9}{
		\begin{tabular}{c c c c c c c c c c}
			\toprule
			\textbf{Materials} &  \textbf{Exp} & \textbf{DFT} & \textbf{\npjmodel{}} & \textbf{CGCNN} & \textbf{MTCGCNN} & \textbf{GATGNN} & \textbf{MEGNet} & \textbf{ElemNet}  \\
			\hline
			\midrule
			GaSb  & 0.72 & 0.36 & \textbf{0.77} (0.9\textbf{*}) & 1.01 (4.27)  & 0.26 (0.06) & 3.78(1.58)& 2.26(1.31) & 1.09 (0.33) \\
			\hline
			GaP  & 2.26 & 1.69 &  \textbf{2.10} (1.86) & 1.95\textbf{*} (1.49) & 0.73 (0.64) & 2.77(0.08)& 1.21(2.51) & 2.80 (1.29) \\
			\hline
			GaAs  & 1.42 & 0.18 & 1.54\textbf{*} (1.56) & 2.51 (\textbf{1.42}) & 1.83 (1.90) & 2.73(3.50)& 0.77(0.98) &0.83 (0.76)\\
			\hline
			InN  & 1.97 & 0.47 & \textbf{1.92} (1.85\textbf{*}) & 1.30 (2.90) & 1.77 (2.30) & 2.79(0.08) &1.33(2.16) &1.64 (1.43) \\
			\hline
			GaN  & 3.2 & 1.73 & 2.11 (1.47) & 3.51\textbf{*} (1.55) & 0.28 (0.16)&  2.27(0.56)& 1.59(2.66) &\textbf{3.69} (1.44)\\
			\hline
			NiO  & 4.3 & 2.214 & \textbf{2.45} (2.08) & 0.96 (1.12) & 0.08 (0.05)& 2.12(1.36) & 2.01(2.71) &2.31\textbf{*} (1.88)\\
			\hline
			Si  & 1.12 & 0.85 & \textbf{1.08} (0.95\textbf{*}) & 1.56 (1.64) & 0.39 (0.22) & 3.60(0.31)& 1.86(1.69) &0.33 (0.17)\\
			\hline
			ZnO  & 3.37 & 1.05 & 3.42\textbf{*} (2.1) & \textbf{3.32} (1.45) & 0.83 (0.56) & 2.74(1.36)& 2.09(2.53) &2.55 (2.01)\\
			\hline
           FeO  & 2.4 & 0 & \textbf{2.25} (1.72) & 2.16\textbf{*} (2.85) & 1.12 (0.96) & 1.92(1.02)& 2.93(2.81) &1.44 (1.27) \\
			\hline
			MnO  & 4 & 0.20 & \textbf{2.31} (1.81) & 1.51 (1.22) & 1.04 (0.77) & 2.44(2.35)& 1.73(2.11) &1.98\textbf{*} (1.44) \\
			\hline
			\bottomrule
		\end{tabular}
		
	}
	\caption{ Experiment (Exp) and predicted value for Band Gap for 10 crystals calculated by DFT and other machine learning models after fine-tuned by experimental data. The predicted value without fine tuning by experimental data is provided in the bracket.The closest prediction is marked in bold and the second best with  \textbf{*}. \npjmodel{} predicts closest to the ground truth after fine tuning with experimental data.}
	
	\label{tab:band_gap}
\end{table} 
\\
\xhdr{Results (Formation Energy)}
We report the mean absolute error (MAE) of Formation Energy in Table~\ref{tab:fe_dft} achieved by different methods. The DFT prediction of the Formation Energy on the 1,500 crystals 
has an MAE of 0.21  with respect to experimental data and by training our model with DFT data we are performing close to the performance achieved by DFT. The results have a consistent trend for all the methods, whereby we observe that increasing the amount of training data, even if that is error-prone DFT data, helps in minimizing MAE. 
\npjmodel{} performs consistently better by a large margin than CGCNN and MTCGCNN, which takes the graph structure as an explicit input. However, it is interesting to observe that 
 ElemNet performs very close to \npjmodel{} as 
Formation Energy depends more on the composition than that on the explicit connection of atoms. Further, we conduct an experiment where we replace the experimental data with the same amount of DFT data to train our model. We then evaluate the performance of the model using experimental data as test data and find an inferior performance. We report the results in Table~\ref{tab:fe} (last column (in bracket)).
\\\\
\xhdr{Results (Band Gap)}
In Table~\ref{tab:band_gap} we report the experimental value of Band Gap for 10 test instances along with the predicted values by DFT and other machine learning methods. The error margin of DFT with the actual experimental values is quite high.
It is interesting to see that other than a few, DFT prediction is far from experimental data and in most of the cases, it is underestimating the experimental values.
 After fine-tuning DFT trained machine learning models with experimental data, the prediction becomes closer to the experimental value. However, \npjmodel{} performs closest to the experimental result in almost all the cases in comparison to other alternatives.  ElemNet, although second on the average when trained only on DFT (row 2 of Table \ref{tab:mt_mape}), cannot consistently maintain that position, whereas CGCNN performs better.  
 We have also provided results (in bracket) when we do not do any fine tuning. It can be seen even in such a scenario in many of the cases the performance is better than DFT. Further the power of \npjmodel{} in quickly mitigating the bias of DFT  when fine-tuned on minuscule data shows  the usefulness of modeling explicit structural information.
\begin{table}[]
    \centering
    \small
    \setlength{\tabcolsep}{4pt}
    \scalebox{0.9}{
    \begin{tabular}{c c c c c c} 
        \hline
        \multirow{2}{*}{\shortstack[t]{\textbf{Property}\\ \textbf{Name}}} &
        \multirow{2}{*}{\shortstack[t]{\textbf{Ablation }\\ \textbf{Settings}}} &
        \multicolumn{4}{c}{\textbf{Train-Test Split}}  \\ 
        \cline{3-6} & &  {\textbf{20\% - 80\%}} & {\textbf{40\% - 60\%}} & {\textbf{60\% - 40\%}} &{\textbf{80\% - 20\%}}\\
        \hline
        \hline
        
        \multirow{4}{*}{\shortstack[t]{\textbf{Formation}\\ \textbf{Energy}}}
        & Without Global + Local effect  & 0.124 & 0.113 & 0.092 & 0.086  \\
        \cline{2-6}
        & Global effect & 0.112 & 0.092 & 0.085 & 0.077 \\
        \cline{2-6}
        &  Local effect & \textbf{0.079} & \textbf{0.067} & \textbf{0.063} & \textbf{0.061} \\
        \cline{2-6}
        &  \npjmodel{} & 0.086 & 0.082 & 0.076 & 0.067  \\
       \cline{1-6}
       \cline{1-6}
       
       \multirow{4}{*}{\shortstack[t]{\textbf{Band}\\ \textbf{Gap}}}
       & Without Global + Local effect  & 0.502 & 0.482 & 0.452 & 0.408  \\
        \cline{2-6}
        & Global effect & 0.479 & 0.425 & 0.393 & 0.382 \\
        \cline{2-6}
        &  Local effect & 0.471 & 0.419 & 0.387 & 0.375 \\
        \cline{2-6}
        &  \npjmodel{} & \textbf{0.467} & \textbf{0.402} & \textbf{0.383} &  \textbf{0.366} \\
       \cline{1-6}
       \cline{1-6}
       
       \multirow{4}{*}{\shortstack[t]{\textbf{Fermi}\\ \textbf{Energy}}} 
       & Without Global + Local effect  & 0.513 & 0.481 & 0.477 & 0.443  \\
        \cline{2-6}
        & Global effect & 0.495 & 0.476 & 0.441 & 0.437 \\
        \cline{2-6}
        &  Local effect & 0.488 & 0.472 & 0.436 & 0.428 \\
        \cline{2-6}
        &  \npjmodel{} & \textbf{0.471} & \textbf{0.409} & \textbf{0.389} &  \textbf{0.374} \\
       \cline{1-6}
       \cline{1-6}
       
       \multirow{4}{*}{\shortstack[t]{\textbf{Magnetic}\\ \textbf{Moment}}}
       & Without Global + Local effect  & 1.082 & 1.038 & 1.023 & 1.027  \\
        \cline{2-6}
        & Global effect & 1.072 & 1.066 & 1.027 & 1.019 \\
        \cline{2-6}
        &  Local effect & 1.068 & 1.052 & 1.022 & 1.013 \\
        \cline{2-6}
        &  \npjmodel{} & \textbf{1.033} & \textbf{1.024} & \textbf{0.997} & \textbf{0.943}  \\
       \cline{1-6}
       \cline{1-6}
       
       \multirow{4}{*}{\shortstack[t]{\textbf{Bulk}\\ \textbf{Moduli}}} 
       & Without Global + Local effect  & 0.091 & 0.088 & 0.081 &  0.075 \\
        \cline{2-6}
        & Global effect & 0.088 & 0.081 & 0.075 & 0.068 \\
        \cline{2-6}
        &  Local effect & 0.087 & 0.077 & 0.072 & 0.063 \\
        \cline{2-6}
        &  \npjmodel{} & \textbf{0.080} & \textbf{0.072} & \textbf{0.063} & \textbf{0.052}  \\
       \cline{1-6}
       \cline{1-6}
       
       \multirow{4}{*}{\shortstack[t]{\textbf{Shear}\\ \textbf{Moduli}}} 
        & Without Global + Local effect  & 0.122 & 0.119 & 0.106 & 0.098  \\
        \cline{2-6}
        & Global effect & 0.119 & 0.108 & 0.099 & 0.093 \\
        \cline{2-6}
        &  Local effect & 0.117 & 0.102 & 0.097 & 0.091 \\
        \cline{2-6}
        &  \npjmodel{} &\textbf{ 0.105} & \textbf{0.098} & \textbf{0.091} & \textbf{0.089}  \\
       \cline{1-6}
       \cline{1-6}
       
       \multirow{4}{*}{\shortstack[t]{\textbf{Poisson}\\ \textbf{Ratio}}} 
       & Without Global + Local effect  & 0.039 & 0.035 & 0.033 &  0.032 \\
        \cline{2-6}
        & Global effect & 0.037 & 0.034 & 0.032 & 0.031 \\
        \cline{2-6}
        &  Local effect & 0.037 & 0.033 & 0.031 & 0.031 \\
        \cline{2-6}
        
        &  \npjmodel{} & \textbf{0.035} & \textbf{0.032} & \textbf{0.031} & \textbf{0.030}  \\
       \cline{1-6}
       \hline
    \end{tabular}
	}
    \caption{Summary of experiments of ablation study on importance of different reconstruction loss components on CrysAE training and eventually its effect on \npjmodel{} (Mean Absolute Error (MAE) for property prediction).}
    \label{tab:ablation_study}
\end{table}

\begin{table}[]
    \centering
    \small
    \setlength{\tabcolsep}{4pt}
    \scalebox{0.9}{
    \begin{tabular}{c c c c c c} 
        \hline
        \multirow{2}{*}{\shortstack[t]{\textbf{Property}\\ \textbf{Name}}} &
        \multirow{2}{*}{\shortstack[t]{\textbf{Ablation }\\ \textbf{Settings}}} &
        \multicolumn{4}{c}{\textbf{Train-Test Split}}  \\ 
        \cline{3-6} & &  {\textbf{20\% - 80\%}} & {\textbf{40\% - 60\%}} & {\textbf{60\% - 40\%}} &{\textbf{80\% - 20\%}}\\
        \hline
        \hline
        
        \multirow{2}{*}{\shortstack[t]{\textbf{Formation}\\ \textbf{Energy}}}
        & Without $L_{1}$ regularizer & 0.092 & 0.089 & 0.083 & 0.0731  \\
        \cline{2-6}
        &  With $L_{1}$ regularizer & \textbf{0.086} & \textbf{0.082} & \textbf{0.076} & \textbf{0.067}  \\
       \cline{1-6}
       \cline{1-6}
       
       \multirow{2}{*}{\shortstack[t]{\textbf{Band}\\ \textbf{Gap}}}
        & Without $L_{1}$ regularizer  & 0.476 & 0.417 & 0.391 & 0.374  \\
        \cline{2-6}
        & With $L_{1}$ regularizer & \textbf{0.467} & \textbf{0.402} & \textbf{0.383} &  \textbf{0.366} \\
       \cline{1-6}
       \cline{1-6}
       
       \multirow{2}{*}{\shortstack[t]{\textbf{Fermi}\\ \textbf{Energy}}} 
        & Without $L_{1}$ regularizer  & 0.502 & 0.441 & 0.415 & 0.394  \\
        \cline{2-6}
        &  With $L_{1}$ regularizer & \textbf{0.471} & \textbf{0.409} & \textbf{0.389} &  \textbf{0.374} \\
       \cline{1-6}
       \cline{1-6}
       
       \multirow{2}{*}{\shortstack[t]{\textbf{Magnetic}\\ \textbf{Moment}}}
        & Without $L_{1}$ regularizer  & 1.094 & 1.046 & 1.028 & 1.013  \\
        \cline{2-6}
        & With $L_{1}$ regularizer & \textbf{1.033} & \textbf{1.024} & \textbf{0.997} & \textbf{0.943}  \\
       \cline{1-6}
       \cline{1-6}
       
       \multirow{2}{*}{\shortstack[t]{\textbf{Bulk}\\ \textbf{Moduli}}} 
        & Without $L_{1}$ regularizer  & 0.093 & 0.082 &  0.067 & 0.061  \\
        \cline{2-6}
        &  With $L_{1}$ regularizer & \textbf{0.080} & \textbf{0.072} & \textbf{0.063} & \textbf{0.052}  \\
       \cline{1-6}
       \cline{1-6}
       
       \multirow{2}{*}{\shortstack[t]{\textbf{Shear}\\ \textbf{Moduli}}}
        & Without $L_{1}$ regularizer  & 0.128 & 0.115 & 0.099 & 0.095  \\
        \cline{2-6}
        &  With $L_{1}$ regularizer &\textbf{ 0.105} & \textbf{0.098} & \textbf{0.091} & \textbf{0.089}  \\
       \cline{1-6}
       \cline{1-6}
       
       \multirow{2}{*}{\shortstack[t]{\textbf{Poisson}\\ \textbf{Ratio}}}
        & Without $L_{1}$ regularizer  & 0.038 & 0.033 & 0.033 & 0.032  \\
        \cline{2-6}
        &  With $L_{1}$ regularizer & \textbf{0.035} & \textbf{0.032} & \textbf{0.031} & \textbf{0.030}  \\
       \cline{1-6}
       \hline
    \end{tabular}
	}
    \caption{Summary of experiments of ablation study on sparse  feature selection using $L_1$ regularizer, performed on different train test splits across different properties (Mean Absolute Error (MAE)).}
    \label{tab:ablation_study_fs}
\end{table} 

 \begin{table}[]
    \centering
    \small
    \setlength{\tabcolsep}{4pt}
    \scalebox{0.9}{
    \begin{tabular}{c c c c c c} 
        \hline
        \multirow{2}{*}{\shortstack[t]{\textbf{Property}\\ \textbf{Name}}} &
        \multirow{2}{*}{\shortstack[t]{\textbf{Ablation }\\ \textbf{Settings}}} &
        \multicolumn{4}{c}{\textbf{Train-Test Split}}  \\ 
        \cline{3-6} & &  {\textbf{20\% - 80\%}} & {\textbf{40\% - 60\%}} & {\textbf{60\% - 40\%}} &{\textbf{80\% - 20\%}}\\
        \hline
        \hline
        
        \multirow{2}{*}{\shortstack[t]{\textbf{Formation}\\ \textbf{Energy}}}
        & GCN Encoder  & 0.187 & 0.162 & 0.138 & 0.125  \\
        \cline{2-6}
        &  CGCNN Encoder & \textbf{0.086} & \textbf{0.082} & \textbf{0.076} & \textbf{0.067}  \\
       \cline{1-6}
       \cline{1-6}
       
       \multirow{2}{*}{\shortstack[t]{\textbf{Band}\\ \textbf{Gap}}}
        & GCN Encoder   & 0.613 & 0.515 & 0.493 & 0.476  \\
        \cline{2-6}
        &  CGCNN Encoder & \textbf{0.467} & \textbf{0.402} & \textbf{0.383} &  \textbf{0.366} \\
       \cline{1-6}
       \cline{1-6}
       
       \multirow{2}{*}{\shortstack[t]{\textbf{Fermi}\\ \textbf{Energy}}} 
        & GCN Encoder   & 0.513 & 0.489 & 0.450 & 0.436  \\
        \cline{2-6}
        &  CGCNN Encoder & \textbf{0.471} & \textbf{0.409} & \textbf{0.389} &  \textbf{0.374} \\
       \cline{1-6}
       \cline{1-6}
       
       \multirow{2}{*}{\shortstack[t]{\textbf{Magnetic}\\ \textbf{Moment}}}
        & GCN Encoder   & 1.204 & 1.113 & 1.080 & 1.041  \\
        \cline{2-6}
        &  CGCNN Encoder & \textbf{1.033} & \textbf{1.024} & \textbf{0.997} & \textbf{0.943}  \\
       \cline{1-6}
       \cline{1-6}
       
       \multirow{2}{*}{\shortstack[t]{\textbf{Bulk}\\ \textbf{Moduli}}} 
        & GCN Encoder   & 0.171 & 0.133 &  0.114 & 0.101  \\
        \cline{2-6}
        &  CGCNN Encoder & \textbf{0.080} & \textbf{0.072} & \textbf{0.063} & \textbf{0.052}  \\
       \cline{1-6}
       \cline{1-6}
       
       \multirow{2}{*}{\shortstack[t]{\textbf{Shear}\\ \textbf{Moduli}}}
        & GCN Encoder   & 0.183 & 0.172 & 0.168 & 0.141 \\
        \cline{2-6}
        &  CGCNN Encoder &\textbf{ 0.105} & \textbf{0.098} & \textbf{0.091} & \textbf{0.089}  \\
       \cline{1-6}
       \cline{1-6}
       
       \multirow{2}{*}{\shortstack[t]{\textbf{Poisson}\\ \textbf{Ratio}}}
        & GCN Encoder   & 0.041 & 0.037 & 0.036 & 0.034  \\
        \cline{2-6}
        &  CGCNN Encoder & \textbf{0.035} & \textbf{0.032} & \textbf{0.031} & \textbf{0.030}  \\
       \cline{1-6}
       \hline
    \end{tabular}
	}
    \caption{Summary of experiments (Mean Absolute Error (MAE)) of ablation study on effect of GCN as graph encoder in CrysAE and \npjmodel{}.}
    \label{tab:ablation_study_gnn}
\end{table}
\subsection{Ablation Study}
We demonstrate the effectiveness of architectural choices and training strategies for \npjmodel{}, by  designing the following set of ablation studies:
\begin{enumerate}
	\item The importance of explicitly capturing global and local features and understanding their effect on property prediction
	\item The impact of sparse feature selection on property prediction, and 
	\item The choice of GNN models in the autoencoder CrysAE
\end{enumerate}
In the following subsections we will thoroughly discuss these.
\\
\xhdr{Importance of local and global feature understanding}
Here we investigate the importance of different reconstruction loss components on CrysAE training and eventually its effect on property prediction.
 (a). Without Global + Local effect : In this scenario we do not train CrysAE and only train \npjmodel{}.
(b). Global effect : We train CrysAE by minimizing the reconstruction loss of only global features and ignoring local feature losses.
(c). Local effect: Here we focus only minimizing local feature losses.
We report the performance of the model (MAE) in Table \ref{tab:ablation_study} on different train test splits across different properties. We observe that the performance of the model in the setting  Without Global + Local effect, is the worst. We also notice that for all the properties, Local effect individually leads to better performance than  Global effect, except for the Poisson ratio where the effect is similar for both the cases. However, it is found that the impact of local and global effect are somewhat complementary, hence simultaneous reconstruction of 
global and local features (\npjmodel{}) results in the best performance. The only exception is  formation energy where 
addition of global feature leads to performance deterioration..
\\\\
\xhdr{Impact of sparse feature selection} We perform an ablation study  to analyze the impact of sparse feature selection on property prediction.
This is done by removing the $L_{1}$ regularizer term from \npjmodel{} loss function in Eq.\ref{eq:proppred}. We evaluate the performance of the model and report the results (MAE) in Table \ref{tab:ablation_study_fs}. We observe discernible improvement due to the introduction of sparse feature selection using $L_{1}$ regularizer.
\\\\
\xhdr{Effect of other GNN variants as graph encoder} To explore the effectiveness of other GNN variants as graph encoders, we conduct an experiment where we replace the CGCNN encoder with one of the popular GNN variants:  GCN \cite{kipf2017semi} encoder and evaluate the performance of the model. GCN only considers the graph structural information and atom features to learn the graph representation and unlike CGCNN, it does not consider the individual edges weights in the multi-graph representing a crystal. We report the results of the model performance (MAE) in Table \ref{tab:ablation_study_gnn}. We observe that the model performance degrades when trained with GCN. The edge weight calculation, which is a major contribution of CGCNN, is extremely helpful to capture the local structure of the crystal.

\subsection{Explanation through feature selection}
\label{sec:explain}
We have introduced a feature selector that is trained along with the property prediction parameters with available tagged data. The feature selector helps to select the subset of the atomic features contributing to the chemical properties of the crystal which makes the model explainable by design. To demonstrate the effectiveness of the feature selector, we have selected few case studies and provide the feature explanation for formation energy, band gap and magnetic moment.
\begin{figure*}[h!]
	\centering
	\includegraphics[width=\textwidth, height=48mm]{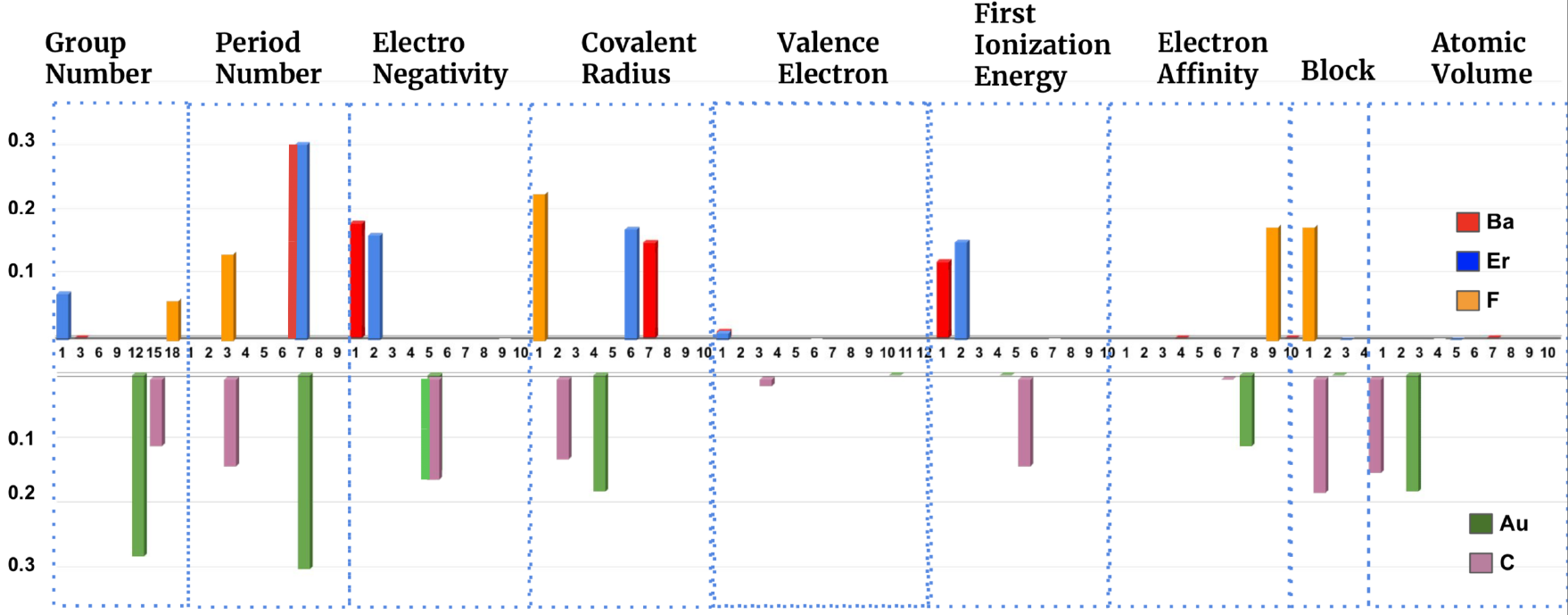}
	\caption{Feature selector values corresponding to atom features after trained on Formation Energy tagged data. The top bar chart represents the feature weights of $BaEr_2F_8$ 
	and the below one represents the feature weights of AuC.  
	}
	\label{fig:feature_mask_cs}
\end{figure*}\\
\xhdr{Formation Energy}
We here report case studies corresponding to two crystals $BaEr_2F_8$ and AuC,  illustrating the important role of feature selector in providing explanation. 
We report the feature selector values corresponding to categorical atomic properties after being trained on Formation Energy tagged data in Fig.\ref{fig:feature_mask_cs}.  The bars represent the weights assigned by the feature selector on the categorical values of atomic features and different colors indicate different atoms. Higher  category denotes higher value of the feature. Fig.\ref{fig:feature_mask_cs} depicts the importance of the atomic features in two extreme cases. One is $BaEr_2F_8$, whose Formation Energy is predicted as -4.41 eV/atom indicating its stability while the other is AuC with predicted Formation Energy 2.2 eV/atom  denoting the material is quite unstable. In both cases we see that Period Number is the most important atomic feature as it has maximum weight. Period and Group Numbers provide the
information to distinguish each element. 
As the Group Numbers and the number of
Valence Electrons are closely related, 
we see that the feature selector only selected the former thus avoiding duplicity.
Electronegativity and Covalent Radius both are another two important features (with non-zero weight) which is evident from the figure.  
Non-zero difference in Electronegativity of atoms indicates
stability in structure.
Both Au and C
have the same Electronegativity (category value 5), and feature selector gives same weight to it, as a result the difference of Electronegativity is zero in the case of unstable AuC. 
While
for the case of stable $BaEr_2F_8$, the feature selector provides different non-zero weights to smaller Electronegative elements and zero weight
to the largest Electronegative atom F.
The Covalent Radius determines the extent of overlap of electron densities of constituents, therefore, it appeared as another important feature. 
Higher the radius means weaker the bond.
It is interesting to note here the
trend of weights is the reverse than that of radius itself (Ba has the largest radius 215 pm and has the
smallest weight) for stable $BaEr_2F_8$. The scenario is reverse for unstable AuC. Ionization Energy plays similar role as Electronegativity and we observe same behavior of feature selector. As can be seen from the example, the feature selector provides elaborate cues for domain experts to reason out the results.
\begin{figure*}[h!]
	\centering
	\includegraphics[width=\textwidth, height=43mm]{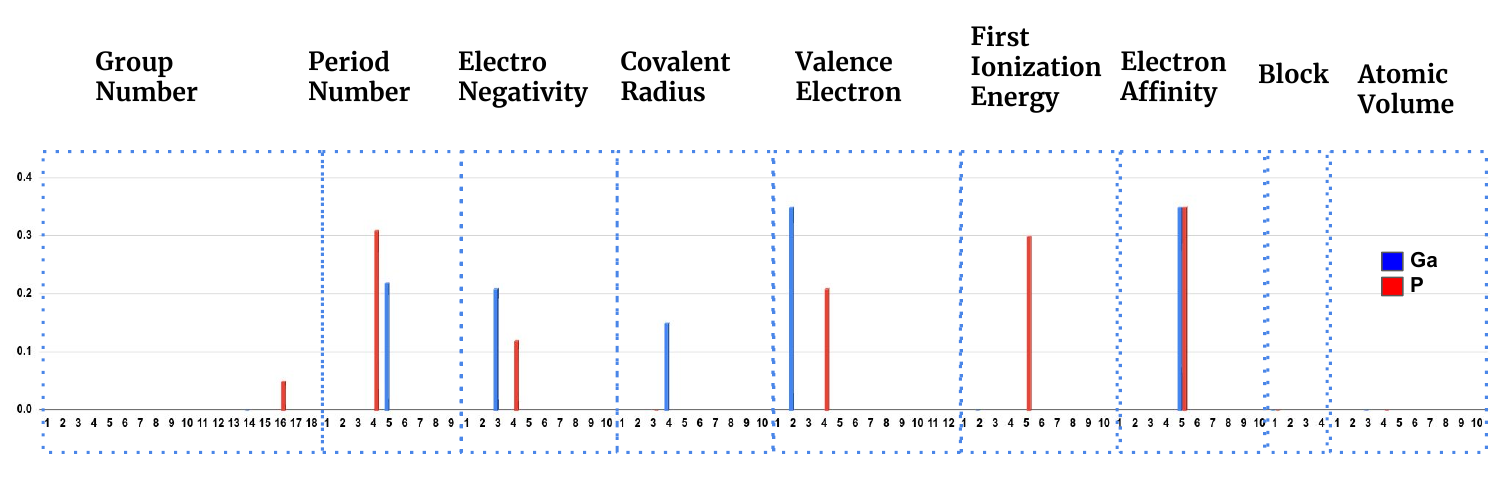}
	\caption{Feature selector values corresponding to atom features after trained on Band Gap tagged data. The top bar chart represents the feature weights of GaP (Band Gap 2.26 eV). }
	\label{fig:feature_mask_cs-g}
\end{figure*}\\
\xhdr{Band Gap}
In Fig.\ref{fig:feature_mask_cs-g}, we show the important features that appear in the case of band gap. It is interesting to see that the Electron Affinity came out to be the most
important atomic feature for the band gap as it determines the location of
conduction band minimum with respect to the vacuum.  Again, as the number of Valence Electrons and the group number are collinear properties, only one (valence electrons) is found to be having the significant weight. The Conduction Band is composed of Ga-states while the valence band is composed of P states with small admixture of Ga-states, that gives Ga-valence electrons more weight. The situation is reversed for the Ionization Energy, which determines the location of the valence band with respect to a vacuum, 
and as the valence band is mostly formed by the  P atom, we see P has  more weight.
\begin{figure*}[h!]
	\centering
	\includegraphics[width=\textwidth, height=48mm]{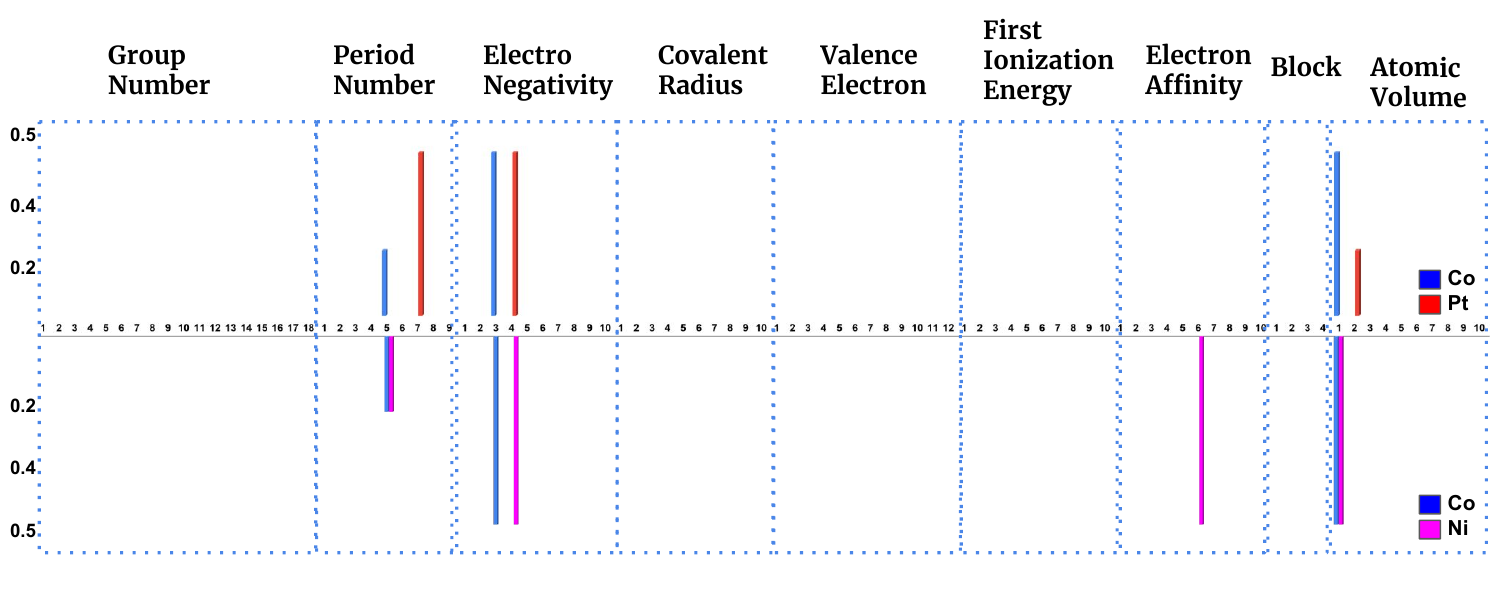}
	\caption{Feature selector values corresponding to atom features after trained on Magnetic Moment tagged data. The top bar chart represents the feature weights of CoPt and the bottom chart  represents the feature weights of CoNi. }
	\label{fig:feature_mask_cs-mm}
\end{figure*}\\
\xhdr{Magnetic Moment}
In order to understand the feature importance in the case of magnetic moment, we compare the results obtained for two Co based alloys, namely, CoPt and CoNi (Fig.\ref{fig:feature_mask_cs-mm}). In both cases  the Atomic Volume, Period Number and  Electronegativity appear to be the three most important features. While in the case of CoNi, Electron Affinity of Ni also appeared to be as additional important feature. It can be seen that in the case of CoPt, the Atomic Volume of Co has a higher weight, while for CoNi, the atomic volume for both the species have the same weight. This is quite intuitive, as for CoPt, the magnetic moment is mostly carried by Co atom, while in the case of CoNi, both the atoms have a significant contribution. The Electronegativity plays an important role in the context of the magnetic moment. For example the magnetic moment of Co in CoPt is slightly higher than its corresponding value in pure Co. The electronegativity difference between Co and Pt causes the electron transfer from Co minority spin band to Pt, which in turn enhances its magnetic moment~\cite{S1,S2}. In the case of Period Number again we see that for CoPt, it is only the period number of the magnetic atom, i.e,  the Co atom that is given visible weight while in the case of NiCo, the period number of the two atoms appears to be important.\\
It is evident from the above analysis that \npjmodel{} is effectively constructing models where the important node features are physically intuitive.
\section{Conclusion}
In conclusion, we propose an explainable property predictor for crystalline materials, \npjmodel{} to predict different crystal state and elastic properties with accurate precision using  small amount of property-tagged data. We address the issue of limited crystal data where the value of a particular property is known, using transfer learning from an encoding module CrysAE; which we train in a property agnostic way with a large amount of untagged crystal data to capture all the important structural and chemical information useful to a specific property predictor. We further find the encoder knowledge is extremely useful in de-biasing DFT error using a meagre instances of experimental results. \npjmodel{} outperforms all the baselines across seven diverse sets of properties. With appropriate case studies, we show that the explanations provided by the feature selection module  are in sync with the domain knowledge. We release the large pretrained model CrysAE so that it could be fine-tuned using a small amount of tagged data by the research community on various applications with restricted data source.


\clearemptydoublepage
\chapter{CrysGNN: Distilling Pre-trained Knowledge to Enhance Property Prediction for Crystalline Materials}
\chaptermark{CrysGNN [AAAI-2023]}
\label{chap: crysgnn}

\noindent\fbox{%
\parbox{\textwidth}{%
Work of this chapter are based on the following publication:  
\\
\textit{CrysGNN: Distilling Pre-trained Knowledge to Enhance Property Prediction for Crystalline Materials.}
\\
\underline{Kishalay Das}, Bidisha Samanta, Pawan Goyal, Seung-Cheol Lee, Satadeep Bhattacharjee and Niloy Ganguly.
\\
AAAI, 2023.}}
\vspace{1em}

\section{Introduction}
\label{intro}
Fast and accurate prediction of different material properties is a challenging and important task in material science.  In recent times there has been an ample amount of data driven works ~\cite{seko2015prediction,pilania2015structure,lee2016prediction,de2016statistical,seko2017representation,isayev2017universal,ward2017including,lu2018accelerated,im2019identifying}
for predicting crystal properties which are as accurate as theoretical DFT (Density functional Theory)~\cite{orio2009density} based approaches, however, much faster than it. The architectural innovations of these approaches towards accurate property predictions come from incorporating specific domain knowledge into a deep encoding module. For example, in order to encode the neighbourhood structural information around a node (atom), GNN based approaches ~\cite{xie2018crystal,chen2019graph,louis2020graph,Wolverton2020,schmidt2021crystal} gained some popularity in this domain. Understanding the importance of many-body interactions, ALIGNN~\cite{choudhary2021atomistic} incorporates bond angular information into its encoder module and became SOTA for a large range of property predictions. However, as different properties expressed by a crystalline material are a complex function of different inherent structural and chemical properties of the constituent atoms, it is extremely difficult to explicitly incorporate them into the encoder architecture. Moreover, data sparsity across properties is a known issue~\cite{das2022crysxpp,jha2019enhancing}, which makes these models difficult to train for all the properties. To circumvent this problem we adopt the concept of self-supervised pre-training \cite{devlin2018bert,trinh2019selfie,chen2020simple,chen2020adversarial,he2020momentum,hu2020pretraining,hu2020gpt,qiu2020gcc,you2020graph} for crystalline materials which enables us to leverage a large amount of untagged material structures to learn the complex hidden features which otherwise are difficult to identify.\\\\
In this work, we introduce a graph pre-training method which captures (a) connectivity of different atoms, (b) different atomic properties and (c) graph similarity from a large set of unlabeled crystal data. To this effect, we curate a new large untagged crystal dataset with 800K crystal graphs and undertake a  pre-training framework (named  \aaaimodel{}) with the dataset. \aaaimodel{}  learns the representation of a crystal graph by initiating self-supervised loss  at both node (atom) and graph (crystal) level. At the node level, we pre-train the GNN model to reconstruct the node features and connectivity between nodes in a self-supervised way, whereas at the graph level, we adopt supervised and contrastive learning to learn structural similarities between graph structures using the space group and crystal system information of the crystal materials respectively.
We subsequently distill important structural and chemical information of a crystal from the pre-trained \aaaimodel{} model and pass it to the property predictor. 
The distillation process  provides wider usage than the conventional pretrain-finetuning framework~\cite{he2022knowledge} as transferring pre-trained knowledge to a property predictor and finetuning it requires a similar graph encoder architecture between the pre-trained model and the property predictor, which limits the knowledge transfer capability of the pre-trained model. On the other hand, using knowledge distillation \cite{romero2014fitnets,hinton2015distilling}, we can retrofit the pre-trained \aaaimodel{} model into any existing state-of-the-art property predictor, irrespective of their architectural design, to improve their property prediction performance. Also experimental results (presented later) show that even in case of similar graph encoder, distillation performs better than finetuning.
\\\\ 
To the best of our knowledge, \aaaimodel{} is the first attempt to develop a deep pre-trained model for the crystalline materials and for that, we curated a new large untagged crystal dataset with 800K crystal graphs. With rigorous experimentation on 19 different properties 
across two popular benchmark materials datasets, we show that distilling necessary information from \aaaimodel{}
\footnote{Source code, pre-trained model, and dataset of CrysGNN is made available at \url{https://github.com/kdmsit/crysgnn}} 
to various state-of-the-art property predictors results in substantial performance gains for  GNN based architectures and the complex ALIGNN model. The improvements range from \emph{\textbf{4.19\% to 16.20\%}} over several highly optimized SOTA models. We also perform ablation studies to investigate the influence of different pre-training losses in enhancing the SOTA model performance and observe significant performance gain employing the both node and graph-level pre-training, compared to node-level or graph-level pre-training separately. Also using both supervised and contrastive graph-level pre-training, we are able to learn more robust and expressive graph representation which enhances the property predictor performance.
This also helps to achieve even better improvements when the dataset is sparse. Moreover, the property-tagged dataset suffers from certain biases as it is theoretically (DFT) derived, hence the property predictor also suffers from such bias.  We found that on being trained with small amount of experimental data, the DFT bias decreases substantially.

\section{Background}
\subsection{Crystal Graph Representation.}
\label{graph_rep}
We realize a crystal material as a multi-graph structure $\mathcal{G}_i =(\mathcal{V}_i, \mathcal{E}_i, \mathcal{X}_i, \mathcal{F}_i )$ as proposed in \cite{xie2018crystal}. $\mathcal{G}_i$ is an undirected weighted multi-graph where $\mathcal{V}_i$ denotes the set of nodes or atoms present in a unit cell of the crystal structure. $\mathcal{E}_i=\{(u,v,k_{uv})\}$ denotes a multi-set of node pairs and $k_{uv}$ denotes number of edges between a node pair $(u,v)$. $\mathcal{X}_i=\{(x_{u} | u \in \mathcal{V}_i )\}$ denotes the node feature set proposed by CGCNN~\cite{xie2018crystal}. It includes different chemical properties like electronegativity, valance electron, covalent radius, etc. Finally, $\mathcal{F}_i=\{\{s^k\}_{(u,v)} |  (u,v) \in \mathcal{E}_i, k\in\{1..k_{uv}\}\}$ denotes the multi-set of edge weights where $s^k$ corresponds to the $k^{th}$ bond length between a node pair $(u,v)$, which signifies the inter-atomic bond distance between two atoms. Next, we formally define \aaaimodel{} pre-training  and knowledge distillation based property prediction strategy.

\subsection{Crystal System and Space group Information}
\label{sec_spec}
By definition crystal material is a periodic arrangement of repeating “motifs”( e.g. atoms, ions). The symmetry of a periodic pattern of repeated motifs is the total set of symmetry operations allowed by that pattern. The total set of such symmetry operations, applicable to the pattern is the pattern's symmetry, and is mathematically described by a so-called Space Group. The Space Group of a Crystal describes the symmetry of that crystal,
and as such, it describes an important aspect of that crystal's internal structure.\\
In turn crystals across certain space groups show
similarities among each other, hence they are divided among seven Crystal Systems: Triclinic, Monoclinic, Orthorhombic, Tetragonal, Trigonal, Hexagonal, and Cubic or Isometric system ~\cite{kittel2018kittel}.
\\\\
\textbf{Crystallographic Axes: } The crystallographic axes are imaginary lines that we can draw within the crystal lattice.  These will define a coordinate system within the crystal. We refer to the axes as a,b,c and the angle between them as $\alpha$, $\beta$ and $\gamma$. By using these crystallographic axes we can define six large groups or crystal systems.

\begin{itemize}
  \item \textbf{Cubic or Isometric system :} The three crystallographic axes are all equal in length and intersect at right angles to each other.
  \begin{equation}
        \boxed{a=b=c\;; \;\; \alpha = \beta =\gamma = 90^{\circ}} \notag
   \end{equation}
  \item \textbf{Tetragonal system :} Three axes, all at right angles, two of which are equal in length (a and b)
    and one (c) which is different in length (shorter or longer).
    \begin{equation}
        \boxed{a=b\neq c\;; \;\; \alpha = \beta =\gamma = 90^{\circ}} \notag
    \end{equation}
   \item \textbf{Orthorhombic system :} Three axes, all at right angles, all three have different length.
   \begin{equation}
        \boxed{a\neq b\neq c\;; \;\; \alpha = \beta =\gamma = 90^{\circ}} \notag
    \end{equation}
   \item \textbf{Monoclinic system :} Three axes, all unequal in length, two of which (a and c) intersect at an oblique angle (not 90 degrees), the third axis (b) is perpendicular to the other two axes.
   \begin{equation}
        \boxed{a\neq b\neq c\;; \;\; \alpha  =\gamma = 90^{\circ} \neq \beta} \notag
    \end{equation}
   \item \textbf{Triclinic system :} The three axes are all unequal in length and intersect at three different angles (any angle but 90 degrees).
   \begin{equation}
        \boxed{a\neq b\neq c\;; \;\; \alpha \neq \beta \neq \gamma \neq 90^{\circ}} \notag
    \end{equation}
  \item \textbf{Hexagonal system :} Here we have four axes. Three of the axes fall in the same plane and intersect at the axial cross at 120°. These 3 axes, labeled $a_1$, $a_2$, and $a_3$, are the same length. The fourth axis, c, may be of {\em a} different length than the axes set. The c axis also passes through the intersection of the {\em a} axes
set at right angle to the plane formed by the {\em a} set.
\end{itemize}

\section{Methodology}
\label{methodology}
\begin{figure*}
	\centering
	\vspace*{-1mm}
	\subfloat[Node-level decoding]{
		\boxed{\includegraphics[width=\columnwidth]{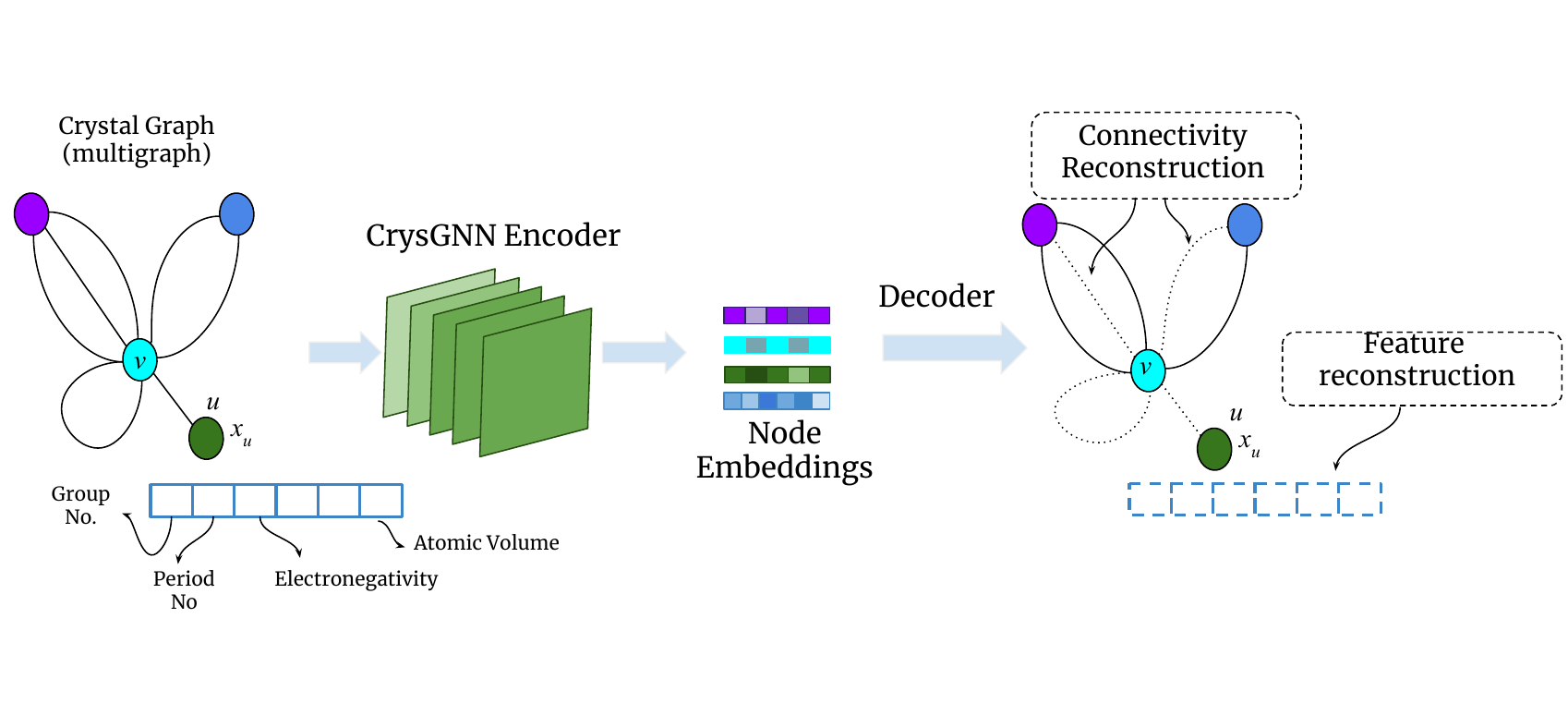}}} \\
	\subfloat[Graph-level decoding]{
		\boxed{\includegraphics[width=\columnwidth]{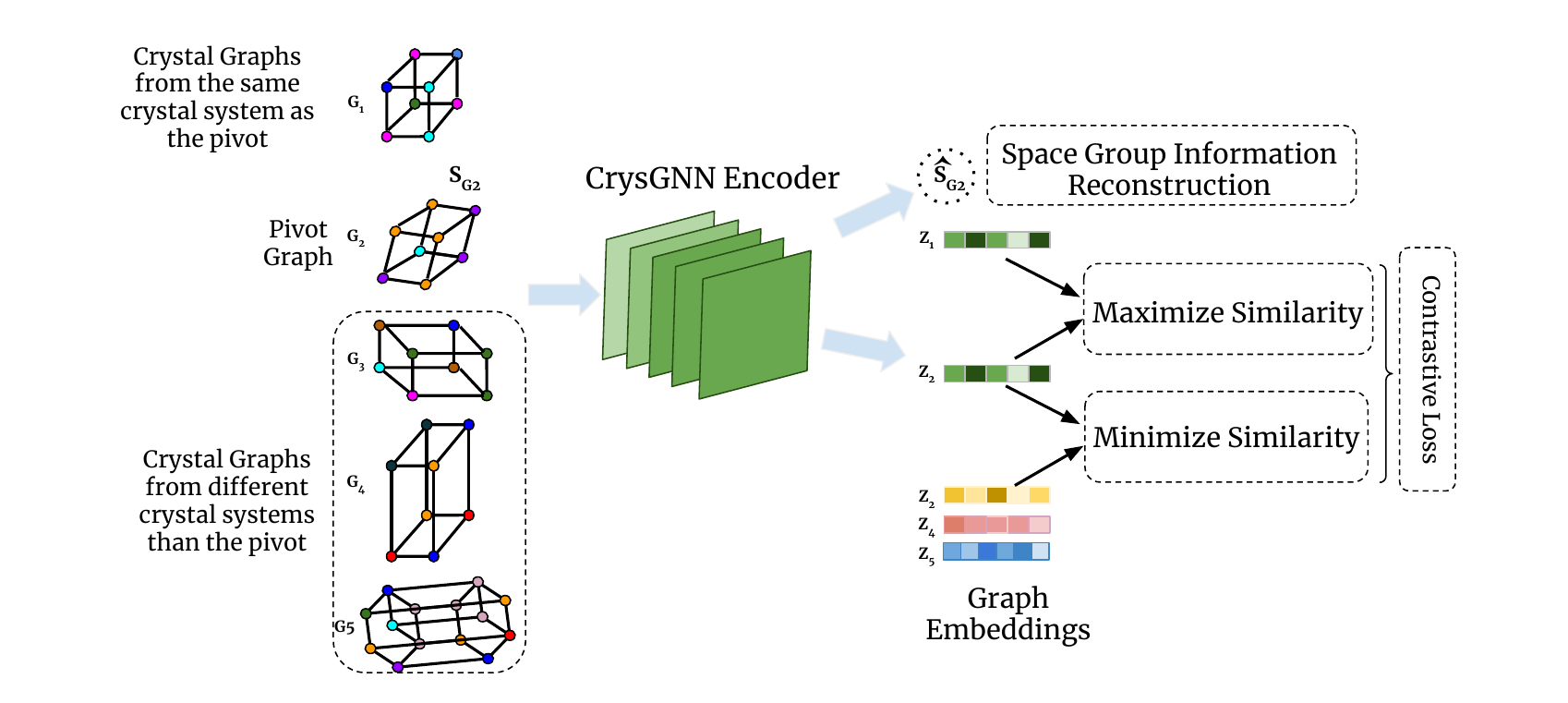}}}
	\caption{Overview of both node and graph-level decoding methods for \aaaimodel{}. (a) In node-level decoding, node feature attributes and connectivity between nodes are reconstructed in a self-supervised way. (b) In graph-level decoding, $G_2$ is the pivot graph and $G_1$ is from the same crystal system (Cubic), whereas $G_3, G_4, G_5$ are from different crystal systems. First we reconstruct space group information of $G_2$, then through contrastive loss, \aaaimodel{} will maximize similarities between positive pair ($G_2,G_1$) and minimize similarities between negative pairs ($G_2, G_3$), ($G_2, G_4$) and ($G_2, G_5$) in embedding space.}
	\label{fig:CrysGNN}
\end{figure*}
Formally, we first curate a huge amount of property un-tagged crystal graphs $\mathcal{D}_{u} = \{\mathcal{G}_i\}$ from various materials datasets to pre-train a deep GNN model $f_{\theta}$, that learns intrinsic structural and chemical patterns of the crystal graphs.
Further, we use a training set of property-tagged crystal graphs $\mathcal{D}_{t} =\{\mathcal{G}_i,y_i\}$ for property prediction, which is smaller in volume and may or may not be disjoint from the original untagged set $\mathcal{D}_{u}$. We train any supervised property predictor $\mathcal{P}_{\psi}$ using $\mathcal{D}_{t}$ to predict the property value given the crystal graph structure. While training the property predictor, we incorporate the idea of knowledge distillation to distill important structural and chemical information from the pre-trained model. This knowledge may prove to  be useful to a property predictor which now need not learn from scratch, but be armed with distilled  knowledge from the pre-trained model.  Hence in this section, we first describe the \aaaimodel{} pre-training strategy, followed by the knowledge distillation and property prediction process. 
\subsection{CrysGNN Pre-training} 
\label{pretraining}
We build a deep auto-encoder architecture \aaaimodel{}, which comprises a graph convolution based encoder followed by an effective decoder. The autoencoder is (pre)trained end to end, using a large amount of property un-tagged crystal graphs $\mathcal{D}_{u} = \{\mathcal{G}_i\}$, where via node and graph-level self-supervised losses, the model can capture the structural and chemical information of the crystal graph data. First, we formalize the representation of a crystal 3D structure into a multi-graph structure, which will be an input to the encoder module.

\begin{figure*}
	\centering
	\vspace*{-1mm}
	\boxed{\includegraphics[width=\columnwidth]{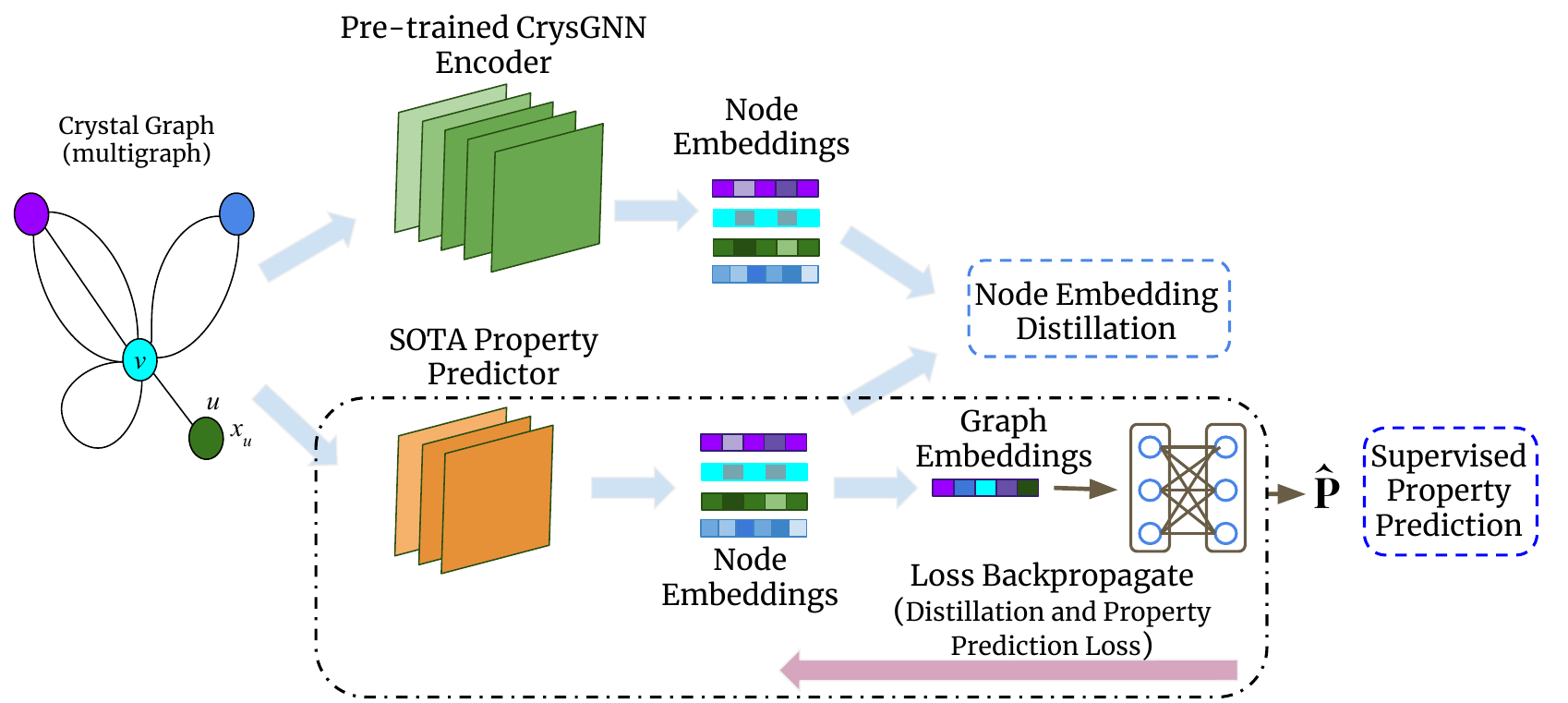}}
	\caption{\rev{Overview of Property Prediction using Knowledge Distillation from \aaaimodel{}. A crystal multigraph is encoded using a pre-trained CrysGNN to obtain node embeddings, which act as a teacher representation. A state-of-the-art (SOTA) property predictor learns from both supervised property labels and distilled node/graph embeddings, producing graph-level representations for accurate property prediction. The model is trained end-to-end via joint distillation and property prediction loss backpropagation.}}
	\label{fig:dist_prop}
\end{figure*}
\subsubsection{Self Supervision.}
\label{self_sup}
We first develop a graph convolution  \cite{xie2018crystal} based encoding module, which takes crystal multi-graph structure $\mathcal{G} =(\mathcal{V}, \mathcal{E}, \mathcal{X}, \mathcal{F} )$ as input and encodes structural semantics of the crystal graph into lower dimensional space. Each layer of convolution, follows an iterative neighbourhood aggregation (or message passing) scheme to capture the structural information within node's (atom's) neighbourhood. After $L$-layers of such aggregation, the encoder returns the final set of node embeddings $\mathcal{Z} = \{z_1,..., z_{|\mathcal{V}|}\}$, where $z_u :=z_u^L$ represents the final embedding of node $u$.  Details of the GNN architecture is in \ref{gnn_arch}.
Next, we design an effective decoding module, which takes node embeddings $\mathcal{Z}$ as input and learns local chemical features and global structural information through node and graph-level decoding, respectively. Decoding node-level information will enable \aaaimodel{} to learn local domain specific chemical features and connectivity information around an atom, while decoding graph-level features  will help \aaaimodel{} capture global structural knowledge. \\\\
\textbf{Node-Level Decoding.} For node-level decoding (Fig-\ref{fig:CrysGNN}(a)), we propose two self-supervised learning methods, where we reconstruct two important features that induce the local chemical environment of the crystal around a node (atom). For a given node $u$, we first reconstruct its node features $x_u$, 
which represent different chemical properties of atom $u$. Given a node embedding $z_u$, which is encoded based on neighbouring structure around atom $u$, we apply a linear transformation on top of $z_u$ to reconstruct the node attributes. In crystalline graphs, node features correspond to different chemical properties associated with the constituent atoms 
through reconstructing these features \aaaimodel{}  captures local chemical semantics around that atom.\\ 
Further, we reconstruct local connectivity around an atom, where given node embeddings of two nodes $u$ and $v$, we apply a bi-linear transformation module to generate combined transformed embedding of two nodes $z_{uv}$, which we pass through a feed forward network to predict the strength of association between two atoms. Through reconstructing this local connectivity around an atom, \aaaimodel{}  encodes the periodicity of the node i.e. the number of neighbours around it along with the relative position of its neighbours and their bond length.
\\\\
\textbf{Graph-level Decoding.} 
We aim to capture periodic structure of a crystal material through graph-level decoding (Fig-\ref{fig:CrysGNN}(b)). We specifically leverage two concepts in doing so. (a). {\bf Space group} which is used to describe the symmetry of a unit cell of the crystal material. In materials science literature there are  230 unique space groups and each crystal (graph) has a unique space group number. 
(b). {\bf Crystal system.} As discussed in section \ref{sec_spec}, the space group level information can classify a crystal graph into 7 broad groups of crystal systems like Triclinic, Monoclinic, Orthorhombic, Tetragonal, Trigonal, Hexagonal, and Cubic. Several electronic and optical properties such as band gap, dielectric constant depend on the space-group and the crystal structure justifying its usage.
\\\\
Given the set of node embeddings $\mathcal{Z} = \{z_1,..., z_{|\mathcal{V}|}\}$, we use a symmetric aggregation function to generate graph-level representation $\mathcal{Z}_{\mathcal{G}}$. First, we pass $\mathcal{Z}_{\mathcal{G}}$ through a feed-forward neural network to predict the space group number of graph $\mathcal{G}$.
Further, we develop a contrastive learning framework for pre-training of \aaaimodel{}, where pre-training is performed by maximizing (minimizing) similarity between two crystal graphs belonging to the same (different) crystal system via contrastive loss in graph embedding space.  
A mini-batch of $N$ crystal graphs is randomly sampled and processed through contrastive learning to 
align the positive pairs $\mathcal{Z}_{\mathcal{G}_i},\mathcal{Z}_{\mathcal{G}_j}$ of graph embeddings, which belong to the same crystal system and contrast the negative pairs which are from different crystal systems. Here we adopt the
normalized temperature-scaled cross-entropy loss (NT-Xent)\cite{sohn2016improved,van2018representation,wu2018unsupervised} and NT-Xent for the $i^{th}$ graph is defined:
\begin{equation}
    \label{eq:contrastive}
        \mathcal{L}_{i}=-log\frac{exp(sim(\mathcal{Z}_{\mathcal{G}_i},\mathcal{Z}_{\mathcal{G}_j})/\tau)}{\sum_{k=1}^{K} exp(sim(\mathcal{Z}_{\mathcal{G}_i},\mathcal{Z}_{\mathcal{G}_k})/\tau)}
\end{equation}
where $\tau$ denotes the temperature parameter and  $sim(\mathcal{Z}_{\mathcal{G}_i},\mathcal{Z}_{\mathcal{G}_j})$ denotes cosine similarity function.
The final loss $\mathcal{L}_{NTXent}$ is computed across all positive pairs in the minibatch. 
Overall we pre-train this deep auto-encoder architecture \aaaimodel{} end to end to optimize the following loss :
\begin{equation}
    \label{eq:pretrain_loss}
        \mathcal{L}_{pretrain}=\alpha\mathcal{L}_{FR}+\beta\mathcal{L}_{CR}+\gamma\mathcal{L}_{SG}+\lambda\mathcal{L}_{NTXent}
\end{equation}
where $\mathcal{L}_{FR},\mathcal{L}_{CR}$ are the reconstruction losses for node feature, and local connectivity, $,\mathcal{L}_{SG}$ is the space group supervision loss, $\mathcal{L}_{NTXent}$ is the contrastive loss and $\alpha$, $\beta$, $\gamma$, $\lambda$ are the weighting coefficients of each loss. We denote the set of parameters in \aaaimodel{} model as $\theta$ and the pre-trained \aaaimodel{} as $f_{\theta}$.

\subsection {Details of GNN Encoder in CrysGNN}
\label{gnn_arch}
We develop an graph convolution \cite{xie2018crystal} based encoding module, which takes crystal multi-graph structure $\mathcal{G} =(\mathcal{V}, \mathcal{E}, \mathcal{X}, \mathcal{F} )$ as input and encode structural semantics of the crystal graph into lower dimensional space. Considering $L$ graph convolution layers, neighbourhood aggregation of the $l$-th layer is represented as :
\begin{equation}
    \label{eq:cgcnn}
    z_u^{(l+1)}  = z_u^{(l)} + \sum\limits_{v,k} \sigma ( {h^{(l)}_{(u,v)}}_k W_c^{(l)} +  b_c^{(l)}) \odot g({h^{(l)}_{(u,v)}}_k W_s^{(l)}+b_s^{(l)}) 
\end{equation}
where $h_{(u,v)_k}^{(l)} = z_u^{(l)} \oplus z_v^{(l)} \oplus {s^{(k)}_{(u,v)}}$. Here $s^{(k)}_{(u,v)}$ represents inter-atomic bond length of the $k$-th bond between $(u,v)$ and $z_u^{(l)}$ denotes embedding of node $u$ after $l$-th layer, which is initialized to a transformed node feature vector $z_u^0 := x_u  W_x$ where $W_x$ is the trainable parameter and $ x_u $ is the input node feature vector (Table. \ref{tab:feature_desc}). In Equation \ref{eq:cgcnn}, $W_c^{(l)},W_s^{(l)},b_c^{(l)},b_s^{(l)}$ are the convolution weight matrix, self weight matrix, convolution bias, and self bias of the $l$-th layer, respectively, $\oplus$ operator denotes concatenation, $\odot$ denotes element-wise multiplication, $\sigma$ is the sigmoid function indicating the edge importance and $g$ is a feed forward network. After $L$-layers of such propagation, $\mathcal{Z} = \{z_1,..., z_{|\mathcal{V}|}\}$ represents the set of final node embeddings, where $z_u :=z_u^L$ represents final embedding of node $u$.

\subsection{Distillation and Property Prediction}
We aim to retrofit the pre-trained \aaaimodel{} model into any SOTA property predictor to enhance its learning process and improve performance (Fig-\ref{fig:dist_prop}). Hence we incorporate the idea of knowledge distillation to distill important structural and chemical information from the pre-trained model, which is useful for the downstream property prediction task, and feed it into the property prediction process.
Formally, given the pre-trained \aaaimodel{} model $f_{\theta}$, any SOTA property predictor $\mathcal{P}_{\psi}$ and set of property tagged training data $\mathcal{D}_{t} =\{\mathcal{G}_i,y_i\}$, we aim to find optimal parameter values ${\psi^*}$ for $\mathcal{P}$. We train $\mathcal{P}_{\psi}$ using dataset $\mathcal{D}_{t}$ to optimize the following multitask loss:
\begin{equation}
    \label{eq:finetune_loss}
        \mathcal{L}_{prop}=\delta\mathcal{L}_{MSE}+(1-\delta)\mathcal{L}_{KD}
\end{equation}
where $\mathcal{L}_{MSE} = (\hat{y_i}-y_i)^2$ denotes the discrepancy between predicted and true property values by $\mathcal{P}_{\psi}$ (property prediction loss). We define knowledge distillation loss $\mathcal{L}_{KD}$ to match intermediate node feature representation between the pre-trained \aaaimodel{} model and the SOTA property predictor $\mathcal{P}_{\psi}$ as follows:
\begin{equation}
\label{eq:kd_loss}
                \mathcal{L}_{KD}=  \lVert {\mathcal{Z}^{T}_{i}} - {\mathcal{Z}^{S}_{i}} \rVert^2 
\end{equation}
where $\mathcal{Z}^{T}_{i}$ and $\mathcal{Z}^{S}_{i}$ denote intermediate node embeddings of the pre-trained \aaaimodel{} and the property predictor $\mathcal{P}_{\psi}$ for crystal graph $\mathcal{G}_i$, respectively. Note, both $\mathcal{Z}^{T}_{i}$ and $\mathcal{Z}^{S}_{i}$ are projected on the same latent space. Finally, $\delta$ signifies relative weightage between two losses, which is a hyper-parameter to be tuned on validation data. During property prediction the pre-trained network is frozen and we backpropagate $\mathcal{L}_{prop}$ through the predictor $\mathcal{P}_{\psi}$ end to end.

\subsection{Comparison with Prior works}
Although conceptually similar to the work done by ~\cite{hu2020pretraining}, our work differs in the following three key aspects: (1) pre-training strategy proposed by ~\cite{hu2020pretraining} is very effective for molecular dataset, but it is difficult to extend directly to crystalline material because structural semantics are different between molecules and materials \cite{xie2021crystal}. 
Molecules have non-periodic and finite structures,  solid materials’ structures are infinite and periodic in nature. (2) For graph-level pre-training, ~\cite{hu2020pretraining} adapted supervised graph-level property prediction using a huge amount of labelled dataset from chemistry and biology domain, which makes it less effective in several other domains like material science where property labeled data is extremely scarce. Also, a crucial step in graph-level prediction is to find graph structural similarity between two sets of graphs, which they do not explore but mention as a future work. We do not make use of supervised pre-training which requires a large amount of property tagged material data. Instead, we leverage the idea of structural similarity of materials belonging to the similar space group, and via contrastive loss and space group classification loss, we try to capture this similarity. (3) Finally they follow conventional pre-train finetuning framework, whereas in \aaaimodel{}  we incorporate the idea of knowledge distillation \cite{romero2014fitnets,hinton2015distilling} to distill important information from the pre-trained model and inject it into the property prediction process. By design, this knowledge distillation based approach is more robust and independent of the underlying architecture of the property predictor, thus it can enhance the performance of a diverse set of SOTA models.

\begin{table}
  \centering
    \setlength{\tabcolsep}{2 pt}
    \resizebox{0.8\textwidth}{!}{
    \begin{tabular}{c | c | c | c | c | c }
    \toprule
    Task & Datasets & Graph Num. & Structural Info.  & Properties Count & Data Type\\
    \midrule
     \multirow{2}{*}{\shortstack{Pre-training}}
     & OQMD & 661K & \checkmark  & x & DFT Calculated\\
     & MP & 139K & \checkmark  & x & DFT Calculated\\
     \midrule
    \multirow{3}{*}{\shortstack{Property \\ Prediction}} & MP 2018.6.1 & 69K & \checkmark  & 2 & DFT Calculated\\
     & JARVIS-DFT & 55K & \checkmark  & 19 & DFT Calculated\\
     & OQMD-EXP & 1.5K & \checkmark  & 1 & Experimental\\
    \bottomrule
  \end{tabular}
  }
  \caption{\rev{Dataset details used for pre-training and property prediction tasks: The table summarizes the datasets employed in this work, grouped by task. For pre-training, large-scale DFT-calculated crystal structure datasets: OQMD (661K graphs) and Materials Project (139K graphs) are used, containing only structural information without explicit property labels. For property prediction, three datasets are considered: MP 2018.6.1 (69K graphs, 2 target properties), JARVIS-DFT (55K graphs, 19 target properties), and OQMD-EXP (1.5K graphs, 1 experimentally measured property). All datasets include crystal structural information, while the data type distinguishes between DFT-calculated and experimentally obtained properties.}}
  \label{tbl-dataset}
\end{table}

\section{Experimental Results}
\label{exp_results}
In this section, we evaluate how the distilled knowledge from \aaaimodel{} enhances the performance of different state of the art property predictors on a diverse set of crystal properties from two popular benchmark materials datasets. We first briefly discuss the datasets used both in pre-training and downstream property prediction tasks. Then we report the results of different SOTA property predictors on the downstream property prediction tasks. 
Next, we illustrate the effectiveness of our knowledge distillation method compared to the conventional fine-tuning approach.
We further conduct some ablation studies to show the influence of different pre-training losses in predicting different crystal properties and the performance of the system to sparse dataset. Finally, we demonstrate how distilled knowledge from the pre-trained model aids the SOTA models to remove DFT error bias, using very little experimental data.
    
\subsection{Datasets}
\label{sec_crysgnn_dataset}
\textbf{OQMD:} The Open Quantum Materials Database (OQMD) is a high-throughput database based on DFT calculations of chemicals from the Inorganic Crystal Structure Database (ICSD) and embellishments of regularly occurring crystal structures \cite{OQMD}. Comprehensive computations for significant structure types in materials research, such as Heusler and perovskite compounds, are among the distinguishing aspects of this database. The database is public and  a free resource.\\\\
\textbf{Materials project:} This is another free and public resource database that contains crystal structures and calculated materials properties~\cite{MP}. The dataset is made from the results obtained with density functional theory based calculations. The dataset is composed of electronic structure, thermodynamic, mechanical, and dielectric properties. It also provides  a visual web-based search interface.\\\\
\textbf{JARVIS:} JARVIS (Joint Automated Repository for Various Integrated Simulations) is a data repository that incorporates not only DFT-based calculations, but also data from classical force-fields and machine learning approaches~\cite{JARVIS}. The databases are free and public as well.
\\\\
We curated 800K untagged crystal graph data from two popular materials databases, Materials Project (MP) and OQMD, to pre-train \aaaimodel{} model. Further to evaluate the performance of different SOTA models with distilled knowledge from \aaaimodel{}, we select MP 2018.6.1 version of Materials Project and  2021.8.18 version of JARVIS-DFT, another popular materials database, for property prediction as suggested by \cite{choudhary2021atomistic}.  Please note, MP 2018.6.1 dataset is a subset of the dataset used for pre-training, whereas JARVIS-DFT is a separate dataset which is not seen during the pre-training.
MP 2018.6.1 consists of 69,239 materials with two properties bandgap and formation energy, whereas JARVIS-DFT consists of 55,722 materials with 19  properties which can be broadly classified into two categories : 1) properties like formation energy, bandgap, total energy, bulk modulus, etc. which depend greatly on crystal structures and atom features, and 2) properties like $\epsilon_x$, $\epsilon_y$, $\epsilon_z$, n-Seebeck, n-PF, etc. which depend on the precise description of the materials’ electronic structure. \\
\rev{Note, although MP 2018.6.1 a subset of the MP dataset used in pre-training is also used for the property prediction tasks (formation energy and band gap), the pre-training and property prediction tasks are fundamentally different. The pre-training stage is entirely self-supervised and uses only unlabeled data, whereas the property prediction stage is a supervised, property-specific task that relies on labeled data. Therefore, the results on the MP dataset are reliable.}

In the following section, we will evaluate effectiveness of \aaaimodel{} on both class of properties. Moreover, all these properties in both Materials Project and JARVIS-DFT datasets are based on DFT calculations of chemicals. Therefore, to investigate how pre-trained knowledge helps to mitigate the DFT error, we also take a small dataset OQMD-EXP~\cite{kirklin2015open}, containing 1,500 available experimental data of formation energy. Details of each of these datasets are given in Table \ref{tbl-dataset}. More detail about dataset, different crystal properties and experimental setup is in Appendix

\subsection{Baseline Models.}
To evaluate the effectiveness of \aaaimodel{}, we choose following four diverse state of the art algorithms for crystal property prediction.
\begin{enumerate}
	\item \textbf{CGCNN}~\cite{xie2018crystal}: This work generates crystal graphs from inorganic crystal materials and builds a graph convolution based supervised model for predicting various properties of the crystals.
	\item \textbf{GATGNN}~\cite{louis2020graph}: In this work, authors have incorporated a graph neural network with multiple graph-attention layers (GAT) and a global attention layer, which can learn efficiently the importance of different complex bonds shared among the atoms within each atom’s local neighborhood.
	\item \textbf{ALIGNN}~\cite{choudhary2021atomistic}: In this work, authors adopted line graph neural networks to develop an alternative way to
    include angular information into convolution layer which alternates between message passing on the bond graph and its bond-angle line graph.
	\item \textbf{CrysXPP}~\cite{das2022crysxpp}: In this work,  the authors train an autoencoder (CrysAE) on a volume of un-tagged crystal graphs and then the learned knowledge is (transferred to)  used to initialize the encoder of  CrysXPP, which is fine-tuned with property specific tagged data.Also they design a feature selector that helps to interpret the model’s prediction. 
\end{enumerate}
\subsection{Baseline Implementations.} We used the available PyTorch implementations of all the baselines, viz, CGCNN\footnote{https://github.com/txie-93/cgcnn.git}, GATGNN\footnote{https://github.com/superlouis/GATGNN.git}, CrysXPP\footnote{https://github.com/kdmsit/crysxpp.git} and ALIGNN\footnote{https://github.com/usnistgov/alignn.git}. In order to retrofit \aaaimodel{} to each SOTA model, we modified their supervised loss into the multi-task loss proposed in Equation \ref{eq:finetune_loss} (main manuscript) and train each model. To ensure a fair comparison between all the baselines and perform the knowledge distillation between node embeddings, we have ensured node embedding dimension for each SOTA and \aaaimodel{} as 64.

\subsection{Training Setup / Hyper-parameter Details}
We use five convolution layers of the encoder module to train \aaaimodel{} and train it for 200 epochs using Adam \cite{kingma2014adam} for optimization with a learning rate of 0.03. We keep the embedding dimension for each node as 64, batch size of data as 128, and equal weightage (0.25) for $\alpha$, $\beta$, $\gamma$, and $\lambda$ of Equation \ref{eq:pretrain_loss}. During property prediction, we used the default configuration for hyper-parameters for all four vanilla SOTA models. However, while predicting properties by the distilled version of the SOTA models we keep node embedding dimension for each SOTA as 64, and train each model for 500 epochs using Adam~\cite{kingma2014adam} optimization and keep $\delta$ as 0.5. We perform the experiments in shared servers having Intel E5-2620v4 processors which contain 16 cores/thread  and four GTX 1080Ti 11GB GPUs each. \\
\rev{\textbf{Sensitivity to temperature $\tau$:} We tuned the temperature parameter $\tau$ over the range $ \{0.06 \text{-} 0.1 \}$ and observed no noticeable differences in either the pretraining loss or downstream task performance. Consequently, we fix $\tau = 0.07$, which is used as the default value in NTXent.
\\
\textbf{Sensitivity to embedding dimension:} We fix the node embedding dimension to 64, consistent with most state-of-the-art property prediction models. Increasing the embedding size to 128 or 256 yields only marginal improvements in downstream performance, while significantly increasing the number of parameters and training cost. Therefore, we adopt an embedding dimension of 64 to balance performance and computational efficiency.}

\begin{table*}
  \centering
    \setlength{\tabcolsep}{11 pt}
    \resizebox{1\textwidth}{!}{
      \begin{tabular}{c | c c | c c| c c | c c}
        \toprule
        Property &  CGCNN & CGCNN & CrysXPP & CrysXPP & GATGNN & GATGNN & ALIGNN & ALIGNN \\
        & & (Distilled) &  & (Distilled) &  & (Distilled) &  & (Distilled)\\
        \midrule
        Formation Energy & 0.039 & \textbf{0.032} & 0.041 & \textbf{0.035} & 0.096 & \textbf{0.091} & 0.026 & \textbf{0.024}  \\
        Bandgap (OPT)     & 0.388 & \textbf{0.293} & 0.347 & \textbf{0.287} & 0.427 & \textbf{0.403} & 0.271 & \textbf{0.253}  \\
        \midrule
        Formation Energy  & 0.063 & \textbf{0.047} & 0.062 & \textbf{0.048} & 0.132 & \textbf{0.117} & 0.036 & \textbf{0.035}  \\
        Bandgap (OPT)     & 0.200 & \textbf{0.160} & 0.190 & \textbf{0.176} & 0.275 & \textbf{0.235} & 0.148 & \textbf{0.131} \\
        Total Energy     & 0.078 & \textbf{0.053} & 0.072 & \textbf{0.055} & 0.194 & \textbf{0.137} & 0.039 & \textbf{0.038}  \\
        Ehull             & 0.170 & \textbf{0.121} & 0.139 & \textbf{0.114} & 0.241 & \textbf{0.203} & 0.091 & \textbf{0.083} \\
        Bandgap (MBJ)     & 0.410 & \textbf{0.340} & 0.378 & \textbf{0.350} & 0.395 & \textbf{0.386} & 0.331 & \textbf{0.325} \\
        Spillage     & 0.386 & \textbf{0.374} & 0.363 & \textbf{0.357} & 0.350 & \textbf{0.348} & 0.358 & \textbf{0.356} \\
        SLME (\%)     & 5.040 & \textbf{4.790} & 5.110 & \textbf{4.630} & 5.050 & \textbf{4.950} & 4.650 & \textbf{4.590} \\
        Bulk Modulus (Kv)     & 12.45 & \textbf{12.31} & 13.61 & \textbf{12.70} & 11.64 & \textbf{11.53} & 11.20 & \textbf{10.99} \\
        Shear Modulus (Gv)     & 11.24 & \textbf{10.87} & 11.20 & \textbf{10.56} & 10.41 & \textbf{10.35} & 9.860 &\textbf{9.800}  \\
         \bottomrule
      \end{tabular} 
      }
   \caption{Summary of the prediction performance (MAE) of different properties (belong to first class) in Materials project (Top)  and JARVIS-DFT (Bottom). Model M is the vanilla variant of a SOTA model and M  (Distilled) is the distilled variant using the pretrained \aaaimodel{}. The best performance is highlighted in bold.}
  \label{tbl-full-data-1}
\end{table*}
\begin{table*}
  \centering
  \small
    \setlength{\tabcolsep}{11 pt}
    \resizebox{1\textwidth}{!}{
      \begin{tabular}{c | c c | c c| c c | c c}
        \toprule
        Property  & CGCNN & CGCNN & CrysXPP & CrysXPP & GATGNN & GATGNN & ALIGNN & ALIGNN\\
        &  & (Distilled) &  & (Distilled) &  & (Distilled) &  & (Distilled)\\
        \midrule
        $\epsilon_x$ (OPT)    & 24.93 & \textbf{23.86} & 24.47 & \textbf{23.33} & 26.79 & \textbf{25.23}  & 21.42 & \textbf{21.37}\\
        $\epsilon_y$ (OPT)    & 25.06 & \textbf{24.08} & 24.84 & \textbf{23.98} & 26.77 & \textbf{25.02}  & 21.66 & \textbf{21.25}   \\
        $\epsilon_z$ (OPT)    & 24.99 & \textbf{23.85} & 24.13 & \textbf{23.76} & 26.09 & \textbf{24.81}  & \textbf{20.51} & 21.03 \\
        $\epsilon_x$ (MBJ)     & 27.18 & \textbf{26.64} & 28.40 & \textbf{26.03} & \textbf{27.56} & 27.61 & 24.76 & \textbf{24.32}\\
        $\epsilon_y$ (MBJ)     & 27.14 & \textbf{25.80} & 26.92 & \textbf{25.34} & 27.21 & \textbf{26.89}  & 23.99 & \textbf{23.83}\\
        $\epsilon_z$ (MBJ)     & 28.38 & \textbf{26.08} & 26.76 & \textbf{25.97} & 25.50 & \textbf{24.84}  & \textbf{23.93} & 24.59\\
        n-Seebeck  & 50.32 & \textbf{46.66} & 48.79 & \textbf{46.53} & 52.03 & \textbf{49.46}  & 42.85 & \textbf{42.13}\\
        n-PF  & 528.25 & \textbf{506.41} & 502.79 & \textbf{498.43} & 511.56 & \textbf{505.22} & 477.09 & \textbf{474.88} \\
      p-Seebeck  & 53.01 & \textbf{49.15} & 50.13 & \textbf{48.67} & 54.99 & \textbf{50.72}  & 46.04 & \textbf{45.19}\\
        p-PF  & 513.57 & \textbf{501.11} & 493.35 & \textbf{489.54} & 516.84 & \textbf{494.92}  & 460.11 & \textbf{447.81}\\
         \bottomrule
      \end{tabular}
      }
  \caption{Summary of the prediction performance (MAE) of different properties (belong to second class) in JARVIS-DFT dataset where performance improvement is modest. Model M is the vanilla variant of a SOTA model and M  (Distilled) is the distilled variant using the pretrained \aaaimodel{}. All the models are trained on $80\%$  data, validated on $10\%$ and evaluated on $10\%$ of the data. The best performance is highlighted in bold.}
  \label{tbl-full-data-2}
\end{table*}

\subsection{Downstream Task Evaluation}
\label{main_results}
To evaluate the effectiveness of \aaaimodel{}, we choose four diverse state of the art algorithms for crystal property prediction, CGCNN~\cite{xie2018crystal}, GATGNN~\cite{louis2020graph}, CrysXPP~\cite{das2022crysxpp} and ALIGNN~\cite{choudhary2021atomistic}. To train these models for any specific property, we adopt the multi-task setting discussed in equation \ref{eq:finetune_loss} ,where we distill relevant knowledge from the pre-trained \aaaimodel{} to each of these algorithms to predict different properties. 
We report mean absolute error (MAE) of the predicted and actual value of a particular property to compare the performance of different participating methods. For each property, we trained on $80\%$ data, validated on $10\%$ and evaluated on $10\%$ of the data. We compare the results of distilled version of each SOTA model with its vanilla version (version reported in the respective papers), to show the effectiveness of the proposed framework.\\\\
\noindent\textbf{Results.} 
As mentioned in section \ref{sec_crysgnn_dataset}, properties of crystalline materials can be broadly classified into two categories: 1) properties like formation energy, bandgap, total energy, bulk modulus, etc. which depend greatly on crystal structures and atom features, and 2) properties like $\epsilon_x$, $\epsilon_y$, $\epsilon_z$, n-Seebeck, n-PF, etc. which depend on the precise description of the materials’ electronic structure. First, we evaluate  performance of SOTA models with distilled knowledge from \aaaimodel{} for the first class of properties. In Table \ref{tbl-full-data-1}, 
we report MAE of different crystal properties of Materials project and JARVIS-DFT datasets.   In the distilled version of the SOTA models, while training the model, we distill information from the pre-trained \aaaimodel{} model. We observe that the  distilled version of any state-of-the-art model outperforms the vanilla model across all the properties.  In specific, average improvement in CGCNN, CrysXPP, GATGNN and ALIGNN are 16.20\%, 12.21\%, 8.02\% and 4.19\%, respectively. 
These improvements are particularly significant as in most of the cases, the MAE is already low for SOTA models, still pretraining enables improvement over that.  In fact, lower the MAE, higher the improvement. 
We calculate Spearman's Rank Correlation between MAE for each property across different SOTA models and their improvement due to distilled knowledge and found it to be very high (0.72), which supports  the aforementioned observations.
The average relative improvement across all properties for ALIGNN (4.19\%) and GATGNN (8.02\%) is lesser compared to CGCNN (16.20\%) and CrysXPP (12.21\%). A possible reason could be that ALIGNN and GATGNN are more complex models (more number of parameters) than the pre-trained \aaaimodel{} framework.
Hence designing a deeper pre-training model or additionally incorporating angle-based information (ALIGNN) or attention mechanism (GATGNN) as a part of pre-training framework may help to improve further.  This requires further investigation and we keep it as a scope of future work. \\\\
We further evaluate the effectiveness of CrysGNN for the second class of properties for both the vanilla and distilled model and report the MAE in Table \ref{tbl-full-data-2}. We observe, for this class of properties all the SOTA models had a higher MAE and though pre-training enhances the performance, the improvement is modest. In specific, average improvement in CGCNN, CrysXPP, GATGNN and ALIGNN are  6\%, 4\%, 3.6\% and 0.8\%, respectively. A potential reason is, electronic dielectric constant, Seebeck coefficients, and power factors all depend greatly on the precise description of the materials' electronic structure, which is neither captured by the SOTA models nor by the pre-trained \aaaimodel{} framework explicitly. Hence error is high for these properties by SOTA models and injecting distilled structural information from the pre-trained model is able to achieve only modest improvements.

\subsection{Comparison with Existing Pre-trained Models.} We further demonstrate the effectiveness of the knowledge distillation method vis-a-vis the conventional fine-tuning approaches. Note that the encoding architecture is same for \aaaimodel{}, CGCNN, and CrysXPP. CrysXPP is very similar to a {\em pretrained-finetuned} version of CGCNN. 
Thus we compare distilled version of CGCNN  with finetuned version of \aaaimodel{} and  CrysXPP. 
Additionally, we consider  Pretrain-GNN~\cite{hu2020pretraining} which is a popular pre-training algorithm for molecules. We pre-train all the baseline models on our curated 800K untagged crystal data and fine-tune on seven properties in JARVIS dataset and report the MAE in Table \ref{tbl-distill-finetune}. 
We feed multi-graph structure of the crystal material (as discussed in ``Crystal Graph Representation") in Pretrain-GNN and try different combinations of node-level pre-training strategy along with the graph-level supervised pre-training (as suggested in ~\cite{hu2020pretraining}) and report the minimum MAE for any specific property. For finetuned \aaaimodel{}, we take the pre-trained encoder of \aaaimodel{} and feed a multilayer perceptron to predict a specific property. We observe that distilled CGCNN outperforms finetuned version of \aaaimodel{} and both the baselines with a significant margin for all the properties. Pretrain-GNN performs the worst and the potential reason is - it is designed considering simple two-dimensional structure of molecules with a minimal set of node and bond features, which is hard to generalize for crystal materials which have very complex structure with a rich set of node and edge features.

\begin{table*}
  \centering
    \setlength{\tabcolsep}{20 pt}
    \resizebox{0.95\textwidth}{!}{
      \begin{tabular}{c | c c c c}
        \toprule
        Property & CGCNN & \aaaimodel{} & CrysXPP & Pretrain - GNN \\
        & (Distilled) & (Finetuned) & & \\
        \midrule
        Formation Energy  & \textbf{0.047} & 0.056 & 0.062 &  0.764 \\
        Bandgap (OPT)     &  \textbf{0.160} & 0.183 & 0.190 &  0.688 \\
        Total Energy     &  \textbf{0.053} & 0.069 & 0.072 &  1.451 \\
        Ehull             &  \textbf{0.121} & 0.130 & 0.139 &  1.112 \\
        Bandgap (MBJ)     &  \textbf{0.340} & 0.371 & 0.378 &  1.493 \\
        Bulk Modulus (Kv) & \textbf{12.31}  & 13.42 & 13.61 &  20.34 \\
        Shear Modulus (Gv)& \textbf{10.87}  & 11.07 & 11.20 &  16.51 \\
        SLME (\%)         & \textbf{4.791}   & 5.452  & 5.110  &  9.853 \\
        Spillage          & \textbf{0.354}  & 0.374 & 0.363 &  0.481 \\
        \bottomrule
      \end{tabular} 
      }
   \caption{Comparison of the prediction performance (MAE) of seven properties in JARVIS-DFT between \aaaimodel{} and existing pretrain-finetune models, the best performance is highlighted in bold.}
  \label{tbl-distill-finetune}
\end{table*}

\subsection{Effectiveness on sparse training dataset.}
To demonstrate the effectiveness of the pre-training in limited data settings, we conduct additional set of experiments under different training data split. In specific, we vary available training data from 20 to 60 \%, train different SOTA models and check their performance on test dataset. We report the MAE values of different baselines and their distilled version in Table \ref{tbl-limited-data} for five different properties, for which available data is very limited. We observe that the distilled version of any SOTA model consistently outperforms its vanilla version even in the limited training data setting, which illustrates the robustness of our pre-training framework. 
Specifically, for CGCNN and GATGNN, the improvements are more for 20\% training data, because these deep models suffer from data scarcity issue and  with distilled knowledge from the pre-trained model they are able to mitigate the issue. ALIGNN achieves moderate improvements as before but surprisingly, we found the least improvement (additional improvement for 20\% training data compared to other settings) for CrysXPP. CrysXPP itself pre-trains an autoencoder and transfers the learned information to the property predictor. Hence, the vanilla version performs well in the limited data settings.  \\
Moreover, in some cases, we observe that the MAE values of the distilled version of a model using lesser training data is better than vanilla version of the model using more training data. For example, while predicting Bandgap (MBJ), CGCNN (Distilled) with 20\% and 40\% training data outperforms CGCNN with 40\% and 60\% training data, respectively. These results verify that even with lesser training data, the distilled information from the large pre-trained model gives performance comparable to using larger training data by the original model.

\begin{table*}
  \centering
  \small
    \setlength{\tabcolsep}{6 pt}
    \resizebox{\textwidth}{!}{
    \begin{tabular}{cccccccccc}
    \toprule
    Property & Train-Val  & CGCNN & CGCNN & CrysXPP & CrysXPP & GATGNN & GATGNN & ALIGNN & ALIGNN\\
     & -Test(\%) &  & (Distilled) &  & (Distilled) &  & (Distilled) &  & (Distilled)\\
    \midrule
     \multirow{3}{*}{\shortstack{Bandgap \\ (MBJ)}}
     & 20-10-70  & 0.588 & 0.453\textbf{*} (23.04) & 0.598 & 0.450\textbf{*} (24.82) & 0.541 & 0.521 (3.70) & 0.497 & 0.485 (2.53) \\
     & 40-10-50  & 0.532 & 0.419\textbf{*} (21.41) & 0.496 & 0.405\textbf{*} (18.40) & 0.462 & 0.448\textbf{*} (2.81) & 0.404 & 0.395 (2.20)\\
     & 60-10-30  & 0.449 & 0.364 (19.08) & 0.435 & 0.360 (17.36) & 0.449 & 0.439 (2.29) & 0.387 & 0.380 (1.98)\\
     \midrule
     \multirow{3}{*}{\shortstack{Bulk Modulus \\ (Kv)}}
     & 20-10-70  & 16.91 & 16.26 (3.80) & 15.42 & 14.25\textbf{*} (7.59) & 14.80 & 14.19 (4.12) & 14.70 & 14.06 (4.35)\\
     & 40-10-50  & 14.81 & 14.46 (2.36) & 15.13 & 14.02\textbf{*} (7.34) & 12.98 & 12.59 (3.00) & 12.47 & 12.11 (2.89)\\
     & 60-10-30  & 14.23 & 14.05 (1.26) & 14.76 & 13.73 (6.98) & 12.01 & 11.75 (2.16) & 11.23 & 11.01 (1.96)\\
     \midrule
     \multirow{3}{*}{\shortstack{Shear Modulus \\(Gv)}}
     & 20-10-70  & 13.89 & 12.50 (10.01) & 13.39 & 12.07\textbf{*} (9.86) & 12.83 & 12.42 (3.20) & 12.71 & 12.31 (3.15)\\
     & 40-10-50  & 12.04 & 11.54\textbf{*} (4.15) & 12.16 & 11.01\textbf{*} (9.46)  & 11.43 & 11.23 (1.75) & 10.98 & 10.67 (2.82)\\
     & 60-10-30  & 11.75 & 11.31 (3.74) & 11.77 & 10.67 (9.35) & 10.65 & 10.47 (1.69) & 10.24 & 10.04 (1.95)\\
     \midrule
     \multirow{3}{*}{\shortstack{SLME (\%)}}
     & 20-10-70  & 7.13 & 6.62 (7.17) & 7.05 & 5.90\textbf{*} (16.40) & 6.35 & 6.02 (5.20) & 6.36 & 6.27 (1.43)\\
     & 40-10-50  & 6.14 & 5.78 (5.90) & 6.89 & 5.81 (15.67) & 5.78 & 5.63 (2.60) & 5.65 & 5.57 (1.42)\\
     & 60-10-30  & 5.55 & 5.24 (5.68) & 5.41 & 4.84 (10.54) & 5.48 & 5.34 (2.55) & 4.88 & 4.82 (1.33)\\
     \midrule
     \multirow{3}{*}{\shortstack{Spillage}}
     & 20-10-70  & 0.424 & 0.389\textbf{*} (8.43) & 0.422 & 0.403 (4.45) & 0.406 & 0.392\textbf{*} (3.5) & 0.402 & 0.392 (2.49)\\
     & 40-10-50  & 0.408 & 0.391\textbf{*} (4.28) & 0.398 & 0.384 (3.57) & 0.396 & 0.388 (1.94) & 0.378 & 0.372 (1.72)\\
     & 60-10-30  & 0.395 & 0.380 (3.80) & 0.379 & 0.368 (3.00) & 0.382 & 0.377 (1.36) & 0.362 & 0.357 (1.13)\\
     
    \bottomrule
  \end{tabular}
  }
  \caption{MAE values of five different properties in JARVIS-DFT dataset with the increase in training instances from 20 to 60\%. Model M is the vanilla variant of the SOTA model, M (Distilled) is distilled variant of it and relative improvement is mentioned in bracket. Distilled knowledge from \aaaimodel{} improves the performance of all the baselines consistently.  In few cases, we observe that MAE values of the distilled version of a model using even lesser training data is better than vanilla version of the same model using more training data and highlight it with *.}
  \label{tbl-limited-data}
\end{table*}

\subsection{Analysis of Different Pre-training Losses}
We perform an ablation study to investigate the influence of different pre-training losses in enhancing the SOTA model performance. While pre-training \aaaimodel{} (Eq. \ref{eq:pretrain_loss}), we capture both local chemical and global structural information via node and graph-level decoding, respectively. 
Further, we are curious to know the influence of each of these decoding policies independently in the downstream property prediction task.  
In specific, we conduct the ablation experiments, where we pre-train \aaaimodel{} with (a) only node-level decoding ($\mathcal{L}_{FR}$, $\mathcal{L}_{CR}$), (b) only graph-level decoding ($\mathcal{L}_{SG}$, $\mathcal{L}_{NTXent}$). Further, we perform ablations with individual graph-level losses, and pretrain with (c) removing $\mathcal{L}_{NTXent}$ (node-level with $\mathcal{L}_{SG}$ (space group)) and (d) removing $\mathcal{L}_{SG}$ (node-level with $\mathcal{L}_{NTXent}$(crystal system)). We train two baseline models, CGCNN and ALIGNN, with distilled knowledge from all the aforementioned variants of the pre-trained model and evaluate the performance on four crystal properties. \\

\begin{figure}[!thb]
	\centering
	\boxed{\includegraphics[width=\columnwidth]{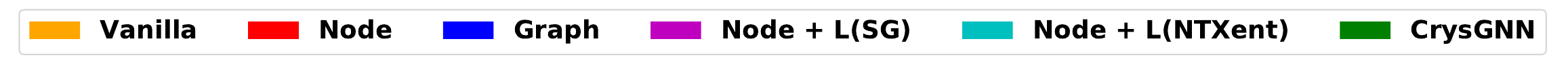}}
	\\
	\vspace*{-1mm}
	\subfloat[Formation Energy]{
		\boxed{\includegraphics[width=0.5\columnwidth]{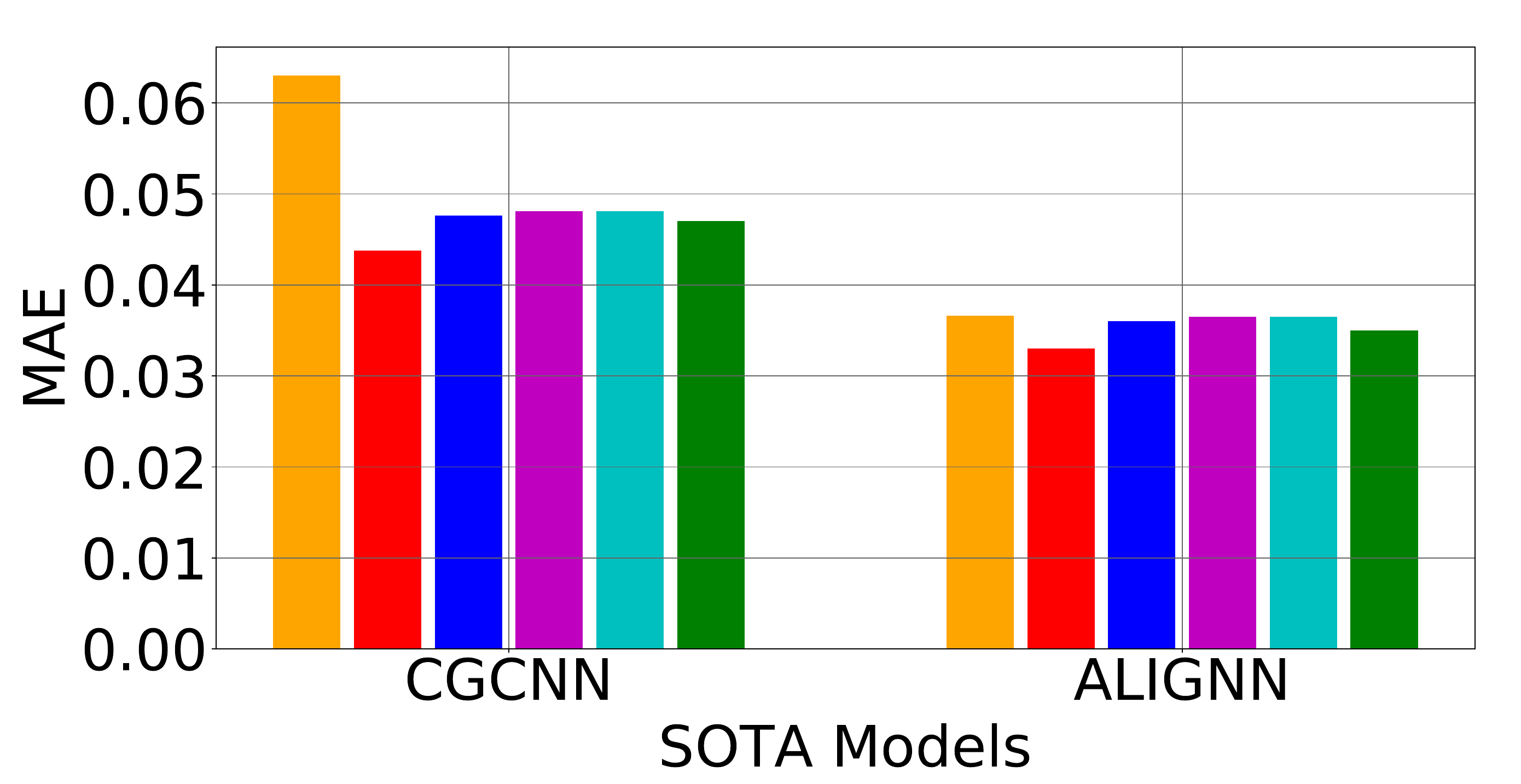}}}
	\subfloat[Bandgap (MBJ)]{
		\boxed{\includegraphics[width=0.5\columnwidth]{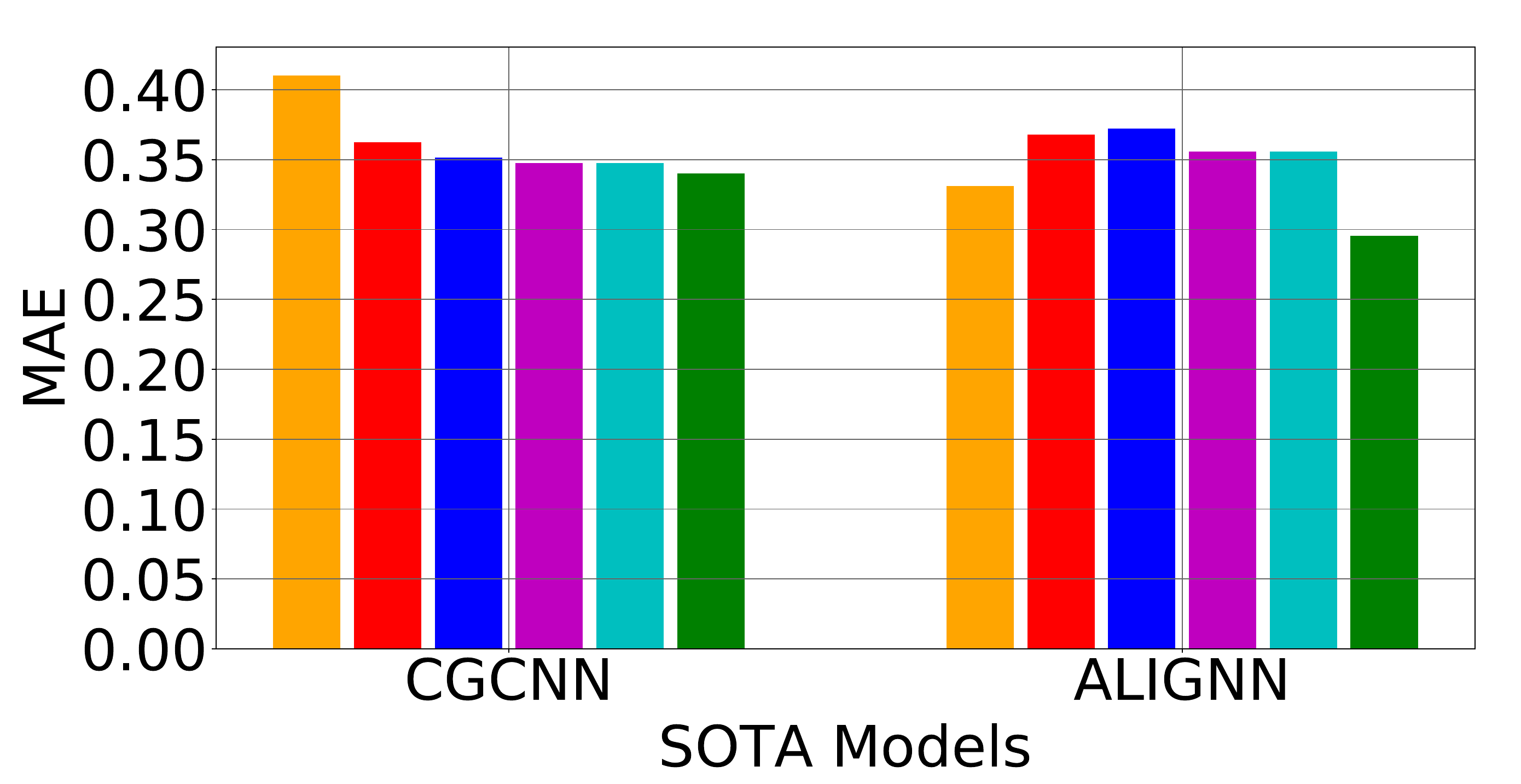}}}\\
	\subfloat[Total Energy]{
		\boxed{\includegraphics[width=0.5\columnwidth]{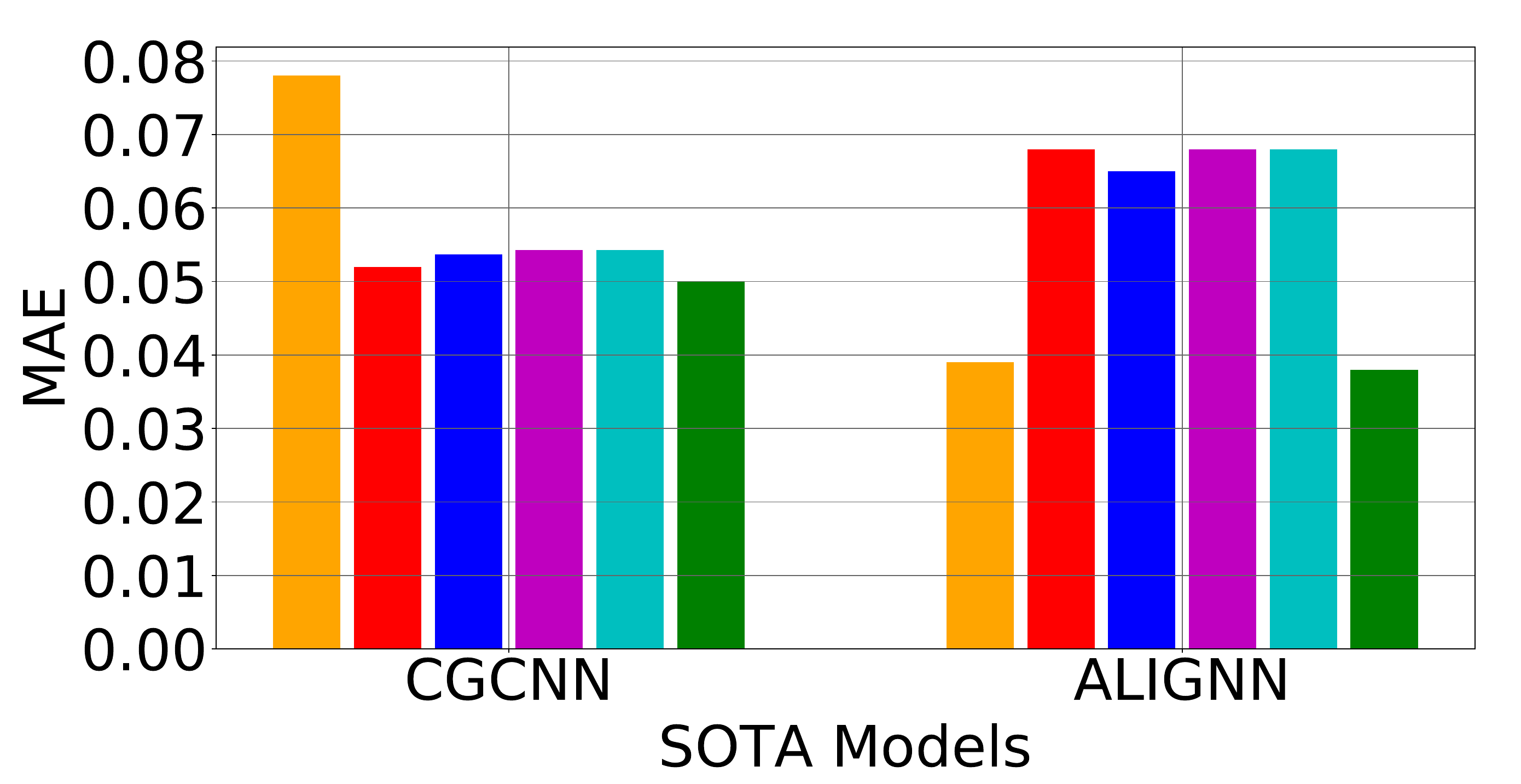}}}
	\subfloat[Bulk Modulus (Kv)]{
		\boxed{\includegraphics[width=0.5\columnwidth]{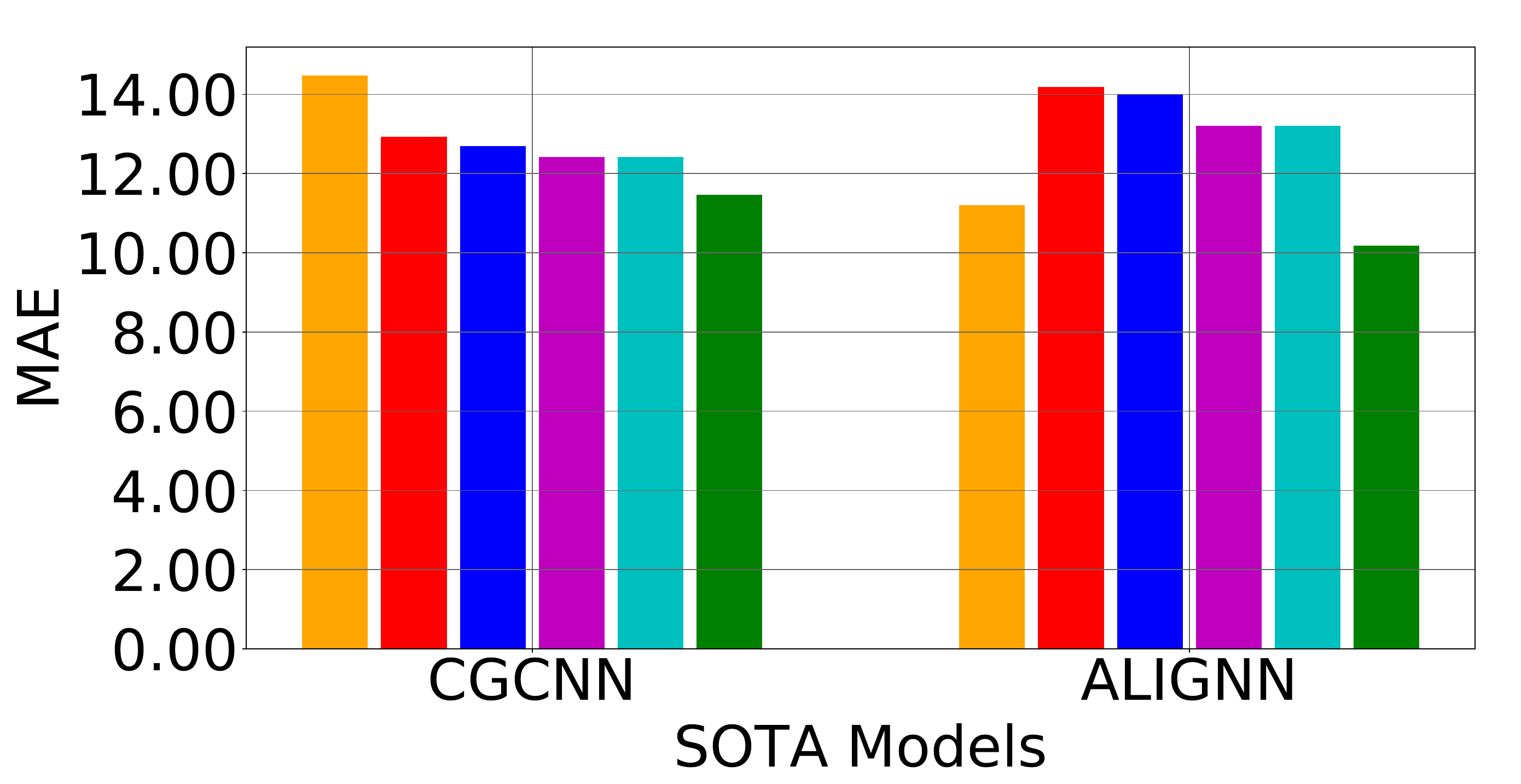}}}
	\caption{Summary of experiments of ablation study on importance of different pre-training loss components on \aaaimodel{} training and eventually its effect on CGCNN and ALIGNN models on four different properties (MAE for property prediction). (i) Vanilla: SOTA based model (without distillation) and all the other cases are SOTA models (distilled) from different pre-trained version of \aaaimodel{}.
	(ii) Node : only node-level pre-training, (iii) Graph : only graph-level pre-training, (iv) Node + L(SG) : node-level and $\mathcal{L}_{SG}$, (v) Node + L(NTXent) : node-level and $\mathcal{L}_{NTXent}$ and (vi) \aaaimodel{}~: both node and graph-level pre-training.}
	\label{fig:ablation_diff_loss}
\end{figure}

Experimental results are presented in Fig.~\ref{fig:ablation_diff_loss}. 
We can observe clearly that all the variants offer significant performance gain in all four properties using the combined node and graph-level pre-training, compared to node-level or graph-level pre-training separately. \rev{Only exception is formation energy, where only node-level pre-training produces less error compared to other variants, in both the baseline. Formation energy of a crystal is defined as the difference between the energy of a unit cell comprised of $N$ chemical species and the sum of the chemical potentials of all the $N$ chemical species. Hence pre-training at the node-level (node features and connection) is adequate for enhancing performance of formation energy prediction and incorporating graph-level information works as a noisy information, which degrades the performance.} We also observe improvement in performance using both supervised and contrastive graph-level losses ($\mathcal{L}_{SG}$ and $\mathcal{L}_{NTXent}$), compared to using only one of them, which proves the learned representation via supervised and contrastive learning is more expressive that using any one of them.
Moreover, in ALIGNN, with either node or graph-level pre-training separately, performance degrades across different properties. ALIGNN explicitly captures the three body interactions which drive its performance, to replicate that inclusion of both node and graph information is necessary.
\begin{table*}[!thb]
    \centering
    \setlength{\tabcolsep}{6pt}
    \resizebox{1.0\textwidth}{!}{
    \begin{tabular}{l c c c c c c c c}
    \toprule
	\textbf{Experiment Settings}  & CGCNN & CGCNN & CrysXPP & CrysXPP & GATGNN & GATGNN & ALIGNN & ALIGNN\\
	 &  & (Distilled) &  & (Distilled) &  & (Distilled) &  & (Distilled)\\
	\midrule
	\textbf{\vtop{\hbox{\strut Train on DFT }\hbox{\strut Test on Experimental}}} 
    & 0.265 & 0.244 (7.60) & 0.243 & 0.225 (7.40) & 0.274 & 0.232 (15.3) 	& 0.220 & 0.209 (5.05)  \\
	\midrule
	\textbf{\vtop{\hbox{\strut Train on DFT and 20 \% Experimental }\hbox{\strut Test on 80 \% Experimental}}}  
	 & 0.144 & 0.113 (21.7) & 0.138 & 0.118 (14.2) & 0.173 &  0.168 (2.70) & 0.099 & 0.094 (5.60) \\
	\midrule
	\textbf{\vtop{\hbox{\strut Train on DFT and 80 \% Experimental }\hbox{\strut Test on 20 \% Experimental}}} 
	 & 0.094 & 0.073 (22.7) & 0.087 & 0.071 (18.4) & 0.113  & 0.109 (3.40) & 0.073 & 0.069 (5.90) \\
    \bottomrule
    \end{tabular}
	}
	\caption{MAE of predicting experimental values by different SOTA models and their distilled versions with full DFT data and  different percentages of experimental data for formation energy in OQMD-EXP dataset. Relative improvement in the distilled model is mentioned in bracket.}
    \label{tab:fe}
\end{table*}
\subsection{Removal of DFT error bias using experimental data}
One of the fundamental issues in material science is that experimental data for crystal properties are very rare~\cite{kubaschewski1993materials,bracht1995properties,turns1995understanding}. Hence existing SOTA models rely on DFT calculated data to train their parameters. 
However, mathematical approximations in DFT calculation lead to erroneous predictions (error bias) compared to the actual experimental values of a particular property. Hence 
DFT error bias is prevalent in all SOTA models. 
\cite{das2022crysxpp} has shown that pre-training helps to remove DFT error bias when fine-tuned with experimental data. Hence, 
we investigate whether SOTA models can remove the DFT error with distilled knowledge from pre-trained model, using a small amount of available experimental data. In specific, we consider OQMD-EXP dataset to conduct an experiment, where we train SOTA models and their distilled variants with available DFT data and different percentages of experimental data for formation energy. We report the MAE of different SOTA models and their distilled variant in Table \ref{tab:fe}.  We observe, with more amount of experimental training data, all the SOTA models are minimizing the error consistently. Moreover, with distilled knowledge from pre-trained \aaaimodel{}, all SOTA models are reducing MAE further and we observe consistently larger degree of improvement with more amount of experimental training data in almost all the models. 

\section{Conclusion}
\label{conclusion}
In this work, we present a novel but simple pre-trained GNN framework, \aaaimodel{}, for crystalline materials, which captures both local chemical and global structural semantics of crystal graphs. To pre-train the model, we curate a huge dataset of  800k unlabelled crystal graphs. 
Further, while predicting different crystal properties, we distill important knowledge from \aaaimodel{} and inject it into different state of the art property predictors and enhance their performance. Extensive experiments on multiple popular datasets and diverse set of SOTA models show that with distilled knowledge from the pre-trained model, all the SOTA models outperform their vanilla versions.
Extensive experiments show its superiority over conventional fine-tune models and its inherent ability to remove DFT-induced bias.
The pretraining framework can be extended beyond structural graph information in a multi-modal setting to include other important (text and image) information about a crystal which would be our immediate future work.


\clearemptydoublepage
\chapter{CrysMMNet: Multimodal Representation for Crystal Property Prediction}
\chaptermark{CrysMMNet [UAI-2023]}
\label{chap: crysmmnet}

\noindent\fbox{%
\parbox{\textwidth}{%
Work of this chapter are based on the following publication:  
\\
\textit{CrysMMNet: Multimodal Representation for Crystal Property Prediction.}
\\
\underline{Kishalay Das}, Pawan Goyal, Seung-Cheol Lee, Satadeep Bhattacharjee and Niloy Ganguly.
\\
UAI, 2023.}}
\vspace{1em}

\section{Introduction}
\label{uai_intro}
In the recent past, we have witnessed a surge of interest in developing machine learning models ~\cite{seko2015prediction,pilania2015structure,lee2016prediction,de2016statistical,seko2017representation,isayev2017universal,ward2017including,lu2018accelerated,im2019identifying} for fast and accurate property prediction of crystalline materials. 
Crystalline materials are typically modeled by a minimal unit cell containing all the constituent atoms in different coordinates, repeated infinite times in 3D space on a regular lattice, which makes material structures periodic in nature. A key challenge in learning crystal representation is how to capture accurately global periodic structural information along with local chemical semantics. Recent state-of-the-art models ~\cite{xie2018crystal,chen2019graph,louis2020graph, Wolverton2020,schmidt2021crystal,choudhary2021atomistic,hsu2021efficient,das2022crysxpp,yan2022periodic} construct multi-edge graphs for a 3D material structure where they create edges between nearby atoms within a pre-specified distance threshold in 3D space and apply GNN model to learn representations of crystal structures that are optimized for downstream property prediction tasks. Although existing variants of GNN models predict different crystal properties with high precision, they rely on a single modality of crystal data i.e crystal graph structure which limits the expressive power of these models. The architectural innovations of these approaches are primarily based on incorporating specific domain knowledge of the local bonding environment, such as explicitly encoding bond angle ~\cite{choudhary2021atomistic}, dihedral angle ~\cite{hsu2021efficient}, etc. but they fail to incorporate crucial global periodic structural information like lattice constraint, space group number, crystal symmetry, rotational information, component 3D orientation, heterostructure information, etc, which will enrich its representation and subsequently aid the property prediction accuracy.
\\\\
 In this work, we propose to learn a more robust and enriched representation by using multi-modal data i.e graph structure and textual description of materials. One of the major advantages of using the textual description of materials is it provides a diverse set of periodic structural information which is useful to estimate different crystal properties but difficult to incorporate explicitly into a graph structure. Leveraging textual modalities beyond graph structures of materials remains unexplored by the research community and to the best of our knowledge, there is no existing dataset containing textual descriptions of the materials. Hence, we first curate the textual dataset of two popular materials databases (Graph-based), Material Project (MP) and JARVIS, containing textual descriptions of each material of those databases. We used a popular tool robocrystallographer ~\cite{ganose2019robocrystallographer} to generate descriptions for global crystal structures, which looks at the structural symmetry, local environment, and extended connectivity to generate a description that includes space group number, crystal symmetry, rotational information, component orientations, heterostructure information, etc.\\
 Further, we propose, \uaimodel{} (\textbf{Crys}tal \textbf{M}ulti-\textbf{M}odal \textbf{Net}work), a simple multi-modal framework for crystalline materials, which has two components: Graph Encoder and Text Encoder. Given a material, Graph Encoder uses its graph structure and applies GNN based approach to encode the local neighborhood structural information around a node (atom), and subsequently learn graph (crystal) representation. On the contrary, Text Encoder is a transformer-based model, which encodes the global structural knowledge from the textual description of the material and generates a textual representation. Finally, both graph structural and textual representation are fused together to generate a more enriched multimodal representation of materials, which captures both global and local structural knowledge and subsequently improves property prediction accuracy.
 \\\\
 To show the merit of our proposed algorithm, we performed comprehensive experiments on two popular benchmark datasets, Materials Project and JARVIS-DFT, across ten diverse sets of properties and compare the results with popular state-of-the-art models. We observe that for all the properties \uaimodel{} can achieve the lowest error in comparison with other baseline models. In addition, our results demonstrate that multi-modal representation learning  helps to achieve even better improvements when the dataset is sparse. We also perform some ablation studies to investigate the expressiveness of textual representation and robustness of multimodal representation on different GNN architectural choices. Result shows, textual representations alone are not expressive enough to learn the structure-property relationship of the materials.
 Moreover, fusing both graph structural and textual representation together leads to substantial performance improvements for all the state-of-the-art GNN models compared to their vanilla versions. We also investigate the influence of local compositional information and global material structural knowledge encoded through
textual representation and found for all crystal properties both local and global knowledge improves the downstream property prediction accuracy. 
We have shared the textual dataset, that we have curated for both the benchmark material databases with the community for future use.\footnote{Source code and dataset of \uaimodel{} is made available at \url{https://github.com/kdmsit/crysmmnet}}
\begin{figure*}
	\centering
	\includegraphics[width=\columnwidth]{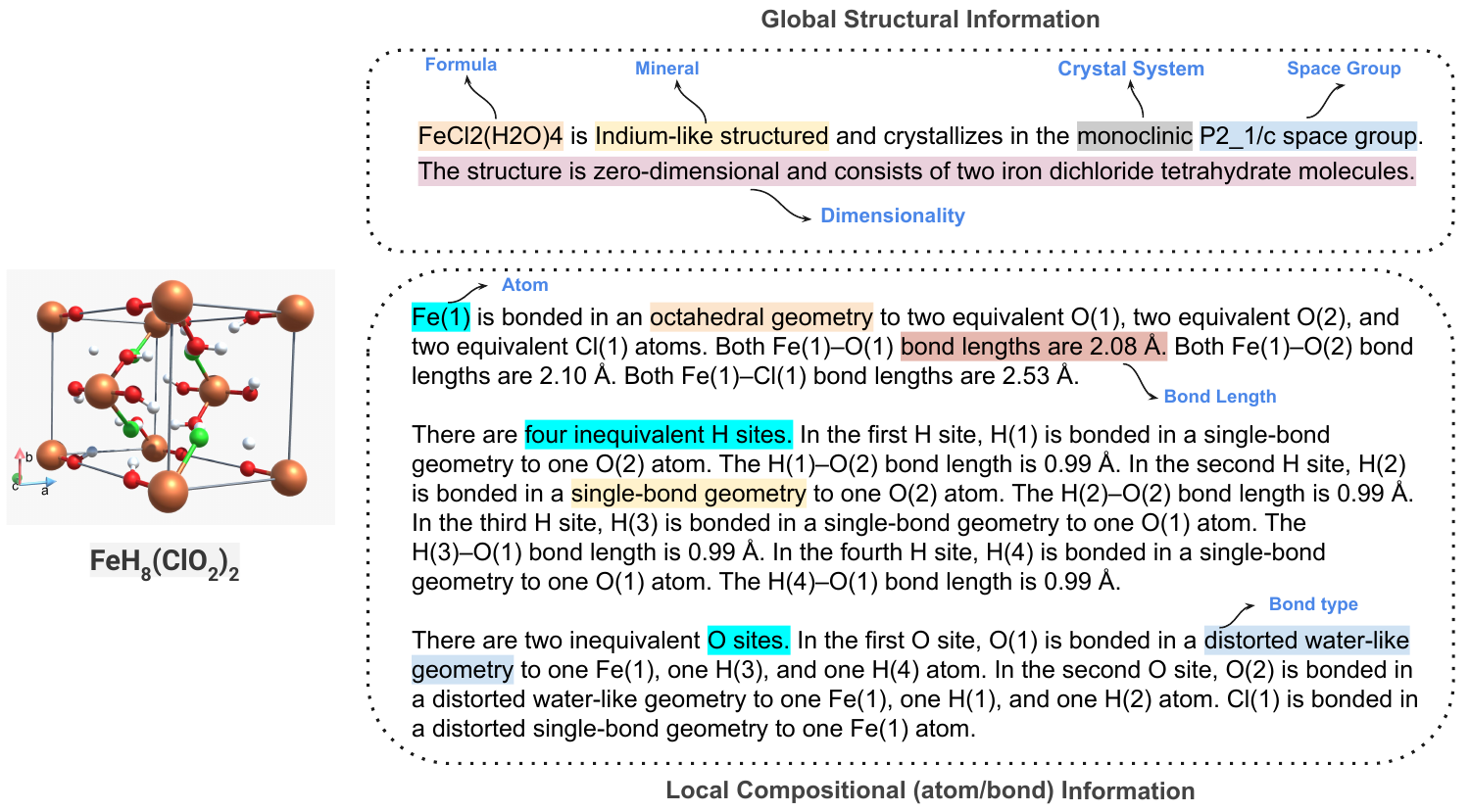}
	\caption{Textual description of $\mathbf{FeH_8(ClO_2)_2}$ material from JARVIS dataset generated by Robocrystallographer~\cite{ganose2019robocrystallographer}. The generated text contains both local chemical compositional information related to  atom/bonds (like site coordination, geometry, polyhedral connectivity, and tilt angles) and global structural knowledge (like mineral type, space group information, symmetry, and dimensionality).}
	\label{fig:crysmmnet_text}
\end{figure*}
\subsection{Crystal Representation}
The structure of a crystalline material can be modeled by a minimum unit cell, repeated infinite times in three-dimensional (3D) euclidean space on a regular lattice, which makes the crystalline structure periodic in nature. As mentioned in ~\cite{xie2021crystal,yan2022periodic}, for a given crystal we can describe its unit cell by two matrices: Feature Matrix ($\mathbf{X}$) and Coordinate Matrix ($\mathbf{C}$). Feature Matrix $\mathbf{X} =[\mathbf{x}_1,\mathbf{x}_2,...,\mathbf{x}_n]^T \in R^{n \times d}$ denotes atomic feature set of the material, where $\mathbf{x}_i \in R^{d}$ corresponds to the d-dimensional feature vector of i-th atom. On the other hand, Coordinate Matrix $\mathbf{C}=[\mathbf{c}_1,\mathbf{c}_2,...,\mathbf{c}_n]^T \in R^{n \times 3}$ denotes atomic coordinate positions, where $\mathbf{c}_i \in R^{3}$ corresponds to cartesian coordinates of i-th atom in the unit cell. Further, there is an additional lattice matrix $\mathbf{L}=[\mathbf{l}_1,\mathbf{l}_2,\mathbf{l}_3]^T \in R^{3 \times 3}$, which describes how a unit cell repeats itself in the 3D space towards $\mathbf{l}_1,\mathbf{l}_2$ and $\mathbf{l}_3$ direction to form the periodic 3D structure of the material. Formally, a given crystal can be defined as $\mathbf{M}=(\mathbf{X},\mathbf{C},\mathbf{L})$ and we can represent its infinite periodic structure as 
\begin{equation}
    \label{eq:crystal}
    \begin{split}
    & \mathbf{\hat{C}}  = \{ \mathbf{\hat{c}}_i |  \mathbf{\hat{c}}_i = \mathbf{c}_i + \sum_{j=1}^{3} k_j\mathbf{l}_j \}; \:
    \mathbf{\hat{X}}  = \{ \mathbf{\hat{x}}_i |  \mathbf{\hat{x}}_i = \mathbf{x}_i \}
    \end{split}
\end{equation}
where $k_1,k_2,k_3, i \in Z, 1 \leq i \leq n$.

\subsection{Crystal Property Prediction using GNNs}
Graph neural networks have emerged as highly promising models in various domains of computer science, showcasing significant potential in many real-world applications including social networks \cite{hamilton2017inductive,chen2018fastgcn,dai2018learning}, recommender systems \cite{berg2017graph,ying2018graph}, hyper-networks \cite{yadati2019hypergcn,bandyopadhyay2020hypergraph}, chemical and biological networks \cite{duvenaud2015convolutional,gilmer2017neural} etc.
Recently, graph neural network (GNN) based approaches have been very effective to encode structural information of the crystal materials into enriched embedding space so that it can predict different crystal properties with high accuracy. CGCNN \cite{xie2018crystal} is the first proposed model, which represents 3D crystal structure as an undirected weighted multi-edge graph $\mathcal{G} =(\mathcal{V}, \mathcal{E}, \mathcal{X}, \mathcal{F})$ where $\mathcal{V}$ denotes the set of nodes (atoms) in the unit cell of material and $\mathcal{E}=\{(u,v,k_{uv})\}$ denotes a multi-set of node pairs and $k_{uv}$ denotes number of edges between a node pair $(u,v)$. $\mathcal{X}=\{(x_{u} | u \in \mathcal{V})\}$ denotes the node feature set, which includes different chemical properties like electronegativity, valance electron, covalent radius, etc. Finally, $\mathcal{F}_i=\{\{s^k\}_{(u,v)} |  (u,v) \in \mathcal{E}, k\in\{1..k_{uv}\}\}$ denotes the multi-set of edge weights where $s^k$ corresponds to the $k^{th}$ bond length between a node pair $(u,v)$, which signifies the inter-atomic bond distance between two atoms. Further, CGCNN develops a graph convolution neural network to update node features based on their local chemical and structural environment. \\
Following CGCNN, there are a lot of subsequent studies ~\cite{chen2019graph,louis2020graph, Wolverton2020,schmidt2021crystal}, where authors proposed different variants of GNN architectures for effective crystal representation learning. Through multiple layers of graph convolutions, these models can implicitly encode many-body interactions. Further, ALIGNN ~\cite{choudhary2021atomistic} explicitly captures many-body interactions by incorporating bond angles and local geometric distortions into the GNN encoding module to enhance property prediction accuracy and became SOTA for a large range of properties. 
Recently, transformer-based architecture Matformer ~\cite{yan2022periodic} is proposed to learn the periodic graph representation of the material, which is invariant to periodicity and can capture repeating patterns explicitly. Matformer marginally improves the performance compared to ALIGNN, however, is much faster than it.
Moreover, scarcity of labeled data makes these models difficult to train for all the properties, and recently, some key studies \cite{jha2019enhancing,das2022crysxpp,das2023crysgnn} have shown promising results to mitigate this issue using transfer learning, pre-training, and knowledge distillation respectively.
\section{Methodology}
\label{uai_methodology}
\begin{figure*}
	\centering
        \includegraphics[width=\columnwidth]{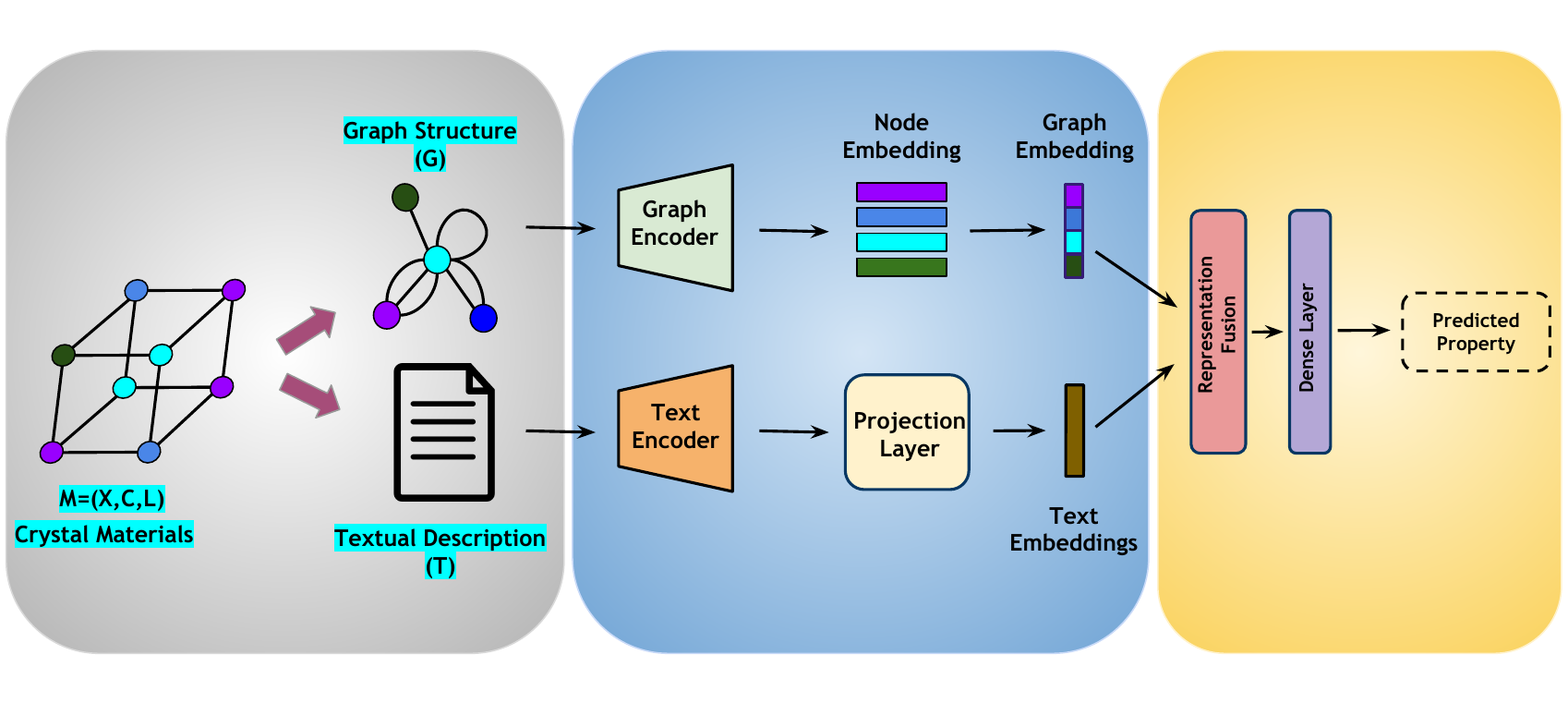}
	\caption{Overview of our adopted methodology \uaimodel{}. Given Crystal Material (M), we use two modalities - Graph Structure $(\mathcal{G})$ and Textual Description $(\mathcal{T})$. Graph structure $(\mathcal{G})$ is passed through a graph encoder to generate graph embedding $(\mathcal{Z}_\mathcal{G})$. Textual Description $(\mathcal{T})$ is fed through a text encoder followed by a projection layer to generate text embedding $(\mathcal{Z}_\mathcal{T})$. Finally, both the representations are fused together and predict the crystal properties based on joint modeling of the input modalities.}
	\label{fig:crysmmnet}
\end{figure*}
In this section, we first discuss the insights on the textual dataset that we have curated for two popular crystalline databases and explain the local compositional and global periodic information we are able to encode using textual representation, which is difficult to incorporate explicitly into a graph structure. Then we give a detailed overview of our proposed muti-modal framework \uaimodel{} that generates joint embedding for materials, which facilitates accurate property prediction. 
\subsection{Textual Dataset}
\label{textual}
Leveraging textual modalities beyond the conventional graph structure of the materials to capture both local atomic and global periodic knowledge remains largely unexplored by the research community. To the best of our knowledge, there is no existing dataset containing textual descriptions of the materials. Hence, we first curate the textual dataset for two popular material databases JARVIS and Material Project(MP), containing textual descriptions for each material of those databases. Conventionally, in these databases, the periodic structure of the materials is represented in Crystallographic Information File (CIF File). We use Robocrystallographer~\cite{ganose2019robocrystallographer}, which is a free utility, to generate a textual description of the material from the CIF file. Robocrystallographer decomposes crystal structures into local compositional (site coordination, geometry, polyhedral connectivity, and tilt angles) and global structural (mineral type, space group information, symmetry, dimensionality) components (Figure \ref{fig:crysmmnet_text}) and output this information in three formats: JSON for machine use, human-readable text, and machine learning format. In this work, we use human-readable text for collecting textual datasets, which are easily interpretable and resembles a human description of the crystal structure.\\
\textbf{Local compositional} information describes local chemical environments around different atoms and inter-atomic bonds in a unit cell. It provides a detailed description of different sites of the materials, like atomic compositions of different sites, site coordination, inter-atomic connectivity through chemical bonds, bond type, and length. Further, the geometry of each site is mentioned and the presence of corner-sharing tetrahedra connectivity is specified. On the contrary, \textbf{global structural} information illustrates the global environment i.e. periodic structure and orientation of the material in 3D space. The most useful information it provides is regarding crystal symmetry, which includes the specific space group  and crystal system the material belongs to. Space group is used to describe the symmetry of a unit cell of the crystal material in 3D space. In materials science literature there are 230 unique space groups and each crystal (graph) has a unique space group number. Further based on the space group level information can classify a crystal graph into 7 broad groups of crystal systems like Triclinic, Monoclinic, Orthorhombic, Tetragonal, Trigonal, Hexagonal, and Cubic. Moreover, it contains the mineral type of the material and the dimensionality of the crystal structure. \\
Minerals are naturally occurring, inorganic substances with a specific chemical composition and a crystalline structure. The most common types include silicates (which contain silicon and oxygen), carbonates (which contain carbon and oxygen), sulfates (which contain sulfur and oxygen), halides (which contain a halogen element), oxides (which contain oxygen and one or more other elements), and sulfides (which contain sulfur and one or more other elements). Examples of minerals in each category include quartz, calcite, gypsum, halite, hematite, and pyrite. The chemical composition and crystal structure of a mineral determine its properties, such as its hardness, color, and cleavage.
Further, dimensionality of a material is a significant global feature that refers to the number of dimensions that a particular component of the material spans. The dimensionality of a bonded cluster of atoms can be determined by calculating the rank of the subspace spanned by the central atom and its periodically connected neighbors.\\
A comprehensive understanding of both local and global environments is necessary for robust prediction of material properties. For example, in the case of formation energy, the local chemical environment, such as atom composition, bond length, and bond angles, plays a crucial role in determining the electronic and geometric structure of the material, which directly affects its formation energy. A slight variation in the local environment can result in significant changes in the electron density and, subsequently, the energy required to form the materials. Similarly, the global chemical environment, like the space group, has a profound impact on the formation energy by controlling the arrangement of atoms within the material. Different space groups are associated with different crystal structures and packing arrangements, which can lead to different formation energies.  Moreover, the study by Larsen \textit{et. al}~\cite{Larsen} showed that the formation energies of layered materials can be related to their dimensionality, highlighting the importance of considering this feature in the investigation of materials.
\subsection{Multi-Modal Framework}
Next, we propose a simple, yet effective multi-modal framework, \uaimodel{}, for graph and textual embedding of materials, which realizes material dataset  as $D =\{ (\mathcal{G}, \mathcal{T}),\mathcal{Y} \}$, where $\mathcal{G}$,$\mathcal{T}$ and $\mathcal{Y}$ denote multi-graph structure, textual description and property value the material respectively.  In our multi-modal architecture, the goal is to learn a
function $f_{\theta}(\mathcal{G}, \mathcal{T})$
\begin{equation}
    \label{eq:func}
    \begin{split}
    f_{\theta} : (\mathcal{G}, \mathcal{T}) \rightarrow \mathcal{Y}
    \end{split}
\end{equation}
By design, \uaimodel{} (as shown in Figure \ref{fig:crysmmnet}) is composed of three modules: graph encoder $M_V(\mathcal{G}) \rightarrow \mathcal{Z}_{\mathcal{G}}$, text encoder $M_L(\mathcal{T}) \rightarrow \mathcal{Z}_{\mathcal{T}}$, and joint embedding model $E(\mathcal{Z}_{\mathcal{G}},\mathcal{Z}_{\mathcal{T}}) \rightarrow \mathcal{Z}$, where $\mathcal{Z}_\mathcal{G}$,$\mathcal{Z}_\mathcal{G}$ and $\mathcal{Z}$ are graph-level, textual and multimodal embedding respectively. Next, we explain each part of the \uaimodel{} framework in detail. 
\subsubsection{Graph Encoder: }\uaimodel{} adopts a GNN architecture inspired by ALIGNN \cite{choudhary2021atomistic} as Graph Encoder, to encode the chemical, structural, and bond angular information of a crystal graph $\mathcal{G}$. We derive additional line graph $\mathcal{L}(\mathcal{G})$ from the crystal graph $\mathcal{G}$ to describe the angles between the edges in $\mathcal{G}$, where nodes and edges in line graph $\mathcal{L}(\mathcal{G})$ correspond to inter-atomic bonds and bond angles. We denote $h^l_i$, $e^l_{i,j}$ and $t^l_{i,j,k}$ as \textit{l}-th layer representation for i-th atom, $\{i,j\}$-th bond, and $\{i,j,k\}$-th angle (triplet) respectively. Graph encoder alternates edge-gated graph convolution layers between $\mathcal{L}(\mathcal{G})$ and $\mathcal{G}$ to propagate bond angular information through inter-atomic bond representation to atom embedding and vice versa. Specifically, at the $(l)$-th layer, given the line graph $\mathcal{L}(\mathcal{G})$, we apply Gated Graph ConvNet (GatedGCN) ~\cite{dwivedi2020benchmarkgnns} to update triplet representation and generate bond messages m as follows :
\begin{equation}
    \label{eq:linegraph}
    \begin{split}
    & {t}^{l+1}_{i,j,k} = {t}^{l}_{i,j,k} + \gamma \biggl(BN\biggl( A^{l}_{lg}e_{i,j}^{l} + B^{l}_{lg}e_{j,k}^{l} + C^{l}_{lg}{t}^{l}_{i,j,k}\biggr)\biggr) \\ 
    & \hat{t}^{l+1}_{i,j,k} = \frac{\sigma (t^{l+1}_{i,j,k})}{\sum_{\substack{(j,m) \in N_{i,j}}} \sigma (t^{l+1}_{i,j,m}) + \epsilon }\\
    & m_{i,j}^{l} = e_{i,j}^{l} + \gamma \biggl(BN\biggl((W^{l}_{lg}e_{i,j}^{l} +\sum_{\substack{(j,k) \\ \in N_{i,j}}}\hat{t}^{l+1}_{i,j,k} \odot V^{l}_{lg}e_{j,k}^{l}\biggr)\biggr)
    \end{split}
\end{equation}
Further, we apply another GatedGCN on the crystal graph $\mathcal{G}$ and update bond and atom features as follows : 
\begin{equation}
    \label{eq:crystalgraph}
    \begin{split}
    & {e}^{l+1}_{i,j} = {e}^{l}_{i,j} + \gamma \biggl(BN\biggl( A^{l}_{g}h_{i}^{l} + B^{l}_{g}h_{j}^{l} + C^{l}_{g}{m}^{l}_{i,j}\biggr)\biggr) \\
    & \hat{e}^{l+1}_{i,j} = \frac{\sigma (e^{l+1}_{i,j})}{\sum_{\substack{k \in N_{i}}} \sigma (e^{l+1}_{i,k}) + \epsilon }\\
    & h^{l+1}_{i} = h^{l}_{i} + \gamma \biggl(BN\biggl((W^{l}_{g}h^{l}_{i} +\sum_{\substack{j \in N_{i}}}\hat{e}^{l+1}_{i,j} \odot V^{l}_{g}h_{j}^{l}\biggr)\biggr)
    \end{split}
\end{equation}
where $\sigma$ is the sigmoid function, $\epsilon$ is a small fixed constant for numerical stability, $\odot$ is the Hadamard product, BN is batch normalization and $\gamma$ is the activation function where we use Sigmoid Linear Unit (SiLU). $A^{l}_{lg},B^{l}_{lg},C^{l}_{lg},V^{l}_{lg},W^{l}_{lg}$ are learnable parameters of GatedGCN applied on $\mathcal{L}(\mathcal{G})$ and $A^{l}_{g},B^{l}_{g},C^{l}_{g},V^{l}_{g},W^{l}_{g}$ are learnable parameters of GatedGCN applied on $\mathcal{G}$. We apply $L$ such layers of aggregation and update in Graph Encoder and return the final set of node embeddings $\mathcal{H} = \{h_1,..., h_{|\mathcal{V}|}\}$, where $h_i:=h_i^L \in R^{d}$ represents the final embedding of node $i$. We subsequently use a symmetric aggregation function (AvgPool) to generate graph-level representation $\mathcal{Z}_{\mathcal{G}} \in R^{d'}$ (we set $d'$ as 256) of the crystal material M.
\begin{equation}
    \label{eq:aggregation}
    \begin{split}
    & \mathcal{Z}_{\mathcal{G}} = \sum_{i=1}^{|\mathcal{V}|}  h_{i}^{L}
    \end{split}
\end{equation}
\subsubsection{Text Encoder: }As a text encoder, we adopt a pre-trained MatSciBERT model, which is a domain-specific language model for materials science, followed by a projection layer. MatSciBERT is effectively a pre-trained SciBERT model on a scientific text corpus of 3.17B words, which is further trained on a huge text corpus of materials science containing around 285 M words, using domain adaptive pretraining objective proposed by \cite{gururangan2020don}. We feed textual description of material $\mathcal{T}$ and extract embedding of [CLS] token $\mathcal{Z}_{CLS} \in R^{768}$ as a representation of the whole text. Further. we pass $\mathcal{Z}_{CLS}$ through a projection layer (two-layer neural network) to generate the textual embedding for the material $\mathcal{Z}_{\mathcal{T}} \in R^{d}$
\begin{equation}
    \label{eq:projection_head}
    \begin{split}
    & \mathcal{Z}_{\mathcal{T}} = W_2(g(W_1\mathcal{Z}_{CLS}))
    \end{split}
\end{equation}
We use standard non-linear function ReLU(·) as g(·), $W_1 \in R^{768 \times 128}$ and $W_2 \in R^{128 \times d}$ are parameter matrix that project $\mathcal{Z}_{CLS}$ to embedding space  $R^{d}$. In our experiment, we set d as 64.
\subsubsection{Joint Embedding Model: } The graph encoder encodes local structural and chemical semantics around atoms in a unit cell of the material, whereas the text encoder captures global periodic knowledge from the textual description. Further, in the joint embedding model, we fuse both the representations $(\mathcal{Z}_{\mathcal{G}},\mathcal{Z}_{\mathcal{T}})$ together into a single multi-modal representation $\mathcal{Z}:=  (\mathcal{Z}_{\mathcal{G}} \oplus \mathcal{Z}_{\mathcal{T}}) \in R^{(d'+d)}$, which can now capture both local and global structural semantics of the material. We tried different ways to fuse both the embeddings like sum, average, concatenation $(\oplus)$ and found concatenation performs best.\\
Further, we pass this multi-modal representation $\mathcal{Z}$ through a multi-layer perceptron which predicts the
value of the properties. We train \uaimodel{} end to end to optimize the following mean square error(MSE) loss :
\begin{equation}
\label{eq:mse_loss}
                \mathcal{L}_{MSE}=  \lVert {\hat{\mathcal{Y}}} - {\mathcal{Y}} \rVert^2
\end{equation}
where $\hat{\mathcal{Y}}$ and $\mathcal{Y}$ are predicted and true property values respectively. Note, while training \uaimodel{} we freeze the weights of MatSciBERT and don't tune it further. Fine-tuning MatSciBERT with \uaimodel{} training for a specific property will add more computational overhead as it will increase the number of parameters significantly. This provides scope for further investigation and we keep it as future work.
\section{Experimental Results}
In this section, we begin by describing the experimental setup which includes the benchmark datasets used for evaluation, alternate baseline approaches, and implementation details.  Then we evaluate the performance of \uaimodel{} in comparison with different SOTA property predictors on the downstream property prediction tasks using two popular material benchmark datasets. Next, we present the empirical evaluation results of our proposed framework in limited training data settings. Further, we conduct some ablation studies to demonstrate the expressiveness and robustness of textual representation and the importance of global and local knowledge encoded in textual embedding. Finally, we perform a qualitative analysis of the attention layer of MatSciBert to visualize attention in different tokens in the material description.
\subsection{Experimental Setup}
To evaluate the effectiveness of \uaimodel{}, we conduct experiments on two benchmark material datasets, Materials Project ~\cite{MP} (MP 2018.6.1) and JARVIS-DFT ~\cite{choudhary2020joint} (2021.8.1), which comprises some important physical properties obtained with high-throughput DFT calculations. MP 2018.6.1 consists of 69,239 materials whereas JARVIS-DFT consists of 55,722 materials. We curated textual datasets for both datasets using robocrystallographer with a textual description of each material as described in subsection \ref{textual}. 
We choose seven state of the art algorithms for crystal property prediction CIFID ~\cite{choudhary2018machine}, CGCNN ~\cite{xie2018crystal}, SchNet ~\cite{schutt2017schnet}, MEGNET ~\cite{chen2019graph}, GATGNN ~\cite{louis2020graph}, ALIGNN ~\cite{choudhary2021atomistic} and Matformer ~\cite{yan2022periodic}. To avoid any deterioration of the performance of the baseline algorithms due to insufficient hyperparameter tuning, we report the property prediction results from the respective papers of the baseline models.\\
We use four convolution layers of the graph encoder module and pre-trained MatSciBERT followed by a two-layer neural network (projection layer) as the text encoder module in \uaimodel{}. We train it for 1000 epochs using AdamW  \cite{loshchilov2017decoupled} optimizer with normalized weight decay of $10^{-5}$ and keep the batch size as 64. We schedule the learning rate according to the one-cycle policy ~\cite{smith2018disciplined} with a maximum learning rate of 0.001. We keep embedding dimensions of  the graph and text encoder as 64 and 256 respectively. We perform the experiments in shared servers having Intel E5-2620v4 processors which contain 16 cores/thread and four GTX 1080Ti 11GB GPUs each.
\subsection{Downstream Task Evaluation}
\begin{table*}
  \centering
    \setlength{\tabcolsep}{3 pt}
    \resizebox{1.0\textwidth}{!}{
      \begin{tabular}{c c | c c c c c c c c c}
        \toprule
        Property &  Unit & CIFID & CGCNN & SchNet & MEGNET & GATGNN & ALIGNN  & Matformer & \uaimodel{} \\
        \midrule
        Formation Energy  & eV/atom & 0.140   & 0.063  & 0.045  & 0.047 & 0.047  & 0.033           & 0.033\textbf{*} &\textbf{0.028}\\
        Bandgap(OPT)      & eV      & 0.301   & 0.200  & 0.192  & 0.145 & 0.170  & 0.142           & 0.137\textbf{*}  &\textbf{0.128}\\
        Bandgap(MBJ)      & eV      & 0.532   & 0.413  & 0.433  & 0.344 & 0.513  & 0.310           & 0.302\textbf{*}  & \textbf{0.278}\\
        Total Energy      & eV/atom & 0.244   & 0.078  & 0.047  & 0.058 & 0.056  & 0.037           & 0.035\textbf{*} & \textbf{0.034}\\
        Bulk Moduli(Kv)   & GPa     & 14.12   & 14.47  & 14.33  & 15.11 & 14.32  & 10.40\textbf{*} & 11.21 & \textbf{9.625}\\
        Shear Moduli(Gv)  & GPa     & 11.98   & 11.75  & 10.67  & 13.09 & 12.48  & 9.481\textbf{*}  & 10.76 & \textbf{8.471}\\
         \bottomrule
      \end{tabular} 
      }
   \caption{Summary of the prediction performance (MAE) of \uaimodel{} and different state-of-the-art models for different properties in JARVIS-DFT Dataset. The best performance is highlighted in bold and the second-best results are highlighted with \textbf{*}.}
  \label{tbl-jarvis}
\end{table*}
\subsubsection{JARVIS-DFT Dataset}
To evaluate \uaimodel{}, we first conduct experiments on the JARVIS-DFT dataset, which is a widely used large-scale material benchmark containing 55,722 crystals. Following previous state-of-the-art works, we choose six crystal properties including formation energy, bandgap (OPT), bandgap (MBJ), total energy, bulk moduli, and shear moduli for the downstream property prediction task. We use 80\%,10\%, and 10\% train, validation, and test split for all the properties as used by ALIGNN. We report mean absolute error (MAE) of the predicted and actual value of a particular property for test data in table \ref{tbl-jarvis} to compare the performance of \uaimodel{} and different participating methods.
We observe that \uaimodel{} outperforms every baseline model across all the properties with a significant margin. In specific, we observe 13.84\%, 2.85\%, 6.56\%, 7.33\%, 7.40\%, and 10.65\%  improvements compared to the competing second-best baseline model for formation energy, total energy, bandgap (OPT), bandgap (MBJ), bulk moduli, and shear moduli respectively, which shows the effectiveness of multimodal representation capturing both local chemical semantics and global periodic structural knowledge towards crystal property prediction.
\begin{table*}
  \centering
    \setlength{\tabcolsep}{4.5 pt}
    \resizebox{1.0\textwidth}{!}{
      \begin{tabular}{c c | c c c c c c c }
        \toprule
        Property &  Unit  & CGCNN & SchNet & MEGNET & GATGNN & ALIGNN & Matformer &\uaimodel{} \\
        \midrule
        Formation Energy  & eV/atom  &  0.031 & 0.033  & 0.030 & 0.033 & 0.022 & 0.021\textbf{*} & \textbf{0.020}\\
        Bandgap           & eV       &  0.292 & 0.345  & 0.307 & 0.280 & 0.218 & 0.211\textbf{*} & \textbf{0.197}\\
        Bulk Moduli(Kv)   & log(GPa) &  0.047 & 0.066  & 0.060 & 0.045 & 0.051 & 0.043\textbf{*} & \textbf{0.038}\\
        Shear Moduli(Gv)  & log(GPa) &  0.077 & 0.099  & 0.099 & 0.075 & 0.078 & 0.073\textbf{*} & \textbf{0.062}\\
         \bottomrule
      \end{tabular} 
      }
   \caption{Summary of the prediction performance (MAE) of \uaimodel{} and different state-of-the-art models for different properties in The Materials Project dataset. The best performance is highlighted in bold and the second-best results are highlighted with \textbf{*}.}
  \label{tbl-mp}
\end{table*}
\subsubsection{Materials Project (MP) Dataset}
We further use another benchmark material dataset, Materials Project-2018.6.1, comprises 69,239 materials. Here we evaluate \uaimodel{} with all state-of-the-art models using four crystal properties namely formation energy, bandgap, bulk moduli, and shear moduli. For formation energy and bandgap, we use 60000, 5000, and 4239 crystals as train, validation, and test split as used by ALIGNN, whereas use 4664, 393, and 393 crystals as train, validation, and test split for bulk and shear moduli as used by GATGNN. We report the mean absolute error (MAE) of the predicted and actual property value for test data in table \ref{tbl-mp} to compare the performance of \uaimodel{} with different participating methods. Note, to maintain consistency with the results reported by baseline works, we report (GPa) values of bulk and shear moduli in table \ref{tbl-jarvis}, whereas log(GPa) values in table \ref{tbl-mp}. We observe that \uaimodel{} outperforms every baseline model across all the properties with a significant margin. In specific, we observe 4.76\%, 6.63\%, 6.9\%, and 15.06\%  improvements compared to the competing second-best baseline model for formation energy, bandgap, bulk moduli, and shear moduli respectively. Overall, the superior performances show the effectiveness of multi-modal representation in \uaimodel{}.
\subsubsection{Results on Limited Training Data}
\uaimodel{} performs well in limited data settings as well. With 4664 training samples only \uaimodel{} achieves 6.9\% and 15.06\% improvements for bulk moduli and shear moduli respectively in the Materials Project dataset. Further, in JARVIS-DFT Dataset, we conduct an additional set of experiments for three different properties including bandgap (MBJ), bulk moduli, and shear moduli, where we have limited labeled data.  More specifically, we take 20-10-10\% training-validation-test data split and evaluate the performance of CGCNN, ALIGNN, and \uaimodel{} in the table \ref{tbl-lim-data}. We observe, \uaimodel{} achieves improvement for all three properties compared to CGCNN and ALIGNN. Overall, these superior performances indicate the robustness of our model and its adaptive ability to tasks of various data scales.
\begin{table*}
  \centering
  \small
    \setlength{\tabcolsep}{3 pt}
    \resizebox{0.6\columnwidth}{!}{
      \begin{tabular}{c c | c c c}
        \toprule
        Property & Train-set &CGCNN & ALIGNN & \uaimodel{} \\
         & Size & &  &  \\
        \midrule
        Bandgap(MBJ) & 3634 & 0.522  & 0.483 & \textbf{0.456} \\
        Bulk Moduli  & 3936 &  14.98 & 14.13 & \textbf{13.04} \\
        Shear Moduli & 3936 &  13.07 & 12.61 & \textbf{11.57} \\
        \bottomrule
      \end{tabular} 
}
   \caption{: MAE values of CGCNN, ALIGNN and \uaimodel{} for three different properties in the JARVIS-DFT dataset with 20\% training instances. The best performance is highlighted in bold.}
  \label{tbl-lim-data}
\end{table*}
\subsection{Ablation Study} In this subsection, We demonstrate the robustness of multimodal representation on different GNN architecture choices and the influence of textual modality on \uaimodel{} performance, by designing the following set of ablation studies:
\begin{enumerate}
    \item Is only textual information sufficient to infer better property prediction accuracy?
    \item How robust is the multimodal representation on different GNN architecture choices for graph encoder?
    \item What are the influences of global structural and local compositional knowledge from the textual datasets on property prediction performance?
\end{enumerate}

\begin{table*}
  \centering
  \small
    \setlength{\tabcolsep}{6 pt}
    \resizebox{0.6\textwidth}{!}{
      \begin{tabular}{c  c c c}
        \toprule
        Property & CrysTextNet & CGCNN & ALIGNN \\
        \midrule
        Formation Energy  & 0.447  &  0.063 & 0.033 \\
        Total Energy      & 0.352  &  0.078 & 0.037 \\
        \midrule
        Bandgap(OPT)      & 0.595  &  0.201 & 0.142 \\
        Bandgap(MBJ)      & 0.849  &  0.411 & 0.311 \\
        \midrule
        Bulk Moduli(Kv)   & 21.98  &  14.47 & 10.42 \\
        Shear Moduli(Gv)  & 14.76 &   11.75 & 9.483 \\
         \bottomrule
      \end{tabular} 
      }
      \caption{Summary of experiments for the ablation study on the effectiveness of Textual Representation.}
  \label{tbl-onlytext}
\end{table*}
In the following subsections, we will thoroughly discuss these. 
\subsubsection{Expressiveness of Textual Representation}
First, we are interested to understand whether textual representations are alone expressive enough, to encode atomic chemical and periodic structural semantics from the curated textual data and predict different properties precisely. We conduct an ablation experiment, where we consider only the text embeddings $\mathcal{Z}_{\mathcal{T}}$ of \uaimodel{} (output of projection layer) and pass it alone through a multi-layer perceptron to predict the property value. We denote this model as CrysTextNet and compare it with state-of-the-art graph-based models on different properties of the JARVIS-DFT dataset. We report the MAE for test data in table \ref{tbl-onlytext}. We observe, for all the properties, test MAE is higher for the CrysTextNet model compared to state-of-the-art graph-based models like CGCNN and ALIGNN. In specific, for mechanical properties like  bulk modulus and shear modulus, CrysTextNet works better (closer test MAE with competing GNN baselines) than properties like formation energy, band-gap, and total energy. This is because properties, such as formation energy, band gap, and total energy rely on microscopic chemical information which textual representation fails to encode. Instead, graph encodings include node features ${\bf x}_u$ ($u \in \mathcal{V}$) that are high-dimensional vectors with meaningful chemical quantities like electronegativity, group number, covalent radius, number of valence electrons, first ionization energy, etc. Furthermore, through message passing and aggregation in the graph convolution layer, GNN models capture many-body interactions among atoms in the material. On the other side, mechanical properties like bulk modulus and shear modulus are more dependent on structural information like lattice structure and symmetry of the material, which textual representations are able to capture.\\
Overall, though textual representations can capture many useful local and global information about the materials, unlike graph structural models, they are not alone expressive enough to capture atomic chemical features and the structural connectivity between different atoms in the materials.  Message passing and neighborhood aggregation between atoms through the GNN model are still very fundamental in learning the structure-property relationship of the materials.

\begin{table*}
  \centering
  \small
    \setlength{\tabcolsep}{11 pt}
    \resizebox{1\textwidth}{!}{
      \begin{tabular}{c | c >{\columncolor{green!20}}c | c >{\columncolor{green!20}}c | c >{\columncolor{green!20}}c}
        \toprule
        Property & CGCNN & CrysMMNet & MEGNET & CrysMMNet & GATGNN & CrysMMNet\\
         &  & (CGCNN) &  & (MEGNET) &  & (GATGNN)\\
        \midrule
        Formation Energy  & 0.063  &  0.046 & 0.076 & 0.060 & 0.077 & 0.064 \\
        Bandgap(OPT)      & 0.200  &  0.163 & 0.184 & 0.165 & 0.169 & 0.157 \\
        Bandgap(MBJ)      & 0.413   &  0.339 & 0.369 & 0.339	& 0.343 & 0.331 \\
        Total Energy      & 0.078  &  0.059 & 0.058 &  0.057     & 0.056 &  0.053     \\
        Bulk Moduli(Kv)   & 14.47  &  12.98 & 15.11 & 13.29 & 14.32 & 13.73 \\
        Shear Moduli(Gv)  & 11.75 &   10.71 & 13.09 & 11.86 & 12.48 & 12.04 \\
         \bottomrule
      \end{tabular} 
      }
   \caption{Summary of the prediction performance (MAE) of different state-of-the-art GNN models with textual representation for six different properties in The JARVIS-DFT Dataset. Model M is the SOTA baseline model and \uaimodel{}(M) is a variant where we replace graph encoder with M.}
  \label{tbl-all-gnn}
\end{table*}
\subsubsection{Robustness of Textual Representation}
Further, we investigate the robustness of textual representations on different crystal GNN encoders. We conduct an ablation study where we replace graph encode of \uaimodel{} with popular crystal GNN variants, e.g, CGCNN, MEGNET, GATGNN  and evaluate the performance. We set up the experiments with six properties of the JARVIS dataset and report the MAE values in table \ref{tbl-all-gnn} for the baseline GNN models and different variants of \uaimodel{} with different GNN architectural choices as graph encoder. We observe all these variants outperform corresponding vanilla GNN models with a good margin for all the properties, which shows textual representations are rich enough to encode global structural knowledge which aids the property prediction accuracy of any state-of-the-art GNN models. 

\subsubsection{Importance of Local and Global Knowledge}
Finally, we are curious to understand the importance of local (atom/bond) compositional information and global material structural knowledge encoded through textual representation in \uaimodel{}. Specifically, we conduct an ablation study, where we train \uaimodel{} in two additional setups along with the conventional (Global+Local) setup for \uaimodel{}. (a) Only Global: In this scenario, we take only global knowledge about the periodic structure of the materials as textual data to train \uaimodel{}. (b) Only Local: In this scenario, we take only local compositional information about atoms and inter-atomic bonds as textual data to train \uaimodel{}. We use six properties of JARVIS-DFT dataset for the experiment and report MAE in Table \ref{tbl-text-ablation}. We observe performance gain across all the properties using both global and local information combined as textual knowledge, compared to only global or local knowledge separately. 
\begin{table}
  \centering
  \small
    \setlength{\tabcolsep}{3 pt}
      \begin{tabular}{c c c c}
        \toprule
        Property & Global+Local & Only Global & Only Local \\
        \midrule
        Formation Energy  & \textbf{0.028}  &  0.039 & 0.039 \\
        Total Energy      & \textbf{0.034}  &  0.042 & 0.046 \\
        \midrule
        Bandgap(OPT)      & \textbf{0.114}  &  0.191 & 0.147 \\
        Bandgap(MBJ)      & \textbf{0.209}  &  0.216 & 0.218 \\
        \midrule
        Bulk Moduli(Kv)   & \textbf{6.860}  &  6.910 & 6.870 \\
        Shear Moduli(Gv)  & \textbf{6.440} &   6.730 & 6.880 \\
         \bottomrule
      \end{tabular} 
   \caption{Summary of experiments for the ablation study on the importance of Local and Global Material Knowledge. We observe performance gain across all the properties using both global and local information combined as textual knowledge, compared to only global or local knowledge separately.}
  \label{tbl-text-ablation}
\end{table}

\subsection{Qualitative Analysis of Attention Layers}Finally, to visualize and understand attentions in different tokens in the material description, we perform a qualitative analysis of the attention layer in MatSciBert.  We utilized the standard BertViz tool ~\cite{vig-2019-multiscale} \footnote{https://github.com/jessevig/bertviz} to analyze and visualize the attention scores in the MatSciBert Model. We present a case study of  the textual data of $FeH_8(ClO_2)_2$ in \textit{\textbf{Figure 3 \& 4 }}, where we have examined the attention score of the [CLS] token at the 5th layer of MatSciBert. \\
We observe MatSciBert allocates higher attention scores to tokens that defines global features of the crystal, such as   \textit{`Formula', `Mineral', `Crystal System', `Space Group Number', and `Dimensionality'}. Further, we investigate attention weights for local information corresponding to Fe, H, and O atoms. MatScibert provides more attention score to tokens related to \textit{bond types (octahedral geometry, equivalent bond, distorted water-like geometry, etc)} and \textit{bond lengths (2.08 Å, 2.10 Å, and 2.53 Å bond length)}. It is evident from these observations that MatSciBert is attending the important tokens related to global and local material information, to generate more expressive multimodal representation.
\begin{figure*}
	\centering
        \subfloat{\includegraphics[scale=0.8]{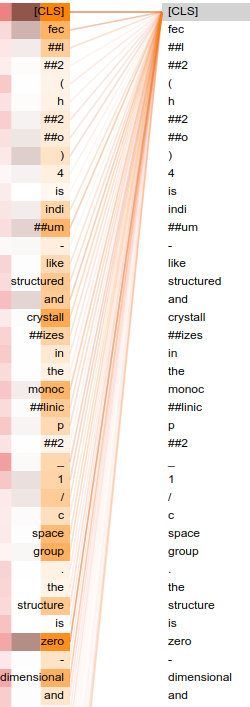}}
	\subfloat{\includegraphics[scale=0.8]{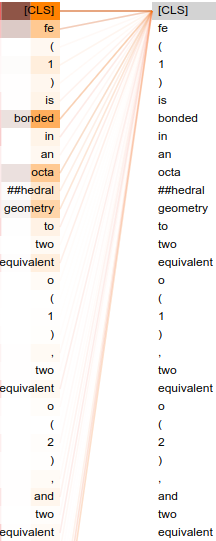}}
	\caption{\rev{Attention weights of 5th Layer at MatSciBert between [CLS] token and other tokens in material's description for $\mathbf{FeH_8(ClO_2)_2}$. We observe MatSciBert allocates higher attention scores to tokens that defines global features of the crystal, such as   \textit{`Formula', `Mineral', `Crystal System', `Space Group Number', and `Dimensionality'}}}
	\label{fig:attn_analysis_1}
\end{figure*}
\begin{figure*}
	\centering
         \subfloat{\includegraphics[scale=0.8]{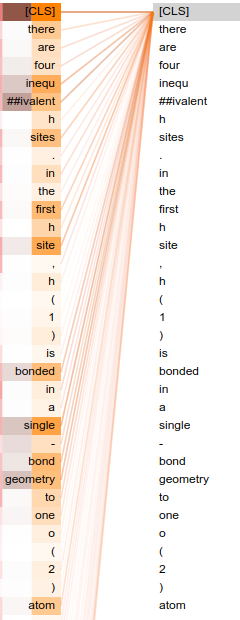}}
         \subfloat{\includegraphics[scale=0.8]{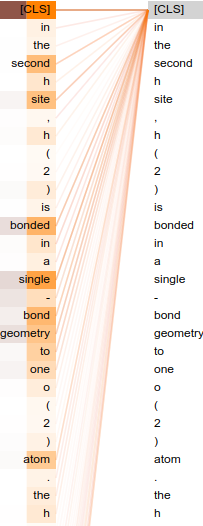}}
	\caption{\rev{Attention weights of 5th Layer at MatSciBert between [CLS] token and other tokens in material's description for $\mathbf{FeH_8(ClO_2)_2}$. We observe MatSciBert allocates higher attention scores to tokens that defines global features of the crystal, such as   \textit{`Formula', `Mineral', `Crystal System', `Space Group Number', and `Dimensionality'}}}
	\label{fig:attn_analysis_2}
\end{figure*}

\section{Conclusions}
In this work, we address the limitation of state-of-the-art GNN models for crystal property prediction  to capture global periodic structural information and leverage textual modalities beside graph structures to resolve the issue. To this end, we curate textual datasets of two popular benchmark databases containing textual descriptions of each material containing both local compositional and global structural information of a material. Further,  we propose a simple yet effective multi-modal framework, \uaimodel{}, for crystalline materials, which fuse both graph structural and textual representation together to generate a more enriched and robust multimodal representation for materials, which subsequently improves property prediction accuracy. Extensive experiments show \uaimodel{} outperforms all the popular state-of-the-art baselines across ten diverse sets of properties on two popular datasets. Further, we conduct ablation studies to demonstrate the expressiveness and robustness of textual representation on different crystal GNN encoders and show performance gain across all the properties using both global and local information combined as textual knowledge, compared to only global or local knowledge separately. Finally, we visualize attention weights between [CLS] token and other tokens in the material's description to understand
the important tokens. that the text encoder is attending to generate more expressive multimodal representation

\clearemptydoublepage
\chapter{TGDMat: Periodic Materials Generation using Text-Guided Joint Diffusion Model}
\chaptermark{TGDMat [ICLR-2025]}
\label{chap: textgudedgen}

\noindent\fbox{%
\parbox{\textwidth}{%
Work of this chapter are based on the following publication:  
\\
\textit{Periodic Materials Generation using Text-Guided Joint Diffusion Model.}
\\
\underline{Kishalay Das}, Subhojyoti Khastagir, Pawan Goyal, Seung-Cheol Lee, Satadeep Bhattacharjee and Niloy Ganguly.
\\
ICLR, 2025.}}
\vspace{1em}

\section{Introduction}
\label{iclr_intro}
Screening 3D periodic structures and their atomic compositions to identify novel crystal materials with specific chemical properties remains a long-standing challenge in the materials design community. These materials have been fundamental to key innovations such as the development of batteries, solar cells, semiconductors etc.~\cite{ butler2018machine,desiraju2002cryptic}. Historically, there have been attempts to generate novel materials by conducting resource-intensive and time-consuming simulations based on Density Functional Theory (DFT)~\cite{kohn1965self}.  Recently, the equivariant diffusion models~\cite{jiao2023crystal,luo2023towards,xie2021crystal} have demonstrated great potential to generate stable 3D periodic structures of new crystal materials. 
\\\\
\begin{figure}[ht!]
	\centering
	\includegraphics[width=\columnwidth]{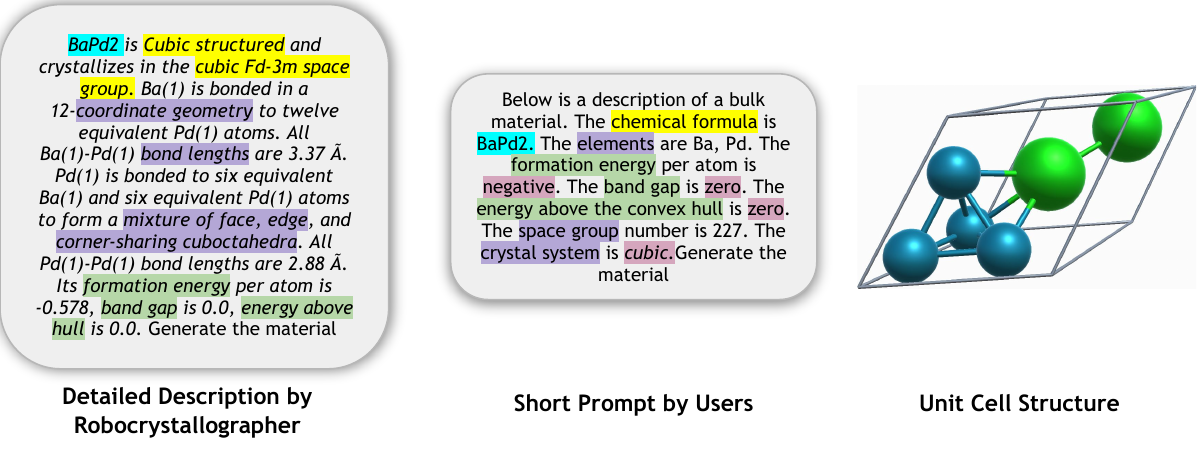}
	\caption{Detailed textual description generated by Robocrystallographer, short/less-detailed prompts by experts, and crystal unit cell structure of $\mathbf{BaPd_2}$ from Material Projects dataset. Text generated by Robocrystallographer contains both local chemical compositional information related to atom/bonds (like site coordination, geometry, polyhedral connectivity, and tilt angles) and global structural knowledge (like mineral type, space group information, symmetry, and dimensionality).The shorter prompt encodes minimal information about the material like its chemical formula, constituent elements, crystal system, and few chemical properties.}
	\label{fig:tgdmat_text}
\end{figure}
However, these models possess several inherent limitations. 1) None of these existing SOTA models learns the joint distribution of atom coordinates, types, and lattice structure of the material through an end-to-end diffusion network. Existing models like CDVAE~\cite{xie2021crystal} and SyMat~\cite{luo2023towards} learn lattice parameters and atom types separately using a VAE model and further use a score network to learn the conditional distribution of atom coordinates given atom types and lattice. DiffCSP~\cite{jiao2023crystal}, on the other hand, focuses primarily on structure prediction task where it assumes atom types are given and predict the stable crystal structure (lattice and coordinates). 
2) Furthermore, these models use SE(3)-equivariant GNNs as backbone denoising network, which largely relies on messages passing around the local neighborhood of the atoms. Hence they fail to incorporate global structural knowledge into the diffusion process, which can enhance the diffusion performance. 3) Finally, these models are unconditional by design. From initial noisy structures without any external constraints, they generate stable crystal structures, which are distributionally similar to structures of the training dataset. 
This setup may have limited utility
in real-world scenarios, as it lacks a mechanism for users to specify a criteria for the material to be generated.  In a realistic setup, users would want to specify certain key details about the target material, like the chemical formula, space group, crystal symmetry, bond lengths, chemical properties, etc as input to the diffusion model, which the generated structure must then match.
\\\\
In this paper, we propose, \iclrmodel{}, a novel \textit{\underline{T}ext-\underline{G}uided \underline{D}iffusion Model for \underline{Mat}erial Generation} that mitigates the limitations mentioned above and enhances the generation capability. Though Text Guided Diffusion Models (TGDMs) produce impressively high-quality data in the form of images~\cite{nichol2021glide,ramesh2022hierarchical,rombach2022high,saharia2022photorealistic}, audio~\cite{kreuk2022audiogen,yang2207discrete}, video~\cite{du2024learning}, molecules~\cite{gong2024text,luo2023text} etc, it remains largely unexplored in periodic material generation.
Text-guided diffusion model for new material generation has some key benefits. First, we can leverage popular tools like Robocrystallographer~\cite{ganose2019robocrystallographer} to generate a textual description of the material which provides a rich and diverse set of global structural knowledge like chemical formula, lattice constraint, space group number, crystal symmetry, chemical properties, etc. We believe this additional information is helpful for diffusion models in learning underlying crystal geometry. Second, it provides end users the flexibility to use custom prompts to guide the material generation process, ensuring that the resulting material aligns with the user's provided description. Towards that goal, we first develop a diffusion model that jointly generates the atom coordinates, atom types, and lattice structure of crystal materials using a periodic E(3)-equivariant denoising model, satisfying periodic E(3) invariance properties of learned data distribution. Subsequently, we fuse textual information into the reverse diffusion process, which guides the denoising process in predicting material structure as specified by the textual description.\\\\
To sum up, our novel contributions in this work are as follows:
\begin{itemize}
    \item To the best of our knowledge, we are the first to explore text-guided diffusion for material generation. Our proposed \iclrmodel{} bridges the gap between natural language understanding and material structure generation.
    \item Unlike prior models, \iclrmodel{} conducts joint diffusion on lattices, atom types, and coordinates, enhancing its ability to accurately capture the crystal geometry. Additionally, incorporating global structural knowledge through textual descriptions at each denoising step improves TGDMat’s ability to generate plausible materials with valid and stable structures.
    \item Extensive experiments using popular datasets on benchmark tasks show \iclrmodel{} outperforms baseline models with a good margin. Notably, in CSP task, with just one generated sample, \iclrmodel{} outperforms all baseline models, highlighting the importance of text-guided diffusion. Moreover, in the generation task, \iclrmodel{} outperforms all baselines and their text-fusion variants, showcasing the effectiveness of the joint diffusion paradigm.
    \item Fusing textual knowledge reduces the overall computational cost for both training and inference of the diffusion model. Moreover, when applied to real-world custom text prompts by experts, \iclrmodel{} demonstrates rich generative capability under general textual conditions.
\end{itemize}

\section{Preliminaries : Crystal Structure Representation}
\label{prelim}
Crystal material can be modeled by a minimal \textit{unit cell}, which gets repeated infinite times in 3D space on a regular lattice to form the periodic crystal structure. Given a material with $N$ number of atoms in its unit cell, we can describe the unit cell by two matrices: \textit{Atom Type Matrix (\textbf{\textit{A}})} and \textit{Coordinate Matrix ($\textbf{\textit{X}}$)}. Atom Type Matrix $\textbf{\textit{A}} =[\textbf{\textit{a}}_1,\textbf{\textit{a}}_2,...,\textbf{\textit{a}}_N]^T \in \mathbb{R}^{N \times k}$ denotes set of atomic type in one hot representation (k: maximum possible atom types) \rev{of all the atoms.} On the other hand, Coordinate Matrix $\textbf{\textit{X}}=[\textbf{\textit{x}}_1,\textbf{\textit{x}}_2,...,\textbf{\textit{x}}_N]^T \in \mathbb{R}^{N \times 3}$ denotes atomic coordinate positions, where $\textbf{\textit{x}}_i \in \mathbb{R}^{3}$ corresponds to coordinates of $i^{th}$ atom in the unit cell. Further, there is an additional \textit{Lattice Matrix} $\textbf{\textit{L}}=[\textbf{\textit{l}}_1,\textbf{\textit{l}}_2,\textbf{\textit{l}}_3]^T \in \mathbb{R}^{3 \times 3}$, which describes how a unit cell repeats itself in the 3D space towards $\textbf{\textit{l}}_1,\textbf{\textit{l}}_2$ and $\textbf{\textit{l}}_3$ direction to form the periodic 3D structure of the material. Formally, a given material can be defined as $ \mathbf{\textit{M}}=(\textbf{\textit{A}},\textbf{\textit{X}},\textbf{\textit{L}})$ and we can represent its infinite periodic structure as 
\begin{equation}
    \label{eq:crystal_rep}
    \begin{split}
    \mathbf{\hat{X}}  = \{ \hat{x}_i |  \hat{x}_i = \textit{x}_i + \sum_{j=1}^{3} k_j{l}_j \}; \\
    \mathbf{\hat{A}}  = \{ \hat{a}_i |  \hat{a}_i = {a}_i \}
    \end{split}
\end{equation}
where $k_1,k_2,k_3 \in \mathbb{Z};  i \in \mathbb{Z}, 1 \leq i \leq N$.
\subsection{Invariances in Crystal Structure}
\label{prelim_inv}
The basic idea of using generative models for crystal generation is to learn the underlying data distribution of material structure $p(\textbf{\textit{M}})$. Since crystal materials satisfy physical symmetry properties ~\cite{dresselhaus2007group,zee2016group}, one of the major challenges here is the learned distribution must satisfy periodic E(3) invariance i.e. invariance to permutation, translation, rotation, and periodic transformations.
\begin{itemize}
    \item \textbf{\textit{Permutation Invariance :}} If we permute the indices of constituent atoms it will not change the material. Formally, given any material $ \textbf{\textit{M}}=(\textbf{\textit{A}},\textbf{\textit{X}},\textbf{\textit{L}})$, using any permutation matrix $\mathbf{P}$ if we permute $\textbf{\textit{A}}$ and $\textbf{\textit{X}}$ as $\mathbf{P}(\textbf{\textit{A}})$ and $\mathbf{P}(\textbf{\textit{X}})$, then new material $\textbf{\textit{M}}_\textbf{\textit{P}}=(\mathbf{P}(\textbf{\textit{A}}),\mathbf{P}(\textbf{\textit{X}}),\textbf{\textit{L}})$ will remains unchanged. Hence the underlying distribution is also the same i.e $p( \textbf{\textit{M}}) = p(\textbf{\textit{M}}_\textbf{\textit{P}})$.
    
    \item \textbf{\textit{Translation Invariance :}} If we translate the atom coordinates by a random vector it will not change the structure of the material. Formally, given any material $ \textbf{\textit{M}}=(\textbf{\textit{A}},\textbf{\textit{X}},\textbf{\textit{L}})$, if we translate $\textbf{\textit{X}}$ by an arbitrary translation vector $\mathbf{u} \in \mathbb{R}^{3}$, new generated material $\textbf{\textit{M}}_\textbf{\textit{P}} =(\textbf{\textit{A}},\textbf{\textit{X}}+\mathbf{u}\mathbf{1}^T,\textbf{\textit{L}})$ will be the same as $ \mathbf{\textit{M}}$. Hence $p( \textbf{\textit{M}}) = p(\textbf{\textit{M}}_\textbf{\textit{T}})$ must satisfy.
    
    \item \textbf{\textit{Rotational Invariance :}} If we rotate the atom coordinates and lattice matrix, the material remains unchanged. Formally, using any orthogonal rotational matrix $\mathbf{Q} \in R^{3 \times 3}$ (satisfying $\mathbf{Q}^T\mathbf{Q}= \mathbf{I}$), if we rotate $\textbf{\textit{X}}$ and $\textbf{\textit{L}}$ of any material $ \textbf{\textit{M}}$ and generate new $\textbf{\textit{M}}_\textbf{\textit{R}}=(\textbf{\textit{A}},\textbf{\textit{QX}},\textbf{\textit{QL}})$, then actually different representations of the same material. Hence $p( \textbf{\textit{M}}) = p(\textbf{\textit{M}}_\textbf{\textit{R}})$ must satisfy.
    
    \item \textbf{\textit{Periodic Invariance :}} Finally, since the atoms in the unit cell can periodically repeat itself infinite times along the lattice vector, there can be many choices of unit cells and coordinate matrices representing the same material. Formally, given coordinates $\textbf{\textit{X}}$, after applying periodic transformation using random matrix $\textbf{\textit{K}} \in R^{n \times 3}$, new coordinates $\mathbf{X'}=\textbf{\textit{X}}+\textbf{\textit{K}}\textbf{\textit{L}}$ are periodically equivalent. Hence $\mathbf{\textit{M}}=(\textbf{\textit{A}},\textbf{\textit{X}},\textbf{\textit{L}})$ and $\textbf{M'}=(\textbf{\textit{A}},\mathbf{X'},\textbf{\textit{L}})$ are same material and $p( \textbf{\textit{M}}) = p(\textbf{M'})$ must hold.
\end{itemize}\
\section{Textual Dataset}
\label{text_data}
Leveraging textual information to guide the reverse diffusion process remains unexplored in the material design community. To the best of our knowledge, there is currently no dataset available that includes textual descriptions of the materials present in standard benchmark databases (Section \ref{results_expsetup}) used for material generation. In specific, we propose two methods for generating textual descriptions of materials. Hence, we first curate the textual dataset containing textual descriptions of these materials to train our model. \\\\
\textbf{\textit{Long Detailed Textual Description:}} First, we utilize a freely available utility tool known as \textit{Robocrystallographer}~\cite{ganose2019robocrystallographer} to generate detailed textual descriptions about the periodic structure of crystal materials encoded in Crystallographic Information Files (CIF Files). Robocrystallographer breaks down crystal structures into two main components: local compositional details such as atomic coordination, geometry, polyhedral connectivity, and tilt angles, as well as global structural aspects like crystal formula, mineral type, space group information, symmetry, and dimensionality. This information is presented in three formats: JSON for machine processing, human-readable text for easy comprehension akin to descriptions provided by humans, and machine learning format for specialized analysis. We choose the human-readable text format to compile textual datasets, which closely resemble descriptions given of the crystal structure by humans. \\\\
\textbf{\textit{Short Custom Prompts:}} Secondly, we utilized shorter and less detailed prompts that are more easily interpretable by users. We extend the prompt template proposed by ~\cite{gruver2024fine}, which encodes minimal information about the material like its chemical formula, constituent elements, crystal system it belongs to, and its space group number. Further, we specify a few chemical properties, and instead of mentioning their actual values, we provide generic information like negative/positive formation energy, zero/nonzero band gaps, etc.  We used the Pymatgen tool~\cite{ong2013python} to extract this information from the Crystallographic Information Files (CIF Files) and curate the textual prompts.\\\\
An illustrative example of both these textual descriptions and the unit cell structure is provided in Figure \ref{fig:tgdmat_text}.  We have publicly shared the textual datasets for both benchmark material databases with the community for future use.

\section{Proposed Methodology : \iclrmodel{}}
\label{method}
\begin{figure*}
	\centering
	\boxed{\includegraphics[width=0.9\textwidth]{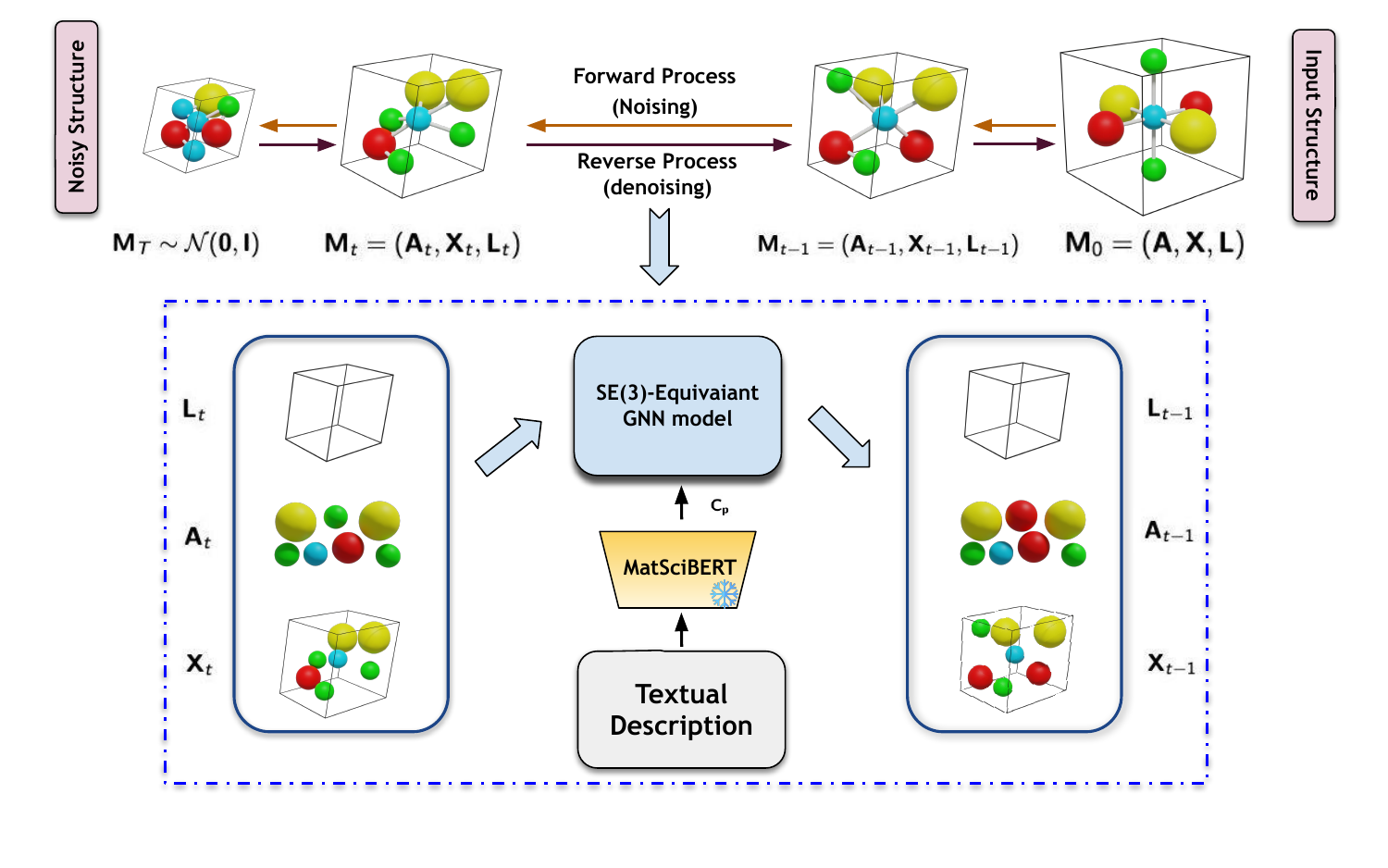}}
	\caption{Model Architecture of our proposed text guided diffusion model \iclrmodel{}. At $t^{th}$ step of reverse diffusion, given $ \textbf{\textit{M}}_t=(\textbf{\textit{A}}_t,\textbf{\textit{X}}_t,\textbf{\textit{L}}_t)$, we use periodic-E(3)-equivariant GNN model guided by contextual representation of the textual prompts ($\textbf{\textit{C}}_\textbf{\textit{p}}$) to generate $ \textbf{\textit{M}}_{t-1}=(\textbf{\textit{A}}_{t-1},\textbf{\textit{X}}_{t-1},\textbf{\textit{L}}_{t-1})$}
	\label{fig:architecture}
\end{figure*}

\xhdr{Problem Formulation} In this work, given the textual description, we focus on generating a stable crystal structure that aligns with the provided textual description. Formally, given a dataset $\mathcal{M}=\{ \textbf{\textit{M}}_i,\textbf{\textit{T}}_i \}$, containing crystal structure {$\textbf{\textit{M}}_i = (\textbf{\textit{A}}_i,\textbf{\textit{X}}_i,\textbf{\textit{L}}_i)$} and its text description ($\textbf{\textit{T}}_i$), the goal of text guided crystal generation problem is to capture the underlying conditional data distribution $p(\textbf{\textit{M}} | \textbf{\textit{T}})$ via learning a generative model $f_{\theta} (\textbf{\textit{M}} | \textbf{\textit{T}})$, where $\theta$ is a set of learnable parameters. While training, we need $f_{\theta}$ to ensure that the learned distribution is invariant to different symmetry transformations mentioned in Section \ref{prelim_inv}. Once trained, given a text description of a plausible material, the learned generative model can sample a valid and stable structure of the material, that is invariant to different symmetry transformations.
\\\\
\xhdr{Model Overview} Our proposed model, \iclrmodel{} (Fig. \ref{fig:architecture}), uses an equivariant diffusion model guided by contextual representation of the textual description ($\textbf{\textit{C}}_\textbf{\textit{p}}$) to generate a new crystal structure $ \textbf{\textit{M}}=(\textbf{\textit{A}},\textbf{\textit{X}},\textbf{\textit{L}})$. Unlike prior methods~\cite{jiao2023crystal,luo2023towards,xie2021crystal}, our method jointly diffuses \textbf{\textit{A}}, \textbf{\textit{X}}, \textbf{\textit{L}} to learn the underlying data distribution of crystal structure $p (\mathbf{\textit{M}} | \textbf{\textit{C}}_\textbf{\textit{p}})$. Diffusion models~\cite{ho2020denoising,song2019generative,song2020improved} are popular generative models that are formulated using a T steps Markov Chain. Given an input crystal material $ \textbf{\textit{M}}_0=(\textbf{\textit{A}}_0,\textbf{\textit{X}}_0,\textbf{\textit{L}}_0)$, the forward process gradually add noise to $\textbf{\textit{A}}_0,\textbf{\textit{X}}_0,\textbf{\textit{L}}_0$ independently over T steps and the reverse denoising process samples a noisy structure $ \textbf{\textit{M}}_T=(\textbf{\textit{A}}_T,\textbf{\textit{X}}_T,\textbf{\textit{L}}_T)$ from a prior distribution and reconstruct back $ \textbf{\textit{M}}_0$ using some GNN model. At each $t^{th}$ step of denoising $( T \geq t \geq 0)$, the contextual representation of the crystal textual description ($\textbf{\textit{C}}_\textbf{\textit{p}}$) will guide the diffusion process so that the intermediate structure $ \textbf{\textit{M}}_t$ aligns the target 3D structure constrained on textual conditions. Moreover, the learned distribution of material structure must satisfy periodic E(3) invariance. It is well studied in the literature~\cite{xu2022geodiff} that if the prior distribution $p(x)$ is invariant to a group and the transition probabilities of a Markov chain $y \sim p(y|x)$ exhibit equivariance, the marginal distribution of $y$ at any given time step also remains invariant to group transformations. Hence the learned distribution $p(\textbf{\textit{M}}_0)$ of the denoising model will satisfy periodic E(3) invariance if the prior distribution $p(\textbf{\textit{M}}_T)$ is invariant and the neural network used to parameterize the transition probability $q(\textbf{\textit{M}}_{t-1} | \textbf{\textit{M}}_t)$ is equivariant to permutational, translation, rotational, and periodic transformations. To satisfy that, we use periodic-E(3)-equivariant GNN model as a backbone denoising network to guide the denoising process. Next in this section, we first explain diffusion on \textbf{\textit{M}} in Section \ref{diffusion}, then demonstrate the text-guided denoising network in Section \ref{text_guided_diffusion} and finally training details in Section \ref{train_sample}.

\subsection{Joint Equivariant Diffusion on \textbf{\textit{M}}}\label{diffusion}
Given an input crystal material $ \textbf{\textit{M}}_0= (\textbf{\textit{A}}_0,\textbf{\textit{X}}_0,\textbf{\textit{L}}_0)$, we define a forward diffusion process through a Markov chain over T steps to defuse  \textbf{\textit{A}}, \textbf{\textit{X}}, \textbf{\textit{L}} independently as follows :
\begin{equation}
    \label{eq:joint_dist}
    \begin{split}
        q(\textbf{\textit{A}}_{t},\textbf{\textit{X}}_{t},\textbf{\textit{L}}_{t} | \textbf{\textit{A}}_{t-1},\textbf{\textit{X}}_{t-1},\textbf{\textit{L}}_{t-1}) = \\
        q(\textbf{\textit{A}}_{t} | \textbf{\textit{A}}_{t-1}) q(\textbf{\textit{X}}_{t} | \textbf{\textit{X}}_{t-1}) 
        q(\textbf{\textit{L}}_{t} | \textbf{\textit{L}}_{t-1}) \: \ t=1,2,...T
    \end{split}
\end{equation}
\subsection{Diffusion on Lattice (\textbf{\textit{L}})} Lattice Matrix $\textbf{\textit{L}}=[{l}_1,{l}_2,{l}_3]^T \in \mathbb{R}^{3 \times 3}$ is a global feature of the material which determines the shape and symmetry of the unit cell structure. Since \textbf{\textit{L}} is in continuous space, we leverage the idea of the Denoising Diffusion Probabilistic Model (DDPM) for diffusion on \textbf{\textit{L}}. In specific, given input lattice matrix $\textbf{\textit{L}}_0 \sim p(\textbf{\textit{L}})$, the forward diffusion process iteratively diffuses it over T timesteps to a noisy lattice matrix $\textbf{\textit{L}}_T$ 
through a transition probability $q(\textbf{\textit{L}}_{t} | \textbf{\textit{L}}_{0})$ at each $t^{th}$ step, which can be derived as follows :
\begin{equation}
    \label{eq:lattice_forward}
        q(\textbf{\textit{L}}_{t} \ | \ \textbf{\textit{L}}_{0}) = \mathcal{N} \bigg(\textbf{\textit{L}}_{t} \ | \ \sqrt{\bar{\alpha}_{t}}\textbf{\textit{L}}_{0}, \ (1 \ - \ \bar{\alpha}_{t})\mathbf{I} \bigg)
\end{equation}
where, $\bar{\alpha}_{t} = \prod_{k=1}^{t} \alpha_{k}$, $\alpha_{t} = 1-\beta_{t}$ and $\{ \beta_{t} \in (0,1) \}^{T}_{t=1}$ controls the variance of diffusion step following certain variance scheduler. 
By reparameterization, we can rewrite equation \ref{eq:lattice_forward} as:
\begin{equation}
    \label{eq:lattice_forward_reparam}
        \textbf{\textit{L}}_{t} = \sqrt{\bar{\alpha}_{t}}\textbf{\textit{L}}_{0} +  \sqrt{1-\bar{\alpha}_{t}}\bm{\epsilon}^{\textbf{\textit{L}}}
\end{equation}
where, $\bm{\epsilon}^{l}$ is a noise, sampled from $\mathcal{N}(\mathbf{0},\mathbf{I})$, added with original input sample $\textbf{\textit{L}}_{0}$  at $t^{th}$ step to generate $\textbf{\textit{L}}_{t}$. 
After T such diffusion steps, noisy lattice matrix $\textbf{\textit{L}}_T $ is generated from prior noise distribution $\sim \mathcal{N}(\mathbf{0},\mathbf{I})$. In the reverse denoising process, given noisy $\textbf{\textit{L}}_T \sim \mathcal{N}(\mathbf{0},\mathbf{I})$ we reconstruct true lattice structure $\textbf{\textit{L}}_0$ thorough iterative denoising step via learning reverse conditional distribution, which we formulate as follows :
\begin{equation}
    \label{eq:lattice_backward}
        p(\textbf{\textit{L}}_{t-1} | \textbf{\textit{M}}_{t},\textbf{\textit{C}}_\textbf{\textit{p}}) = \mathcal{N} \big \{ \textbf{\textit{L}}_{t-1} \ | \ \mu^{\textbf{\textit{L}}}(\textbf{\textit{M}}_{t},\textbf{\textit{C}}_\textbf{\textit{p}}),\beta_{t}\frac{(1 - \bar{\alpha}_{t-1})}{(1 -\bar{\alpha}_{t})}\mathbf{I} \big \}
\end{equation}
where $\mu^{\textbf{\textit{L}}}(\textbf{\textit{M}}_{t},\textbf{\textit{C}}_\textbf{\textit{p}}) = \frac{1}{\sqrt{\alpha_{t}}}\big(\textbf{\textit{L}}_{t}-\frac{1 - \alpha_{t}}{\sqrt{1 - \bar{\alpha}_{t}}} \ \hat{\bm{\epsilon}}^{\textbf{\textit{L}}} (\textbf{\textit{M}}_{t},\textbf{\textit{C}}_\textbf{\textit{p}},t)\big)$.
Intuitively, $\hat{\bm{\epsilon}}^{l}$ is the denoising term that needs to be subtracted from $\textbf{\textit{L}}_{t}$ to generate $\textbf{\textit{L}}_{t-1}$ and textual representation $\textbf{\textit{C}}_\textbf{\textit{p}}$ will steer this reverse diffusion process. We use a text-guided denoising network $\Phi_\theta(\textbf{\textit{A}}_{t},\textbf{\textit{X}}_{t},\textbf{\textit{L}}_{t},t,\textbf{\textit{C}}_\textbf{\textit{p}})$ to model the noise term $\hat{\bm{\epsilon}}^{\textbf{\textit{L}}} (\textbf{\textit{M}}_{t},\textbf{\textit{C}}_\textbf{\textit{p}},t)$. Following the simplified training objective proposed by ~\cite{ho2020denoising}, we train the aforementioned denoising network using $l_2$ loss between $\hat{\bm{\epsilon}}^{\textbf{\textit{L}}}$ and $\bm{\epsilon}^{\textbf{\textit{L}}}$
\begin{equation}
    \label{eq:appendix_lattice_loss}
        \mathcal{L}_{lattice} = \mathbb{E}_{\bm{\epsilon}^{\textbf{\textit{L}}},t \sim \mathcal{U}(1,T)}
        \lVert  \bm{\epsilon}^{\textbf{\textit{L}}} \ - \ \hat{\bm{\epsilon}}^{\textbf{\textit{L}}} {\rVert}^2_2 
\end{equation}

\subsection{Diffusion on Atom Types (\textbf{\textit{A}})} 
Prior studies ~\cite{jiao2023crystal,xie2021crystal} consider Atom Type Matrix $\textbf{\textit{A}}$ as the logits/probability distribution for k classes $\in$ $\mathbb{R}^{N \times k}$ (continuous variable in real space) and apply DDPM to learn the distribution. However for discrete data these models are inappropriate and produce suboptimal results ~\cite{austin2021structured,campbell2022continuous,hoogeboom2021argmax}. Hence we consider $\textbf{\textit{A}}$ as N discrete variables belonging to k classes and leverage discrete denoising diffusion probabilistic model (D3PM) ~\cite{austin2021structured} for diffusion on \textbf{\textit{A}}. In specific, denoting row vector \textbf{\textit{a}} as a one-hot representation of an atom \textit{a}, we can write transition probability for forward process as:
\begin{equation}
        q(\textbf{\textit{a}}_t | \textbf{\textit{a}}_{t-1}) = Cat(\textbf{\textit{a}}_t;\textbf{\textit{p}} = \textbf{\textit{a}}_{t-1}\textbf{\textit{Q}}_{t})
\end{equation}
where $Cat(\textbf{\textit{a}};\textbf{\textit{p}})$ is a categorical distribution over the one-hot row vector \textbf{\textit{a}} with probabilities given by the row vector \textbf{\textit{p}} and $\textbf{\textit{Q}}_{t}$ is the Markov transition matrix at time step t defined as $[\textbf{\textit{Q}}_{t}]_{i,j} = q(a_t = i \ | \ a_{t-1} = j)$. Different choices of $\textbf{\textit{Q}}_{t}$ and corresponding stationary distributions are proposed by ~\cite{austin2021structured} which provides flexibility to control the data corruption and denoising process. We adopted the absorbing state diffusion process, introducing a new absorbing state [MASK] in $\textbf{\textit{Q}}_{t}$. At each time step t, we can formally define the transition matrix as:
\begin{equation}
        [\textbf{\textit{Q}}_{t}]_{i,j} = 
        \begin{cases}
        1,           & \text{if $i=j=[MASK]$}.\\
        1-\beta_{t}, & \text{if $i=j \neq [MASK]$}\\
        \beta_{t}, & \text{if $i=j = [MASK]$}.
        \end{cases}
\end{equation}
Intuitively, at each time step t, an atom either stays in its type state with probability $1-\beta_{t}$ or moves to [MASK] state with probability $\beta_{t}$ and once it moves to [MASK] state, it stays in that state. Hence, the stationary distribution
of this diffusion process has all the mass on the [MASK] state. During reverse denoising process, given textual representation $\textbf{\textit{C}}_\textbf{\textit{p}}$, we first sample noisy $\textbf{\textit{a}}_T$ and obtain $\textbf{\textit{a}}_0$ thorough iterative denoising step via learning reverse conditional transition:
\begin{equation}
        p_{\theta}(\textbf{\textit{a}}_{t-1} | \textbf{\textit{a}}_t, \textbf{\textit{C}}_\textbf{\textit{p}}) \propto \sum_{\textbf{\textit{a}}_0} q(\textbf{\textit{a}}_{t-1}, \textbf{\textit{a}}_t | \textbf{\textit{a}}_0)p_{\theta}(\textbf{\textit{a}}_0 | \textbf{\textit{a}}_t, \textbf{\textit{C}}_\textbf{\textit{p}})
\end{equation}
We use the text-guided denoising network $\Phi_\theta(\textbf{\textit{A}}_{t},\textbf{\textit{X}}_{t},\textbf{\textit{L}}_{t},t,\textbf{\textit{C}}_\textbf{\textit{p}})$ to model this backward denoising process, which is trained using the following loss function as proposed by ~\cite{austin2021structured} :
\begin{equation}
    \label{eq:appendix_type_loss}
        \mathcal{L}_{type} = \mathcal{L}_{VB}+\lambda\mathcal{L}_{CE}
\end{equation}
where $\mathcal{L}_{VB}$ is the variational lower bound loss defined as follows:
\begin{equation}
    \label{eq:variational_loss}
    \begin{aligned}
        \mathcal{L}_{VB} = 
        \mathbb{E}_{\substack{q(\textbf{\textit{a}}_0)}} 
        \bigg[ 
        \underbrace{D_{KL} \{q(\textbf{\textit{a}}_{T}| \textbf{\textit{a}}_0) || p(\textbf{\textit{a}}_{T})\}}_{L_T} \ + \ 
        \sum^T_{t=2} \mathbb{E}_{\substack{q(\textbf{\textit{a}}_t | \textbf{\textit{a}}_0)}}
        \underbrace{[ D_{KL} \{ q(\textbf{\textit{a}}_{t-1}| \textbf{\textit{a}}_t,\textbf{\textit{a}}_0) || p_{\theta}(\textbf{\textit{a}}_{t-1}| \textbf{\textit{a}}_t)\}]}_{L_{t-1}} \\
        - \underbrace{\mathbb{E}_{\substack{q(\textbf{\textit{a}}_1 | \textbf{\textit{a}}_0)}}[\text{log} \ p_{\theta}(\textbf{\textit{a}}_0| \textbf{\textit{a}}_1)\}]}_{L_0}
        \bigg]
    \end{aligned}
\end{equation}
and $\mathcal{L}_{CE}$ is the cross-entropy loss defined as follows:
\begin{equation}
    \label{eq:ce_loss}
    \begin{aligned}
        \mathcal{L}_{CE} = 
        \mathbb{E}_{\substack{q(\textbf{\textit{a}}_0)}} 
        \bigg[
        \sum^T_{t=2} \mathbb{E}_{\substack{q(\textbf{\textit{a}}_t | \textbf{\textit{a}}_0)}}[\text{log} \ p_{\theta}(\textbf{\textit{a}}_0| \textbf{\textit{a}}_t)\}]
        \bigg]
    \end{aligned}
\end{equation}
and $\lambda$ is a hyperparameter.
\subsection{Diffusion on Atom Coordinates (\textbf{\textit{X}})}
Coordinate Matrix $\textbf{\textit{X}}=[\textbf{\textit{x}}_1,\textbf{\textit{x}}_2,...,\textbf{\textit{x}}_N]^T \in \mathbb{R}^{N \times 3}$ denotes atomic coordinate positions, where ${x}_i \in \mathbb{R}^{3}$ corresponds to coordinates of $i^{th}$ atom in the unit cell. We can diffuse the atom coordinates in two ways: either by diffusing cartesian coordinates or fractional coordinates. Prior works like CDVAE~\cite{xie2021crystal} and SyMat~\cite{luo2023towards} diffuse cartesian coordinates whereas DiffCSP~\cite{jiao2023crystal} diffuse fractional coordinates. In our setup, as we are jointly learning atom coordinates and lattice matrix simultaneously, we follow the line of work by DiffCSP and diffuse fractional coordinates. Atomic fractional coordinates in crystal material lives in quotient space $\mathbb{R}^{N \times 3} / \mathbb{Z}^{N \times 3}$ induced by the crystal periodicity. Since the Gaussian distribution used in DDPM is unable to model the cyclical and bounded domain of \textbf{\textit{X}}, it is not suitable to apply DDPM to model \textbf{\textit{X}}. Hence at each step of forward diffusion, we add noise sample from Wrapped Normal (WN) distribution ~\cite{de2022riemannian} to \textbf{\textit{X}} and during backward diffusion leverage Score Matching Diffusion Networks ~\cite{song2019generative,song2020improved} to model underlying transition probability $q(\textbf{\textit{X}}_{t} \ | \ \textbf{\textit{X}}_{0}) = \mathcal{N}_W (\textbf{\textit{X}}_{t} \ | \ \textbf{\textit{X}}_{0}, \sigma_t^2 \mathbf{I})$. In specific, at each $t^{th}$ step of diffusion, we derive $\textbf{\textit{X}}_{t}$ as : 
$\textbf{\textit{X}}_{t} = f_w(\textbf{\textit{X}}_{0} + \bm{\sigma_t}\bm{\epsilon}^{\textbf{\textit{X}}})$
where, $\bm{\epsilon}^{\textbf{\textit{X}}}$ is a noise, sampled from $\mathcal{N}(\mathbf{0},\mathbf{I})$, $\bm{\sigma_t}$ is the noise scale following exponential scheduler and $f_w(.)$ is a truncation function. Given a fractional coordinate matrix X, truncation function $f_w(\textbf{\textit{X}}) = (\textbf{\textit{X}} - \lfloor \textbf{\textit{X}} \rfloor)$ returns the fractional part of each element of \textbf{\textit{X}}. \\
As argued in ~\cite{jiao2023crystal}, $q(X_t|X_0)$ is periodic translation equivariant, and approaches uniform distribution $\mathcal{U}(0,1)$ for sufficiently large values of $\sigma_T$. Hence during the backward denoising process, we first sample $\textbf{\textit{X}}_{T} \sim \mathcal{U}(0,1)$ and iteratively denoise via score network for T steps to recover back the true fractional coordinates $\textbf{\textit{X}}_{0}$. We use the text-guided denoising network $\Phi_\theta(\textbf{\textit{A}}_{t},\textbf{\textit{X}}_{t},\textbf{\textit{L}}_{t},t,\textbf{\textit{C}}_\textbf{\textit{p}})$ to model the backward diffusion process, which is trained using the following score-matching objective function :
\begin{equation}
    \label{eq:appendix_coord_loss}
        \mathcal{L}_{coord} = \mathbb{E}_{\substack{\textbf{\textit{X}}_{t} \sim q(\textbf{\textit{X}}_{t} | \textbf{\textit{X}}_{0}) \\
        t \sim \mathcal{U}(1,T)}}
        \lVert  \nabla_{\textbf{\textit{X}}_{t}} \text{log} q(\textbf{\textit{X}}_{t} | \textbf{\textit{X}}_{0}) - \hat{\bm{\epsilon}}^{\textbf{\textit{X}}}(\textbf{\textit{M}}_{t},\textbf{\textit{C}}_\textbf{\textit{p}},t) {\rVert}^2_2
\end{equation}
where $\nabla_{\textbf{\textit{X}}_{t}} \text{log} q(\textbf{\textit{X}}_{t} | \textbf{\textit{X}}_{0}) \propto \sum_{\textbf{\textit{K}} \in \mathbb{Z}^{N \times 3}} \text{exp}(- \ \frac{\lVert \textbf{\textit{X}}_{t} - \textbf{\textit{X}}_{0} + \textbf{\textit{K}} {\rVert}^2_F }{2\bm{\sigma_t}^2})$ is the score function of transitional distribution and $\hat{\bm{\epsilon}}^{\textbf{\textit{X}}}(\textbf{\textit{M}}_{t},\textbf{\textit{C}}_\textbf{\textit{p}},t)$ denoising term.
\subsection{Text Guided Denoising Network} 
\label{text_guided_diffusion}
In this subsection, we will illustrate the detailed architecture of our proposed Text Guided Denoising Network $\Phi_\theta(\textbf{\textit{A}}_{t},\textbf{\textit{X}}_{t},\textbf{\textit{L}}_{t},t,\textbf{\textit{C}}_\textbf{\textit{p}})$, which we used to denoise \textbf{\textit{A}}, \textbf{\textit{X}} and \textbf{\textit{L}}. As mentioned in \ref{prelim_inv}, the learned distribution of material structure $p(\textbf{\textit{M}})$  must satisfy periodic E(3) invariance. Hence we leverage an periodic-E(3)-equivariant Graph Neural Network (GNN) integrated with a pre-trained textual encoder to model the denoising process. In particular, as a text encoder, we adopt a pre-trained MatSciBERT ~\cite{gupta_matscibert_2022} model, which is a domain-specific language model for materials science, followed by a projection layer. MatSciBERT is effectively a pre-trained SciBERT model on a scientific text corpus of 3.17B words, which is further trained on a huge text corpus of materials science containing around 285 M words.  We feed textual description of material $\mathcal{T}$ and extract embedding of [CLS] token $\textbf{\textit{h}}_{CLS}$ as a representation of the whole text. Further. we pass $\textbf{\textit{h}}_{CLS}$ through a projection layer to generate the contextual textual embedding for the material $\textbf{\textit{C}}_\textbf{\textit{p}} \in \mathbb{R}^{d}$, which we pass to the equivariant GNN model to guide the denoising process. Practically, as the backbone network for the backward diffusion process, we extend CSPNet architecture~\cite{jiao2023crystal}, originally developed for crystal structure prediction (CSP) task. CSPNet is built upon EGNN~\cite{satorras2021n}, satisfying periodic E(3) invariance condition on periodic crystal structure. At the $k^{th}$ layer message passing, the Equivariant Graph Convolutional Layer (EGCL) takes as input the set of atom embeddings $\textbf{\textit{h}}^{k}=[\textbf{\textit{h}}^{k}_1,\textbf{\textit{h}}^{k}_2,...,\textbf{\textit{h}}^{k}_N]$, atom coordinates $\textbf{\textit{x}}^{k}=[\textbf{\textit{x}}^{k}_1,\textbf{\textit{x}}^{k}_2,...,\textbf{\textit{x}}^{k}_N]$ and Lattice Matrix \textbf{\textit{L}} and outputs a transformation on $\textbf{\textit{h}}^{k+1}$. Formally, we can define the $k^{th}$ layer message passing operation as follows :
\begin{equation}
\label{eq:appendix_msg_pass_1}
    \textbf{\textit{m}}_{i,j} = \rho_m \{\textbf{\textit{h}}^{k}_i, \ \textbf{\textit{h}}^{k}_j, \ \textbf{\textit{L}}^T\textbf{\textit{L}}, \ 
    \psi_{FT}(\textbf{\textit{x}}^{k}_i - \textbf{\textit{x}}^{k}_j ) \} ; \;
\end{equation}
\begin{equation}
\label{eq:appendix_msg_pass_2}
    \textbf{\textit{m}}_{i} = \sum_{j=1}^{N} \textbf{\textit{m}}_{i,j}
\end{equation}
\begin{equation}
\label{eq:appendix_msg_pass_3}
    \textbf{\textit{h}}^{k+1}_{i} = \textbf{\textit{h}}^{k}_{i} + \rho_h \{ \textbf{\textit{h}}^{k}_{i}, \textbf{\textit{m}}_{i} \}
\end{equation}
where $\textbf{\textit{m}}_{i} = \sum_{j=1}^{N} \textbf{\textit{m}}_{i,j}$,  $\rho_m, \rho_h$ are multi-layer perceptrons and $\psi_{FT}$ is a Fourier Transformation function applied on relative difference between fractional coordinates $\textbf{\textit{x}}^{k}_i$, $ \textbf{\textit{x}}^{k}_j$. Fourier Transformation is used since it is invariant to periodic translation and extracts various frequencies of all relative fractional distances that are helpful for crystal structure modeling.\\
We fuse textual representation $\textbf{\textit{C}}_\textbf{\textit{p}}$ into input atom feature $\textbf{\textit{h}}^{0}_{i}$ as 
\begin{equation}
\label{eq:appendix_text_fusion}
   \textbf{\textit{h}}^{0}_{i} = \rho \ \{ \ f_{atom}(\textbf{\textit{a}}_i) \ \lvert \rvert \  f_{pos}(t) \ \lvert \rvert  \ \textbf{\textit{C}}_\textbf{\textit{p}} 
\end{equation}
where t is the timestamp of the diffusion model, $f_{pos}(.)$ is sinusoidal
positional encoding ~\cite{ho2020denoising,vaswani2017attention}, $f_{atom}(.)$ learned atomic embedding function and $\lvert \rvert$ is concatenation operation. Input atom features $\textbf{\textit{h}}^{0}$ and coordinates $\textbf{\textit{x}}^{0}$ are fed through $\mathcal{K}$ layers of EGCL to produce $\hat{\bm{\epsilon}}^{\textbf{\textit{L}}}$, $p(\textbf{\textit{A}}_{t-1} \ | \ \textbf{\textit{M}}_{t})$ and $\hat{\bm{\epsilon}}^{\textbf{\textit{X}}}$ as follows :
\begin{equation}
    \label{eq:appendix_msg_pass_4}
    \begin{split}
        \hat{\bm{\epsilon}}^{\textbf{\textit{L}}} = \textbf{\textit{L}} \rho_L (\frac{1}{N} \sum^{i=1}_{N} \textbf{\textit{h}}^{\mathcal{K}}) ; \; \\
        p(\textbf{\textit{A}}_{t-1} \ | \ \textbf{\textit{M}}_{t}) = \rho_A (\textbf{\textit{h}}^{\mathcal{K}}) ; \; \\
        \hat{\bm{\epsilon}}^{\textbf{\textit{X}}} = \rho_X (\textbf{\textit{h}}^{\mathcal{K}})
    \end{split}
\end{equation}
where $\rho_L, \rho_A, \rho_X$ are multi-layer perceptrons on the final layer embeddings. Intuitively, we feed global structural knowledge about the crystal structure into the network by injecting contextual representation $\textbf{\textit{C}}_\textbf{\textit{p}}$ into input atom features. This added signal will participate through message-passing operations in Eq. \ref{eq:appendix_msg_pass_1} and guides in denoising atom types, coordinates, and lattice parameters such that it can capture the global crystal geometry and aligned with the input stable structure specified by a textual description.

\begin{algorithm}
\caption{Training Algorithm}\label{alg:training}
\begin{algorithmic}[1]
\State \textbf{Input:} Atom type Matrix $\textbf{\textit{A}}_0$ (One hot Vector Representation), Coordinate Matrix $\textbf{\textit{X}}_0$, Lattice matrix $\textbf{\textit{L}}_0$, Markov Transition Matrix $[\textbf{\textit{Q}}_{t}]^T_{t=1}$, Textual Representation $\textbf{\textit{C}}_\textbf{\textit{p}}$, Number of diffusion step T and hyperparameters $\lambda_{\textbf{\textit{A}}}$, $\lambda_{\textbf{\textit{X}}}$, $\lambda_{\textbf{\textit{L}}}$.

\Repeat
\State Sample $t \sim  \mathcal{U}(\mathbf{0},\mathbf{T})$
\State Sample Noise $\epsilon^\textbf{X}, \epsilon^\textbf{L} \sim N(0,I)$
\State $\textbf{\textit{L}}_{t} = \sqrt{\bar{\alpha}_{t}}\textbf{\textit{L}}_{0} +  \sqrt{1-\bar{\alpha}_{t}}\bm{\epsilon}^{\textbf{\textit{L}}} $
\State $\textbf{\textit{X}}_{t} = f_w(\textbf{\textit{X}}_{0} + \bm{\sigma_t}\bm{\epsilon}^{x})$
\State $\textbf{\textit{A}}_{t} = Cat(\textbf{\textit{A}}_t;\textbf{\textit{p}} = \textbf{\textit{A}}_{t-1}\textbf{\textit{Q}}_{t})$
\State $\hat{\epsilon}^{\textbf{L}},\hat{\epsilon}^{\textbf{X}},\textbf{A}'_{t}  \leftarrow \Phi_\theta(\textbf{A}_{t},\textbf{X}_{t},\textbf{L}_{t},t,\textbf{C}_\textbf{p})$
\State $\mathcal{L}_{lattice} = \lVert  \bm{\epsilon}^{\textbf{\textit{L}}} \ - \ \hat{\bm{\epsilon}}^{\textbf{\textit{L}}} {\rVert}^2_2 $
\State $\mathcal{L}_{coord} = 
        \lVert  \nabla_{\textbf{\textit{X}}_{t}} \text{log} q(\textbf{\textit{X}}_{t} | \textbf{\textit{X}}_{0}) - \hat{\bm{\epsilon}}^{X} {\rVert}^2_2$
\State $\mathcal{L}_{type} = \mathcal{L}_{VB}+\lambda\mathcal{L}_{CE}$
\State Minimize $\mathcal{L} = \lambda_{L} \mathcal{L}_{lattice} + \lambda_{A} \mathcal{L}_{type} + \lambda_{X} \mathcal{L}_{coord}$ and update parameters of $\Phi_\theta$
\Until{converged}

\end{algorithmic}
\end{algorithm}
\begin{algorithm}
\caption{Sampling Algorithm}\label{alg:sampling}
\begin{algorithmic}[1]
\State Sample $\textbf{\textit{L}}_T \sim  \mathcal{N}(\mathbf{0},\mathbf{I}),\textbf{\textit{X}}_T \sim  \mathcal{U}(0,1)$
\State Randomly sample each atom type between 0 to 99 (Max possible atom type) and form $\textbf{\textit{A}}_T$
\State $\textbf{\textit{C}}_\textbf{\textit{p}} \gets \text{Textual Representation}$
\For{$t \gets T$ to $1$}
    \State $\epsilon^\textbf{A},\epsilon^\textbf{X},\epsilon^\textbf{L} \sim N(0,I) /*Sample*/$
    \State $\hat{\textbf{A}},\hat{\epsilon}^{\textbf{X}},\hat{\epsilon}^{\textbf{L}} \leftarrow \Phi_\theta(\textbf{A}_{t},\textbf{X}_{t},\textbf{L}_{t},t,\textbf{C}_\textbf{p})$
    
    \State $\textbf{L}_{t-1} \leftarrow \frac{1}{\sqrt{\alpha_{t}}} (\textbf{L}_{t} - \frac{\beta_t}{\sqrt{1-\bar{\alpha}_{t}}}\hat{\epsilon}^{\textbf{L}})+\sqrt{\beta_t\frac{1-\bar{\alpha}_{t-1}}{1-\bar{\alpha}_{t}}}\epsilon^{\textbf{L}}$
    \State $\textbf{A}_{t-1} \leftarrow \text{Softmax}(\hat{\textbf{A}}+\bm{\sigma_t}\epsilon^\textbf{A})$
    \State $\textbf{X}_{t-\frac{1}{2}} \leftarrow w(\textbf{X}_{t}+(\sigma^2_t - \sigma^2_{t-1})\hat{\epsilon}^{\textbf{X}}+\frac{\sigma_{t-1}\sqrt{\sigma^2_t - \sigma^2_{t-1}}}{\sigma_{t}}\epsilon^\textbf{X})$
    \State $_,\hat{\epsilon}^{\textbf{X}} \leftarrow \Phi_\theta(\textbf{A}_{t},\textbf{X}_{t-\frac{1}{2}},\textbf{L}_{t-1},t,\textbf{C}_\textbf{p})$
    \State $\eta_t \leftarrow step\_size * \frac{\sigma_{t-1}}{\sigma_t}$
    \State $\textbf{X}_{t-1} \leftarrow w(\textbf{X}_{t-\frac{1}{2}}+\eta_t\hat{\epsilon}^{\textbf{X}}+\sqrt{2\eta_t}\epsilon^\textbf{X})$   
\EndFor
\\
\textbf{Output: }$\textbf{M}_{new}$ = $(\textbf{A}_0, \textbf{X}_0, \textbf{L}_0)$
\end{algorithmic}
\end{algorithm}

\subsection{Training and Sampling}
\label{train_sample}
\iclrmodel{} is trained using the following combined loss:\\
\begin{equation}
    \label{eq:appendix_loss}
        \mathcal{L} = \lambda_{L} \mathcal{L}_{lattice} + \lambda_{A} \mathcal{L}_{type} + \lambda_{X} \mathcal{L}_{coord}
\end{equation}
\\
where $\mathcal{L}_{lattice}$, $\mathcal{L}_{type}$ and $\mathcal{L}_{coord}$ are lattice $l_2$ loss (Eq. \ref{eq:appendix_lattice_loss}), type cross-entropy loss (Eq. \ref{eq:appendix_type_loss}) and coordinate score matching loss (Eq. \ref{eq:appendix_coord_loss}) respectively and $\lambda_{L}$, $\lambda_{A}$, $\lambda_{X}$ are hyperparameters control the relative weightage between these different loss components. During training, we freeze the MatSciBERT parameters and do not tune it further. During sampling, we use the Predictor-Corrector sampling mechanism to sample $\textbf{\textit{A}}_0$, $\textbf{\textit{X}}_0$ and $\textbf{\textit{L}}_0$. Next we explain algorithms for training and sampling.
\section{Experiments}\label{results}
Next, in this section, We provide a comprehensive evaluation of our method against several baselines on two benchmark tasks. 
First, in Section \ref{results_expsetup}, we provide a brief overview of the experimental setup, including benchmark tasks, and datasets. Next, in Section \ref{results_csp}, we demonstrate how textual data enhances the prediction of stable crystal structures. Following that, in Section \ref{results_gen}, we highlight the effectiveness of our proposed joint diffusion paradigm in enhancing its ability to generate novel crystal materials. Additionally, we present the correctness of the generated materials (Section \ref{result_corectness}), performance on shorter custom prompts (Section \ref{sec-shorter-prompt}), computational cost of training and inference (Section \ref{results_time}), smoothness of the model’s generations (Section \ref{results_ablation}), an ablation study on the choice of different text encoders (Section \ref{result_text_enc}), and finally, visualizations of the generated materials (Section \ref{result_visualize}).
\subsection{Experimental Setup}
\label{results_expsetup} 
\begin{figure}
    \centering
	\subfloat[Random Material Generation (Gen): the goal of the generative model is to generate novel stable materials (atom types, fractional coordinates, and lattice structure)]{\includegraphics[width=\columnwidth, height = 300pt]{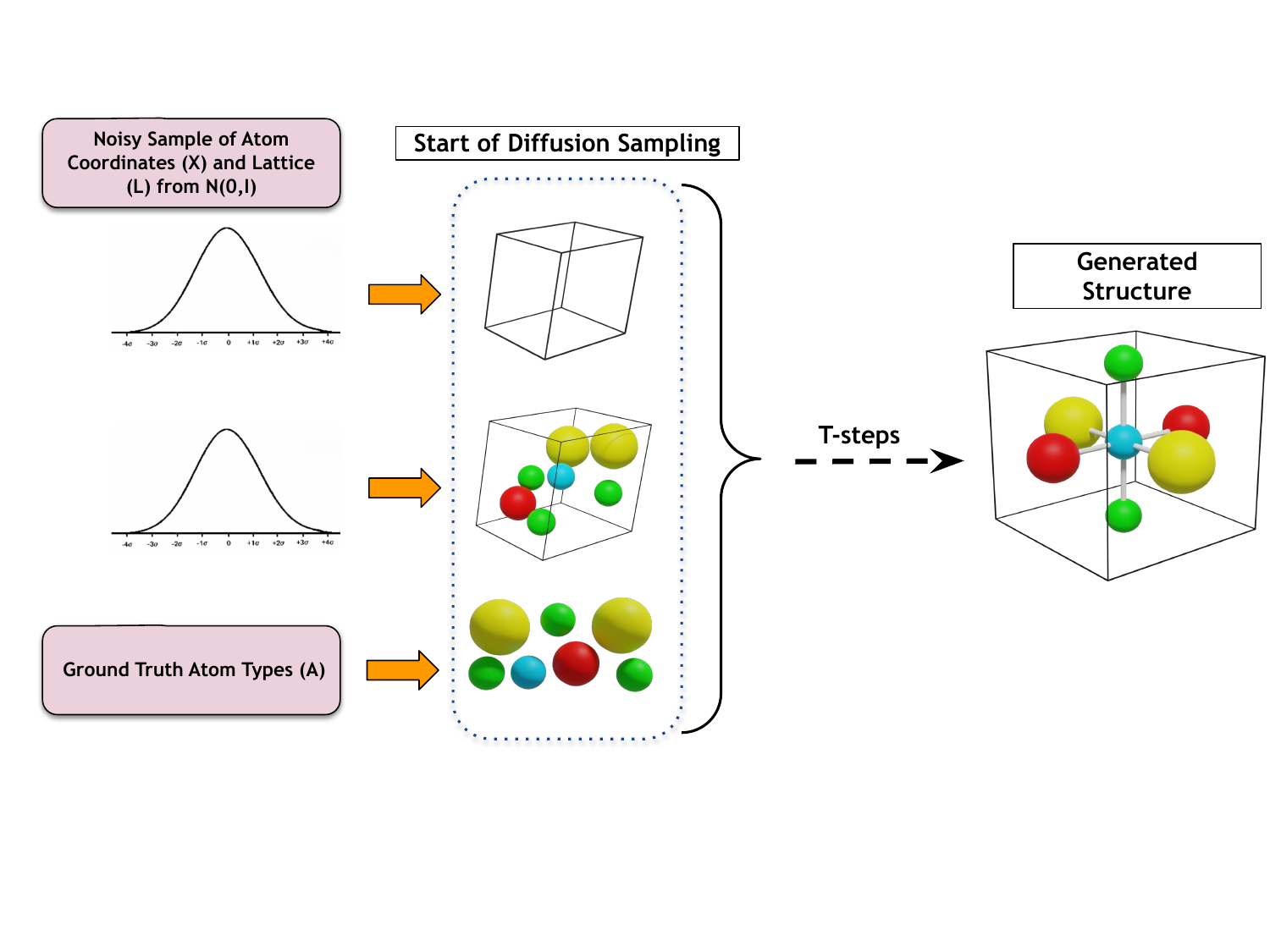}}
    \\
    \subfloat[Crystal Structure Prediction (CSP): atom types of the materials are given and the goal is to predict/match the crystal structure (atom coordinates and lattice).]
    {\includegraphics[width=\columnwidth, height = 300pt]{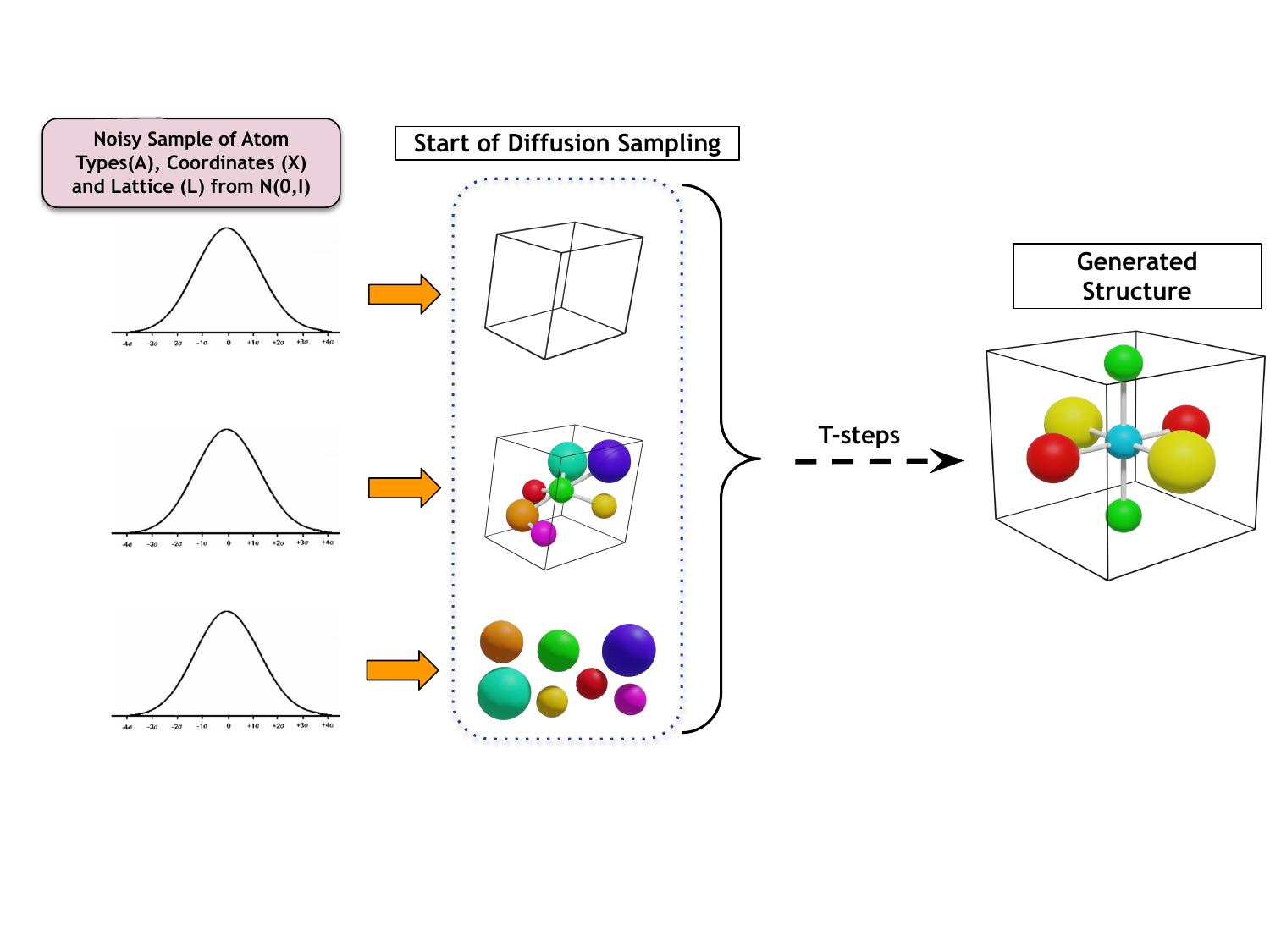}}
	\caption{Different benchmark tasks for crystal material generation.}
	\label{fig:task}
\end{figure}

\xhdr{Benchmark Tasks}
We evaluate our proposed model \iclrmodel{} on two different categories of tasks for material generation, \textit{Random Material Generation (Gen)} and \textit{Crystal Structure Prediction (CSP)}. In \textit{Gen} task, the goal of the generative model is to generate novel stable materials (atom types, fractional coordinates, and lattice structure). In \textit{CSP} task, atom types of the materials are given and the goal is to predict/match the crystal structure (atom coordinates and lattice). In \iclrmodel{} model, by design choice, we use the textual description of crystal materials during each step of the reverse diffusion process to enhance the generation capability in both tasks. A pictorial illustration of both tasks is provided at \ref{fig:task}
\\\\
\xhdr{Dataset} Following Xie et al ~\cite{xie2021crystal} we evaluate our model on three baseline datasets: \textbf{Perov-5}, \textbf{Carbon-24} and \textbf{MP-20}. \textbf{Perov-5} ~\cite{castelli2012new,castelli2012computational} dataset consists of 18,928 perovskite materials, each with 5 atoms in a cell. They generally can be denoted by $\mathbf{ABX_3}$ indicating the three different types of atoms usually observed in such materials. \textbf{Carbon-24} ~\cite{carbon} dataset has 10,153 materials with 6 to 24 atoms of carbon in the crystal lattice. Finally, \textbf{MP-20}~\cite{jain2013materials} dataset has 45,231 materials curated from the Materials Project library ~\cite{10.1063/1.4812323}, where each material has at most 20 atoms in the lattice. Crystals from \textbf{Perov-5} dataset share the same structure but differ in composition, whereas Crystals from \textbf{Carbon-24} share the same composition but differ in structure.  Crystals from \textbf{MP-20}  differs in both structure and composition. We curated textual data for these datasets with a textual description of each material. Specifically, we generate both long detailed textual descriptions and shorter prompts using approaches mentioned in \ref{text_data}. \\
The structures in all three datasets are derived from quantum mechanical simulations and are all at local energy minima. Most materials in \textbf{Perov-5} and \textbf{Carbon-24} are hypothetical, whereas \textbf{MP-20} represents a realistic dataset that includes many experimentally known inorganic materials, each with a maximum of 20 atoms in the unit cell, most of which are globally stable. A model that performs well on MP-20 could potentially generate novel materials that can be synthesized experimentally. While training \iclrmodel{}, we split the datasets into the train, test, and validation sets following the convention of 60:20:20 as done by Xie et al ~\cite{xie2021crystal}.
\\\\
\xhdr{Hyper-Parameters Details}
In our \iclrmodel{} model, we adopted 4 layers CSPNet as message passing layer with hidden dimension set as 512. Further, we use pre-trained MatSciBERT~\cite{gupta_matscibert_2022} followed by a two-layer projection layer (projection dimension 64) as the text encoder module. We keep the dimension of time embedding at each diffusion timestep as 64. We train it for 500 epochs using the same optimizer, and learning rate scheduler as DiffCSP and keep the batch size as 512. We perform all the experiments in the Tesla P100-PCIE-16GB GPU server.

\subsection{Crystal Structure Prediction (CSP)}
\label{results_csp}
\begin{table}[t] 
\centering
\setlength{\tabcolsep}{8 pt}
\resizebox{1.0\textwidth}{!}{
\begin{tabular}{c | c | c c | c c| c c }
\toprule
\multirow{2}*{Method} &  \multirow{2}*{\# Samples} & \multicolumn{2}{c}{Perov-5} & \multicolumn{2}{c}{Carbon-24} & \multicolumn{2}{c}{MP-20}  \\
& & Match Rate $\uparrow$ & RMSE $\downarrow$ & Match Rate $\uparrow $ & RMSE $\downarrow$ & Match Rate $\uparrow $ & RMSE $\downarrow$ \\
\midrule
\multirow{2}*{\shortstack{P-cG-SchNet}} 
& 1  & 48.22 & 0.4179 & 17.29 & 0.3846 & 15.39  & 0.3762\\
& 20 & 97.94 & 0.3463 & 55.91 & 0.3551 & 32.64  & 0.3018\\
\midrule
\multirow{2}*{CDVAE} 
& 1  & 45.31 & 0.1138 & 17.09 & 0.2969 & 33.90 & 0.1045\\
& 20 & 88.51 & 0.0464 & 88.37 & 0.2286 & 66.95 & 0.1026\\
\midrule
\multirow{2}*{DiffCSP} 
& 1  & 52.02 & 0.0760 & 17.54 & 0.2759 & 51.49 & 0.0631 \\
& 20 & \textbf{98.60}   & \underline{0.0128} & 88.47 & 0.2192 & 77.93 & 0.0492 \\
\midrule
\multirow{2}*{\shortstack{\iclrmodel{}\\(Short)}} 
& 1 & \underline{56.54} & \underline{0.0583} & \underline{24.13} & \underline{0.2424} & \underline{52.22} & \underline{0.0597} \\
& 20 & 98.25 & 0.0137 & 88.28 & 0.2252 & \underline{80.97} & \underline{0.0443}  \\
\midrule
\multirow{2}*{\shortstack{\iclrmodel{}\\(Long)}} 
& 1 & \textbf{90.46} & \textbf{0.0203} & \textbf{44.63} & \textbf{0.2266} & \textbf{55.15} & \textbf{0.0572} \\
& 20 & \underline{98.59} & \textbf{0.0072} & \textbf{95.27} & \textbf{0.1534} & \textbf{82.02} & \textbf{0.0483} \\
\bottomrule
\end{tabular} 
}
\caption{\rev{Summary of results on \textit{CSP} task. We highlight the best and second-best performances in bold and underlined, respectively. Both variants of \iclrmodel{} consistently outperform all baselines across three datasets. Unlike prior diffusion models, which require generating many samples per material to achieve good performance, \iclrmodel{} achieves strong Match Rates and low RMSE even with a single generated sample, significantly reducing computational cost. Incorporating textual guidance during reverse diffusion improves alignment with realistic 3D crystal geometries. While \iclrmodel{} remains superior with 20 samples on Carbon-24 and MP-20, its performance is comparable to DiffCSP on Perov-5.}}
  \label{tbl-recons}
\end{table}
\xhdr{Setup} In \textit{CSP} task, the goal is to predict the crystal structure (atom coordinates and lattice) given atom types. In text guidance setup, \iclrmodel{} utilizes textual descriptions during the denoising process and jointly predicts atom coordinates and lattice parameters from randomly sampled noise. To assess \iclrmodel{}'s effectiveness, we choose three SOTA generative models: \textbf{P-cG-SchNet}~\cite{gebauer2022inverse}, \textbf{CDVAE}~\cite{xie2021crystal}, and \textbf{DiffCSP}~\cite{jiao2023crystal}. 
\\\\
\xhdr{Evaluation Metrics} We evaluate the performance of \iclrmodel{} and baseline models on stable structure prediction using standard metrics proposed by the prior works ~\cite{jiao2023crystal,xie2021crystal}, by matching the generated structure and the input ground truth structure in the test set. In Specific, for each material structure in the test set, we generate k samples given the textual description and then identify the matching if at least one of the samples matches the ground truth structure. We calculate the \textbf{Match Rate} and \textbf{RMSE} metrics using the StructureMatcher class in Pymatgen, which identifies the best match between two structures while accounting for all material invariances. Match rate indicates the percentage of the matched structures over the test set satisfying thresholds stol=0.5, angle\_tol=10, ltol=0.3. RMSE is computed between the ground truth and the best-matching candidate, normalized by $\sqrt[3]{V/N}$ where V is the volume of the lattice, and averaged over the matched structures. For baselines and \iclrmodel{}, we evaluate using $k = 1$ and $k = 20$.
\\\\
\xhdr{Results and Discussions} We report the Match Rate and RMSE of all the baselines and \iclrmodel{} for three benchmark datasets in Table \ref{tbl-recons}.
We trained \iclrmodel{} using both detailed textual descriptions and short prompts, as outlined in \ref{text_data} and report them as \textbf{\iclrmodel{}(Long)} and \textbf{\iclrmodel{}(Short)} respectively in Table \ref{tbl-recons}. We observe that both variants of \iclrmodel{} surpass all the baseline models with a good margin across three datasets, which shows the rich capability of text-guided joint diffusion to predict stable crystal structure. Notably, while prior diffusion models demonstrate improved Match Rates and lower RMSE when generating 20 samples per test material, they largely fail in both metrics when generating only one sample per test material. However, generating 20 samples per test material to match the structures is unrealistic and computationally burdensome. This highlights the importance of text-guided diffusion: incorporating textual knowledge during reverse diffusion aids in aligning the noisy structure with the 3D geometry of stable realistic materials. Specifically, with just one generated sample ($k = 1$) per test material, both the variants of \iclrmodel{} outperform all baseline models, thereby reducing computational overhead. Moreover, even with 20 generated samples ($k = 20$) performance of \iclrmodel{} is significantly better for Carbon-24 and MP-20, whereas comparable with DiffCSP for Perov-5.  In general, due to the extensive metadata provided by the detailed description about the crystal structure, the performance enhancement in the long variant surpasses that of the short one. However, these findings collectively demonstrate the effectiveness of text-guided diffusion in the stable crystal structure prediction task.

\subsection{Random Material Generation (Gen)}\label{results_gen} 
\begin{table}[t]
\centering
\footnotesize
\setlength{\tabcolsep}{7 pt}
\renewcommand{\arraystretch}{0.9}

\resizebox{1.0\textwidth}{!}{
\begin{tabular}{c | c | c c | c c| c c c}
\toprule
\multirow{2}*{Dataset} &  \multirow{2}*{Method} & \multicolumn{2}{c|}{Validity $\uparrow$} & \multicolumn{2}{c|}{Coverage $\uparrow$} & \multicolumn{3}{c}{Property Statistics (EMD) $\downarrow$}  \\
& & Compositional & Structural & COV-R & COV-P  & \# Element  & $\rho$ & $\mathcal{E}$ \\
\midrule
\multirow{8}*{Perov-5} 
& CDVAE   & 98.59 & \textbf{100} & 99.45 & 98.46 & 0.0628 & 0.1258 & 0.0264 \\
& CDVAE+  & 98.45 & 99.8 & 99.53 & 99.09 & 0.0609 & 0.1276 & 0.0223 \\
& SyMat   & 97.40 & \textbf{100} & 99.68 & 98.64 & 0.0177 & 0.1893 & 0.2364 \\
& SyMat+  & 97.88 & \underline{99.9} & 99.70 & 98.79 & 0.0172 & 0.1755 & 0.2566 \\
& DiffCSP & \textbf{98.85} & \textbf{100} & \underline{99.74} & 98.27 & 0.0128 & 0.1110 & 0.0263 \\
& DiffCSP+ & 98.44 & \textbf{100} & 99.85 & 98.53 & 0.0119 & 0.1070 & \underline{0.0241} \\
& \iclrmodel{}(Short) & 98.28 & \textbf{100} & 99.7 & \underline{99.24} & \underline{0.0108} & \underline{0.0947} & 0.0257 \\
& \iclrmodel{}(Long) & \underline{98.63} & \textbf{100} & \textbf{99.83} & \textbf{99.52} & \textbf{0.0090} & \textbf{0.0497} & \textbf{0.0187} \\

\midrule
\multirow{8}*{Carbon-24} 
& CDVAE    & - & \textbf{100} & \underline{99.8} & 83.08 & - & 0.1407 & 0.285 \\
& CDVAE+   & - & \textbf{100} & \underline{99.8} & 84.76 & - & 0.1377 & 0.266 \\
& SyMat    & - & \textbf{100} & \textbf{99.9} & \underline{97.59} & - & 0.1195 & 3.9576 \\
& SyMat+   & - & \textbf{100} & \textbf{99.9} & \textbf{97.63} & - & 0.1171 & 3.862 \\
& DiffCSP  & - & \textbf{100} & \textbf{99.9} & 97.27 & - & 0.0805 & \underline{0.082} \\
& DiffCSP+ & - & \textbf{100} & \textbf{99.9} & 97.33 & - & 0.0763 & 0.085 \\
& \iclrmodel{}(Short) & - & \textbf{100} & \underline{99.8} & 91.77 & - & \underline{0.0681} & 0.087 \\
& \iclrmodel{}(Long) & - & \textbf{100} & \textbf{99.9} & 92.43 & - & \textbf{0.043} &  \textbf{0.063} \\

\midrule
\multirow{8}*{MP-20} 
& CDVAE  &  86.70 & \textbf{100} & 99.15 &  99.49 &  1.432  & 0.6875 & 0.2778 \\
& CDVAE+ & 87.42 & \textbf{100} & 99.57 & 99.81 & 0.972 & 0.6388 & 0.2977 \\
& SyMat  & 88.26 & \textbf{100} & 98.97 & \textbf{99.97} &0.5067  & 0.3805 & 0.3506 \\
& SyMat+  & \underline{88.47} & \underline{99.9} & 99.01 & \underline{99.95} & 0.4865 & 0.3879 & 0.3489 \\
& DiffCSP & 83.25 & \textbf{100} & 99.71 & 99.76 &  0.3398 & 0.3502 & 0.1247 \\
& DiffCSP+ & 85.07 & \textbf{100} & \underline{99.8} & 99.89 & \underline{0.3122} & 0.3799 & 0.1355 \\
& \iclrmodel{}(Short) & 86.60 & \textbf{100} & 99.79 & 99.88 & 0.3337 & \textbf{0.3296} & \textbf{0.1154} \\
& \iclrmodel{}(Long) & \textbf{92.97} & \textbf{100} & \textbf{99.89} & \underline{99.95} & \textbf{0.2890} & \underline{0.3382} & \underline{0.1189}  \\
\bottomrule
\end{tabular} 
}
\caption{Summary of results on \textit{Gen} task, with the best and second-best performances in bold and underlined, respectively. The table contains "-" values for metrics that don't apply to certain datasets.}
\label{tbl-gen}
\end{table}
\xhdr{Setup} In \textit{Gen} task, the goal is to generate novel stable materials (both structure and atom types), that are distributionally similar to the materials in the test dataset. In \iclrmodel{}, by design choice, we use the textual description of crystal materials in the test dataset during the reverse diffusion process to enhance the generation capability. To evaluate performance of \iclrmodel{} in this task, we choose three popular state-of-the-art generative models: \textbf{CDVAE}~\cite{xie2021crystal}, \textbf{SyMat}~\cite{luo2023towards}, and \textbf{DiffCSP}~\cite{jiao2023crystal}. For a fair comparison, we also consider their text-guided variants such as \textbf{CDVAE+}, \textbf{SyMat+} and \textbf{DiffCSP+} respectively, where we fuse the contextual representation of (Long) text data into those models using the same process described in \ref{text_guided_diffusion}. 
\\\\
\xhdr{Evaluation Metrics} Following CDVAE~\cite{xie2021crystal}, we evaluate the performance of \iclrmodel{} and baseline models on generating novel material structure using seven metrics under three broad categories: \textbf{Validity}, \textbf{Coverage}, and \textbf{Property Statistics}. Under \textbf{Validity}, following the prior line of work ~\cite{court20203,xie2021crystal}, we measure structural and compositional validity, representing the percentages of generated crystals with valid periodic structures and atom types, respectively. A structure is valid as long as the shortest distance between any pair of atoms is larger than 0.5\textup{~\AA}  whereas the composition is valid if the overall charge is neutral as computed by SMACT~\cite{davies2019smact}. In \textbf{Coverage}, we consider two coverage metrics, COV-R (Recall) and COV-P (Precision). COV-R measures the percentage of the test set materials being correctly predicted, whereas COV-P measures the percentage of generated materials that cover at least one of the test set materials. (More detailed discussions can be found in  ~\cite{xie2021crystal} and ~\cite{ganea2021geomol}). Finally, we evaluate the similarity between the generated materials and those in the test set using various \textbf{Property Statistics}, where we compute the earth mover’s distance (EMD) between the distributions in element number (\# Elem), density ($\rho$, unit g/cm3), and formation energy ($\mathcal{E}$, unit eV/atom) predicted by a GNN model.
\\\\
\xhdr{Results and Discussions} We report the result of \textbf{\iclrmodel{} (Long and Short)} and all the baseline models in Table \ref{tbl-gen}. We observe that both variants of \iclrmodel{} consistently enhances performance across almost all metrics across the benchmark datasets. Particularly on the Perov-5 dataset, \iclrmodel{} outperforms all baseline models across all metrics except for compositional validity, where its performance is on par with state-of-the-art results. In the Carbon-24 and MP-20 datasets, \iclrmodel{} exhibits performance improvements across all metrics except for COV-P. Additionally, our experiments indicate that utilizing shorter prompts results in a slight decrease in overall performance compared to the longer variant. Nonetheless, the performance remains comparable to baseline models. Overall, \iclrmodel{} exhibits promising performance in the Random Material Generation task, indicating its effectiveness in integrating global textual knowledge into the reverse diffusion process to generate more stable periodic structures of 3D crystal materials. Moreover \iclrmodel{ } outperforms text guided variant of baseline models CDVAE+, SyMat+ and DiffCSP+ as well, which highlights the effectiveness of using joint diffusion to learn  \textbf{\textit{A}},\textbf{\textit{X}} and \textbf{\textit{L}}.

\subsection{Correctness of Generated Materials}
\label{result_corectness}
\begin{table*}[ht]
  \centering
  \small
    \setlength{\tabcolsep}{10 pt}
    \scalebox{1}{
\begin{tabular}{c | c | c c c }
\toprule
Method & Global Features  & \multicolumn{3}{c}{\% of Matched Materials}  \\
& in Text Prompt   & Perov-5 & Carbon-24 & MP-20 \\
\midrule
\multirow{4}*{\iclrmodel{}(Long)} 
& Formula           & 97.50 & 98.20 & 70.54\\
& Space Group       & 87.00 & 80.79 & 67.88\\
& Crystal System    & 92.60 & 91.55 & 73.54\\
& Formation Energy  & 95.49 & - & 92.88\\
& Band Gap          & - & 98.61 & 96.73\\
\midrule
\multirow{4}*{\iclrmodel{}(Short)} 
& Formula                   & 90.70 & 92.56 & 65.22\\
& Space Group               & 86.51 & 80.50 & 58.77\\
& Crystal System            & 83.19 & 81.64 & 72.77 \\
& Formation Energy          & 90.33 & - & 91.00\\
& Band Gap                  & - & 95.90 & 93.33\\

\bottomrule
\end{tabular} 
}
\caption{Summary of results on \% of generated materials matching different global features specified by the textual prompts.
}
  \label{tbl-match}
\end{table*}
\xhdr{Setup} In this section, we investigate whether the generated material matches different features specified by the textual prompts. \iclrmodel{} has the capability to process textual prompts given by the user, enabling it to manage global attributes about crystal materials such as Formula, Space group, Crystal System, and different property values like formation energy, band-gap, etc. To ensure the fidelity of our model's outputs concerning these specified global attributes from the text prompt, We randomly generated 1000 materials (sampled from all three Datasets) based on their respective textual descriptions(both Long and Short) and assessed the percentage of generated materials that matched the global features outlined in the text prompt. In specific, we matched the Formula, Space group, and Crystal System, and Dimensions of generated materials with the textual descriptions. Moreover, we examined whether properties such as formation energy and bandgap matched the specified criteria as per the text prompt (positive/negative, zero/nonzero). \\\\
\xhdr{Results and Discussions} We report the results in Table \ref{tbl-match}. 
In general, using longer text, considering Perov-5 and Carbon-24 datasets, the generated material meets the specified criteria effectively. However, when dealing with the MP-20 dataset, which is more intricate due to its complex structure and composition, performance tends to decline. Additionally, when using shorter prompts, overall performance suffers across all datasets compared to longer text inputs. This is because the longer text, provided by the robocrystallographer, offers a comprehensive range of information, both global and local, thereby enhancing the generation capabilities of \iclrmodel{}.

\begin{table*}[ht]
  \centering
  \small
    \setlength{\tabcolsep}{10 pt}
   \scalebox{0.8}{
\begin{tabular}{ c|c|c|c|c|c|c }
\toprule
\multirow{2}*{Text Encoder} & \multicolumn{2}{c|}{Perov-5} & \multicolumn{2}{c|}{Carbon-24} & \multicolumn{2}{c}{MP-20} \\
& Comp(\%) $\uparrow$ & Struct (\%)  $\uparrow$ & Comp(\%) $\uparrow$ &  Struct (\%) $\uparrow$ & Comp(\%) $\uparrow$ &  Struct (\%) $\uparrow$ \\
\midrule
 Only Formula     & 97.06 & 99.19 & - & 98.76 & 86.16 & 96.01\\
 Only Space Group & 85.91 & 98.97 & - & 95.39 & 84.22 & 96.88\\
 Only Property    & 96.62 & 98.53 & - & 94.21 & 86.53 & 91.73\\
 Full Text & \textbf{98.28} &\textbf{100}   & - & \textbf{100}   & \textbf{86.60} & \textbf{100} \\
 \bottomrule
\end{tabular}
}
\caption{Summary of results on generated materials using more custom/shorter Prompt.}
  \label{tbl-short-prompt}
\end{table*}

\subsection{Performance on More Shorter Prompts}
\label{sec-shorter-prompt}
In this section, we explore the generalizability and robustness of our model by examining potential variability in text description lengths. The goal of this paper is, given the text prompt, to generate specific material, not any generic or class of materials. Hence some minimum essential information about the crystal, like formula, space group, crystal system, property value, etc must be given as input to the pre-trained model. However, to investigate the robustness of our proposed \iclrmodel{} model with more custom and shorter prompts, we did an experiment where we evaluated \iclrmodel{} (trained with full text) with even shorter custom prompts with very little information as follows:
\begin{itemize}
    \item \textbf{Specifying only Formula: }\textit{"The chemical formula is GaSiSO2. The elements are Ga, Si, S, O. Generate the material."}
    \item \textbf{Specifying only Space Group Info: }\textit{"The spacegroup number is 1. Generate the material."}
    \item \textbf{Specifying only Property Info: }\textit{"The formation energy per atom is positive. Generate the material."}
\end{itemize}
We report the results in table \ref{tbl-short-prompt}. We observe that though \iclrmodel{} can handle more custom prompts, but it affects the quality of generated materials. Hence, we conclude some minimum essential information about the crystal must be given as input to \iclrmodel{} to generate high quality crystal materials.

\subsection{Computational Cost for Training and Sampling}\label{results_time}
Integrating textual knowledge during reverse diffusion for crystal material generation offers a key advantage: it accelerates convergence towards realistic structures and reduces computational overhead. We observe, compared to other baseline models, \iclrmodel{} incurs substantially lower computation costs during training and sampling processes. Compared to baseline models, our approach notably cuts down on training time, requiring only 500 epochs compared to 3K or 4K epochs for CDVAE and DiffCSP on Perov-5 and Carbon-24 datasets respectively. Additionally, our method reduces sampling steps, making it faster to generate new structures. While CDVAE and DiffCSP need 5K and 1K steps respectively, our model only requires 500 steps. We compare the performance of  CDVAE and DiffCSP with different \iclrmodel{}(Long) variants with 50, 100, 200, 500, and 1K steps and report the match rate of the predicted crystal structure vs running time (GPU hours in P100 GPU server) for Perov-5 and Carbon-24 datasets in Fig. \ref{fig:ablation}(a). \rev{Note, this computation also involves running MatSciBERT. However, during inference, the text prompt is fed into MatSciBERT only once at the beginning to generate the contextual embedding. This embedding is then used throughout the $\tau$ diffusion steps, limiting the computational cost of running MatSciBERT.} We notice that the inference time for CDVAE is lengthier as it necessitates 5K steps for each generation. However, for Carbon-24, \iclrmodel{} with 200 or 500 steps outperforms DiffCSP with 1K steps. Additionally, for Perov-5, \iclrmodel{} with 500 steps achieves results comparable to DiffCSP with 1K steps.
\subsection{Smoothness of Model's Generation}
\label{results_ablation}
\begin{figure}
    \centering
	\subfloat[Match Rate vs Running time]{\includegraphics[width=0.45\columnwidth,height=3cm]{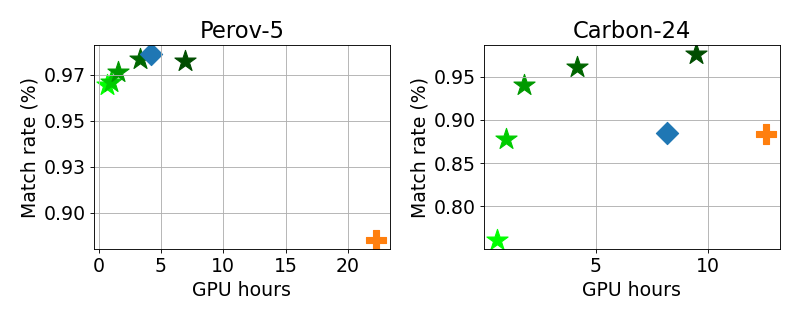}}
    \hspace{10mm}
    \subfloat[Smoothness of \iclrmodel{}]{\boxed{\includegraphics[width=0.30\columnwidth]{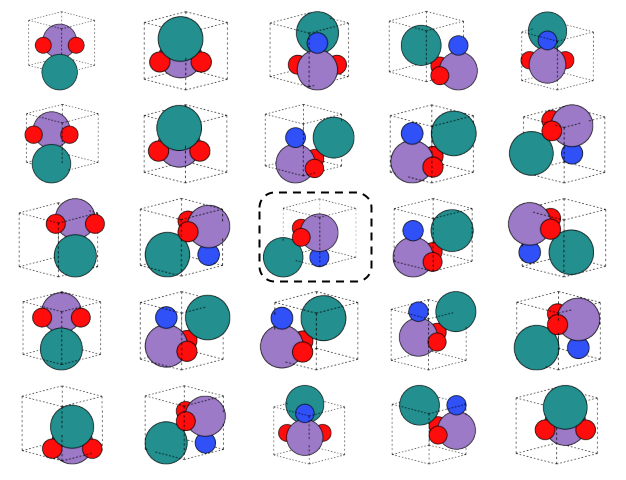}}}
	\caption{(a) Match Rate vs Running time (GPU Hours) for different variants of \iclrmodel{}(Long) \{50 Steps $\textcolor[rgb]{0,1,0,1}{\star}$, 100 Steps $\textcolor[rgb]{0,0.8,0,1}{\star}$, 200 Steps $\textcolor[rgb]{0,0.6,0,1}{\star}$, 500 Steps $\textcolor[rgb]{0,0.4,0,1}{\star}$, 1K Steps $\textcolor[rgb]{0,0.3,0,1}{\star}$ \}, DiffCSP $\textcolor[rgb]{0.12,0.46,0.70}{\blacklozenge}$ and CDVAE \textcolor[rgb]{1,0.498,0.054}{\textcolor[rgb]{1, 0.498, 0.054}{
        $\mathord{\text{\ding{58}}}$}}. (b) Materials sampled given the textual description of the center ground truth material $\boxed{\textbf{M}}$. The sampled materials are structurally similar (rotated or translated) to each other as well as the ground truth.}
	\label{fig:ablation}
\end{figure}

We qualitatively demonstrate the smoothness of the crystal generation process in our model. We provide a textual description of a ground truth material and generate several samples of materials to assess the diversity of the generated structures. Figure \ref{fig:ablation}(b) summarizes the results for one crystal material from the Perov-5 dataset, which shows that the generated materials are structurally similar to each other and the given input material.

\subsection{Ablation Study: Choice of Text Encoder}
\label{result_text_enc}
\begin{table*}[ht]
\centering
  \small
\caption{Ablation study results on different choices of Text Encoders.}
\label{tbl-text-enc}
\renewcommand{\arraystretch}{0.9}
\setlength{\tabcolsep}{6 pt}
\scalebox{0.8}{
\begin{tabular}{ c|c|c|c|c|c|c }
\toprule
\multirow{2}*{Text Encoder} & \multicolumn{2}{c|}{Perov-5} & \multicolumn{2}{c|}{Carbon-24} & \multicolumn{2}{c}{MP-20} \\
& MR $\uparrow$ & RMSE $\downarrow$ & MR $\uparrow $& RMSE $\downarrow$ & MR $\uparrow $& RMSE $\downarrow$ \\
\midrule
BERT & 96.64 & 0.0109 & 72.21 & 0.2679 & 79.53 & 0.057 \\

MatSciBERT & \textbf{98.63} & \textbf{0.0072} & \textbf{95.27} & \textbf{0.1534} & \textbf{82.02} & \textbf{0.039} \\
\midrule
& Comp $\uparrow$ & Struct $\uparrow$ & Comp $\uparrow$ & Struct $\uparrow$ & Comp $\uparrow$ & Struct $\uparrow$ \\
\midrule
 BERT & 97.44 & 99.97 & - & \textbf{100} & 84.73 & 98.37\\
 MatSciBERT & \textbf{98.63} & \textbf{100} & - & \textbf{100} & \textbf{92.97} & \textbf{100} \\
 \bottomrule
\end{tabular}
}
\end{table*}
Further, we investigate the expressiveness of textual representation during the reverse diffusion process. In particular, we are interested in understanding whether there are any benefits we are gaining from using a domain-specific pre-trained text encoder MatSciBERT. We conduct an ablation study where we substitute MatSciBERT with pre-trained BERT~\cite{devlin2018bert} model (which is domain agnostic) as text encoder in \iclrmodel{ and evaluate the performance on both tasks. The results presented in Table \ref{tbl-text-enc} demonstrate that MatSciBERT surpasses BERT~\cite{devlin2018bert}
in performance for both tasks. This highlights the richer expressiveness of contextual representation achieved through the use of a domain-specific pre-trained language model.

\subsection{Visualization of Generated Materials}
\label{result_visualize}
\begin{table*}[ht]
  \centering
    \setlength{\tabcolsep}{1.5 pt}
    \renewcommand{\arraystretch}{1}
    \resizebox{1.0\textwidth}{!}{
    \begin{tabular}{m{4in}|m{3in}|c|cccc| } 
\toprule
Detailed Description & Short Prompt &  Ground truth & \multicolumn{4}{c|}{Generated Samples} \\
\midrule
 
YCoSO2 crystallizes in the orthorhombic Pmm2 space group. Y is bonded in a distorted square co-planar geometry to two equivalent S, two equivalent O, and two equivalent O atoms. Both Y-S bond lengths are 2.74 Å. Both Y-O bond lengths are 2.24 Å. There is one shorter (2.09 Å) and one longer (2.39 Å) Y-O bond length. \dots In the second O site, O is bonded in a distorted square co-planar geometry to two equivalent Y and two equivalent S atoms. & 
Below is a description of a bulk material. The chemical formula is YCoSO2. The elements are Y, Co, S, O. The formation energy per atom is positive. The spacegroup number is 24. The crystal system is orthorhombic. Generate the material:& 
\begin{minipage}{.1\linewidth}
  \includegraphics[width=\linewidth]{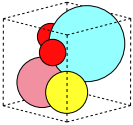}
\end{minipage} & 
\begin{minipage}{.1\linewidth}
  \includegraphics[width=\linewidth]{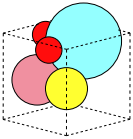}
\end{minipage} & 
\begin{minipage}{.1\linewidth}
  \includegraphics[width=\linewidth]{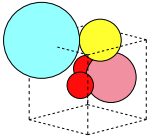}
\end{minipage} & 
\begin{minipage}{.1\linewidth}
  \includegraphics[width=\linewidth]{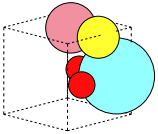}
\end{minipage} & 
\begin{minipage}{.1\linewidth}
  \includegraphics[width=\linewidth]{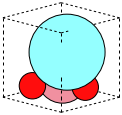}
\end{minipage} 
\\
  
\midrule

C crystallizes in the orthorhombic Cmcm space group. There are two inequivalent C sites. In the first C site, C(1) is bonded to one C(2) and three equivalent C(1) atoms to form a mixture of corner and edge-sharing CC4 trigonal pyramids. \dots There are two shorter (1.54 Å) and one longer (1.56 Å) C(2)-C(2) bond length. The energy per atom is -154.2425.&
Below is a description of a bulk material. The chemical formula is C. The elements are C. The energy per atom is negative. The spacegroup number is 62. The crystal system is orthorhombic. Generate the material.& 
\begin{minipage}{.1\linewidth}
  \includegraphics[width=\linewidth]{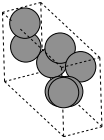}
\end{minipage} & 
\begin{minipage}{.1\linewidth}
  \includegraphics[width=\linewidth]{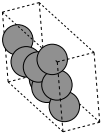}
\end{minipage} & 
\begin{minipage}{.1\linewidth}
  \includegraphics[width=\linewidth]{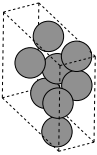}
\end{minipage} & 
\begin{minipage}{.1\linewidth}
  \includegraphics[width=\linewidth]{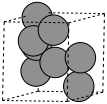}
\end{minipage} & 
\begin{minipage}{.1\linewidth}
  \includegraphics[width=\linewidth]{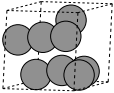}
\end{minipage} 
\\

\midrule
MgNdTl crystallizes in the hexagonal P-62m space group. Mg(1) is bonded in a 4-coordinate geometry to two equivalent Tl(1) and two equivalent Tl(2) atoms. Both Mg(1)-Tl(1) bond lengths are 3.01 Å. \dots In the second Tl site, Tl(1) is bonded in a 9-coordinate geometry to three equivalent Mg(1) and six equivalent Nd(1) atoms. The formation energy per atom is -0.355. The band gap is 0.0. The energy above the convex hull is 0.0. The spacegroup number is 188.& 
Below is a description of a bulk material. The chemical formula is NdMgTl. The elements are Nd, Mg, and Tl. The formation energy per atom is -0.355. The band gap is 0.0. The energy above the convex hull is 0.0. The spacegroup number is 188. The crystal system is hexagonal. Generate the material.& 
\begin{minipage}{.1\linewidth}
  \includegraphics[width=\linewidth]{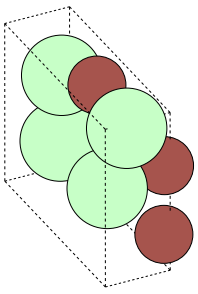}
\end{minipage} & 
\begin{minipage}{.1\linewidth}
  \includegraphics[width=\linewidth]{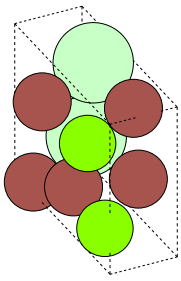}
\end{minipage} & 
\begin{minipage}{.1\linewidth}
  \includegraphics[width=\linewidth]{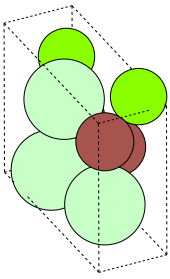}
\end{minipage} & 
\begin{minipage}{.1\linewidth}
  \includegraphics[width=\linewidth]{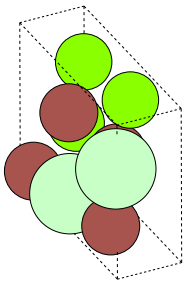}
\end{minipage} & 
\begin{minipage}{.1\linewidth}
  \includegraphics[width=\linewidth]{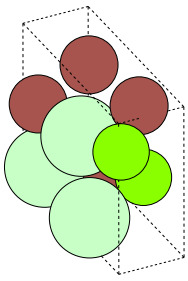}
\end{minipage} 
\\

\bottomrule
\end{tabular}
 }
   \caption{Visualization of the generated structures given textual description. Note that our model produces rotated or translated versions of the ground truth material, owing to the periodic-E(3)invariance.}
  \label{tbl-samples}
\end{table*}
Additionally, we present visualizations of a few generated materials based on the textual descriptions in Table \ref{tbl-samples} and compare them with the ground truth material structure. The generated samples exhibit clear matches to the ground truth structure, highlighting the generation capability of the \iclrmodel{} model given information in text form. 
\section{Conclusion}
\label{iclr_conclusion}
In this work, we explore a practical approach of generating stable crystal materials given a textual description of the material. We propose \iclrmodel{}, which jointly diffuse atom types, fractional coordinates, and lattice structure for crystal materials using a periodic-E(3)-equivariant denoising model. We further integrate textual information into the reverse diffusion process through a pre-trained transformer model, which guides the denoising process in learning the crystal 3D geometry matching the specification by textual description. Extensive experiments conducted on two benchmark generative tasks reveal that \iclrmodel{} surpasses all popular baseline models by a good margin. Furthermore, integrating textual knowledge reduces the overall computational cost for both training and inference of the diffusion model. Moreover, when applied to real-world custom text prompts by experts, \iclrmodel{} demonstrates rich generative capability under general textual conditions. Source code and dataset is available at \url{https://github.com/kdmsit/TGDMat}


\chapter{\textcolor{black}{Conclusion, Future Work and Limitation}}
\chaptermark{\textcolor{black}{Conclusion}}
\label{chap: conclusion}
\color{black}
In conclusion, this thesis presents a robust, efficient AI frameworks for crystal material discovery, focusing on two key tasks: (1) property prediction - accurately estimating chemical properties of materials to validate candidate structures, and (2) material generation - creating stable 3D periodic structures of new materials. The proposed solutions include CrysXPP (an explainable property predictor using transfer learning), CrysGNN (a large pre-trained graph neural network with knowledge distillation), CrysMMNet (a multi-modal framework combining graph structures with textual descriptions), and TGDMat (a text-guided joint diffusion model for generating materials conditioned on natural language descriptions). These methods collectively advance data-efficient, interpretable, and scalable approaches to accelerate materials discovery while addressing limitations like data scarcity, lack of global structural knowledge in existing models, and DFT error bias.

\section{Summary of contributions}
This thesis advances the field of Scalable Material Design and Property Prediction by establishing a suite of robust, interpretable, and scalable AI frameworks for both material property prediction and material generation. By addressing key limitations of existing methods, such as data scarcity, lack of interpretability, unimodal limitations, and uncontrollable generative processes, this thesis advances the frontier of AI for materials design, paving the way for accelerated and controllable discovery of novel materials with tailored properties.

This thesis advances the field of machine learning by exploring subset selection techniques as a foundational tool for building robust and efficient models. Across diverse settings, including federated learning, model compression, in-context learning with large language models, we demonstrate how carefully selected data or model subsets can offer computational efficiency, enhanced robustness, and performance comparable to full-scale datasets or models. The contributions of this thesis can be summarized as follows:

\begin{itemize}
    \item {\it Explainable Property Predictor for Crystalline Materials:} Our first major contribution of this thesis is the development of CrysXPP, an explainable property predictor for crystalline materials. One of the fundamental challenges in materials design lies in the scarcity of labeled property data, the lack of interpretability in deep learning models, and the bias induced by relying heavily on DFT-calculated datasets. To address these, CrysXPP leverages a transfer learning–based framework built on CrysAE, an unsupervised autoencoder trained on large-scale unlabeled crystal graphs. This design enables the model to learn enriched structural and chemical representations, which can then be transferred to the downstream property prediction tasks. CrysXPP achieves high accuracy even in low-data regimes, while a trainable feature selector highlights the most relevant atomic features responsible for a given property, thereby offering interpretability aligned with domain knowledge. Furthermore, fine-tuning with small amounts of experimental data helps mitigate DFT error bias, making predictions more reliable.

    \item {\it GNN Pretraining Model for Crystal Property Prediction:} Our second contribution is the design of CrysGNN, a large-scale pre-trained graph neural network for crystalline materials. While pretraining has become a standard paradigm in NLP and vision, its adaptation to crystalline graph structures had remained underexplored. To fill this gap, we curated a dataset of nearly 800K crystal graphs and proposed a self-supervised pretraining framework that learns both node-level (atomic) and graph-level (crystal) representations. CrysGNN captures rich local and global structural information, which can then be distilled into state-of-the-art property predictors such as CGCNN and ALIGNN. Through knowledge distillation, we show consistent performance improvements ranging from 4\% to 16\% across multiple benchmarks, with notable gains under data-scarce conditions. Additionally, by incorporating small amounts of experimental data during fine-tuning, CrysGNN helps reduce the bias inherited from DFT data, further enhancing robustness and accuracy.

    \item{\it Multi-Modal Representation for Crystal Property Prediction:} Our third contribution is CrysMMNet, a multimodal framework for crystal property prediction that goes beyond the traditional reliance on single-modality graph structures. While graph-based encoders effectively capture local atomic neighborhoods, they struggle to incorporate global structural information such as crystal symmetry, orientation, or space group. To overcome this limitation, we curate textual descriptions for crystals from popular databases (Materials Project and JARVIS-DFT) using Robocrystallographer, and design a framework that fuses graph encodings with textual representations via transformer-based models. This multimodal representation provides a more comprehensive view of the material, leading to consistent improvements across ten property prediction tasks. Extensive ablation studies confirm that combining global and local knowledge via textual and graph modalities results in more robust and enriched material representations compared to using either modality in isolation.

    \item{\it Text-Guided Joint Diffusion Model for Periodic Materials Generation:} Finally, our fourth contribution of this thesis is the proposal of TGDMat, a text-guided joint diffusion model for periodic material generation. Existing generative approaches for crystal materials often treat lattice parameters, atom types, and atomic coordinates separately, rely only on local message passing, and operate in an unconditional manner. TGDMat overcomes these limitations by conducting joint diffusion of all three structural components within a periodic E(3)-equivariant denoising framework, ensuring faithful geometric consistency. More importantly, it introduces text-guided conditioning, where textual prompts are integrated at every denoising step to control the generative trajectory. This innovation bridges natural language understanding with material structure generation, enabling users to specify constraints such as chemical formula, space group, or symmetry. Comprehensive experiments demonstrate that TGDMat significantly outperforms strong baselines in both crystal structure prediction (CSP) and random material generation, while also reducing computational costs. Its ability to handle expert-provided textual prompts showcases strong potential for real-world material discovery.
    
\end{itemize}

\section{Limitations}
\rev{In this thesis, we demonstrate that robust and efficient AI frameworks can substantially advance periodic material generation and property prediction. Nevertheless, the proposed models have certain limitations that need further investigation.}

\begin{itemize}
    \item \rev{Though our proposed pre-trained model CrysGNN is able to enhance the performance of all the state-of-the-art baseline models, improvements for all types of SOTA models are not consistent, and we found lesser improvement in case of complex models like GATGNN and ALIGNN. This provides scope for further investigation on designing a more complex and deeper pretrained model. There are also scopes of exploring different graph representations, loss variations, distillation strategies, etc. Of course, the very fact that the simple model, CrysGNN, can provide substantial improvements, lays the foundation for such future exploration. Moreover, in this present work, we have focused on predicting crystal properties, which is a graph-level regression task. However, the same framework can be used to explore the effect of pre-training on other graph-level tasks (classification or clustering) or node-level tasks, which would be one of our future works.}
    
    \item \rev{Our proposed multimodal framework for crystal materials, CrysMMNet do not fine-tune the MatSciBERT model and instead keep it frozen to generate text embeddings. Jointly tuning both the graph and text encoders would likely require additional cross-modal alignment losses and would be computationally demanding. Furthermore, evaluating the performance of more recent and powerful large language models within this framework remains an important direction for future work.}
    
    \item \rev{One of the major limitations of our generative model TGDMat is the lack of independent textual datasets for material generation tasks. In our experimental setup, our model relied on textual data extracted from existing datasets. Initially, we extracted text data from CIF files of materials in the test sets of Perov-5, Carbon-24, and MP-20, utilizing this data to evaluate our model and all baseline models. While the experimental results show promise, a more robust evaluation could have been achieved with an independent dataset containing only textual prompts. This would enable us to assess how effectively these models can generate the underlying 3D structure of materials through a text-guided diffusion process. Hence curating an independent textual dataset for material generation containing a diverse set of meta-information will be a future scope for research.}
    
    \item \rev{Moreover, given text prompts/descriptions, we generate contextual representation using a text encoder in TGDMat, where we adopted a pre-trained MatSciBERT model, which is a domain-specific language model for materials science. While training TGDMat, we freeze the MatSciBERT parameters and do not tune them further. Moreover, during sampling, the user must follow a specific format (Long/Short) to provide the text description of the target material. This setup limits the expressive power of the textual representation. We investigated the robustness of TGDMat with much shorter prompts to sample from pre-trained TGDMat model  but observed performance degradation across all the benchmark dataset on Gen task. Hence, Exploring state-of-the-art LLMs and further fine-tuning them during training may create more powerful text conditional diffusion models and provide flexibility to process text prompts of different formats. However, that might create computational overhead as it will increase the number of parameters significantly. This provides scope for further investigation and we keep it as scope for future work.
    }
\end{itemize}

\section{Future Research Directions}
In this thesis, we demonstrated how robust and efficient AI frameworks can significantly advance both crystal property prediction and periodic material generation. By introducing explainability, large-scale pretraining, multimodal learning, and text-guided diffusion, we established scalable methods that accelerate and control the discovery of novel materials with desired properties. While the presented contributions offer promising directions, several important avenues remain open for future research.
\begin{enumerate}
    \item \textbf{LLM-Guided Diffusion Models for Material Generation}  Recent generative models have emerged as promising alternatives, falling into two categories: \textbf{Large Language Models (LLMs):} Strong at predicting discrete atomic types and ensuring compositional validity but weak in handling continuous variables like coordinates and lattices. \textbf{Denoising-based frameworks (diffusion, flow matching):} Good at generating continuous variables with high structural validity but poor at capturing discrete atomic compositions. 
    
    To address the complementary weaknesses of existing approaches, a promising future direction is to develop a hybrid framework that integrates large language models (LLMs) with diffusion models. In such a design, the LLM predicts atom types, atomic coordinates, and lattice parameters, after which the atom types are retained while the coordinates and lattices are refined by a pre-trained equivariant diffusion model. This leverages the LLM’s strength in compositional reasoning and the diffusion model’s ability to refine geometry, striking a balance between structural and compositional validity. 
    
    This hybrid approach is expected to produce significantly more stable crystal materials than either standalone LLMs or diffusion models, while also supporting conditional generation such as user-defined compositions or space groups. Moreover, the framework is architecture-agnostic, making it adaptable to future advances in both LLMs and diffusion models, thereby paving the way for more controllable and scalable material discovery pipelines.

    \item \textbf{Accelerating Diffusion Sampling for Scalable Material Generation} A major limitation of current generative models for material generation is their reliance on a large number of sampling steps. For instance, models like CDVAE~\cite{xie2021crystal} and SyMat~\cite{luo2023towards} require up to 5000 steps, while DiffCSP~\cite{jiao2023crystal} and MatterGen~\cite{zeni2023mattergen} still need around 1000 steps. This makes large-scale material generation slow and computationally expensive. My goal is to enhance the efficiency of these models by optimizing sampling strategies, enabling faster and more scalable material discovery.

    \item \textbf{Inverse Material Generation}  An emerging and relatively underexplored direction in material discovery is inverse material generation—designing material structures based on desired properties. Unlike most existing generative models that focus on unconditional generation, producing stable crystal structures from random noise without external constraints, this approach enables goal-driven and data-efficient discovery. Unconditional models typically generate materials that resemble the training distribution but lack a mechanism for users to specify target properties. This limits their practical utility in real-world applications where property-driven design is crucial. A promising direction is text-guided generation, where users specify desired properties in natural language. These textual prompts can be processed by large language models (LLMs) to produce contextual embeddings that guide diffusion models toward generating materials aligned with specified criteria. While there have been some initial efforts~\cite{das2025periodic,park2025exploration}, this remains a challenging and open problem with significant potential for future research.

    \item \textbf{Foundation Models for Periodic Materials} Finally, I have an ambitious vision to develop a general-purpose, large-scale foundational model for periodic crystal systems—capable of performing both generative and discriminative tasks within a unified framework.
    
\end{enumerate}

\subsection{Prioritization of Future Work}
\rev{I prioritize future directions as follows. First, LLM-guided diffusion models for material generation are a crucial direction to explore, as hybrid frameworks combining large language models and diffusion models can exploit the strengths of both to generate more stable, valid, and physically meaningful materials. Next, accelerating diffusion sampling for scalable material generation is essential, for example by leveraging latent diffusion models that operate in low-dimensional, smooth latent spaces to reduce computational cost. Third, inverse material generation conditioned on specific target properties should be explored to enable goal-directed materials design. Finally, developing foundational models for crystal material capable of performing both generative and discriminative tasks within a unified framework represents a long-term but highly impactful research direction.}

\newpage

\begin{center}
    {\Large \textbf{IMMEDIATE EPILOGUE TO THE THESIS}}\\[0.4em]
\end{center}

\vspace{0.5em}

The thesis (viva-voce examination, 
IIT Kharagpur, January 2026) develops a coherent research program spanning both major 
components of the materials-design pipeline: property prediction (\textbf{CrysXPP}, 
\textit{npj Computational Materials} 2022; \textbf{CrysGNN}, AAAI 2023; 
\textbf{CrysMMNet}, UAI 2023) and periodic material generation (\textbf{TGDMat}, 
ICLR 2025). Chapter 7, ``Conclusion, Future Work and Limitation,'' identifies open 
problems and prioritized future directions, including LLM-guided diffusion models and 
latent-space diffusion. Three subsequent works accepted at NeurIPS 2025, ICML 2026, 
and UAI 2026, each with Kishalay Das as an equal-contributing author, form a natural 
immediate epilogue to this dissertation, continuing both its generative and 
property-prediction threads.

\section*{1. CrysLLMGen: LLM Meets Diffusion -- A Hybrid Framework for Crystal Material Generation}

\textbf{Venue:} 39th Conference on Neural Information Processing Systems 
(NeurIPS 2025).

The thesis (Section 7.3, Future Research Direction 1: ``LLM-Guided Diffusion Models 
for Material Generation'') notes that LLMs excel at predicting discrete atomic 
compositions but are weak at continuous variables such as coordinates and lattice 
parameters, while diffusion models show complementary strengths and weaknesses. 
It proposes, as the priority future direction, a hybrid framework where an LLM predicts 
atom types while a pre-trained diffusion model refines the continuous coordinates 
and lattice.

CrysLLMGen~\cite{khastagir2026llm} directly realizes this proposal: a fine-tuned LLM (LLaMA-2-7B) generates 
intermediate atom types, coordinates, and lattice; the atom types are retained, and 
the coordinates and lattice are refined by a pre-trained equivariant diffusion model. 
On standard benchmarks (Perov-5, MP-20), CrysLLMGen achieves balanced gains in 
structural and compositional validity and generates markedly more stable, unique, 
and novel materials than standalone LLM or diffusion-based baselines, while supporting 
strong conditional (text-guided) generation. This work was accepted at NeurIPS 2025 
before the thesis submission and represents a direct realization of Chapter 7's 
top-priority future direction.

\section*{2. CrysLDNet: Latent Diffusion Pretraining for Crystal Property Prediction}

\textbf{Venue:} 43rd International Conference on Machine Learning (ICML 2026).

A second thread of the thesis develops data-efficient, pretrained representations for 
property prediction, beginning with CrysXPP and scaled through CrysGNN, alongside a 
separate thread proposing latent-space diffusion as a future direction (Section 7.3) 
to overcome the limitations of diffusing directly in the high-dimensional feature 
space. Section 7.2 further notes that CrysGNN's gains are inconsistent across 
backbones, with smaller improvements for complex architectures such as GATGNN and 
ALIGNN, and flags scope for ``a more complex and deeper pretrained model.''

CrysLDNet~\cite{mukherjee2026latent} brings these two threads together using a latent diffusion-based pretraining 
framework. It combines a Variational Autoencoder (VAE) with a latent diffusion model: 
crystal structures are first mapped into a smooth, compact latent representation, and 
diffusion-based pretraining is then performed within this latent space. The resulting 
backbone-agnostic encoder yields consistent gains over both training-from-scratch and 
pretrain--finetune baselines on popular benchmark datasets. The framework is also 
particularly effective for structurally complex crystals, with relative improvements 
increasing to 11.10\% for highly complex systems, demonstrating the value of richer 
latent representations for challenging crystal structures. The work was accepted at 
ICML 2026, immediately after the thesis submission.

\section*{3. Model Agnostic Graph Prompt Learning for Crystal Property Prediction}

\textbf{Venue:} Conference on Uncertainty in Artificial Intelligence (UAI 2026).

A recurring theme of the thesis is the need to learn richer representations of crystal 
materials by capturing both local chemical semantics and global structural information. 
CrysGNN advances this objective through large-scale graph pretraining, while CrysMMNet 
further enriches crystal representations by integrating graph structures with textual 
descriptions containing global structural knowledge. The UAI 2026 work extends this 
research direction through a complementary and lightweight approach based on learnable 
soft prompts.

It introduces a model-agnostic, multi-level graph prompting framework~\cite{mukherjee2026model} in which 
node-level prompts learn latent chemical features that are not explicitly encoded in 
the input representation, while graph-level prompts capture global structural symmetry 
associated with different crystal systems. Importantly, this enrichment is achieved 
without redesigning the underlying GNN architecture: the framework introduces only a 
small parameter overhead (0.32\% for Matformer) and can be seamlessly integrated with 
a wide range of crystal graph encoders. Across six state-of-the-art backbones and two 
benchmark datasets, the prompt-enhanced models consistently improve crystal property 
prediction performance by 3--20\%. Accepted at UAI 2026, this work naturally extends 
the thesis's broader objective of developing robust, expressive, and computationally 
efficient representations for crystal property prediction.

\section*{Concluding Perspective}

Taken together, these three works demonstrate the continuity of the research program 
initiated during the PhD. On the generative side, the dissertation progresses from 
text-guided joint diffusion in TGDMat to the explicitly proposed idea of combining 
LLMs with equivariant diffusion models, realized as CrysLLMGen. On the 
property-prediction side, the dissertation's progression from unsupervised representation 
learning (CrysXPP) and large-scale graph pretraining (CrysGNN) to multimodal 
representation learning (CrysMMNet) is further extended by CrysLDNet's 
latent-diffusion pretraining and the UAI 2026 work's lightweight, model-agnostic 
prompt learning.

All three papers were completed by the candidate as first or joint-first author, in 
close temporal proximity to the thesis submission, and represent a direct and coherent 
continuation of the research directions charted in the thesis. These works are not 
retroactive additions to the formally submitted dissertation. Rather, they constitute 
an immediate epilogue to the thesis, illustrating how research questions, methodological 
principles, and future directions developed during the PhD continued to evolve into 
subsequent contributions at NeurIPS, ICML, and UAI.

\clearemptydoublepage
\bibliographystyle{abbrv}
\bibliography{Main.bib}
\clearemptydoublepage

\normalsize \backmatter \singlespace

\chapter{Publications from the Thesis}

The publications are listed in reverse chronological order.

\begin{enumerate}

    \item {\bf Kishalay Das}, Subhojyoti Khastagir, Pawan Goyal, Seung-Cheol Lee, Satadeep Bhattacharjee, Niloy Ganguly. "Periodic Materials Generation using Text-Guided Joint Diffusion Model." International Conference on Learning Representations (ICLR), 2025.

    \item {\bf Kishalay Das}, Pawan Goyal, Seung-Cheol Lee, Satadeep Bhattacharjee, Niloy Ganguly. "CrysMMNet: Multimodal Representation for Crystal Property Prediction." Conference on Uncertainty in Artificial Intelligence (UAI), 2023.

    \item {\bf Kishalay Das}, Bidisha Samanta, Pawan Goyal, Seung-Cheol Lee, Satadeep Bhattacharjee, Niloy Ganguly. "CrysGNN: Distilling pre-trained knowledge to enhance property prediction for crystalline materials." AAAI Conference on Artificial Intelligence (AAAI), 2023.

    \item {\bf Kishalay Das}, Bidisha Samanta, Pawan Goyal, Seung-Cheol Lee, Satadeep Bhattacharjee, Niloy Ganguly. "CrysXPP: An Explainable Property Predictor for Crystalline Material." NPJ Computational Materials Journal, 2022.
	
\end{enumerate}

The following publications represent an immediate continuation of the research directions developed during the thesis and were accepted shortly after the thesis submission:

\begin{enumerate}

    \item \textbf{Latent Diffusion Pretraining for Crystal Property Prediction}\\
        Shrimon Mukherjee$^{*}$, {\bf Kishalay Das$^{*}$}, Partha Basuchowdhuri, 
        Pawan Goyal, and Niloy Ganguly.\\
        \textit{ICML 2026; AI4Mat Workshop @ ICLR 2025}.

    \item \textbf{LLM Meets Diffusion: A Hybrid Framework for Crystal Material Generation}\\
    Subhojyoti Khastagir$^{*}$, {\bf Kishalay Das$^{*}$}, Pawan Goyal, 
    Seung-Cheol Lee, Satadeep Bhattacharjee, and Niloy Ganguly.\\
    \textit{NeurIPS 2025}.

    \item \ textbf {Model-Agnostic Graph Prompt Learning for Crystal Property Prediction}\\
    Shrimon Mukherjee$^{*}$, {\bf Kishalay Das$^{*}$}, Partha Basuchowdhuri, 
    Pawan Goyal, and Niloy Ganguly.\\
    \textit{UAI 2026}.

\end{enumerate}

\noindent
$^{*}$ Equal contribution.
\clearemptydoublepage

\clearemptydoublepage

\phantomsection \addcontentsline{toc}{chapter}{Index}
\renewcommand{\baselinestretch}{1} \small \normalsize

\end{document}